\documentclass{icrc2009}
\pdfoutput=1
\usepackage{graphicx}   
\usepackage[font=footnotesize]{subfig}
\usepackage{fixltx2e}
\usepackage{url}

\usepackage[scaled=.90]{helvet}
\usepackage{amssymb}

\makeatletter
\def\@outputdblcol{%
  \if@firstcolumn
    \global\@firstcolumnfalse
    \global\setbox\@leftcolumn\copy\@outputbox
    \splitmaxdepth\maxdimen
    \vbadness\maxdimen
    \setbox\@outputbox\vbox{\unvbox\@outputbox\unskip}%
    \setbox\@outputbox\vsplit\@outputbox to\maxdimen
    \toks@\expandafter{\topmark}%
    \xdef\@firstcoltopmark{\the\toks@}%
    \toks@\expandafter{\splitfirstmark}%
    \xdef\@firstcolfirstmark{\the\toks@}%
    \ifx\@firstcolfirstmark\@empty
      \global\let\@setmarks\relax
    \else
      \gdef\@setmarks{%
        \let\firstmark\@firstcolfirstmark
        \let\topmark\@firstcoltopmark}%
    \fi
  \else
    \global\@firstcolumntrue
    \setbox\@outputbox\vbox{%
     \hb@xt@\textwidth{%
        \hb@xt@\columnwidth{\box\@leftcolumn \hss}%
        \hfil
        \vrule \@width\columnseprule
        \hfil
       \hb@xt@\columnwidth{\box\@outputbox \hss}}}%
  \@combinedblfloats
    \@setmarks
    \@outputpage
    \begingroup
      \@dblfloatplacement
      \@startdblcolumn
      \@whilesw\if@fcolmade \fi{\@outputpage\@startdblcolumn}%
    \endgroup
  \fi}
\makeatother

\newcommand{\shorttitle}[1]%
{\markboth{Proceedings of the 31\MakeLowercase{$^{st}$} ICRC, {\L}\'{o}d\'{z} 2009}{#1} }
\newcommand{\etal}{\MakeLowercase{\textit{et al. }}} 

\begin{document}
\shorttitle{The ANTARES collaboration}
\begin{onecolumn}
\begin{center}
\begin{figure}

\includegraphics[width=0.20\textwidth]{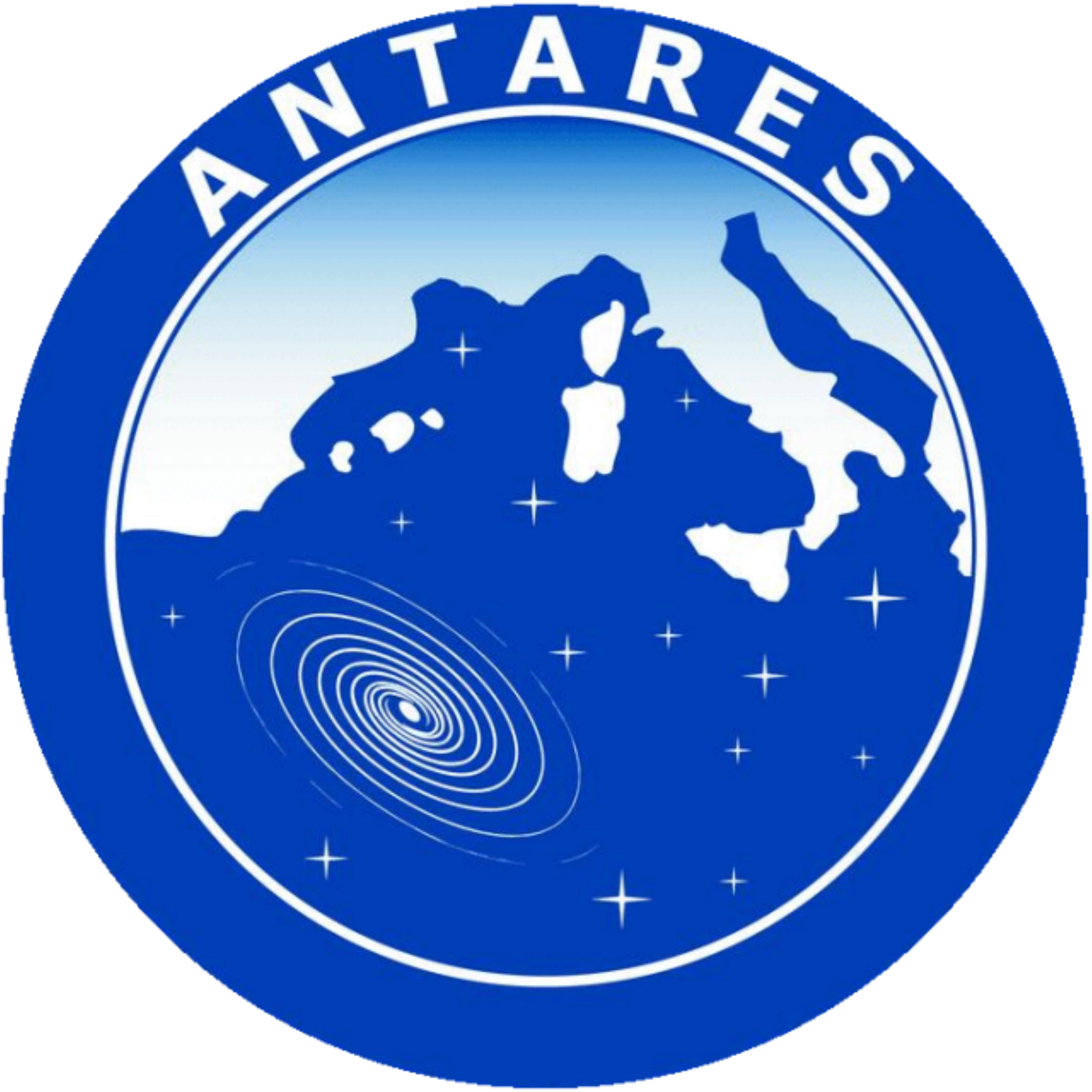} \hfill  \includegraphics[width=0.25\textwidth]{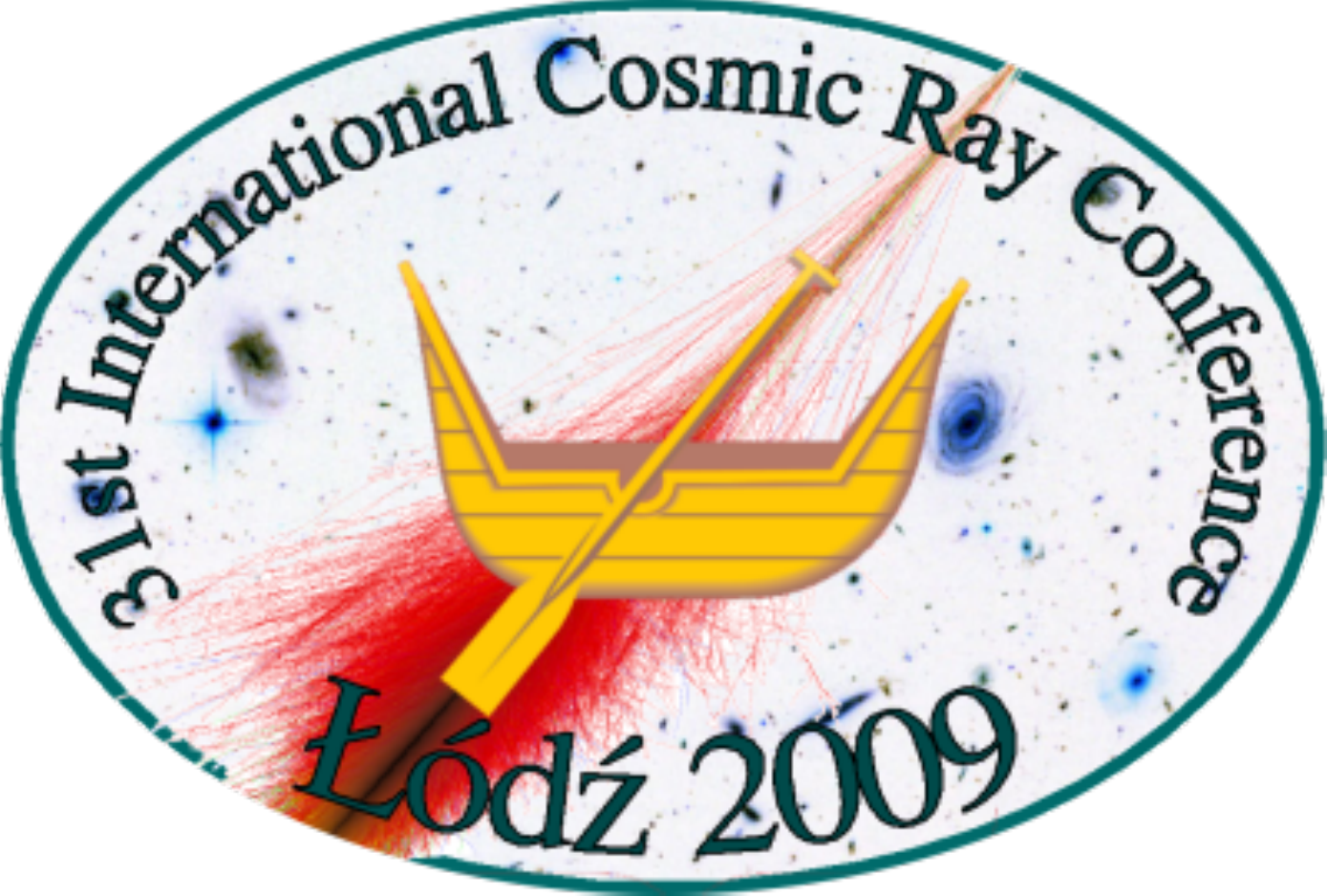}
\end{figure}

{\Huge \bf The ANTARES Collaboration:  \\
\vskip 0.5cm
 \Large contributions to the\\
 31$^{st}$  International Cosmic Ray Conference (ICRC 2009), \\
 Lodz, Poland,\\
 July  2009}
 \end {center}
 \vspace{2cm}
 \begin{center}
 {\large \bf Abstract}
The Antares neutrino telescope, operating at 2.5 km depth in the Mediterranean Sea, 40 km off the Toulon shore, represents the world's largest operational
 underwater neutrino telescope, optimized for the detection of Cerenkov light produced by neutrino-induced muons.  The main goal of Antares is the search  of high energy neutrinos 
 from  astrophysical point  or transient sources. Antares is taking data in its full 12 lines configuration since May 2008: 
 in this paper we collect the 16 contributions by the ANTARES collaboration that were submitted to the 31th International Cosmic Ray Conference ICRC 2009. 
 These contributions includes the detector performances, the first preliminary results on neutrino events and the current physics analysis including 
 the sensitivity to point like sources, the  possibility to detect high energy neutrinos in coincidence with GRB, the search for dark matter or exotic particles.
 \end{center}
\newpage
\begin{center}
{\large \bf ANTARES Collaboration}
\end{center}
{\large

J.A. Aguilar$^{1}$,
I. Al Samarai$^{2}$,
A. Albert$^{3}$,
M. Anghinolfi$^{4}$,
G. Anton$^{5}$,
S. Anvar$^{6}$,
M. Ardid$^{7}$,
A.C. Assis Jesus$^{8}$,
T.~Astraatmadja$^{8,\: a}$,
J.J. Aubert$^{2}$
% ...

$^{1}${\scriptsize{IFIC - Instituto de F\'isica Corpuscular, Edificios Investigaci\'on de Paterna, CSIC - Universitat de Val\`encia, Apdo. de Correos 22085, 46071 Valencia, Spain}} \\
$^{2}${\scriptsize{CPPM - Centre de Physique des Particules de Marseille, CNRS/IN2P3 et Universit\'e de la M\'editerran\'ee, 163 Avenue de Luminy, Case 902, 13288 Marseille Cedex 9, France}} \\
$^{3}${\scriptsize{GRPHE - Institut universitaire de technologie de Colmar, 34 rue du Grillenbreit BP 50568 - 68008 Colmar, France }} \\
$^{4}${\scriptsize{INFN - Sezione di Genova, Via Dodecaneso 33, 16146 Genova, Italy}} \\
$^{5}${\scriptsize{Friedrich-Alexander-Universit\"{a}t Erlangen-N\"{u}rnberg, Erlangen Centre for Astroparticle Physics, Erwin-Rommel-Str. 1, D-91058 Erlangen, Germany}} \\
$^{6}${\scriptsize{Direction des Sciences de la Mati\`ere - Institut de recherche sur les lois fondamentales de l'Univers - Service d'Electronique des D\'etecteurs et d'Informatique, CEA Saclay, 91191 Gif-sur-Yvette Cedex, France}} \\
$^{7}${\scriptsize{Institut d'Investigaci\'o per a la Gesti\'o Integrada de Zones Costaneres (IGIC) - Universitat Polit\`ecnica de Val\`encia. C/ Paranimf, 1. E-46730 Gandia, Spain.}} \\
$^{8}${\scriptsize{FOM Instituut voor Subatomaire Fysica Nikhef, Science Park 105, 1098 XG Amsterdam, The Netherlands}} \\
% ...
}

\newpage
\begin{center}
{\large \bf Table of Contents}
\end{center}
{\large Status Report}
\begin{enumerate}
\item Paschal Coyle on Behalf of the ANTARES Collaboration, {\it "The ANTARES Deep-Sea Neutrino Telescope: Status and First Results."}, pages \pageref{icrc1319:begin}-\pageref{icrc1319:end}.
\end{enumerate}
{\large Detector Operation}
\begin{enumerate}
\item Bruny Baret on behalf of the ANTARES Collaboration, {\it "Charge Calibration of the ANTARES high energy neutrino telescope."}, pages \pageref{icrc1184:begin}-\pageref{icrc1184:end}.
\item Mieke Bouwhuis on behalf of the ANTARES collaboration, {\it "Concepts and performance of the ANTARES data acquisition system"}, pages \pageref{icrc_daq:begin}-\pageref{icrc_daq:end}.
\item Anthony M Brown on behalf of the ANTARES Collaboration {\it "Positioning system of the ANTARES Neutrino Telescope"}, pages \pageref{icrc0178:begin}-\pageref{icrc0178:end}.
\item Juan Pablo G\'omez-Gonz\'alez on behalf of the ANTARES Collaboration {\it "Timing Calibration of the ANTARES Neutrino Telescope"}, pages \pageref{icrc0239:begin}-\pageref{icrc0239:end}.
\end{enumerate}
{\large Reconstruction}
\begin{enumerate}
\item Marco Bazzotti on the behalf of the ANTARES coll., {\it "Measurement of the atmospheric muon flux with the ANTARES detector"}, pages \pageref{icrc_bazzotti:begin}-\pageref{icrc_bazzotti:end}.
\item Aart Heijboer, for the ANTARES collaboration, {\it "Reconstruction of Atmospheric Neutrinos in Antares"}, pages \pageref{icrc1045:begin}-\pageref{icrc1045:end}.
\end{enumerate}
{\large Physics Analyses}
\begin{enumerate}
\item Mieke Bouwhuis on behalf of the ANTARES collaboration, {\it "Search for gamma-ray bursts with the ANTARES\ neutrino telescope"}, pages \pageref{icrc_grb:begin}-\pageref{icrc_grb:end}.
\item Damien Dornic, St\'ephane Basa, Jurgen Brunner, Imen Al Samarai, Jos\'e Busto, Alain Klotz, St\'ephanie Escoffier, Vincent Bertin, Bertrand Vallage, Bruce Gendre, Alain Mazure and Michel Boer on behalf the ANTARES and TAROT Collaboration, {\it "Search for neutrinos from transient sources with the ANTARES telescope and optical follow-up observations"}, pages \pageref{icrc0055:begin}-\pageref{icrc0055:end}.
\item Goulven Guillard, for the A NTARES Collaboration {\it "Gamma ray astronomy with ANTARES"}, pages \pageref{icrc0503:begin}-\pageref{icrc0503:end}.
\item G.M.A. Lim on behalf of the ANTARES collaboration {\it "First results on the search for dark matter in the Sun with the ANTARES neutrino telescope"}, page
s \pageref{icrc0031:begin}-\pageref{icrc0031:end}.
\item Salvatore Mangano, for the ANTARES collaboration, {\it "Skymap for atmospheric muons at TeV energies measured in deep-sea neutrino telescope ANTARES"}, pa
ges \pageref{icrc1131:begin}-\pageref{icrc1131:end}.
\item Gabriela Pavalas and Nicolas Picot Clemente, on behalf of the ANTARES Collaboration, {\it "Search for Exotic Physics with the ANTARES Detector"}, pages \pageref{icrc0695:begin}-\pageref{icrc0695:end}.
\item Francesco Simeone, on behalf of the ANTARES Collaboration. {\it "Underwater acoustic detection of UHE neutrinos with the ANTARES experiment"}, pages \pageref{icrc0471:begin}-\pageref{icrc0471:end}
\item Simona Toscano, for the ANTARES Collaboration {\it "Point source searches with the ANTARES neutrino telescope"}, pages \pageref{icrc0127:begin}-\pageref{icrc0127:end}
\item V\'eronique Van Elewyck for the ANTARES Collaboration, {\it "Searching for high-energy neutrinos in coincidence with gravitational waves with the ANTARES and VIRGO/LIGO detectors"}, pages \pageref{icrc_elewyck:begin}-\pageref{icrc_elewyck:end}
\end{enumerate}
\end{onecolumn}
\begin{twocolumn}






\title{The ANTARES Deep-Sea Neutrino Telescope: \\
Status and First Results}

\author{\IEEEauthorblockN{Paschal Coyle  \IEEEauthorrefmark{1} \\
On Behalf of the ANTARES Collaboration}
                            \\
\IEEEauthorblockA{\IEEEauthorrefmark{1}
coyle@cppm.in2p3.fr\\
 Centre de Physique des Particules de Marseille\\
163 Avenue de Luminy, Case 902,\\
13288 Marseille cedex 09, France}}

\shorttitle{P. Coyle \etal ANTARES status and first results}
\maketitle
\label{icrc1319:begin}

\begin{abstract}

Various aspects of the construction, operation and calibration of the recently completed deep-sea ANTARES neutrino telescope are described. Some first results obtained with a partial five line configuration are presented, including depth dependence of the atmospheric muon rate, the search for point-like cosmic neutrino sources and the search for dark matter annihilation in the Sun.
 
 \end{abstract}

 
\section{Introduction}
The undisputed galactic origin of cosmic rays at energies below the so-called knee implies an existence of a non-thermal population of galactic sources which effectively accelerate protons and nuclei to TeV-PeV energies. The distinct signatures of these cosmic accelerators are high energy neutrinos and gamma rays produced through hadronic interactions with ambient gas or photoproduction on intense photon fields near the source. While gamma rays can be produced also by directly accelerated electrons, high-energy neutrinos provide unambiguous and unique information on the sites of the cosmic accelerators and hadronic nature of the accelerated particles.   

ANTARES (http://antares.in2p3.fr/) is a deep-sea neutrino telescope, designed for the detection of all flavours of high-energy neutrinos emitted by both Galactic (supernova remnants, micro-quasars etc.) and extragalactic (gamma ray bursters, active galactic nuclei, etc.) astrophysical sources. The telescope is also sensitive to neutrinos produced via dark matter annihilation within massive bodies such as the Sun and the Earth. Other physics topics include measurement of neutrino oscillation parameters, the search for magnetic monopoles, nuclearites etc. 

The recently completed ANTARES detector is currently the most sensitive neutrino observatory studying the southern hemisphere and includes the particularly interesting region of the Galactic Centre in its field of view. ANTARES is also a unique deep-sea marine observatory providing continuous, high-bandwidth monitoring from a variety of sensors dedicated to acoustic, oceanographic and Earth science studies. 
 
\section{The ANTARES Detector}

  \begin{figure}[!t]
  \centering
  \includegraphics[width=2.7in]{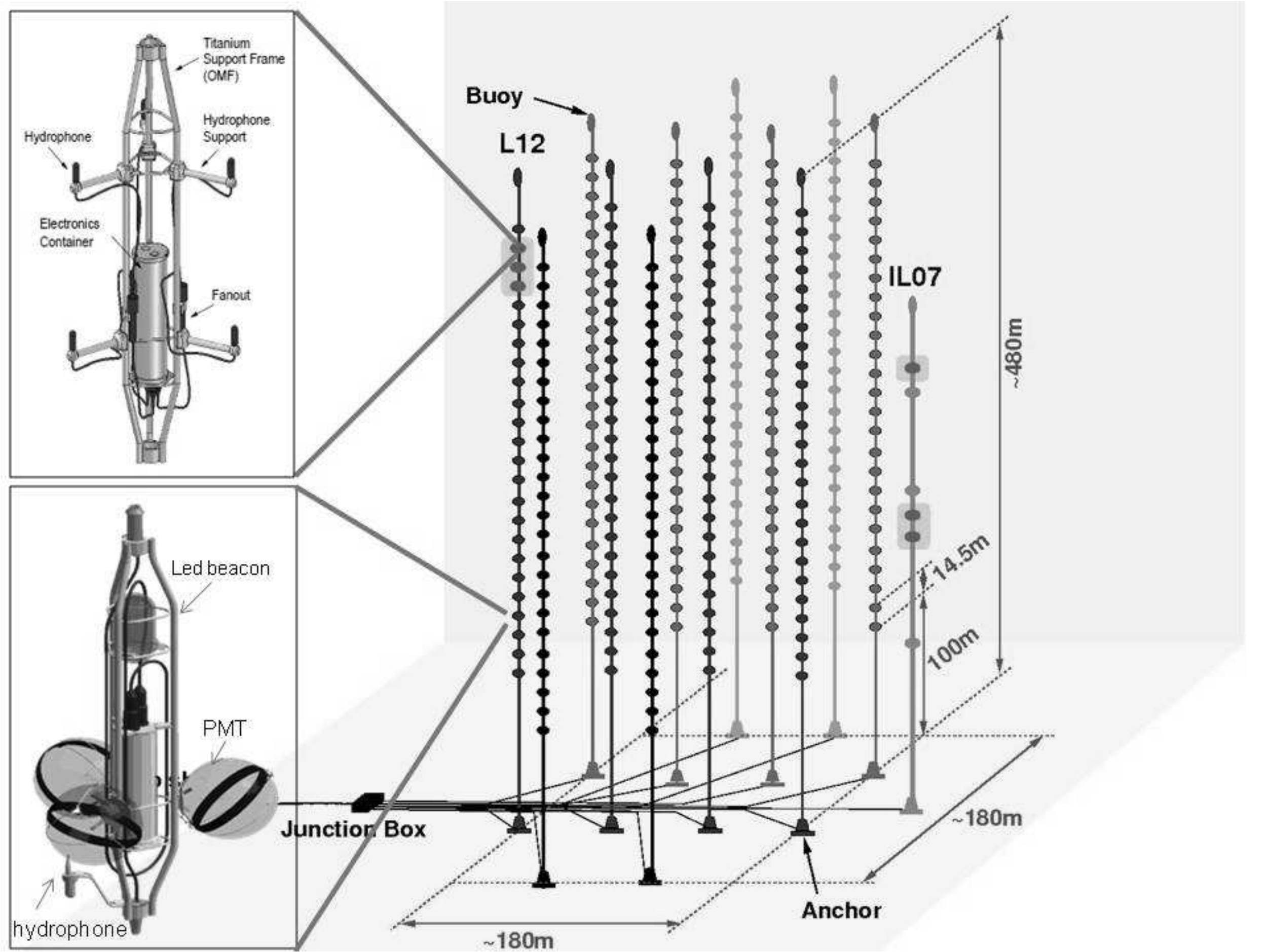}
  \caption{The layout of the completed ANTARES detector. The top insert shows an image of an acoustic storey with 
its six acoustic sensors and the lower insert an image of an optical storey with its three photomultipliers.}
  \label{fig1}
 \end{figure}

The ANTARES detector is located at a depth of 2475~m in the Mediterranean Sea, 42~km from La Seyne-sur-Mer in the South of France ($42^\circ48’ N, 6^\circ10’ E$). A schematic of the detector layout is shown in Figure \ref{fig1}. It is equipped with 885 optical sensors arranged on 12 flexible lines. Each line comprises up to 25 detection storeys each equipped with three downward-looking 10-inch photo-multipliers (PMTs), orientated at $45^\circ$ to the line axis. The lines are maintained vertical by a buoy at the top of the 450~m long line. The spacing between storeys in 14.5~m and the lines are spaced by 60-70~m. An acoustic positioning system provides real-time location of the detector elements to a precision of a few centimeters. A system of optical beacons allows in-situ time calibration. The first detection line was installed in 2006. Five lines have been operating since March 2007. Ten lines were operational in December 2007. With the installation of eleventh and twelfth lines in May 2008, the detector construction was completed. An additional line (IL07) contains an ensemble of oceanographic sensors dedicated measurement of the environmental parameters. The twelfth line and the IL07 also includes hydrophone-only storeys dedicated to the study of the ambient acoustic backgrounds; R\&D for possible acoustic detection of ultra-high energy neutrinos. 

The ANTARES Collaboration currently comprises 29 particle physics, astrophysics and sea science institutes from seven countries (France, Germany, Italy,  Netherlands, Romania, Russia and Spain). 

The three-dimensional grid of photomultiplier tubes is used to measure the arrival time and position of Cherenkov photons induced by the passage of relativistic charged particles through the sea water. The reconstruction algorithm relies on the characteristic emission angle of the light (about 43 degrees) to determine the direction of the muon and hence infer that of the incident neutrino. The accuracy of the direction information allows to distinguish upward-going muons, produced by neutrinos, from the overwhelming background from downward-going muons, produced by cosmic ray interaction in the atmosphere above the detector. Installing the detector at great depths serves to attenuate this background and also allows to operate the PMTs in a completely dark environment.   

  At high energies the large range of the muon allows the sensitive volume of the detector to be significantly greater than the instrumented volume. Although optimised for muon neutrino detection, the detector is also sensitive to the electron and tau neutrinos albeit it with reduced effective area. 

The total ANTARES sky coverage is 3.5$\pi$sr, with an instantaneous overlap of 0.5$\pi$sr with that of the Icecube experiment at the South Pole. Together ANTARES and Icecube provide complete coverage of the high-energy neutrino sky.
Compared to detectors based in ice, a water based telescope benefits from a better angular resolution, due to the absence of light scattering on dust and/or bubbles trapped in the ice. On the other hand, it suffers from additional background light produced by beta decay of  $^{40}$K salt present in the sea water as well as bioluminescent light produced by biological organisms. Furthermore, the continual movement of the detector lines, in reaction to the changing direction and intensity of the deep-sea currents, must be measured and taken into account in the track reconstruction. 

The ANTARES data acquistion \cite{bouwhuis1} is based on the 'all-data-to-shore' concept, in which all hits above a threshold of 0.3 single photon-electrons are digitised and transmitted to shore. Onshore a farm of commodity PCs apply a trigger based on requiring the presence of a 4-5 causally connected local coincidences between pairs of PMTs within a storey. The typical trigger rate is 5-10 Hz, dominated by downgoing muons. In addition, an external trigger generated by the gamma-ray bursts coordinates network (GCN) will cause all the buffered raw data (two minutes) to be stored on disk. This offers the potential to apply looser triggers offline on this subset of the data \cite{bouwhuis2}. 
 
\section{Detector Calibration}

The precision on the neutrino direction is limited at low energies by the kinematics of the neutrino interaction. For neutrino energies above 10~TeV the angular resolution is determined by the intrinisic detector resolution i.e. the timing resolution and accuracy of the location of the PMTs. The energy measurement relies on an accurate calibration of the charge detected by each PMT \cite{baret}.   

\subsection {Acoustic Positioning}
The positions of the PMTs are measured every two minutes with a high-frequency long-baseline acoustic positioning system comprising fixed acoustic emitters-receivers at the bottom of each line and acoustic receivers distributed along a line \cite{brown}. After triangulation of the positions of the moving hydrophones, the shape of each line is reconstructed by a global fit based on a model of the physical properties of the line and additional information from the tiltmeters and compass sensors located on each storey. The relative positions of the PMTs are deduced from this reconstructed line shape and the known geometry of a storey. The system provides a statistical precision of a few mm. The final precision on the PMT locations is a few cm, smaller than the physical extension of the PMTs, and  is limited by the systematic uncertainties on the knowledge of the speed of sound in the sea water.  

 \begin{figure}[!t]
  \centering
  \includegraphics[width=2.7in]{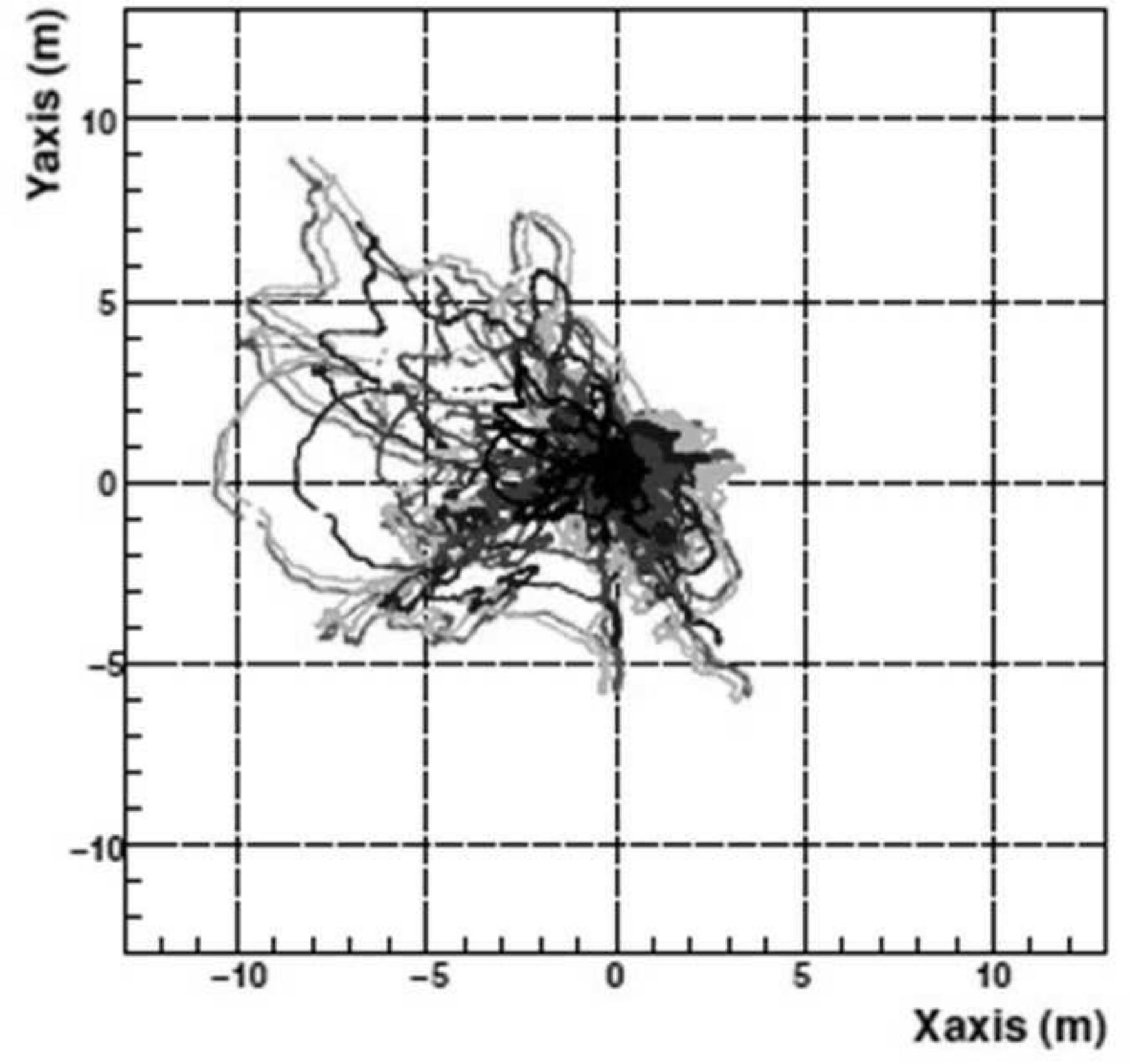}
  \caption{The horizontal movements relative to the bottom of the line, 
of all hydrophones on Line 11 for a 6 month period; black points is the
hydrophone on Storey 1, red is on Storey 8, blue is on Storey 14, green
is on Storey 22 and magenta is on Storey 25.}
  \label{fig2}
 \end{figure}

Figure \ref{fig2} shows the movements of various storeys on a line, relative to its centre axis. The extent of the displacement
depends on the intensity of the sea current. For typical currents of a few centimetres per second, the displacement is a few metres for the topmost storeys.   

\subsection {Time Calibration}
The relative time calibration of ANTARES is performed via a number of independent and redundant systems \cite{Gomez}. 
The master clock system features a method to measure the transit time of the clock signals to the electronics located in each storey of the detector. The determination of the remaining residual time offsets within a storey, due to the delays in the front-end electronics and transit time of the PMTs, are based on the detection of signals from external optical beacons distributed throughout the detector. The presence of $^{40}$K in the sea water also provides a convenient source of calibrated light which is used to verify the time offsets between the triplet of PMTs within a storey as well as study the long term stablity of the PMTs efficiencies.

Every fifth storey of a line contains an optical beacon emitting in the blue. Each beacon illuminates the neighbouring storeys on its line. Comparison of the arrival hit times within a storey provides the relative inter-storey time offsets. Intra-storey time offsets can also be established after corrections are applied for time walk and 'first photon' effects.

\begin{figure}[!t]
  \centering
  \includegraphics[width=2.7in]{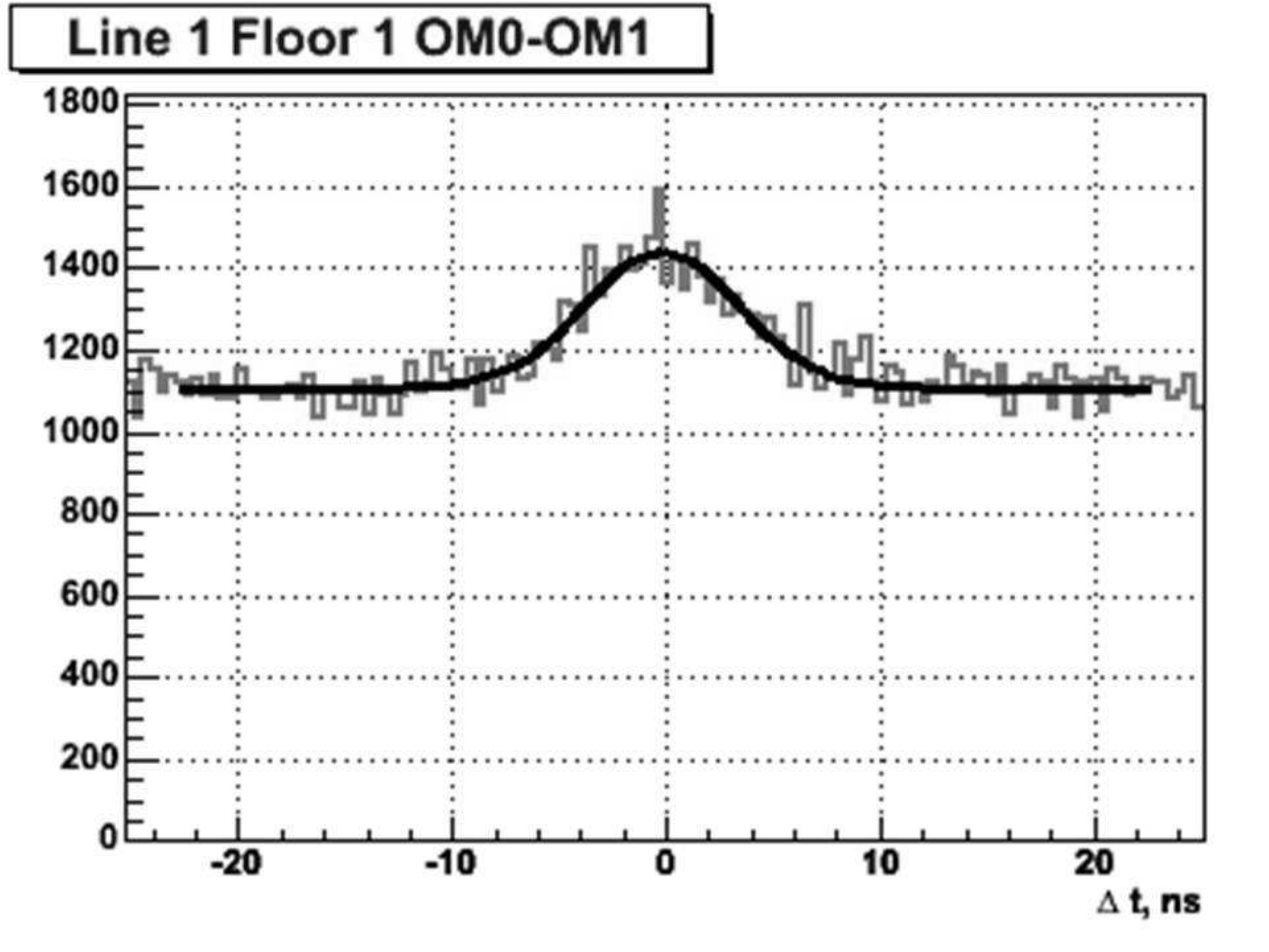}
  \caption{Conicidence peak due to  $^{40}$K decays for a single pair of photomultipliers in one storey.}
  \label{fig3}
 \end{figure}

Potassium-40 is a radioactive isotope naturally present in the sea water. The decay $^{40}K \rightarrow  e^- \nu_e ^{40} Ca$ yields an electron with an energy up to $1.3$~MeV. This energy exceeds the Cherenkov threshold for electrons in water 
(0.26~MeV), and is sufficient to produce up to 150 Cherenkov photons. If the decay occurs in the vincinity of a detector storey, a coincident signal may be recorded by pairs of PMTs on the storey. In Figure \ref{fig3} the distribution of the measured time difference between hits in neighbouring PMTs of the same storey is shown.  The peak around 0~ns is mainly due to single $^{40}$K decays producing coincident signals. The fit of the data is the sum of a Gaussian distribution and a flat background. The full width at half maximum of the Gaussian function is about 9~ns. This width is mainly due to the spatial distribution of the  $^{40}$K decays around the storey. The positions of the peaks of the time distributions for different
pairs of PMTs in the same storey are used to cross-check the time offsets computed by the optical beacon system. This is illustrated in Figure  \ref{fig3} which shows a comparision of the time offsets calculated by the optical beacons and that extracted from the  $^{40}$K analysis; an rms of 0.6~ns is obtained.  

\begin{figure}[!t]
  \centering
  \includegraphics[width=2.7in]{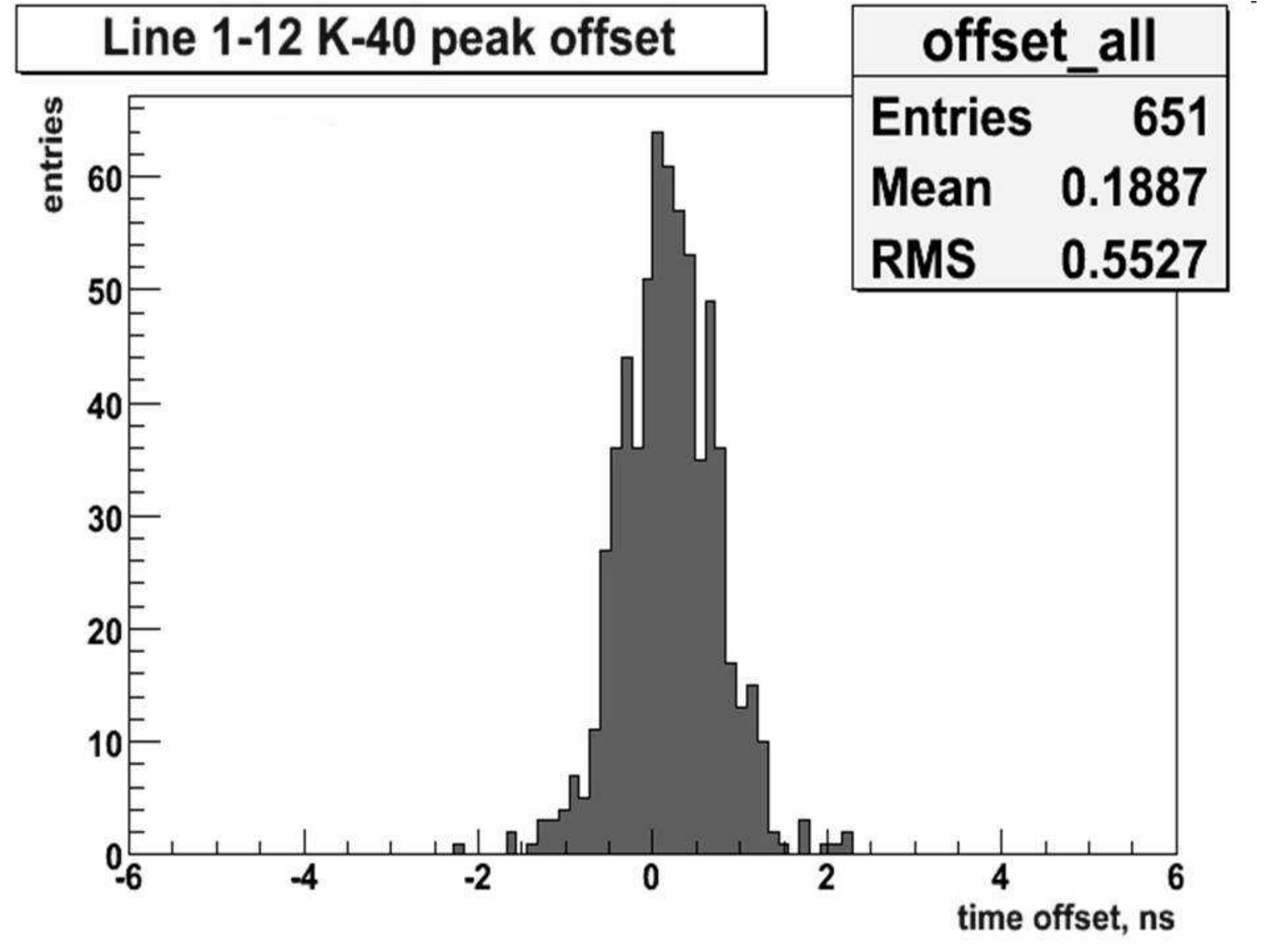}
  \caption{Time offsets for all photomultipliers as extracted using the LED beacons and independently
checked by the K40 conicidence method.}
  \label{fig4}
 \end{figure}

The rate of genuine  $^{40}$K coincidences is given by the integral under the peak of Figure \ref{fig3} and is used to monitor the relative efficiencies of all PMTs and their temporal stability.

\section{Muon Reconstruction}

\begin{figure}[!t]
  \centering
  \includegraphics[width=2.7in]{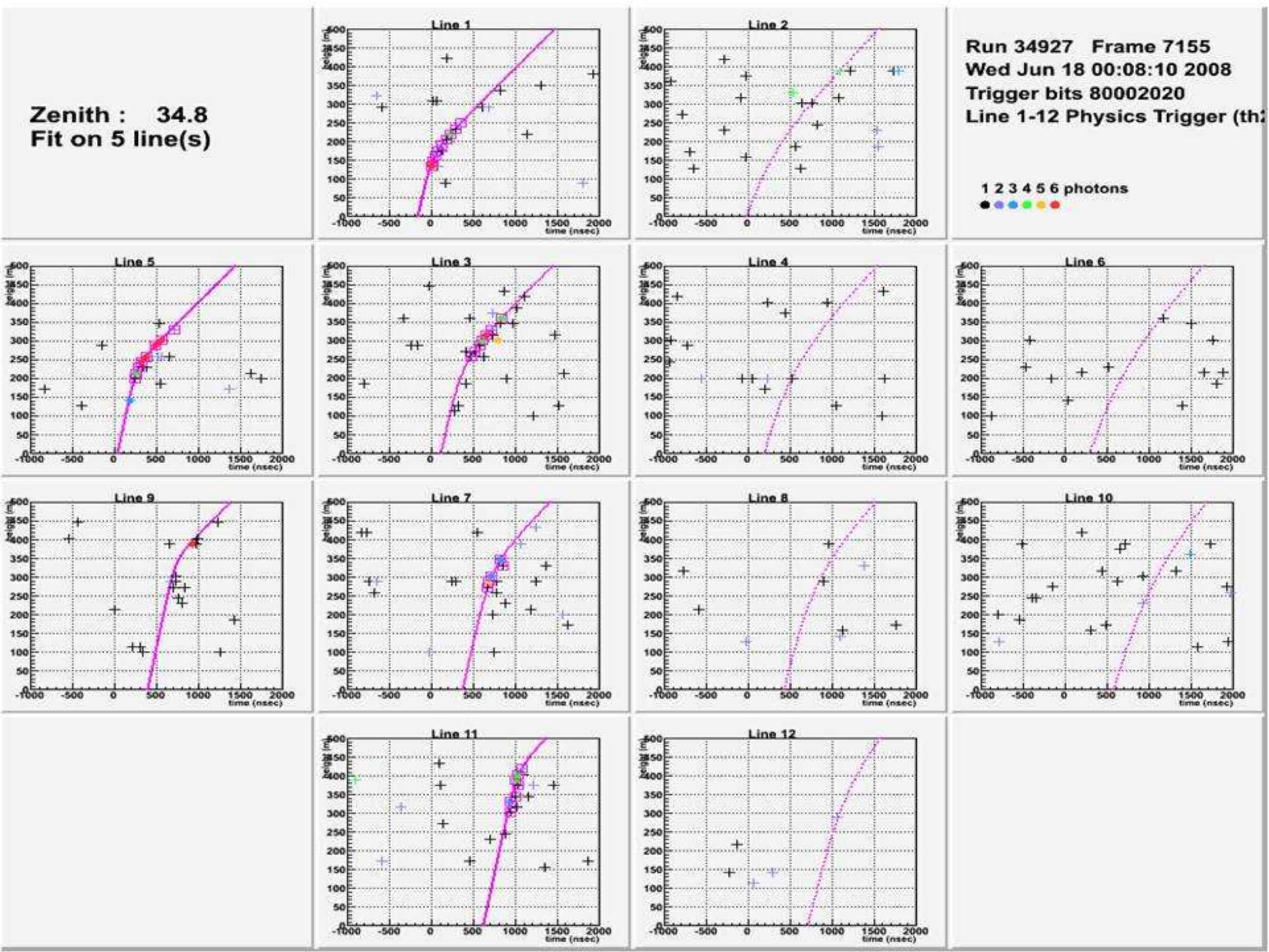}
  \caption{Example display of a neutrino candidate. The 2D plots, one for each line, show on the y axis the vertical position of the PMT with a hit and on the x axis the arrival time of the hit. The fit to the arrival time distribution corresponds to the chi-square algorithm.}
  \label{fig5}
 \end{figure}

Two alternative algorithms for reconstruction of the muon trajectories have been developed \cite{heijboer}. 
In the first approach, a simple chi-square fit is applied to a high purity sample of pre-selected hits. In addition, this algorithm merges the hits observed by the PMTs of the triplet and assumes that they are located on the line axis, i.e. the azimuthal orientation of the storey, measured by the compasses, is ignored. In  Figure \ref{fig5} an upgoing neutrino candidate fitted using this algorithm is shown. This algorithm was initially adopted as a fast reconstruction for online monitoring of the detector. Although it provides an non-optimum angular resolution (typically 1-2 degrees above 10~TeV) it has been used in a number of analyses for which the ultimate angular resolution is not crucial.

In the second approach a full maximum likelihood fit is applied, which uses a detailed parameterisation, derived from simulation, of the probability density function for the track residuals. The fit includes most hits in the event and the PDF takes into account the probability that photons arrive late due to Cherenkov emission by secondary particles or light scattering. A number of increasingly sophisticated prefits are used to aid in the location of the correct maxima of the likelihood. This algorithm makes use of the maximum amount of information, including the line shape and storey orientation, and provides
an angular resolution better than 0.3 degrees above 10~TeV.  
 
\section{Atmospheric Muons}

The dominant signal observed by ANTARES is due to the passage of downgoing atmospheric muons, whose 
flux exceeds that of neutrino-induced muons by several orders of magnitude. They are produced by high
energy cosmic rays interacting with atomic nuclei of the upper atmosphere, producing charged pions and
kaons, which subsequently decay into muons. Although an important background for neutrino detection, they are 
useful to verify the detector response. In particular, with three years of data taking, a deficit in the muon flux in the direction of the moon should allow an important verification of the pointing accuracy of the detector. 

\begin{figure}[!t]
  \centering
  \includegraphics[width=2.7in]{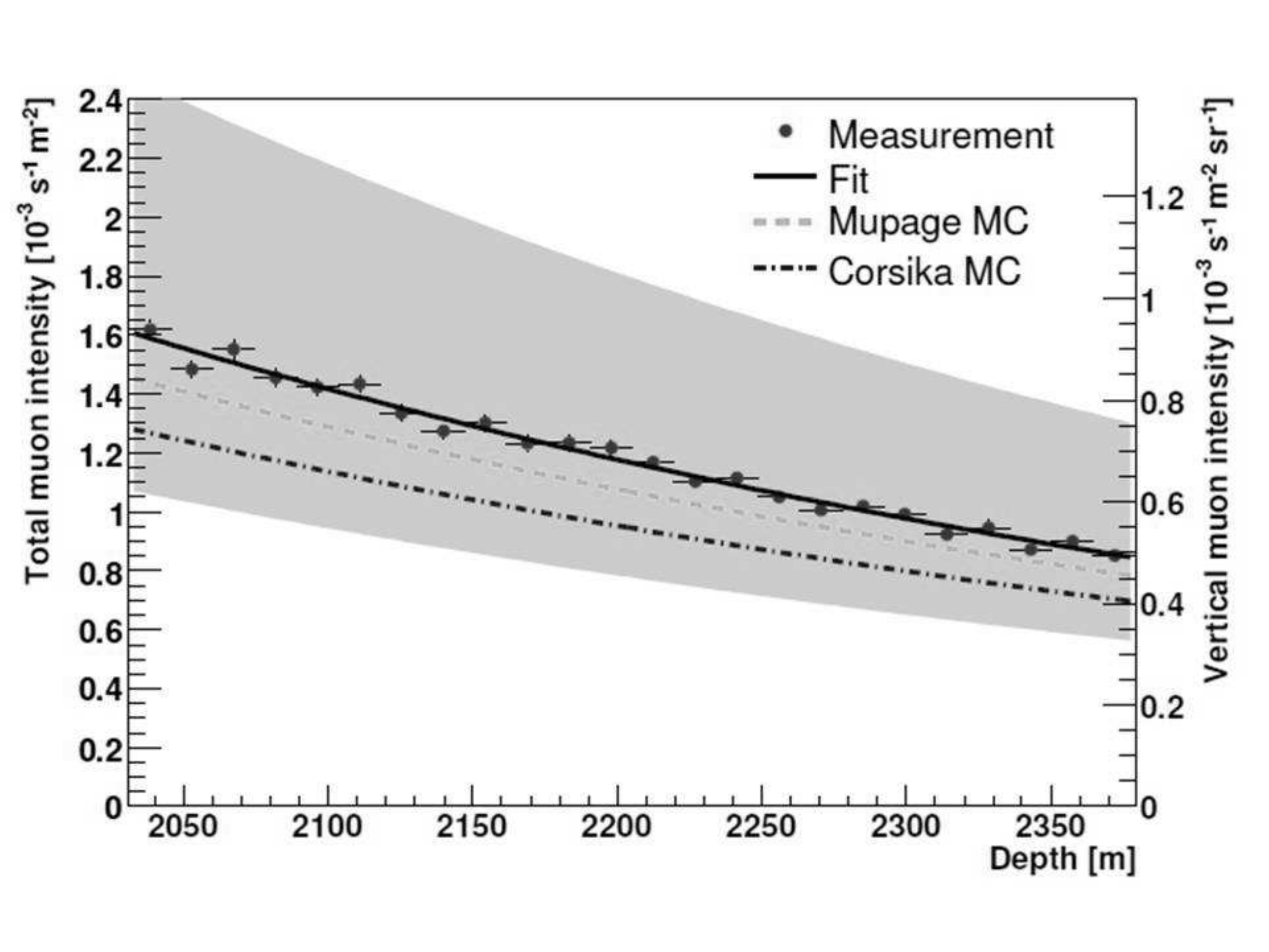}
  \caption{Attenuation of the flux of muons as a function of depth, as extracted using the adjacent storey coincidence method. The shaded band represents the systematic uncertainties due to detector effects. Predictions from MUPAGE and Corsika Monte Carlo simulations are also shown.}
  \label{fig6}
 \end{figure}

\begin{figure}[!t]
  \centering
  \includegraphics[width=3.1in]{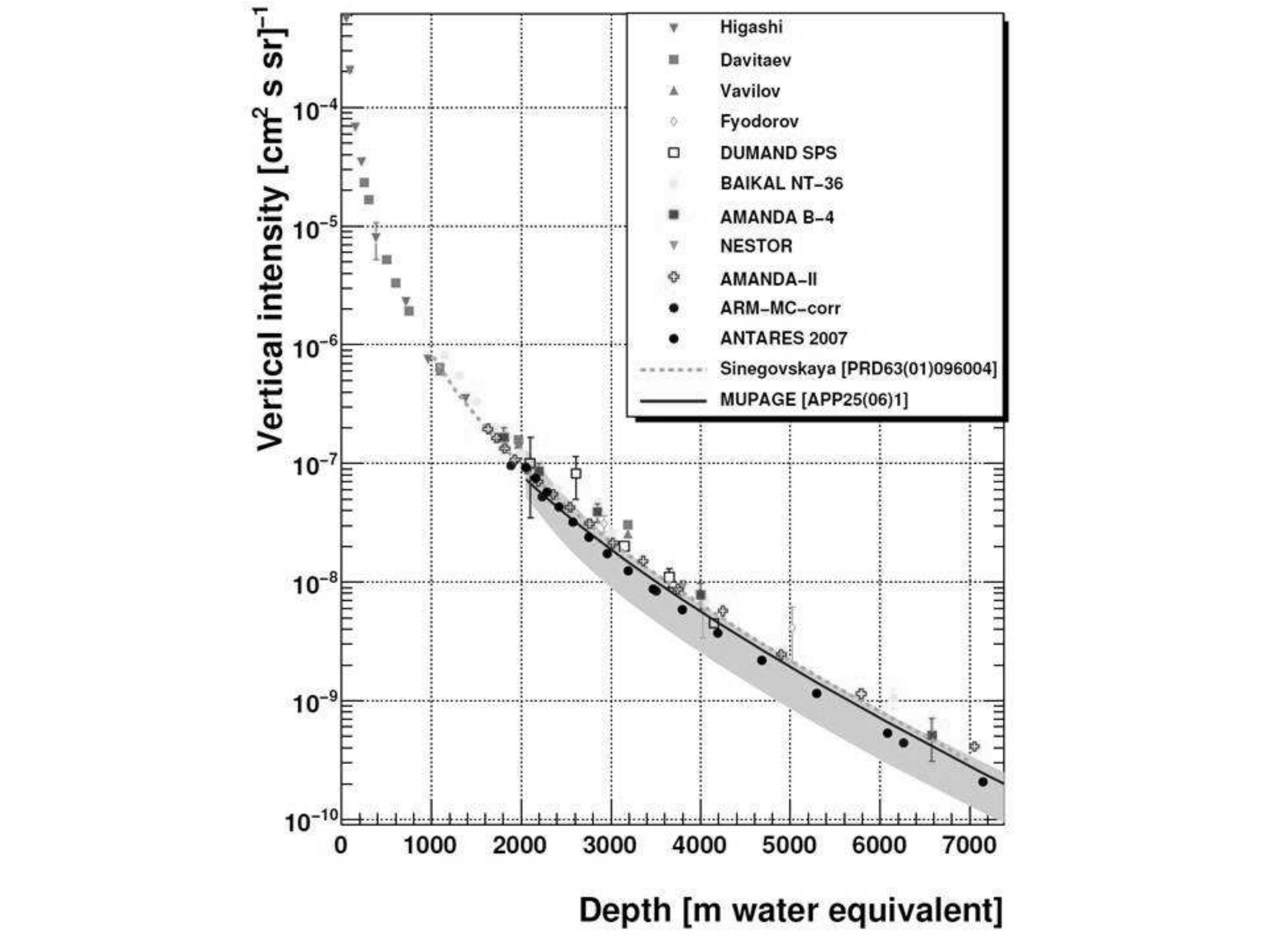}
  \caption{Vertical depth intensity relation of atmospheric muons with $E_\mu > 20$~GeV (black points). The
error band represents systematic uncertainties. A compliation of data from other experiments are also shown.}
  \label{fig7}
 \end{figure}

Two different studies of the vertical depth intensity relation of the muon flux have been performed. In the first, the attenuation of the muon flux as a function of depth is observed as a reduction in the rate of coincidences between adjacent storeys along the length of the detection lines \cite{zaborov}. This method has the advantage that it does not rely on any track reconstruction. In Figure \ref{fig6} the depth dependence of the extracted flux for the 24 inter-storey measurements averaged over all lines is shown.

In the second study, a full track reconstruction is performed and the reconstructed zenith angle is converted to an equivalent slant depth through the sea water \cite{bazotti}. Taking into account the known angular distribution of the incident muons, a depth intensity relation can be extracted (Figure \ref{fig7}). The results ae in reasonable agreement with previous measurements.

\begin{figure}[!t]
  \centering
  \includegraphics[width=2.7in]{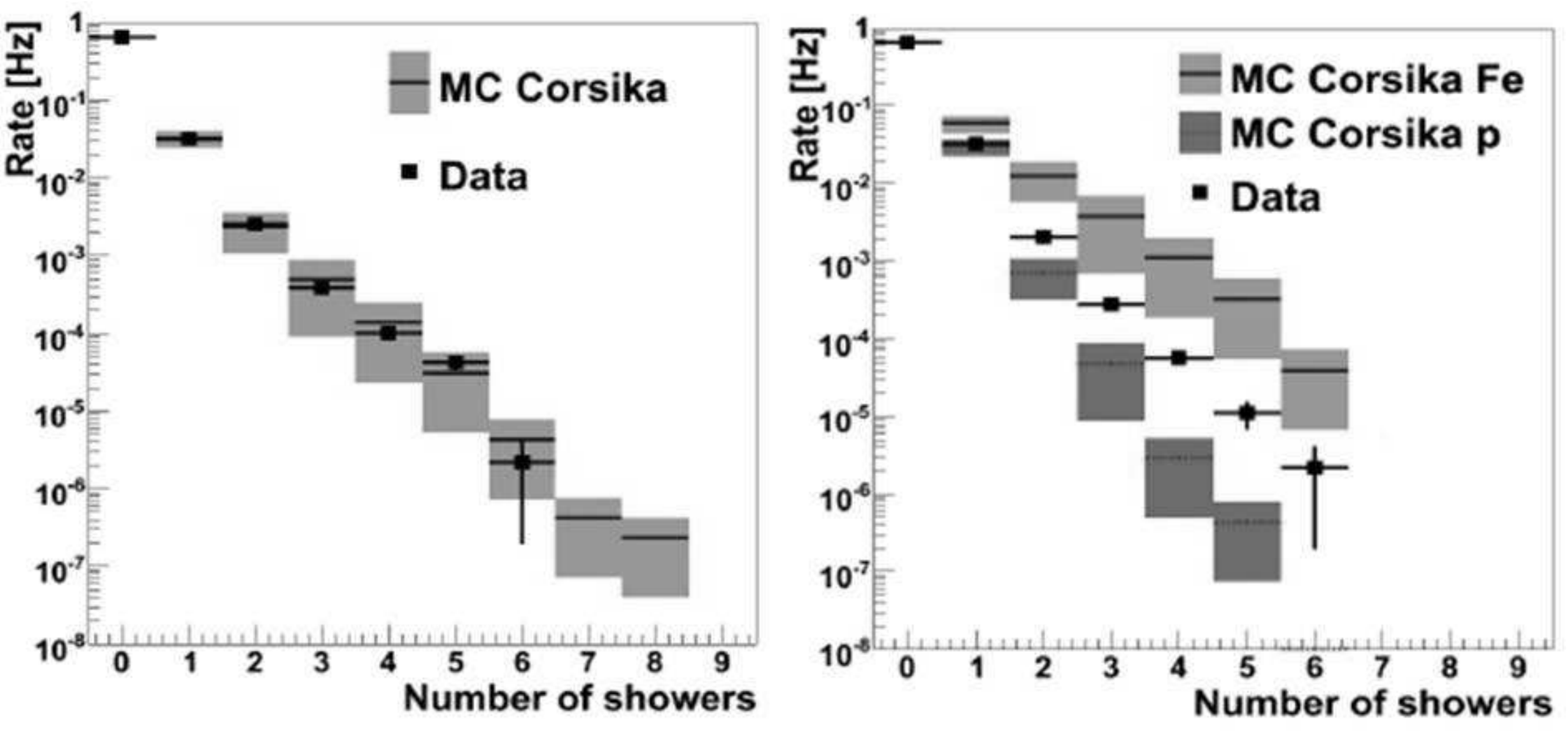}
  \caption{Identification of energetic electromagnetic showers: (left) Data and Monte Carlo comparison of the number of reconstructed electromagnetic showers for the 5-line data, assuming the nominal primary cosmic-ray composition. (right) dependence on primary cosmic-ray composition (proton versus iron).}
  \label{fig8}
 \end{figure}

The composition of the primary cosmic rays in the knee region is of particular interest. As the number of identified electromagnetic showers in an event depends on the muon energy and the number of muons present in a muon bundle, it is sensitive to the primary cosmic ray composition. An algorithm has been developed to  estimate the number of energetic electromagnetic (EM) showers generated along the muon trajectory \cite{mangano1}. This algorithm relies on the fact that the emission point of Cherenkov photons from the muon are uniformly distributed along the muon trajectory whereas Cherenkov photons orginating from an electromagnetic shower will tend to cluster from a single point. The efficiency and purity of the algorithm to identify a shower depends on the shower energy, for example the efficiency to identify a 1~TeV shower is 20\% with a purity of 85\%. In Figure \ref{fig8} (left) the distribution of the number of reconstructed energetic showers per event in the 5-line data is shown. Good agreement with the Corsika Monte Carlo is obtained when the 22Horandel primary composition model is assumed. In Figure \ref{fig8} (right) the data distribution is compared with that obtained assuming a pure proton or a pure iron primary cosmic ray composition. 

A search for a large-scale anistropy in the arrival directions of the atmospheric muons  has been performed but with the current statistics is not yet sensitive to the $0.1\%$ level variations reported by other experiments \cite{mangano2}. The possibilty for detection of gamma ray induced air showers with ANTARES is also under evaluation \cite{guillard}. 

\section{Search for Cosmic Neutrino Point Sources}

\begin{figure}[!t]
  \centering
  \includegraphics[width=2.7in]{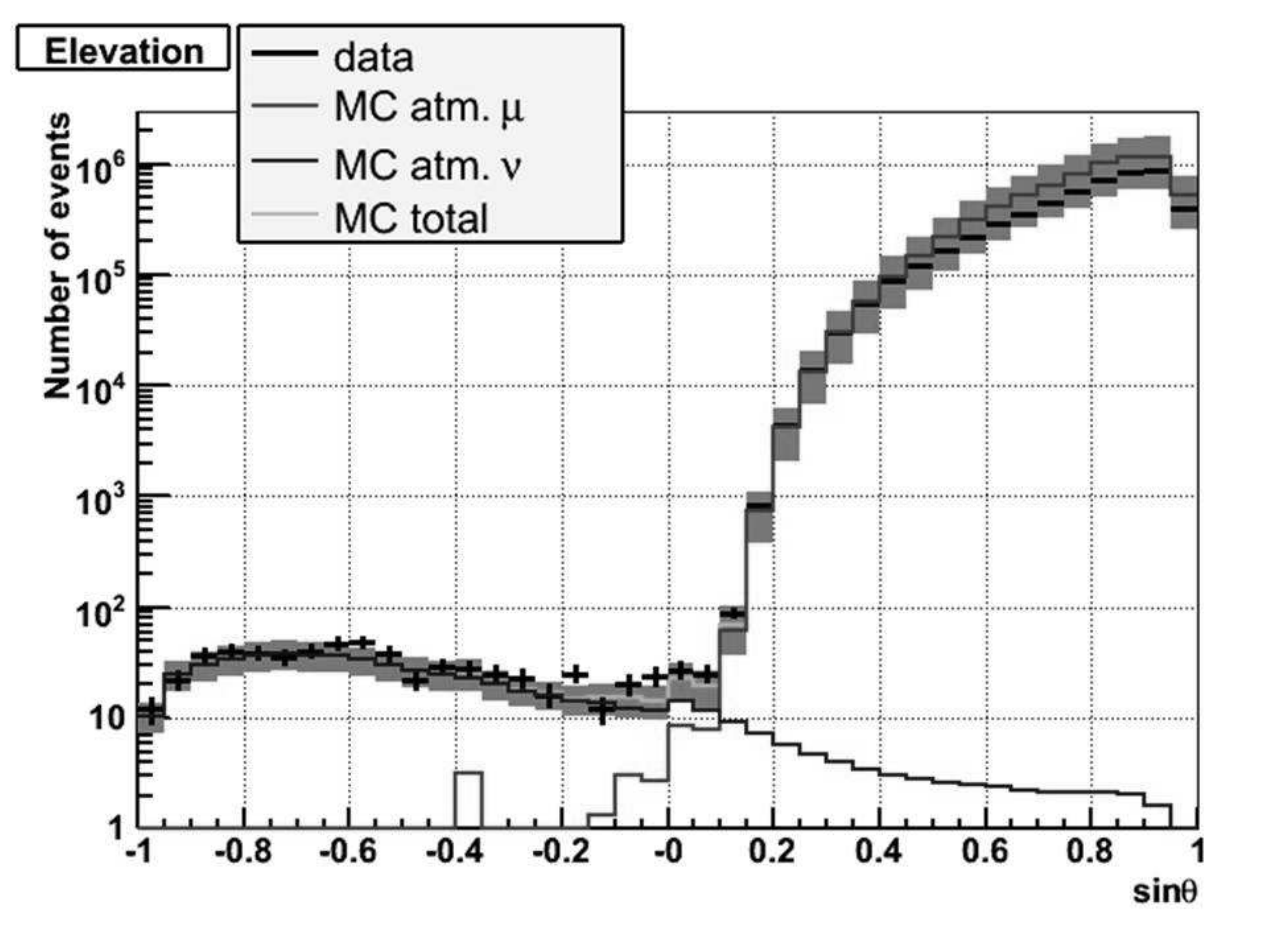}
  \caption{Zenith distribution of reconstructed muons in the 2008 data. The Mone Carlo expectation for the atmospheric muon and atmospheric neutrino backgrounds are indicated.}
  \label{fig9}
 \end{figure}

The muons produced by the interaction of neutrinos can be isolated from the muons generated by the cosmic ray interactions by requiring that the muon trajectory is reconstructed as up-going. In Figure \ref{fig9} the zenith angular distribution of muons in the 2007+2008 data sample by the $\chi^2$ reconstruction algorithm is shown. A total of 750 mulitline up-going neutrinos candidates are found, in good agreement with expectations from the atmospheric neutrino background.

\begin{figure}[!t]
  \centering
  \includegraphics[width=2.7in]{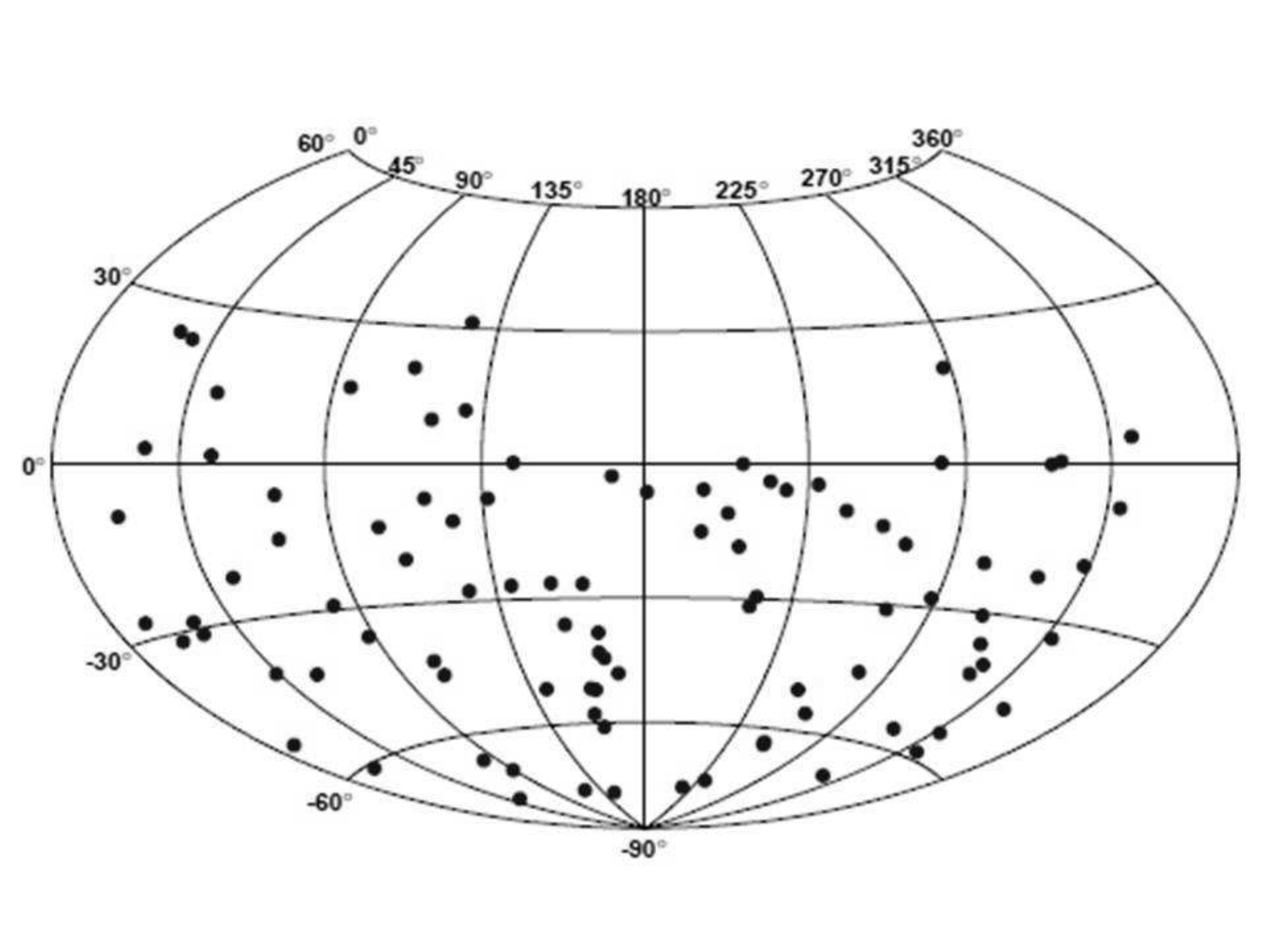}
  \caption{Sky map, in geocentric coordinates, of the upgoing neutrino candidates for the 2007 data.}
  \label{fig10}
 \end{figure}

\begin{table}
 \begin{center}
{\footnotesize

\begin{tabular}{|l|r@{.}l|r@{.}l|c|c|c|} \hline 
Source name &    \multicolumn{2}{c|}{$\delta$ ($^{\circ}$)}  &  \multicolumn{2}{c|}{RA  ($^{\circ}$)} & n$_{{\rm bin.}}$ &  p-value &    $\phi_{90}$ \\ \hline \hline
PSR B1259-63      &    -63&83 &    195&70  & 0 &     -  &    3.1  \\
RCW 86            &    -62&48 &    220&68  & 0 &     -  &    3.3  \\
HESS J1023-575    &    -57&76 &    155&83  & 1 &  0.004 &    7.6  \\
CIR X-1           &    -57&17 &    230&17  & 0 &     -  &    3.3  \\
HESS J1614-518    &    -51&82 &    243&58  & 1 &  0.088 &    5.6  \\
GX 339            &    -48&79 &    255&70  & 0 &     -  &    3.8  \\
RX J0852.0-4622   &    -46&37 &    133&00  & 0 &     -  &    4.0  \\
RX J1713.7-3946   &    -39&75 &    258&25  & 0 &     -  &    4.3  \\
Galactic Centre   &    -29&01 &    266&42  & 1 &  0.055 &    6.8  \\
W28               &    -23&34 &    270&43  & 0 &     -  &    4.8  \\
LS 5039           &    -14&83 &    276&56  & 0 &     -  &    5.0  \\
HESS J1837-069    &    -6&95  &    279&41  & 0 &     -  &    5.9  \\
SS 433            &     4&98  &    287&96  & 0 &     -  &    7.3  \\
HESS J0632+057    &     5&81  &    98&24   & 0 &     -  &    7.4  \\ \hline
ESO 139-G12       &    -59&94 &    264&41  & 0 &     -  &    3.4  \\
PKS 2005-489      &    -48&82 &    302&37  & 0 &     -  &    3.7  \\
Centaurus A       &    -43&02 &    201&36  & 0 &     -  &    3.9  \\
PKS 0548-322      &    -32&27 &    87&67   & 0 &     -  &    4.3  \\
H 2356-309        &    -30&63 &    359&78  & 0 &     -  &    4.2  \\ 
PKS 2155-304      &    -30&22 &    329&72  & 0 &     -  &    4.2  \\
1ES 1101-232      &    -23&49 &    165&91  & 0 &     -  &    4.6  \\
1ES 0347-121      &    -11&99 &    57&35   & 0 &     -  &    5.0  \\
3C 279            &    -5&79  &    194&05  & 1 &   0.030&    9.2  \\
RGB J0152+017     &     1&79  &    28&17   & 0 &     -  &    7.0  \\ \hline
IC22 hotspot      &    11&4  &    153&4  & 0 &     -  &    9.1  \\ \hline

\end{tabular}
}
 \caption{\small Results of the search for cosmic neutrinos correlated
 with potential neutrino sources. The sources are divided into three
 groups: galactic (top), extra-galactic (middle) and the hotspot from
 IceCube with 22 lines (bottom). The source name and location in
 equatorial coordinates are shown together with the number of events
 within the optimum cone for the binned search, the p-value of the
 unbinned method (when different from 1) and the corresponding upper
 limit at 90\% C.L.  $\phi_{90}$ is the value of the normalization
 constant of the differential muon-neutrino flux assuming an $E^{-2}$
 spectrum (i.e. $E^{2} d\phi_{\nu_{\mu}} / dE \le \phi_{90} \times
 10^{-10} $ TeV cm$^{-2}$ s$^{-1}$). The integration energy range is 10
 GeV - 1 PeV.}
 \label{icrc1319:tab:sources}
 \end{center}
\end{table}

For a subset of this data (the 5-line data of 2007) the angular resolution has been improved by applying the pdf based track reconstruction, which makes full use of the final detector alignment. After reoptimisation of the selection cuts for the best upper limits, 94 upgoing neutrino candidates are selected \cite{toscano}. The corresponding sky map for these events is shown in Figure \ref{fig10}. An all sky search, independent of assumption on the source location, has been performed on these data. The most significant cluster was found at ($\delta=-63.7^\circ, RA=243.9^\circ$) with a corresponding p-value of 0.3 ($1\sigma$ excess) and is therefore not significant. 

A search amongst a pre-defined list of 24 of the most promising galactic and extra-galactic neutrino sources (supernova remnants, BL Lac objects, Icecube hot spot, etc.) is reported in Table \ref{icrc1319:tab:sources}. The lowest p-value obtained (a $2.8 \sigma$ excess, pre-trial) corresponds to the location ($\delta=-57.76^\circ, RA=155.8^\circ$). Such a probablity or higher would be expected in $10\%$ of background-only experiments and is therefore not significant. The corresponding flux upper limits, assuming an $E^{-2}$ flux, are plotted in Figure \ref{fig11} and compared to published upper limits from other experiments. Also shown in Figure \ref{fig11} is the predicted upper limit for ANTARES after one full year of twelve line operation.

\begin{figure}[!t]
  \centering
  \includegraphics[width=3.1in]{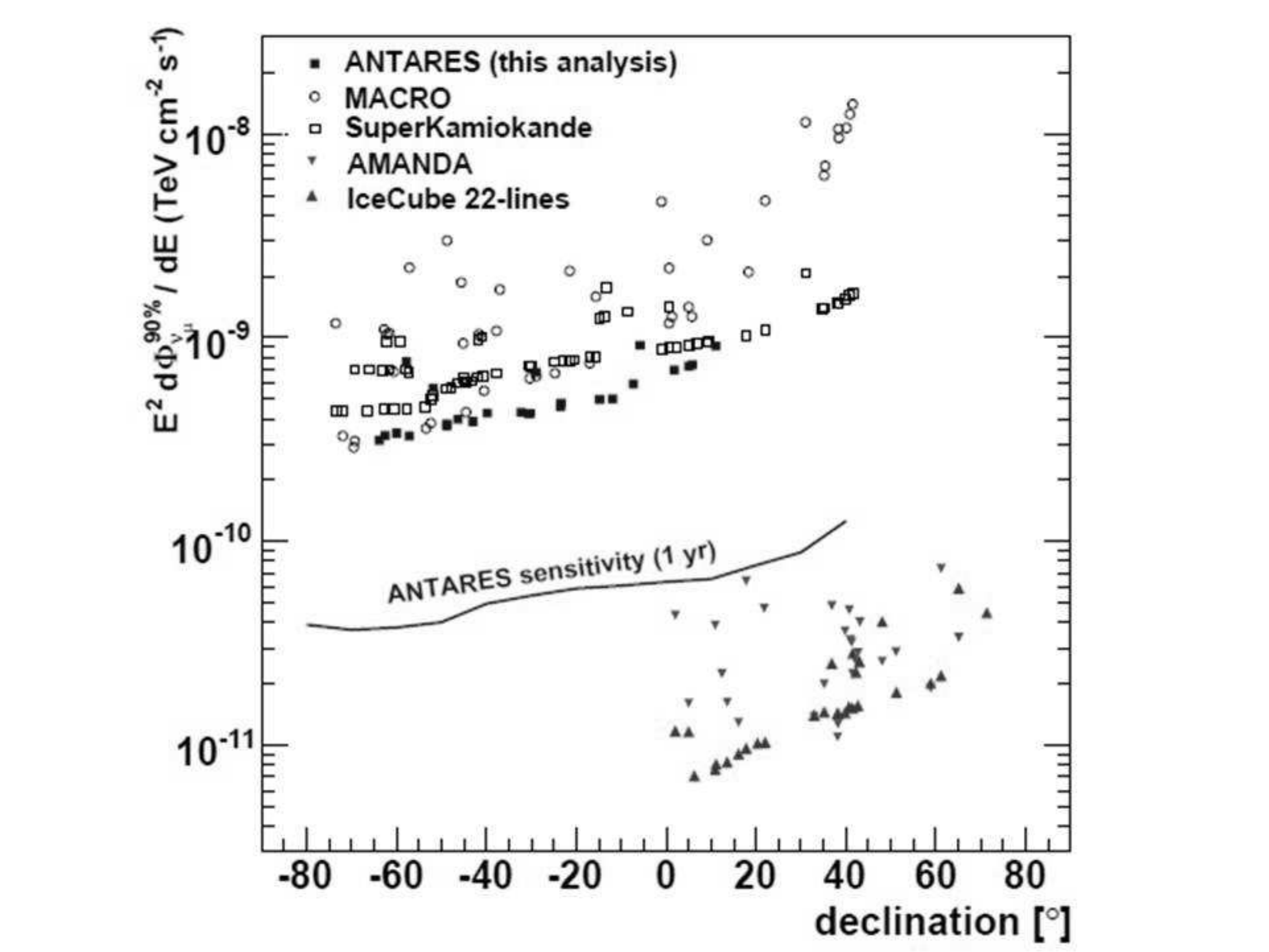}
  \caption{Neutrino flux upper limits at $90\%$ C.L. obtained by this analysis (solid squares), compared with published results from other experiments (IceCube [24], AMANDA [25], SuperKamiokande [26] and MACRO [27]). The expected sensitivity of ANTARES for one year with twelve lines is also shown (solid line). The source spectrum assumed in these results is $E^{−2}$, except for MACRO, for which an $E^{−2.1}$ spectrum was used.}
  \label{fig11}
 \end{figure}
 
\section{Search for Dark Matter}
In many theoretical models a Weakly Interacting Massive Particle (WIMP), a relic from the Big Bang, is proposed to explain the formation of structure in the universe and the discrepancy observed between the measured rotation curves of stars and the associated visible matter distribution in galaxies. A generic property of such WIMPs is that they gravitationally accumulate at the centre of massive bodies such as the Sun or the Earth, where they can self annihilate into normal matter. Only neutrinos, resulting from the decay of this matter, can escape from the body and be detected by a neutrino telescope.

\begin{figure}[!t]
  \centering
  \includegraphics[width=2.7in]{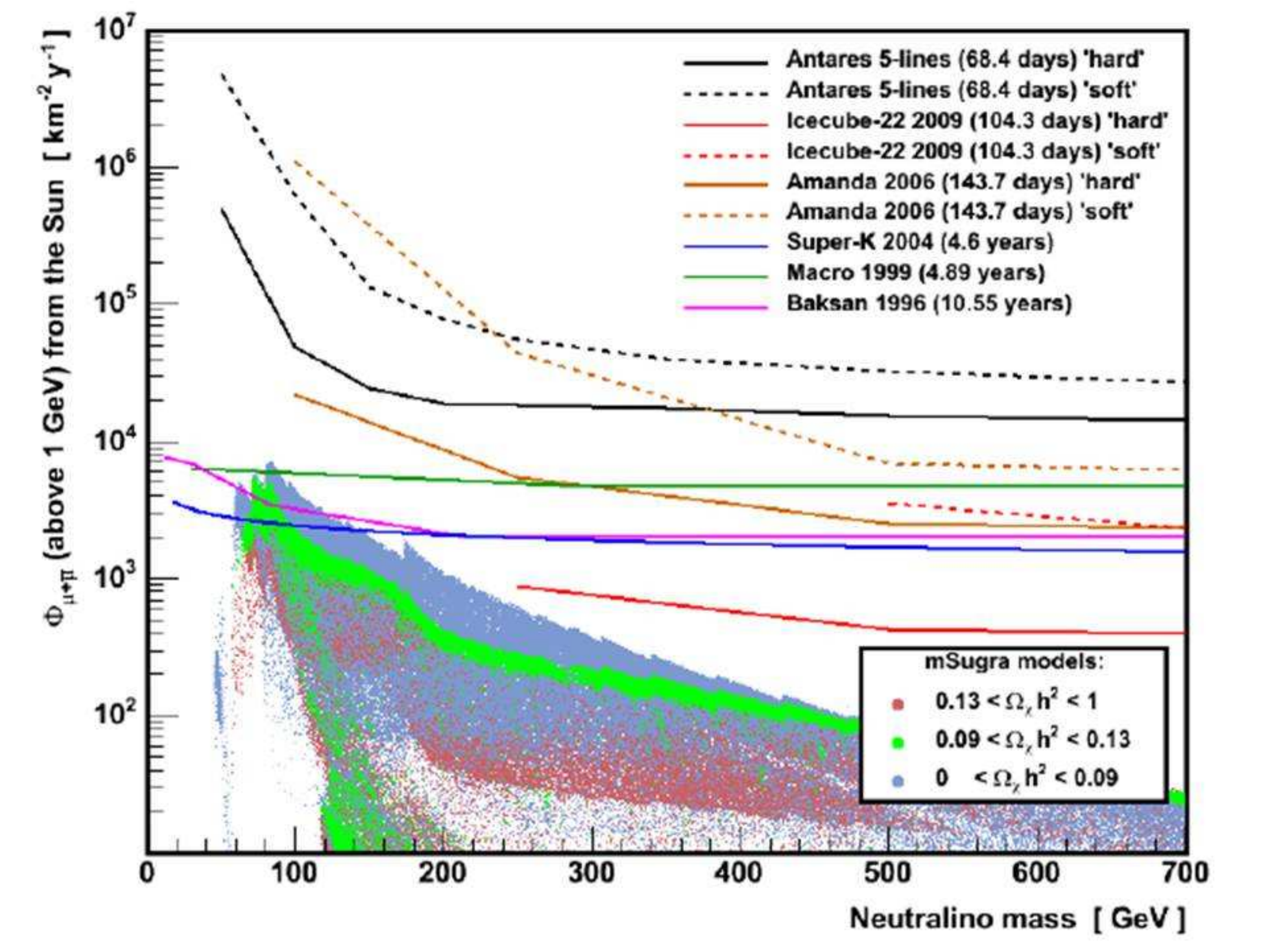}
  \caption{Upper limit on the muon flux from the Sun as a function of neutralino mass. The expected fluxes for a scan of mSUGRA parameters is shown as well as a variety of limits from other experiments.}
  \label{fig12}
 \end{figure}

\begin{figure}[!t]
  \centering
  \includegraphics[width=2.7in]{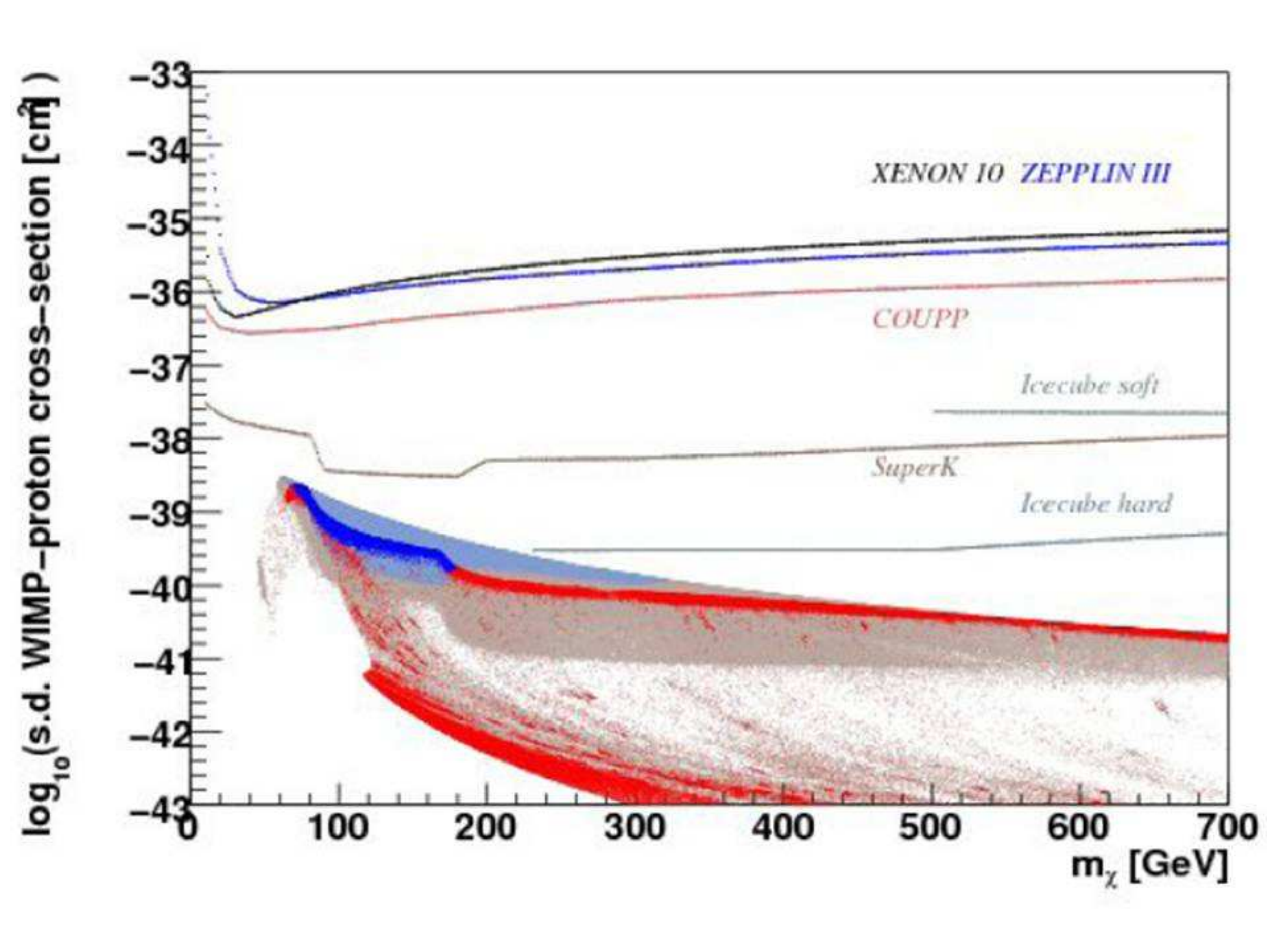}
  \caption{The spin dependent WIMP-proton cross-section versus neutralino mass in the mSUGRA model. The points 
are the results of the scan over the range of model parameters. The points in blue would be excluded at 90\% CL after three years of ANTARES operation. Existing upper limits from a variety of direct and indirect direction experiments are shown. }
  \label{fig13}
 \end{figure}

Within Supersymmetric models with R-parity conservation, the lightest supersymmetric particle (LSP) is stable and is the WIMP candidate. In order to predict the expected solar neutrino fluxes the constrained phenomenological framework of the minimal Supergravity model (mSUGRA, computations using ISASUGRA[5]) has been adopted. Figure \ref{fig10} shows the predicted integrated neutrino fluxes above 10~GeV in ANTARES as a function of neutralino mass for the scan of the model parameters: scalar mass $m_0$ in [0,8000] GeV, gaugino mass $m_{1/2}$ in [0,2000] GeV, tri-linear scalar coupling $A_0$ in [$-3m_0,3m_0$], sign of the Higgsino mixing parameter: $\mu > 0$, ratio of Higgs fields vacuum expectation values $tan \beta$ in [0,60], $m_{top}=172.5~GeV$. The local Dark Matter halo density (NFW-model) was set to $0.3~GeV/cm^3$. 
The most favourable models for neutrino telescopes are in the so-called ‘focus point’ region, for which the decays are mainly via $W^+W^-$ leading to a harder neutrino spectra. Thanks to its low-energy threshold ANTARES is ideally suited to address  low-mass neutralino scenarios.   

A search for neutrinos from the direction of the Sun in the 5-line data \cite{lim}, showed no excess with respect to background expectations. The corresponding derived limit on the neutrino flux is shown in Figure \ref{fig12}. Also shown is the expected limit with 5 years of data taking with the full 12-line detector; a large fraction of the focus point region could be excluded.
  
Due to the capture of the WIMPs inside the Sun, which is mainly hydrogen, neutrino telescopes are particularly sensitive to the spin-dependent coupling of the WIMPs to standard matter. In Figure \ref{fig13} the corresponding limit on the spin-dependent WIMP-proton cross-section after three years of ANTARES operation is shown. For this case, the neutrino telescopes are significantly more sensitive than the direct direction experiments. 

\section{Multi-Messenger Astronomy}

\begin{figure}[!t]
  \centering
  \includegraphics[width=2.7in]{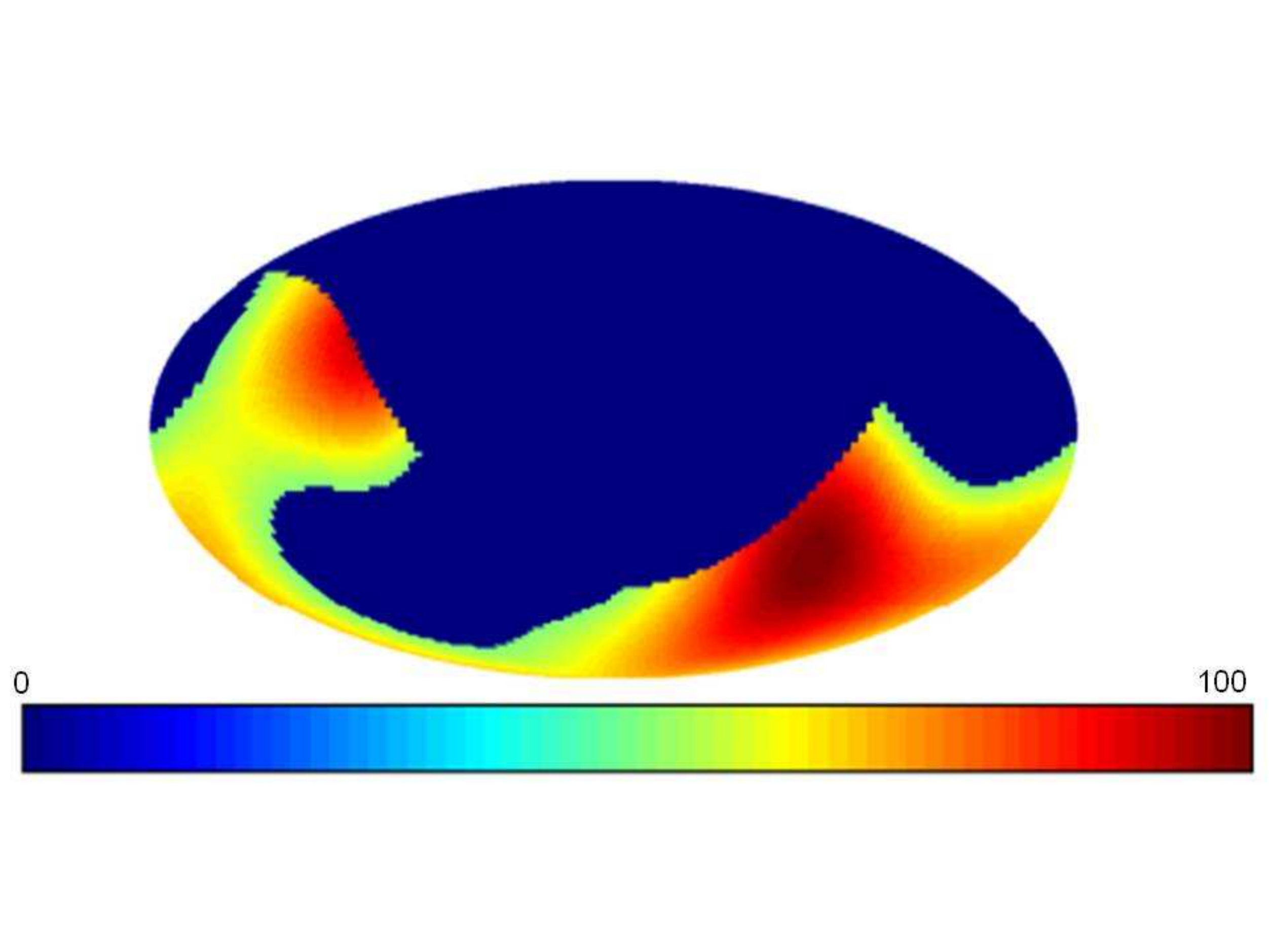}
  \caption{Common sky coverage for VIRGO/LIGO/ANTARES in geocentric coordinates. This map shows the combined antenna pattern for the gravitational wave detector network (above half-minimum), assuming that ANTARES has 100\% visibility in its antipodal hemisphere and 0\% elsewhere.}
  \label{fig14}
 \end{figure}

In order to augment the discovery potential of ANTARES, a program of collaboration with other types of observatory have been established. In this ''multi-messenger'' approach the detection threshold can be lowered in each separate experiment while preserving an acceptable rate of accidental coincidences. One example of such a program is being pursued with the gravitational wave detectors VIRGO and LIGO \cite{elewyck}. Both of these detectors had a data-taking phase during 2007 (VIRGO science run 1 and LIGO S5) which partially coincided with the ANTARES 5-line configuration. A new common science run has also recently started in coincidence with the ANTARES 12-line operation. The common sky coverage for ANTARES-VIRGO+LIGO is signifcant and is shown in Figure \ref{fig14}.
  
In a similar vein, a collaboration with the TAROT optical telescopes has been established \cite{dornic}. The directions of interesting neutrino triggers (two neutrinos within 3 degrees within a time window of 15 minutes or a single neutrino of very high energy) are sent to the Chile telescope in order that a series of optical follow up images can be taken. This procedure is well suited to maximise the sensitivity for transient events such as gamma ray bursters or flaring sources. 

\section {Acoustic Detection}

\begin{figure}[!t]
  \centering
  \includegraphics[width=2.7in]{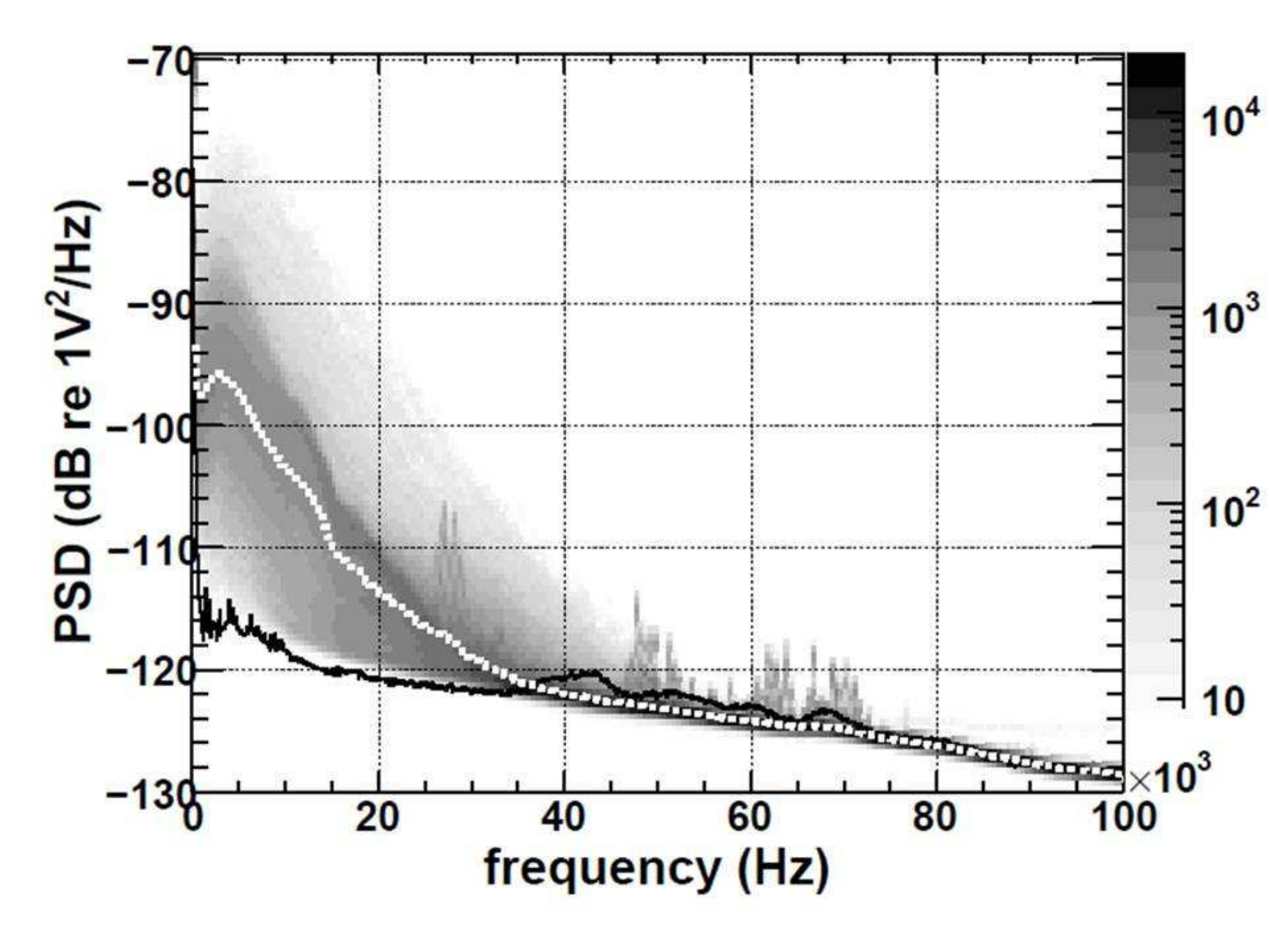}
  \caption{Power spectral density (PSD) of the ambient noise recorded with one sensor. Shown in grey shades of grey is the occurence rate in arbirary units, where dark colours indicate higher occurence. Shown as a white dotted line is the mean value of the in-situ PSD and as a black solid line the noise level recorded in the laboratory before deployment.}
  \label{fig15}
 \end{figure}

\begin{figure}[!t]
  \centering
  \includegraphics[width=2.7in]{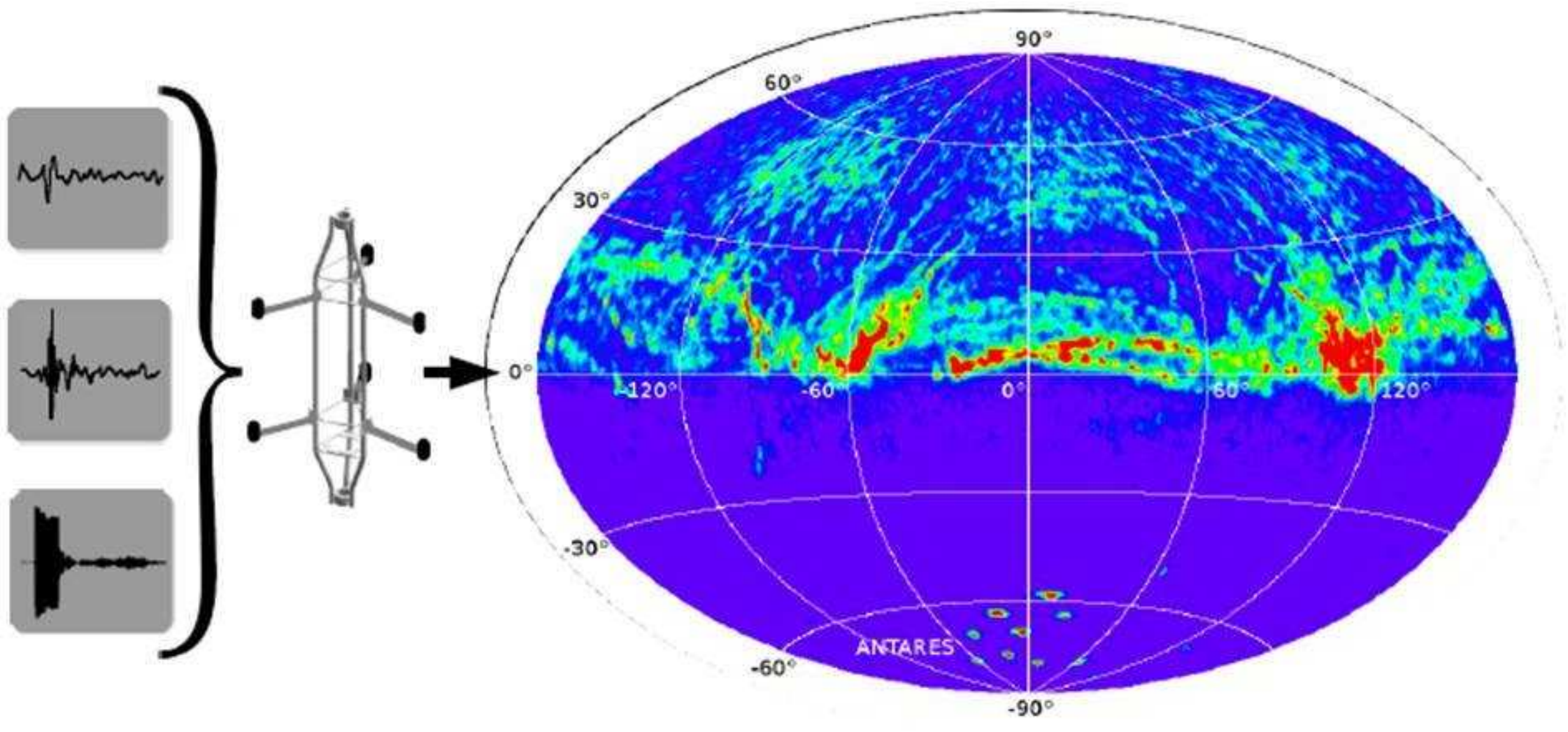}
  \caption{Map of the angular directions of the detected transient acoustic signals at the ANTARES site. }
  \label{fig16}
 \end{figure}

Due to the large attenuation length, $\approx$ 5~km for 10~kHz signals, the detection of bipolar acoustic pressure pulses  in huge underwater acoustic arrays is a possible approach for the identification of cosmic neutrinos with energies exceeding 100~PeV. To this end, the ANTARES infrastructure incorporates the AMADEUS system, an R\&D project intended to evaluate the acoustical backgrounds in the deep sea \cite{simeone}. It comprises a set of piezo-electric sensors for broad-band recording of acoustic signals with frequencies ranging up to 125~kHz with typical sensitivities around 1V/$\mu$Pa. The sensors are distributed in six ''acoustic clusters'', each comprising six acoustic sensors that are arranged at distances roughly 1~m from each other. The clusters are installed along the line 1é and the IL07 line at a horizontal distance of 240~m. The vertical spacing within a line range from 15~m to 125~m (see Figure \ref{fig1}). 

In Figure \ref{fig15} is shown the measured power spectral density of the ambient noise recorded with an acoustic sensor. Due to wind generated surface noise, the observed noise level is larger than that measured in the laboratory. Strong correlation of the measured acoustic noise with the measured surface wind speed are observed. 

In Figure \ref{fig16} an acoustic "sea map" of all transient (not just bipolar) signals detected by the apparatus during a one month period is shown. The acoustic pinger of the ANTARES positioning system are clearly identified in the lower hemisphere. The acoustic activity in the upper hemisphere is presumably due to surface boats and possibly marine mammals.  

\section{Conclusion}

After more than a decade of R\&D and prototyping the construction of ANTARES, the first operating deep-sea neutrino telescope has been completed. Since the deployment of the first line in 2006, data taking has proceeded essentially continuously. During this time the methods and procedures to calibrate such a novel detector have been developed, including in-situ time calibration with optical beacons and acoustic positioning of the detector elements with acoustic hydrophones. The presence of the $^{40}K$ in the sea water has proven particularly useful for monitoring the stability of the time calibration as well as the detector efficiency.

Based on data from the intermediate 5-line configuration, a number of preliminary analyses have been presented;  measurements of the atmospheric muon vertical depth intensity relation, a search for cosmic neutrino point sources in the southern sky, and a search for dark matter annihilation in the Sun. For the latter two analyses no significant signal was observed and competitive upper limits have been obtained. 

The succesful operation of ANTARES, and analysis of its data, is an important step towards KM3NET \cite{icrc1319:km3net}, a future km$^3$-scale high-energy neutrino observatory and marine sciences infrastructure planned for construction in the Mediterranean Sea.

\label{icrc1319:end}


\setcounter{figure}{0}
\setcounter{table}{0}
\setcounter{footnote}{0}
\setcounter{section}{0}
\newpage





\hyphenation{abcdef-ghijklmnoprstuwxyz IEEEtran}

\title{Charge Calibration of the ANTARES high energy neutrino telescope.}

\author{\IEEEauthorblockN{Bruny Baret\IEEEauthorrefmark{1} 
			  on behalf of the ANTARES Collaboration\IEEEauthorrefmark{2}}
                            \\
\IEEEauthorblockA{\IEEEauthorrefmark{1}Laboratoire AstroParticle and Cosmology \\ 10, rue A. Domon et L. Duquet,
75205 Paris Cedex 13, France.}
\IEEEauthorblockA{\IEEEauthorrefmark{2}http://antares.in2p3.fr}
}

\shorttitle{Baret \etal ANTARES Charge calibration}
\maketitle
\label{icrc1184:begin}

\begin{abstract}
ANTARES is a deep-sea, large volume mediterranean neutrino telescope installed off the Coast of Toulon, France. It is taking data in its complete configuration since May 2008 with nearly 900 photomultipliers installed on 12 lines. It is today the largest high energy neutrino telescope of the northern hemisphere. The charge calibration and threshold tuning of the photomultipliers and their associated front-end electronics is of primary importance. It indeed enables to translate signal amplitudes into number of photo-electrons which is the relevant information for track and energy reconstruction. It has therefore a strong impact on physics analysis. We will present the performances of the front-end chip, so-called ARS, including the waveform mode of acquisition. The in-laboratory as well as regularly performed in situ calibrations will be presented together with related studies like the time evolution of the gain of photomultipliers

\end{abstract}

\begin{IEEEkeywords}
neutrino telescope, front-end electronics, calibration.
\end{IEEEkeywords}
 
\section{Introduction}
ANTARES is an underwater neutrino telescope
installed at a depth of 2475 m in the
Mediterranean Sea. The site is at
about 40 km off the coast of Toulon, France. The
control station is installed in Institut Michel Pacha in La Seyne Sur Mer, close to Toulon.
The apparatus consists of an array of 900 photo-multiplier tubes (PMTs) by which the faint light
pulses emitted by relativistic charged particles 
propagating in the water may be detected. Based on such
measurements, ANTARES is capable of
identifying neutrinos of atmospheric as well as of
astrophysical origin. In addition, the detector 
is a monitoring station for geophysics and
sea science investigations.
For an introduction to the scientific aims of the
ANTARES experiment, the reader is referred to
the dedicated presentation at this Conference \cite{ant_sc}.

\section{The ANTARES apparatus}
                                   
The detector consists of an array of 900 large area photomultipliers (PMTs), Hamamatsu
R7081-20, enclosed in pressure-resistant glass
spheres to constitute the optical modules (OMs)\cite{om}, and arranged on 12 detection lines. An
additional line is equipped with
environmental devices.						
Each line is anchored to the sea bed and kept close to 
vertical position by a top buoy. The minimum	
distance between two lines ranges from 60 to 80~m.
Each detection line is composed by 25 storeys,
each equipped with 3 photomultipliers oriented downward at
$45^{\circ}$ with respect to the vertical. The storeys are
spaced by 14.5 m, the lowest one being located	
about 100 m above the seabed. 	
From the functional point of view, each line is	
divided into 5 sectors, each of which consists	
typically of 5 storeys. Each storey is controlled by
a Local Control Module (LCM), and each sector	
is provided with a modified LCM, the Master	
Local Control Module (MLCM), which controls
the data communications between its sector and	
the shore. A String Control Module (SCM),
located at the basis of each line, interfaces the line	
to the rest of the apparatus.
Each of these modules consists of an aluminum
frame, which holds all electronics boards
connected through a backplane and is enclosed in			
a water-tight titanium cylinder.

\section{The front-end electronics}

 \begin{figure*}[!t]
  \centerline{
	\subfloat[]{
  	\includegraphics[width=5.3in,height=2.8in]{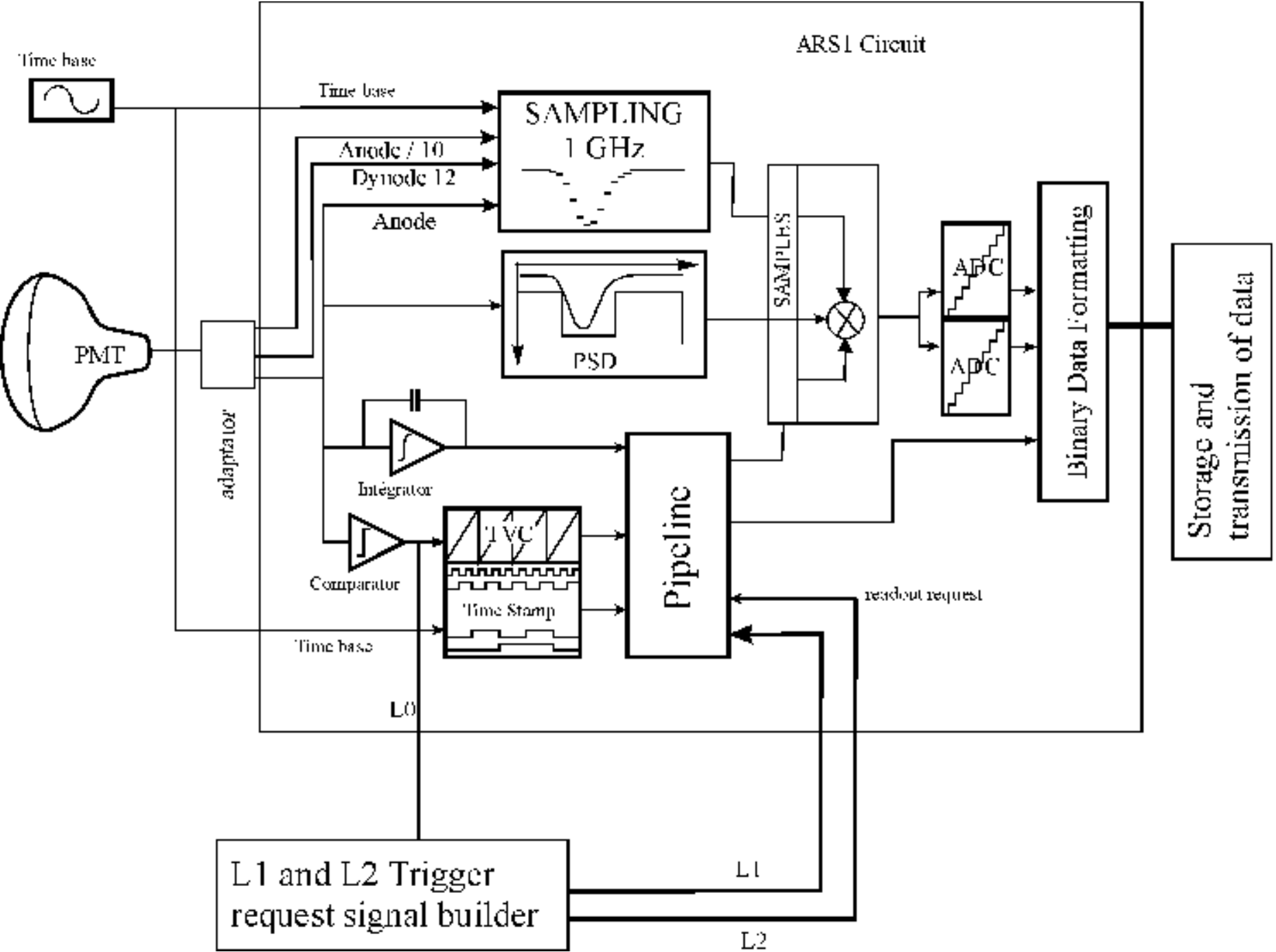}
	}
	}
  \caption{ARS architecture}
  \label{ars_arch}
 \end{figure*}
The full-custom Analogue Ring Sampler (ARS)
has been developed to perform the complex front-end operations \cite{ars}.This chip samples the PMT signal
continuously at a tunable frequency up to 1 GHz
and holds the analogue information on 128
switched capacitors when a threshold level is crossed.
The information is then digitized, in response to a
trigger signal, by means of two integrated dual 8-
bit ADC. Optionally the dynamic range may be
increased by sampling the signal from the last dynode.
A 20 MHz reference clock is used for time
stamping the signals. A Time to Voltage
Converter (TVC) device is used for high-resolution time measurements between clock
pulses. The ARS is also capable of discriminating
between simple pulses due to conversion of single
photoelectrons (SPE) from more complex
waveforms. The criteria used to discriminate
between the two classes are based on the
amplitude of the signal, the time above threshold
and the occurrence of multiple peaks within a
time gate. Only the charge and time information
is recorded for SPE events, while a full waveform
analysis is performed for all other events.
The ARS chips are arranged on a motherboard to
serve the optical modules. Two ARS chips, in a
``token ring'' configuration, perform the charge and
time information of a single PMT. A third chip
provided on each board is used for triggering
purposes.
The settings of each individual chip can be
remotely configured from the shore.

\section{Test bench calibration}
The bare ARSs were calibrated at IRFU-CEA/Saclay. There, the transfer functions of the Amplitude to Voltage Converter (AVC) have been measured. This AVC transfer function is an important parameter for the correction of the walk of the PMT signal and also for measurement of the amplitude of each PMT pulse.
The principal component of this bench is a pulse generator which directly sends signals to a pair of ARSs operating in a flip-flop mode. The generated pulse is a triangle with 4 ns rise time and 14 ns fall, somewhat
similar to the electrical pulse of a PMT with variable amplitude. 
 The tranfer functions of the dynamic range of the ADCs are linear and parametrised by their slope and intercept. The distributions of these two parameters for a large sample of ARS chips are presented on figure \ref{avc_slope} and \ref{avc_intercept} and one can see that they are homogeous which enables to use the same parameters for all ARSs.


 \begin{figure}[!t]
  \centering
  \includegraphics[width=2.5in]{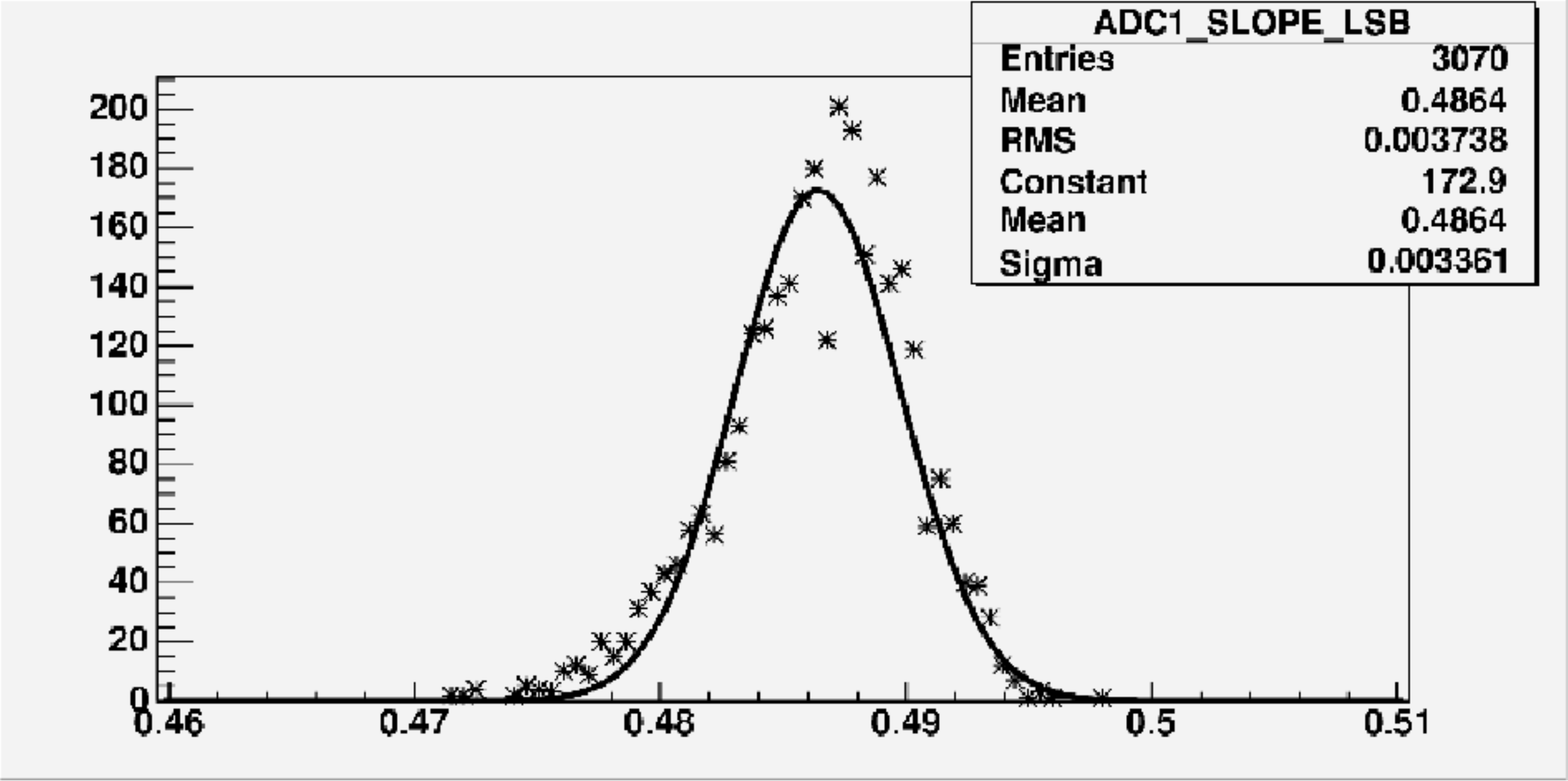}
  \caption{Distribution of the slope (in mV/bit) of the ADC transfer function}
  \label{avc_slope}
 \end{figure}

 \begin{figure}[!t]
  \centering
\includegraphics[width=2.5in]{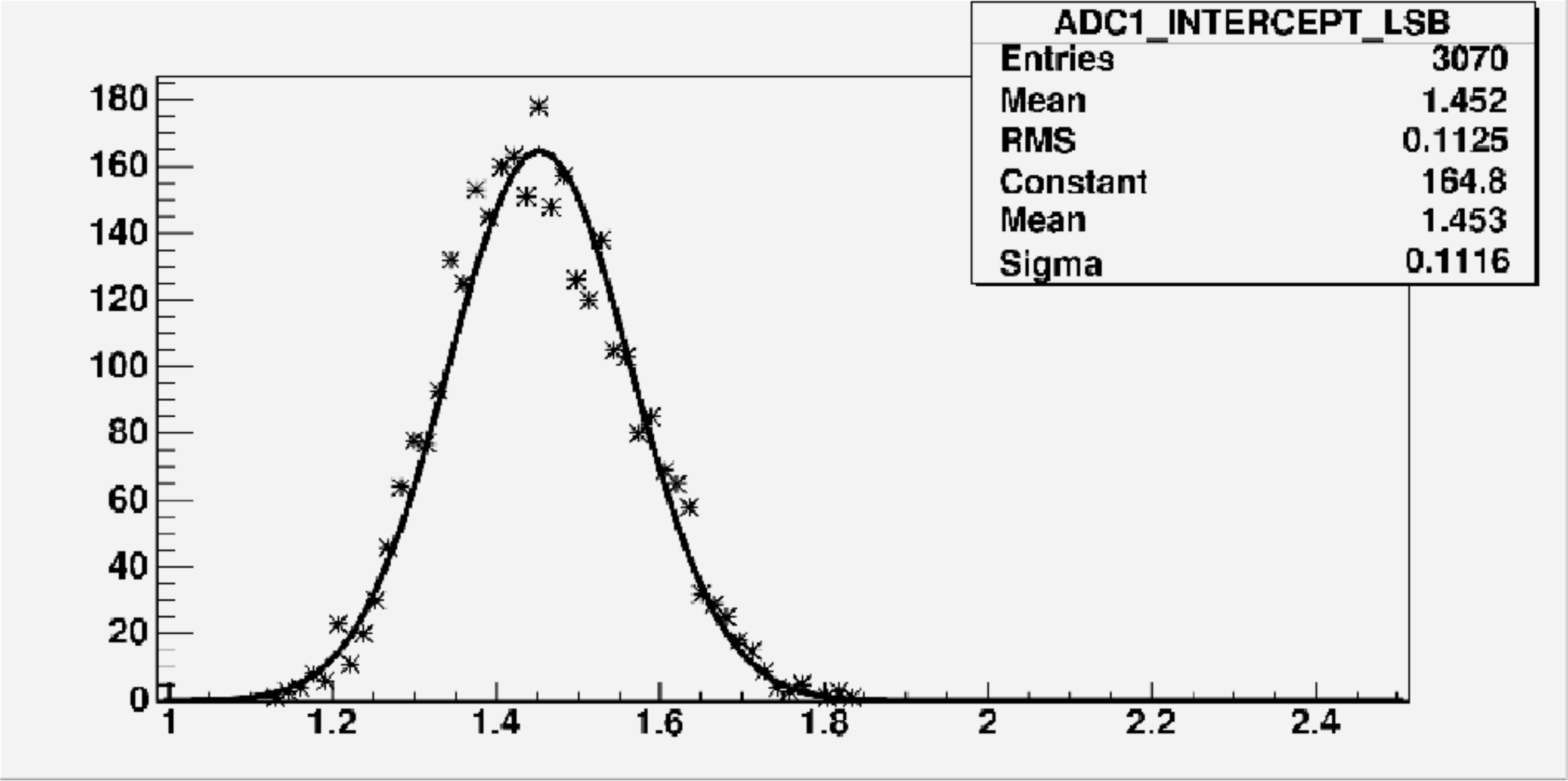} 
  \caption{Distribution of the intercept (in mV) of the ADC transfer function}
\label{avc_intercept}
 \end{figure}
 
\section{In Situ calibration}

 \begin{figure}[!t]
  \centering
  \includegraphics[width=2.6in,height=1.9in]{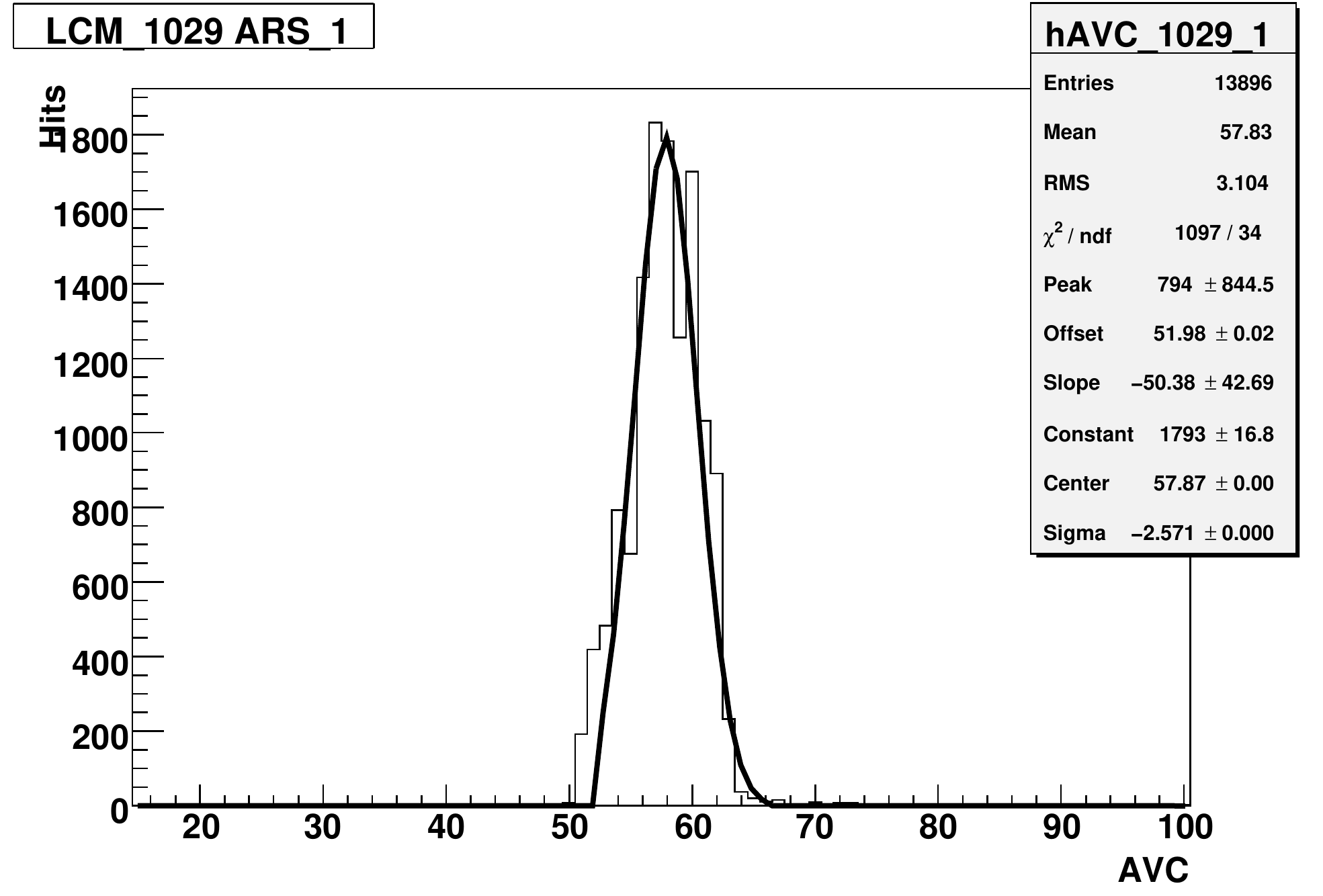}
  \caption{Example of AVC distribution of photoelectron like data with the corresponding fit (see text).}
  \label{1pe}
 \end{figure}

Specials runs reading the PMT current at random times allow to measure the corresponding so-called pedestal value of the AVC channel. Besides, the photoelectron peak can easily be studied with minimum bias events since the optical activity due the ${}^{40}K$ decays and bioluminescent bacteria produces, on average, single photons at the photocathode level. The knowledge of the photoelectron peak and the pedestal is used to estimate the charge over the full dynamical range of the ADC. The integral linearity of the ADC used in the ARS chip has independently been studied using the TVC channel and shows satisfactory results \cite{timecalib}. 
An example of in situ charge distribution for a particular ARS is shown on figure \ref{1pe}.
The values in AVC channel of the pedestal and the photoelectron peak are used to convert individual measurements into photoelectron units. Charge distributions obtained with minimum bias data (based on snapshot of the overall activity of an optical module above a given threshold) can be parameterized using the following simple formula:
\begin{equation}
    Ae^{-\alpha (x-x_{th})}+Be^{\frac{-(x-x_{pe})^2}{2\sigma^2}}
\end{equation}

 \begin{figure}[!t]
  \centering
  \includegraphics[width=2.6in,height=1.9in]{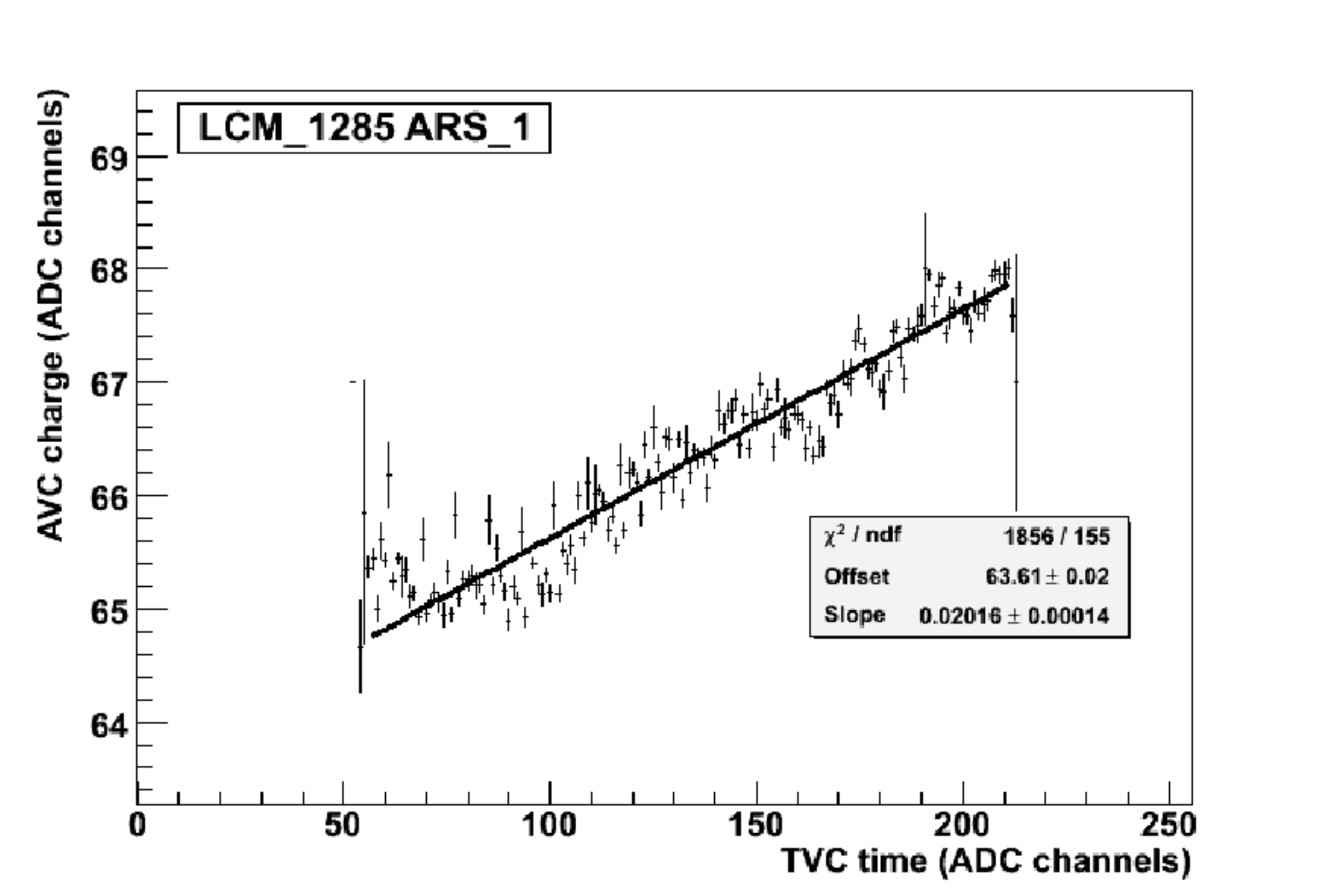}
  \caption{Example of the observed cross talk effect form the time measurement channel affecting the charge measurement channel.}
  \label{Xtalk}
 \end{figure}
The first term accounts for the dark current of the PMT, the second one describes the photoelectron distribution itself. The parameters $x_{th}$ and $x_{pe}$ are respectively the effective thresholds (``offset'') and photoelectron peak (``center'') in AVC units.\\ 
The charge measurements in the AVC channels appear to be influenced by the time measurements in the TVC channel (the inverse effect does not apply). This effect, referred to as ``cross talk effect'' can be, considering the current settings of the ARS, on an event-by-event basis
 as high as 0.2~pe. It is thought to be due to a cross talk of the capacitors inside the ARS pipeline.  It is a linear effect that does not require correction on high statistics basis (when hits populate the full range of the TVC, the effect washes out). Nevertheless a correction has to be applied to the measured charge of a single event. This correction can be inferred with in situ measurements at the level of the photoelectron by plotting the AVC value against the TVC value as can be seen in figure \ref{Xtalk}. After calibration including cross talk correction, minimum bias events recorded by the detector, coming predominantly from ${}^{40}K$ decay and bioluminescence are dominated by single photo-electron charges as is shown on figure \ref{min_bias_charge}\\
 \begin{figure}[!t]
  \centering
  \includegraphics[width=2.8in,height=2.in]{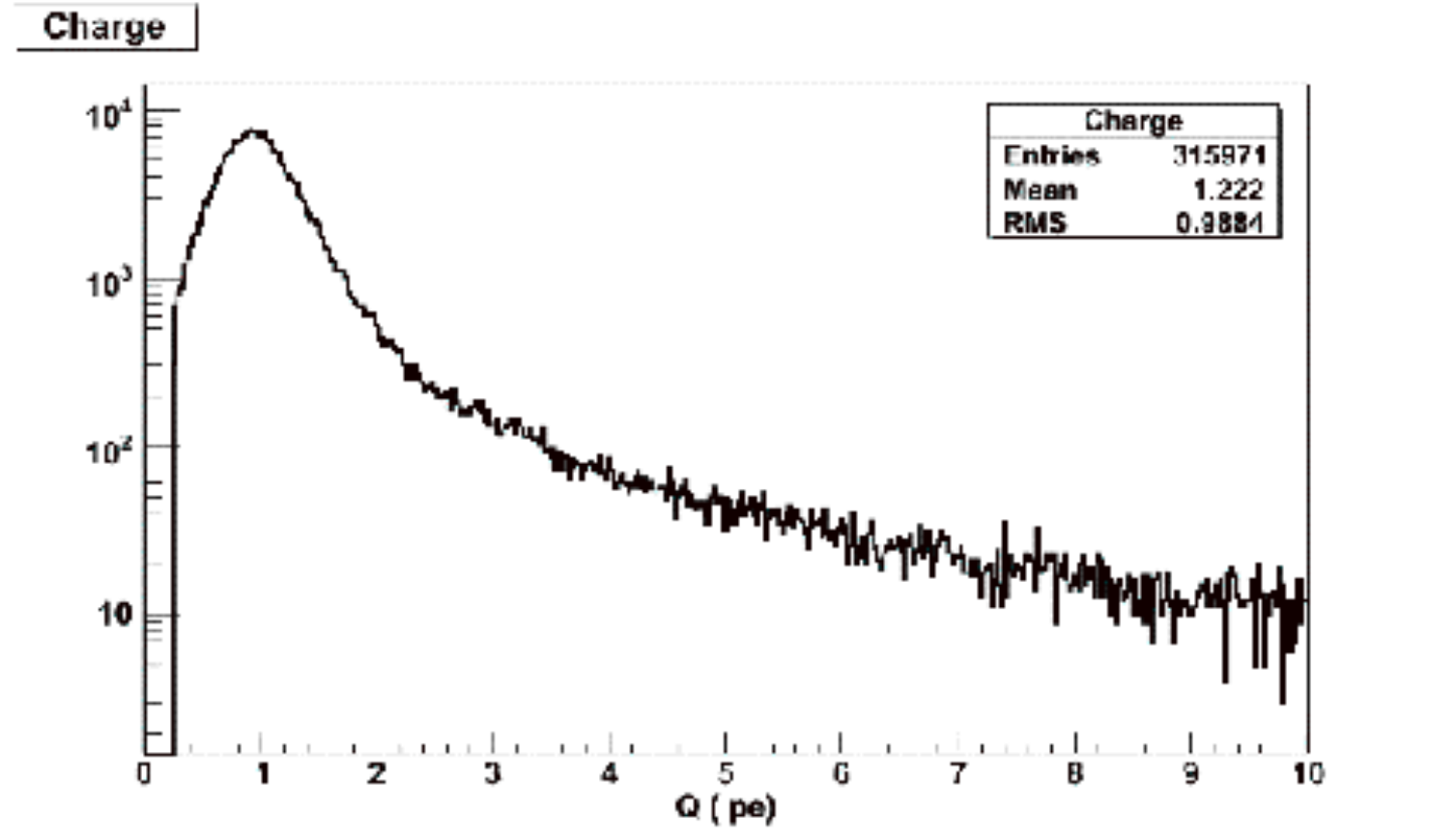}
  \caption{Charge distribution in $pe$ recorded by the detector.}
  \label{min_bias_charge}
 \end{figure}\\
The ARS has also the capability to perform a
full waveform sampling (WF) of the OM signal in addition to the charge measurement
of the PMT pulse and its arrival time. Although this functionality
is mainly used to record double pulses or large
amplitude signals, it is useful to cross-check the computation of the
SPE charge by the integrator circuit of the ARS. In WF mode, 128
digitisations of the OM anode signal are provided, at a sampling rate
of 640 MHz. In order to obtain a precise time stamping of the WF data,
a synchronous sampling of the 50~MHz internal ARS clock is also
performed and read out in addition to the OM data. An example of a WF
record is shown in figure~\ref{WF}.
 \begin{figure}[!t]
  \centering
  \includegraphics[width=2.5in,height=1.8in]{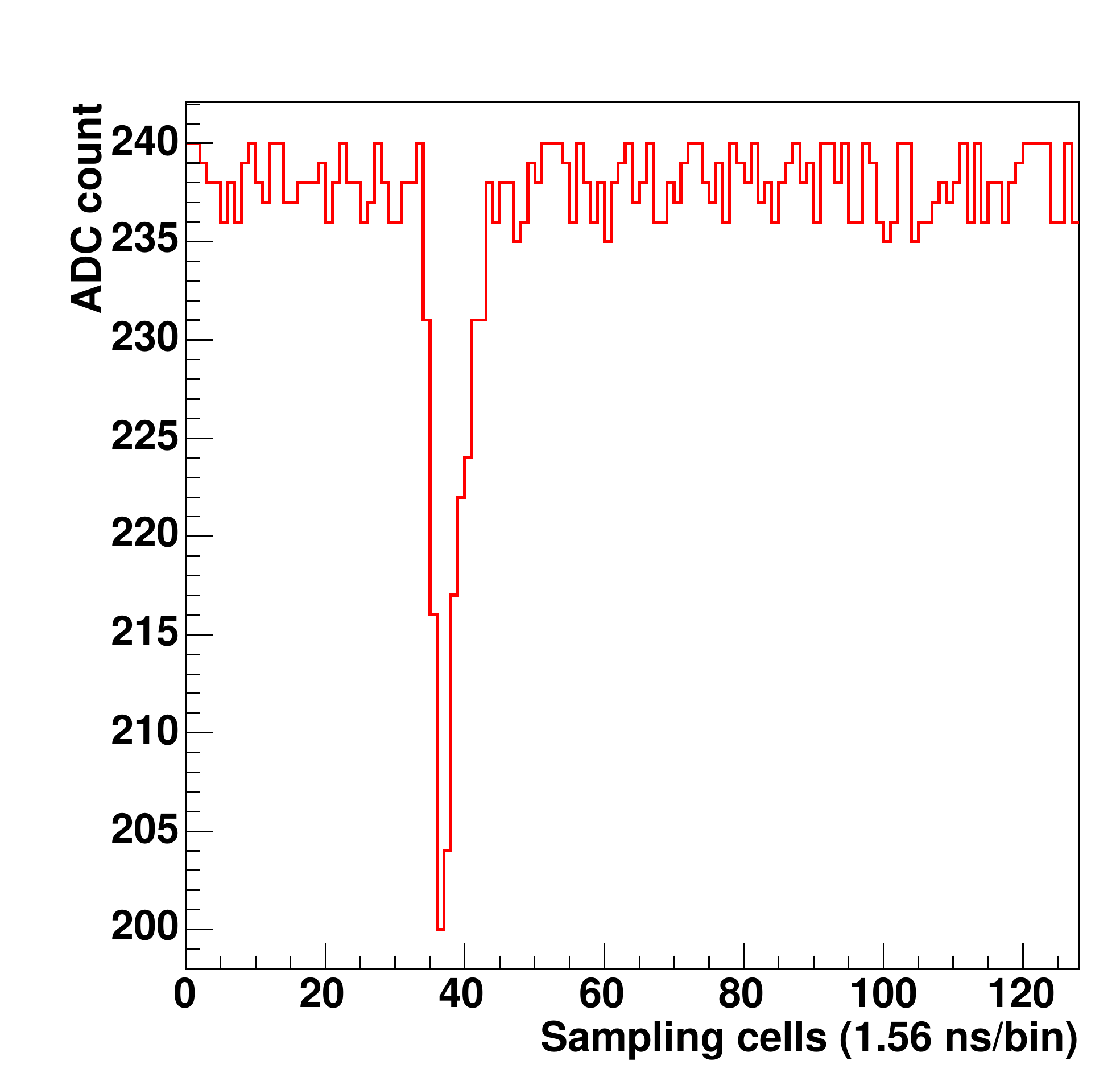}
  \caption{Example of a waveform sampling of an OM signal}
  \label{WF}
 \end{figure}
Figure~\ref{QWF} 
displays the charge distribution of the OM signals
obtained by integrating the WF samples after baseline subtraction. The
single photo-electron peak is clearly identified well above the
electronics noise.
 \begin{figure}[!t]
  \centering
  \includegraphics[width=2.5in,height=1.8in]{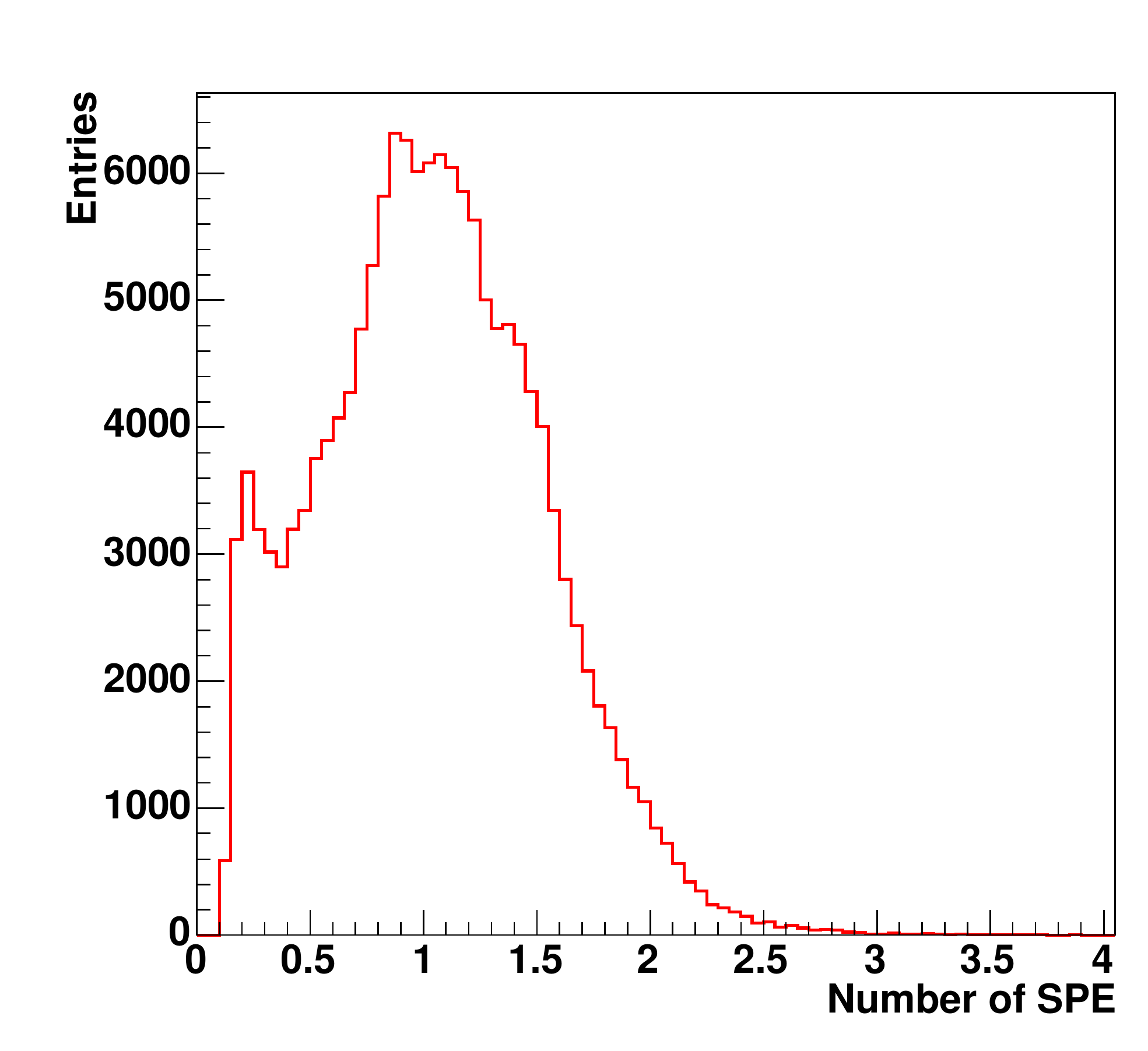}
  \caption{Charge distribution
of the PMT signal obtained by integrating the WF samples }
  \label{QWF}
 \end{figure}
 
When the time over threshold of a pulse is too short, the ARS chip cannot properly generate the time stamp (TS) of the event, which remains null. This happens when the hit amplitude is just above the amplitude threshold.  The charge (AVC) and fine time of the events (TVC) are recorded correctly. This specific behaviour has negligible influence on efficiency but enables to measure the effective threshold in AVC units for different DAC settings by selecting event with TS=0. For different slow control DAC values the mean AVC values of events at the threshold are recorded during these special calibration runs. The result of the linear fit  of the transfer  function gives the intercept (DAC value for null threshold) and slope. An example is shown on figure \ref{thr_insitu}.
 \begin{figure}[!t]
  \centering
  \includegraphics[width=2.5in, height=1.3in]{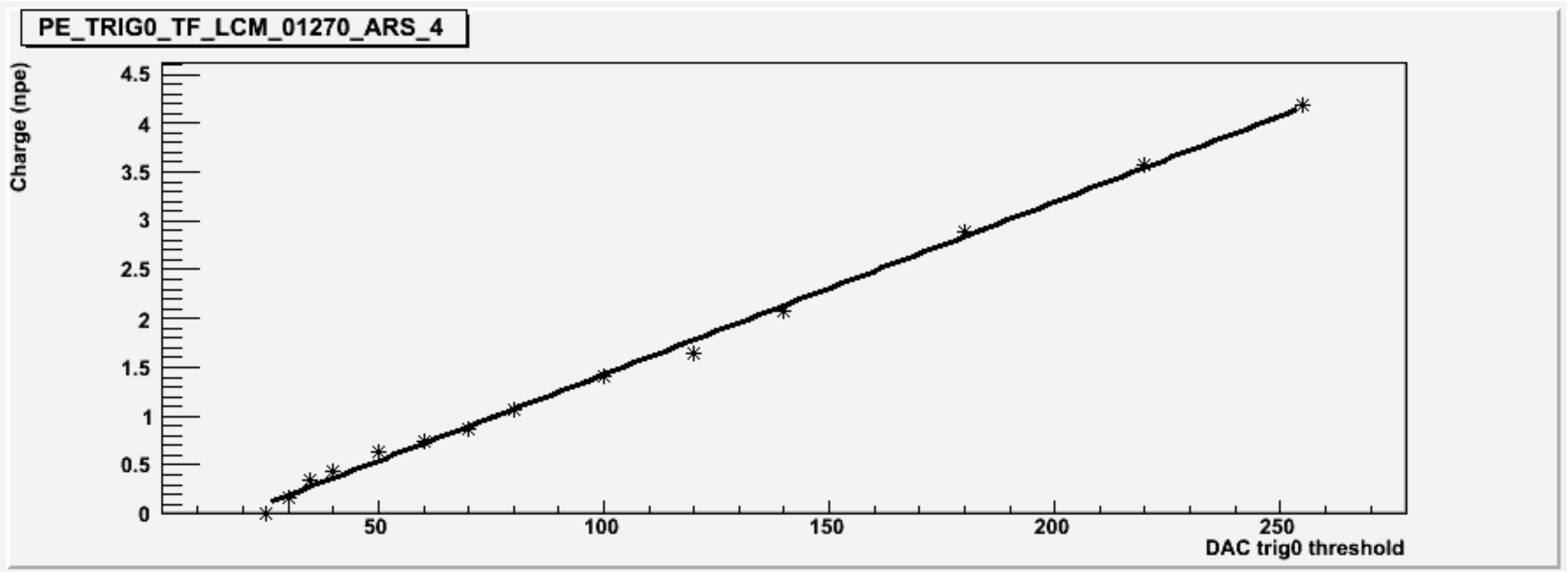}
  \caption{Example of an effective threshold transfer function of an ARS obtained with TS=0 events.}
  \label{thr_insitu}
 \end{figure}\\
This method can be applied to every readout ARS of the detector. There are therefore individual in situ calibration and transfer functions for each ARS. These monitored values are stored in a dedicated database and used for further adjustments of the detector setting. In particular, these effective calibrations are used to homogenize the individual thresholds to a value close to 1/3 pe.  

The desintegration of ${}^{40}K$ present in sea water can be used to monitor the evolution with time of the detector response. Indeed, relativistic electrons produced by the $\beta$ desintegration will produce Cerenkov photons which can trigger two adjacent optical modules in the same storey. 
 \begin{figure}[!t]
  \centering
  \includegraphics[width=3.in]{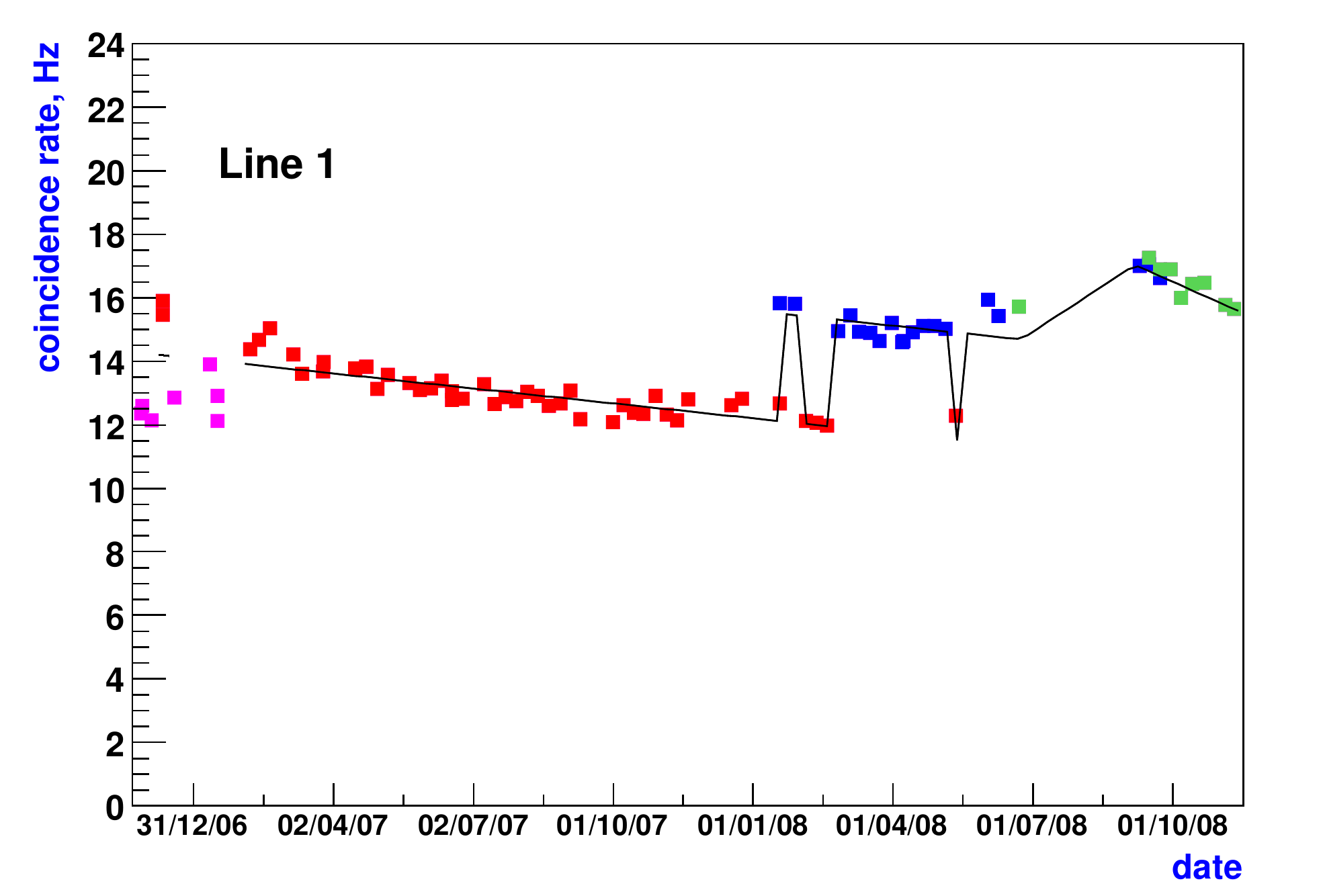}
  \caption{Time evolution of the counting rate due to ${}^{40}K$ desintegration measured with the first line of the detector. Colours account for different threshold setting periods.}
  \label{40K}
 \end{figure}

As can be seen on figure \ref{40K}, ${}^{40}K$ counting rate evolution with time have shown a regular decrease. This PMTs gain drop effect is thought to be due to ageing effect of the photocathode. As can be seen after the period between july and september of 2008 when the detector was off for cable repair, gain seems to be partially recovered when PMTs are off for some time.\\
Since all channels are tunned to have an effective threshold of $0.3~\textrm{pe}$, one has to regularly check, and if necessary correct the value of the effective thresholds. This is done thanks to the TS=0 events as explained earlier. The effect of this procedure can be seen on the effective threshold distribution on figure \ref{th_before} and \ref{th_after}.  
 
 \begin{figure}[!t]
  \centering
  \includegraphics[width=2.5in]{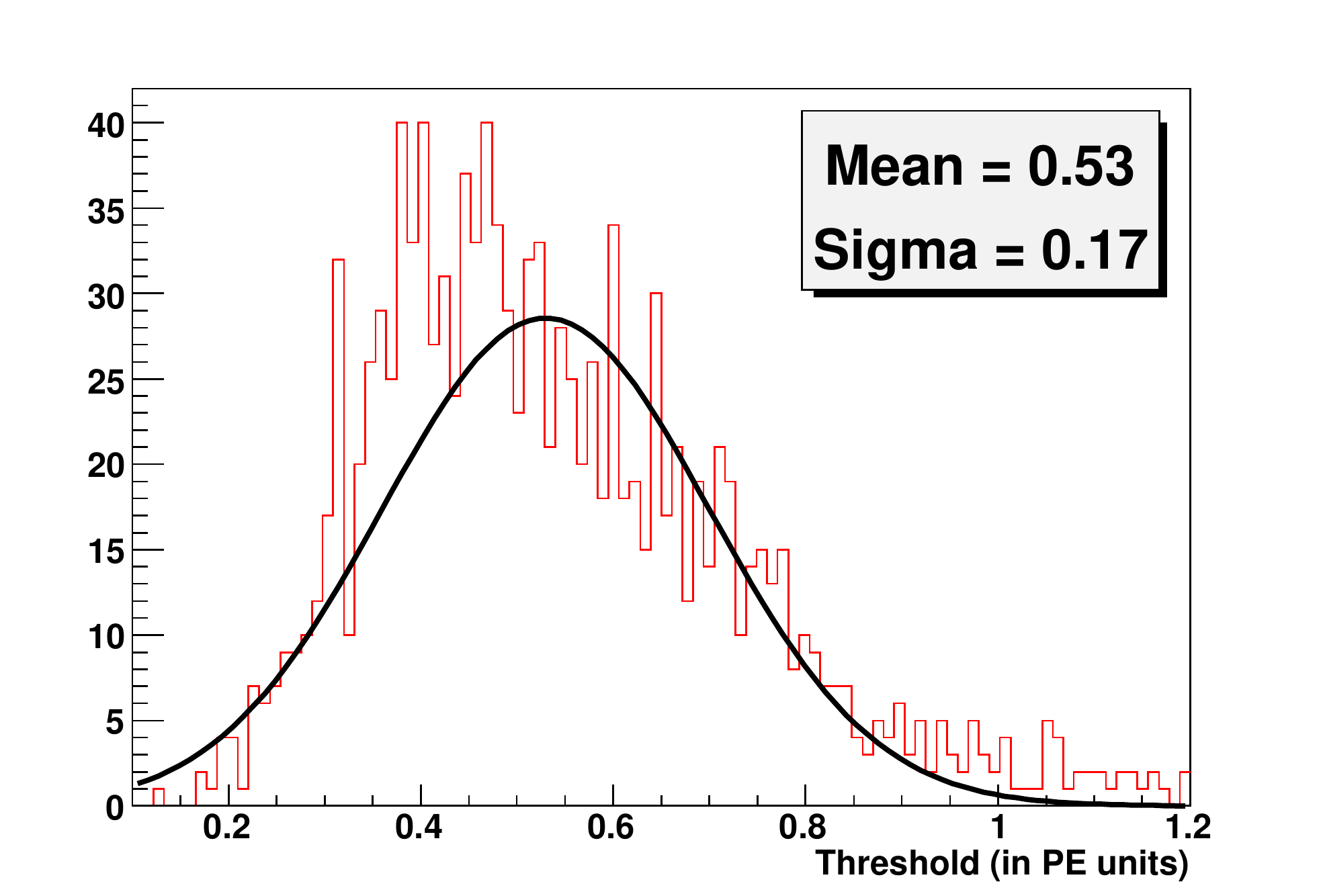}
  \caption{Distributions of the effective thresholds before (12/2007) in situ tuning procedure.}
  \label{th_before}
 \end{figure}
 \begin{figure}[!t]
  \centering
  \includegraphics[width=2.5in]{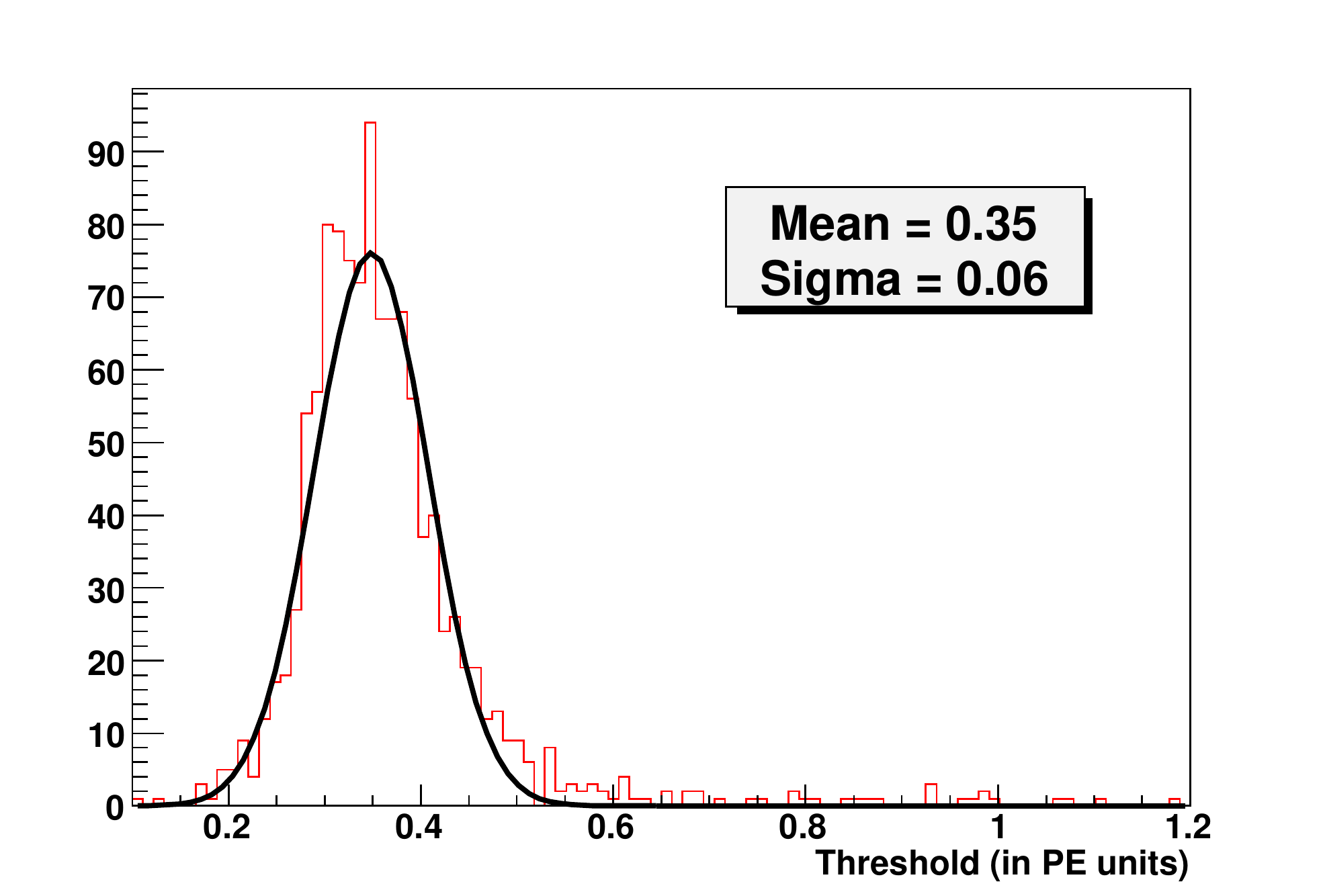}
  \caption{Distributions of the effective thresholds after (03/2008) in situ tuning procedure.}
  \label{th_after}
 \end{figure}

\section{Conclusions}
All the ARSs of the ANTARES neutrino telescope have been calibrated prior to deployment in order to be able to translate the electrical signal from the chips into number of photo-electrons which is the relevant information for event recontruction and physics analysis. Furthermore, in situ calibration procedures have been developped and are regularly performed in order to monitor and control the detector response, especially to take into account PMTs gain evolution.  

\section{Acknowledgement}
This work was in part supported by the French ANR grant ANR-08-JCJC-0061-01.

\label{icrc1184:end}

\setcounter{figure}{0}
\setcounter{table}{0}
\setcounter{footnote}{0}
\setcounter{section}{0}
\newpage




\newcommand{\gc}{Galactic Centre}
\newcommand{\antares}{ANTARES}

\hyphenation{abcdef-ghijklmnoprstuwxyz IEEEtran}

\title{Concepts and performance of the \antares\ data acquisition system}

\author{\IEEEauthorblockN{Mieke Bouwhuis\IEEEauthorrefmark{1}
		on behalf of the \antares\ collaboration}
\IEEEauthorblockA{\IEEEauthorrefmark{1}National Institute for Subatomic
Physics (Nikhef), Amsterdam, The Netherlands}}

\shorttitle{Bouwhuis \etal \antares\ DAQ system}
\maketitle
\label{icrc_daq:begin}

\begin{abstract}

The data acquisition system of the \antares\ neutrino telescope
is based on the unique ``all-data-to-shore" concept.
In this, all signals from the photo-multiplier tubes are digitised,
and all digital data are sent to shore where they are processed in
real time by a PC farm.
This data acquisition system showed excellent stability and
flexibility since the detector became operational in March 2006.
The applied concept enables to operate different physics triggers to the
same data in parallel, each optimised for a specific (astro)physics
signal. 
The list of triggers includes two general purpose muon triggers, a 
\gc\ trigger, and a gamma-ray burst trigger.
The performance of the data acquisition system is evaluated by its
operational efficiency and the data filter capabilities.
In addition, the efficiencies of the different physics triggers are
quantified. 
\end{abstract}

\begin{IEEEkeywords}
neutrino telescope; data acquisition system; triggering
\end{IEEEkeywords}
 
\section{Introduction}

The \antares\ neutrino telescope is situated in the Mediterranean Sea
at a depth of about 2500~m, approximately 40~km south east of the
French town of Toulon.  
Neutrinos are detected through the detection of Cherenkov light
produced by the charged lepton that emerges from a neutrino
interaction in the vicinity of the detector. 
Measurements are focused mainly on muon-neutrinos, since the muon resulting from
a neutrino interaction can travel a distance of up to several kilometres.
Due to the transparency of the sea water (the absorption length is about
50~m), the faint Cherenkov light can be detected at relatively large
distances from the muon track. 
A large volume of sea water is turned into a neutrino
detector by deploying a 3-dimensional array of light sensors in the
water. 

The \antares\ detector consists of thirteen lines, each with up to 25~storeys.
The storeys are connected by cables which provide mechanical
strength, electrical contact and fibre-optic readout. 
Each line is held on the seabed by a dead-weight anchor and 
kept vertical by a buoy at the top of its 450~m length.
Along eleven lines, 25~storeys with three light sensors are placed 
at an inter-storey distance of 14.5~m starting 100~m above the seabed.
On each storey three spherical glass pressure vessels contain 10''
Hamamatsu photo-multiplier tubes (PMTs), which are oriented with their
axes pointing downward at an angle of 45~degrees from the vertical.  
One line consists of 20~such storeys, and one line is equipped with deep-sea
instrumentation.
Each storey in the detector has a titanium cylinder which houses the electronics for
data acquisition and slow control.
This system is referred to as the local control module.
In addition, each line has a line control module that is located at
the anchor. 

Daylight does not penetrate to the depth of the \antares\ site. 
Therefore the telescope can be operated day and night.
But even in the absence of daylight, a ubiquitous background
luminosity is present in the deep-sea due to the decay of radioactive
isotopes (mainly $^{40}$K) and to bioluminescence. 
This background luminosity produces a relatively high count rate of
random signals in the detector (60--150~kHz per PMT).
This background can be suppressed by applying the characteristic
time-position correlations that correspond to a passing muon to the
data. 

\section{Data acquisition system}

The main purpose of the data acquisition (DAQ) system is to convert
the analogue pulses from the PMTs into a readable input for the
off-line reconstruction software. 
The DAQ system is based on the ``all-data-to-shore" concept~\cite{daq}.
In this, all signals from the PMTs that pass a preset threshold
(typically 0.3 photo-electrons) are digitised and all digital data are
sent to shore where they are processed in real-time by a farm of
commodity PCs. 

\subsection{Network architecture}

The network architecture of the off-shore DAQ system has a star topology.
In this, the storeys in a line are organised into separate
sectors, each consisting of 5~storeys, and the detector lines are connected
to a main junction box. 
The junction box is connected to a station on shore via a single
electro-optical cable. 
The network consists of a pair of optical fibres for each detector line,
an 8~channel dense wavelength division \hbox{[de-]multiplexer} (DWDM)
in each line control module (200~GHz spacing), a small Ethernet switch
in each sector and a processor in each local control module.
The Ethernet switch in the sector consists of a combination of the
Allayer AL121 (eight 100~Mb/s ports) and the Allayer AL1022 (two Gb/s
ports). 
One of the 100~Mb/s ports is connected to the processor in the local
control module via its backplane (100Base--CX) and four are connected
to the other local control modules in the same sector via a
bi-directional optical fibre (100Base--SX). 
One of the two Gb/s ports is connected to a DWDM transceiver
(1000Base--CX). 
The DWDM transceiver is then 1--1 connected to an identical
transceiver on shore using two uni-directional optical fibres
(1000Base--LH). 
The line control module has also a processor which is connected to a
DWDM transceiver via its backplane (100Base--CX). 
A similar pair of DWDM transceivers is then used to establish a
100~Mb/s link to shore (100Base--LH). 
The network architecture on-shore consists of an optical
\hbox{[de-]multiplexer} and 6~DWDM transceivers for each detector
line, a large Ethernet switch (192~ports), a data processing farm and
a data storage facility. 
The optical fibres and the Ethernet switches form together a (large)
local area network. 
Hence, it is possible to route the data from any local control module
in the detector to any PC on shore. 

\subsection{Readout}

The front-end electronics consist of custom built analogue ring
sampler (ARS) chips which digitise the charge and the time of the
analogue signals from the PMT. 
The combined data is generally referred to as a hit; it can be a
single photo-electron hit or a complete waveform. 
The arrival time is determined from the signal of the clock system in
the local control module. 
An on-shore clock system (master) drives the clock systems in the
local control modules (slaves). 
The processor in the local control module is a Motorola MPC860P.
It runs the VxWorks real time operating system~\cite{vxworks} and
hosts the DaqHarness and ScHarness processes. 
The DaqHarness and ScHarness are used to handle respectively the data
from the ARS chips and the data from the various deep-sea instruments.
The latter is usually referred to as slow control data.
The processor in the local control module has a fast Ethernet
controller (100~Mb/s) that is connected to the Ethernet switch in the
sector. 
Inside the local control module, two serial ports with either RS485 or
RS232 links and the MODBUS protocol are used to handle the slow
control signals. 
The specific hardware for the readout of the ARS chips is implemented
in a high density field programmable gate array (XilinX Virtex-E
XCV1000E). 
The data are temporarily stored in a high capacity memory (64~MB
SDRAM) allowing a de-randomisation of the data flow.
In this, the data are stored as an array of hits.
The length of these arrays is determined by a predefined time frame of
about 100~ms and the singles rates of the PMTs. It amounts to about
60--200~kB. 
The data are read out from this memory by the DaqHarness process and
sent to shore. 
All data corresponding to the same time frame are sent to a single
data filter process in the on-shore data processing system.

The on-shore data processing system consists of about 50 PCs
running the Linux operating system. 
To make optimal use of the multi-core technology, four data filter
processes run on each PC. 
The physics events are filtered from the data by the data filter
process using a fast algorithm. 
The typical time needed to process 100~ms of raw data amounts to
500~ms. 
The available time is used to apply designated software triggers to
the same data. 
For example, a trigger that tracks the \gc\ is used whenever the count
rates of the PMTs are below 150~kHz (this corresponds to about 80\% of
the time). 
On average, the data flow is reduced by a factor of about 10,000.
The filtered data are written to disk in ROOT format~\cite{ROOT} by a central data
writing process. 
The count rate information of every PMT is stored together with the
physics data. 
The sampling frequency of these rate measurements is about 10~Hz.
The data from the readout of the various instruments are transferred
as an array of parameter values and stored in the database via a
single process. 
The readout of the various deep-sea instruments is scheduled via read
requests that are sent from shore by a designated process. 
The frequency of these read requests is defined in the database.
A general purpose data server based on the tagged data concept is used
to route messages and data~\cite{controlhost}. 
For instance, there is one such server to route the physics
events to the data writer which is also used for online monitoring.
Messages (warnings, errors, etc.) are collected, displayed and written
to disk by a designated GUI. 

\subsection{Operation}

The main control GUI allows the operator to modify the state of the
system. 
In total, the system involves about 750~processes (300~DaqHarness
processes, 300~ScHarness processes, 120~data filters, and various other
processes).
These processes implement the same state machine diagram~\cite{chsm}.
Before the start of a data taking run, the whole system including the
detector is configured. 
In order to archive data efficiently, the main control GUI updates
the run number regularly. 
The database system is used to keep track of the history of the
detector and the data taking. 
It is also used for storing and retrieving configuration parameters of
the whole system. 
The positions of the PMTs are determined using a system of acoustic
transmitters and receivers. 
The corresponding data are recorded at the same time as the physics
data.
The time calibration of the PMTs is obtained using special data taking
runs. 
During these runs, one or more LED beacons (or laser beacon) flash.
The typical flash rate is about 1~kHz.
The time calibration data are recorded using a designated software
trigger. 
All data are archived in the IN2P3 computer centre in Lyon which also
houses the Oracle database system. 

\subsection{External triggers}

The on-shore data processing system is linked to the gamma-ray bursts
coordinates network (GCN)~\cite{GCN}. 
This network includes the Swift and Fermi satellites.
There are about 1--2 GCN alerts per day and half of them correspond to
a real gamma-ray burst. 
For each alert, all raw data are saved to disk during a preset period
(presently 2 minutes). 
The buffering of the data in the data filter processors is used to
save the data up to about one minute before the actual alert~\cite{antares-grb}.

\section{Performance of the DAQ system}  

The performance of the DAQ system can be summarised in terms of the
efficiency to detect neutrinos and the efficiency to operate the
neutrino telescope. 
The efficiency to detect neutrinos is primarily determined by the
capability to filter the physics events from the continuous data
streams. 
With the all-data-to-shore system, different software triggers can be
operated simultaneously. 
At present, two general purpose muon triggers (`standard') and one
minimum bias trigger are used to take data. 
The minimum bias trigger is used for monitoring the data quality.
The standard trigger makes use of the general causality relation:

\begin{eqnarray}
   \left| t_i - t_j \right| \le \left| \bar{x}_i - \bar{x}_j \right|  \times \frac{n}{c} \label{eq:3D}, 
\end{eqnarray}

\noindent
where $t_i$ ($\bar{x}_i$) refers to the time (position) of hit $i$,
$c$ the speed of light and $n$ the index of refraction of the sea water.
In this, the direction of the muon, and hence the neutrino, is not used.
The standard trigger is therefore sensitive to muons covering the full
sky. 
To limit the rate of accidental correlations, the hits have been
preselected. 
This pre-selection (L1) includes coincidences between two neighbouring
PMTs in the same storey and large pulses (number of photo-electrons
typically greater than 3). 
The minimum number of detected photons to trigger an event ranges
between 4--5 L1 hits, depending on the trigger algorithm.
This corresponds to a typical threshold of several 100~GeV.
The purity of the trigger (fraction of events that correspond to a
genuine muon) has been determined using a simulation of the detector
response to muons traversing the detector and a simulation based on
the observed background. 
The purity is found to be better than 90\%.
The 10\% impurity is mainly due to (low-energy) muons that in
combination with the random background produce a trigger.
A small fraction of the events ($\ll 1\%$) is due to accidental
correlations. 
The observed trigger rate is thus dominated by the background of
atmospheric muons and amounts to \mbox{5--10~Hz} (depending on the
trigger conditions). 

In addition to the standard trigger, a trigger that tracks the \gc\ is
used to maximise the detection efficiency of neutrinos coming from the
\gc. 
This trigger makes use of the direction specific causality relation:

\begin{eqnarray}
   (z_i - z_j) - R_{ij}\tan{\theta_C} \le c(t_i - t_j) \nonumber \\
  ~~~~~~~~~\le (z_i - z_j) + R_{ij}\tan{\theta_C} \label{eq:1D},
\end{eqnarray}

\noindent
where $z_i$ refers to the position of hit $i$ along the neutrino
direction, $R_{ij}$ refers to distance between the positions of hits
$i$ and $j$ in the plane perpendicular to the neutrino direction and
$\theta_C$ to the characteristic Cherenkov angle. 
Compared to equation \ref{eq:3D}, this condition is more stringent
because the 2-dimensional distance is always smaller than the
3-dimensional distance. 
Further more, this distance corresponds to the distance travelled by
the photon (and not by the muon). 
Hence, it can be limited to several absorption lengths (e.g.\ 100~m or
so) without loss of detection efficiency. 
This restriction reduces the combinatorics significantly (about a
factor of 10 for each additional hit). 
As a consequence, all hits can be considered and not only preselected
hits (L1) without compromising the purity of the physics events. 
The detection efficiency of the general purpose muon trigger and the
\gc\ trigger are shown in Fig.~\ref{f:volume}. 
The effective volume is defined as the volume in which an interaction
of a muon neutrino would produce a detectable event.
\begin{figure}[t!]
\setlength{\unitlength}{1cm}
\begin{center}
\begin{picture}(8,8)
\put(0,0){\scalebox{0.45}{\includegraphics{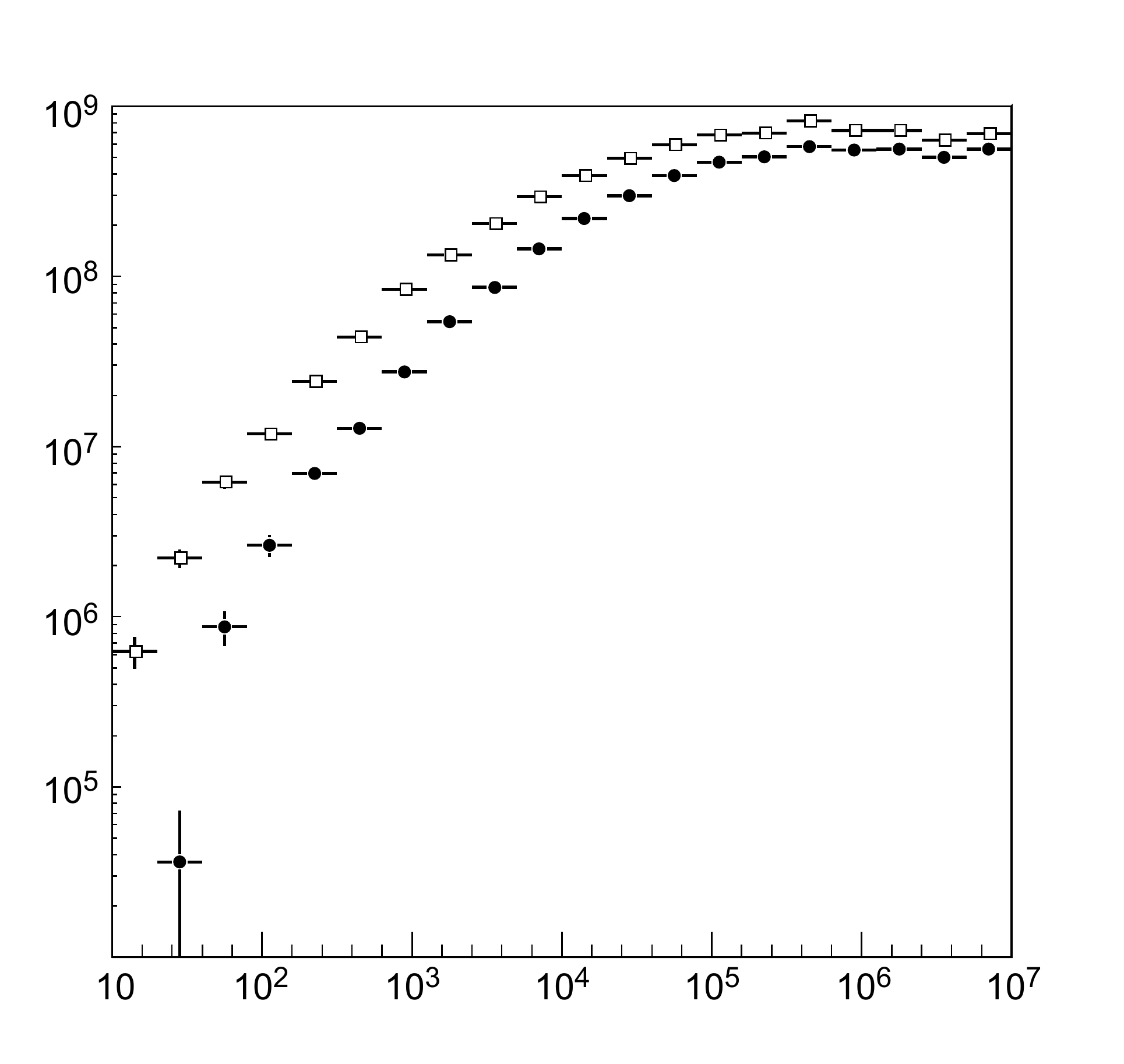}}}%
\put(4.5,0){\makebox(0,0)[b]{neutrino energy [GeV]}}%
\put(0,4.5){\makebox(0,0)[l]{\rotatebox{90}{effective volume [m$^3$]}}}%
\end{picture}
\end{center}
\caption{
Effective volume of the standard trigger (solid circles) and the \gc\
trigger (open squares) as a function of the neutrino energy. 
\label{f:volume}
}
\end{figure}
\noindent
As can be seen from Fig.~\ref{f:volume}, the detection efficiency
obtained with the \gc\ trigger is significantly higher than that
obtained with the standard trigger. 
As a result, the sensitivity to neutrinos from the \gc\ is
greatly improved. 
The field of view for which this improved efficiency is obtained is about 
10~degrees.

Since the start of the operation of the detector (March 2006), the
average data taking efficiency has been better than 90\%.

\enlargethispage{-1.0in}
\label{icrc_daq:end}
\setcounter{figure}{0}
\setcounter{table}{0}
\setcounter{footnote}{0}
\setcounter{section}{0}
\newpage





\hyphenation{abcdef-ghijklmnoprstuwxyz IEEEtran}

\title{Positioning system of the ANTARES Neutrino Telescope}

\author{\IEEEauthorblockN{Anthony M Brown\IEEEauthorrefmark{1}\IEEEauthorrefmark{2} on behalf of the ANTARES Collaboration \IEEEauthorrefmark{3}}
                            \\
\IEEEauthorblockA{\IEEEauthorrefmark{1}Centre de Physique des Particules de Marseille, 164 Av. de Luminy, Case 902, Marseille, France.}
\IEEEauthorblockA{\IEEEauthorrefmark{2}brown@cppm.in2p3.fr}
\IEEEauthorblockA{\IEEEauthorrefmark{3}\url{http://antares.in2p3.fr}}
}

\shorttitle{A.M. Brown - Antares Positioning System}
\maketitle
\label{icrc0178:begin}

\begin{abstract}
 Completed in May 2008, the ANTARES neutrino telescope is located 40 km off the coast of Toulon, at a depth of about 2500 m. The telescope consists of 12 detector lines housing a total of 884 optical modules. Each line is anchored to the seabed and pulled taught by the buoyancy of the individual optical modules and a top buoy. Due to the fluid nature of the sea-water detecting medium and the flexible nature of the detector lines, the optical modules of the ANTARES telescope can suffer from  deviations of up to several meters from the vertical and as such, real time positioning is needed. 

Real time positioning of the ANTARES telescope is achieved by a combination of an acoustic positioning system and a lattice of tiltmeters and compasses. These independent and complementary systems are used to compute a global fit to each individual detector line, allowing us to construct a 3 dimensional picture of the ANTARES neutrino telescope with an accuracy of less than 10 cm. 

In this paper we describe the positioning system of the ANTARES neutrino telescope and discuss its performance during the first year of 12 line data taking.
  \end{abstract}

\begin{IEEEkeywords}
ANTARES neutrino telescope, alignment, acoustic positioning system
\end{IEEEkeywords}
 
\section{Introduction}

Deployed off the coast of Toulon, the ANTARES telescope is, at present, the largest neutrino detector in the Northern hemisphere \cite{pc},\cite{large}. Utilising the Mediterranean Sea as a detecting medium, ANTARES consists of 12 detector lines, spaced approximately 70 meters apart, giving an overall surface area of the order of $~$0.1 km$^2$. The detection principle of ANTARES relies on the observation of Cherenkov photons emitted by charged relativistic leptons, produced through neutrino interactions with the surrounding water and seabed, using a 3 dimensional lattice of photomultiplier tubes (PMTs) \cite{pmt}. For ANTARES this 3 dimensional lattice of PMTs has been optimised for the detection of upward going high energy muon neutrinos and as such, each line of the detector, with the exception of Line 12, consists of 25 storeys separated by a vertical distance of 14.5 metres, each containing a triplet of Optical Modules (OMs) looking downwards at an angle of 45$^\circ$ from the vertical.

An important characteristic of any neutrino telescope is its angular resolution. At low energies the angular resolution is dominated by the angle between the parent neutrino and the resultant relativistic lepton. At larger energies, the angular resolution is dominated by the reconstruction of the relativistic lepton's track. Uncertainty in the reconstruction of the lepton's track is primarily governed by (i) the timing resolution of the individual OMs and (ii) the positional accuracy of where the OMs are located. 

The ANTARES expected angular resolution becomes better than 0.3$^\circ$ for neutrinos above 10 TeV in energy. To achieve this angular resolution, in-situ timing and positioning calibration are needed. ANTARES timing calibration is primarily achieved through the use of LED beacons deployed throughout the detector and is discussed elsewhere \cite{LED},\cite{LED2}. This paper describes the positioning calibration system of the ANTARES neutrino telescope and reviews its performance during the first year of 12 line operation. 
 
\section{ANTARES positioning calibration system.}
  
Each of the ANTARES 12 detector lines are anchored to a bottom string socket (BSS) on the seabed and pulled taught by the buoyancy of the individual OMs and a top buoy. Due to the flexible nature of these detector lines, even a relatively small water current velocity of 5 cm/s can result in the top storeys being displaced by several meters from the vertical. Therefore, real time positioning of each line is needed. This is achieved through two independent systems: an acoustic positioning system and a lattice of tiltmeters-compasses sensors. The shape of each line is reconstructed by performing a global $\chi^{2}$ fit using information from both of these systems. The relative positions of each individual OM is then calculated from this line fit using the known geometry of each individual storey. 

\begin{figure*}[th]
  \centering
  \includegraphics[width=6.0in]{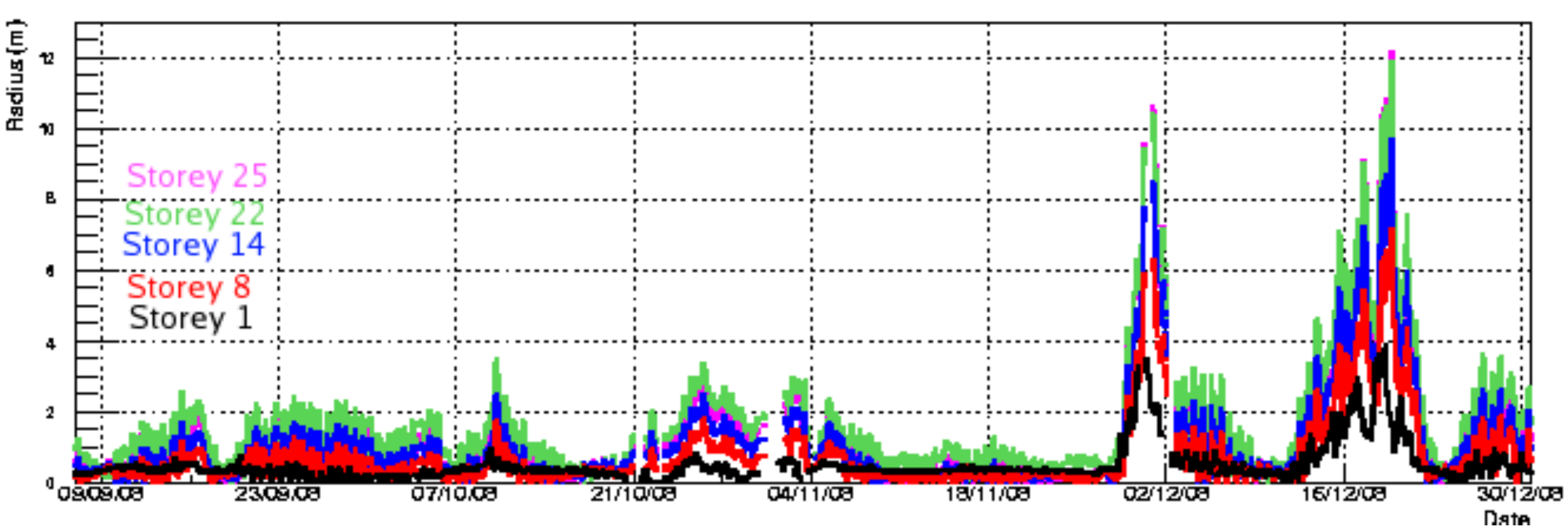}
  \caption{Radial displacement of all hydrophones on Line 11 during a 16 week period.}
  \label{radial}
 \end{figure*}

\subsection{Acoustic Positioning System}

The Acoustic Positioning System (APS) consists of a 3 dimensional array of emitters and receivers exchanging high frequency (40$-$60 kHz) acoustic signals. The use of high frequency signals allows a greater positional accuracy at the expense of a smaller acoustic attenuation length. Nonetheless, the 700$-$1000 meters attenuation length for 40$-$60 kHz signals is sufficient for the ANTARES telescope. The acoustic emitters are located on the BSS of each line, with an additional independent autonomous emitter being located approximately 145 meters from the array. 5 acoustic receivers are non-uniformily distributed along each line, with a higher density towards the top of the line where the line deflection due to the sea current is at its greatest. 

Measurements for the line shape reconstruction are performed every 2 minutes with the completion of an `acoustic run'. At present, each acoustic run consists of 14 cycles; each individual cycle being associated with the emission of an acoustic signal from a single acoustic emitter. For each cycle, the signal transit time to the different receivers is recorded. This information is used, in combination with the sound velocity profile, to calculate the distance between the acoustic emitter and the individual receivers. It should be noted that there are several sound velocity profilers located throughout the ANTARES detector, which measure the sound velocity profile at any given moment. These calculated distances are then used to triangulate the position of each acoustic receiver relative to the acoustic emitters. Figure \ref{radial} shows the radial displacement of all hydrophones on Line 11 during a 16 week period.

\subsection{Tiltmeter-Compass System}

The Tiltmeter-Compass System (TCS) gives the two perpendicular tilt angles, as well as the heading angle, for each storey. This information is given by a single TCM device, containing both tiltmeters and a compass, with a TCM device being installed in the electronics module of each storey. The range of measurement for the tiltmeters is $\pm 20^{\circ}$ on 2 perpendicular axis (roll and pitch), with an accuracy of $0.2^{\circ}$, while the compasses measures $360^{\circ}$ of heading at a resolution of $1^{\circ}$. Data is read out from the TCM device every 2 minutes and is used in conjunction with the APS during the line fit.  

\subsection{Line Shape Fit}

The shape of each detector line is reconstructed based upon a global $\chi^{2}$ fit to a line shape model. This model predicts the mechanical behaviour of the line under the influence of a sea current, taking into consideration the drag and buoyancy coefficients for each individual element along the detector line. At any given point, $i$, along the line, the zenith angle, $\theta_{i}$, from the BSS position, is given by:

   \begin{equation}
    tan(\theta_{i}) = \frac{\sum\limits_{j=i}^{N} F_{j}}{\sum\limits_{j=i}^{N} W_{j}}
    \label{theta1}
   \end{equation}
   
where $F_j$ is the flow resistance, or drag, and $W_j$ is the effective weight of the element in water, given by the weight of the element in air minus the buoyancy of the element. The flow resistance for each individual element on the line is calculated by Eq. \ref{fj}, where $\rho$ and $v$ is the fluid density and velocity respectively, $A_j$ is the cross-sectional area of the element and $C_w$ is the drag coefficient.

   \begin{equation}
    F_j = \frac{1}{2\rho C_{wj}A_jv^2}
    \label{fj}
   \end{equation}
   
Since $tan(\theta_i)$ is also the ratio of the radial displacement of the element from the vertical and the vertical displacement of the element from the base of the line, ($tan(\theta_i) = dr/dz$), then integrating along the line gives us the radial displacement as a function of altitude:

   \begin{equation}
    r(z)= av^2z - bv^2ln[1-cz]
    \label{rz}
   \end{equation} 

where $a$, $b$, and $c$ are known mechanical constants and $v$ is the sea current which is treated as a free parameter during the line fit. A global $\chi^{2}$ fit of this `line shape formula' is then performed on the combined data from the APS and the TCS. By doing so, a complete 3 dimensional positioning of the ANTARES neutrino telescope is performed every 2 minutes. It should be noted that, while a displacement of several meters can occur for the top storeys, the line movements are fairly slow, and therefore completing a full detector positioning every 2 minutes is more than adequate to achieve the 10 cm positional accuracy required for a good angular resolution of the apparatus.

\section{Performance}

Construction of the ANTARES neutrino telescope was completed in May 2008 with the deployment and connection of the last two detector lines. Throughout its construction, the ANTARES neutrino telescope has been operating with an increasing size. During these periods of data taken, the positioning system was extensively studied and proved to be stable in its operation (see \cite{keller}, \cite{vlv}). Here we will concentrate on the results of the positioning systems operation for the first year of data taking in the completed 12 line configuration. 



As mentioned, the global $\chi^{2}$  fit of the line shape formula considers data from both the APS and TCS. However, these two systems are independent, with the data from each system being analysed separately. Firstly, let us consider the contribution from the APS. 

In order to evaluate the stability of the triangulation of the line positions using the APS, the estimates of the distance between each pair of emitter and receiver, are compared before and after triangulation; the spread in this distribution being an indication of the accuracy of the triangulation. For the purpose of illustration, Figure \ref{apscompare} shows that the difference between these two estimated distances is of the order of millimeters and thus, considering acoustic data alone, reconstruction of the individual OMs can easily be achieved with a resolution better than the 10 cm required.  
 
 \begin{figure}[!t]
  \centering
  \includegraphics[width=3.0in]{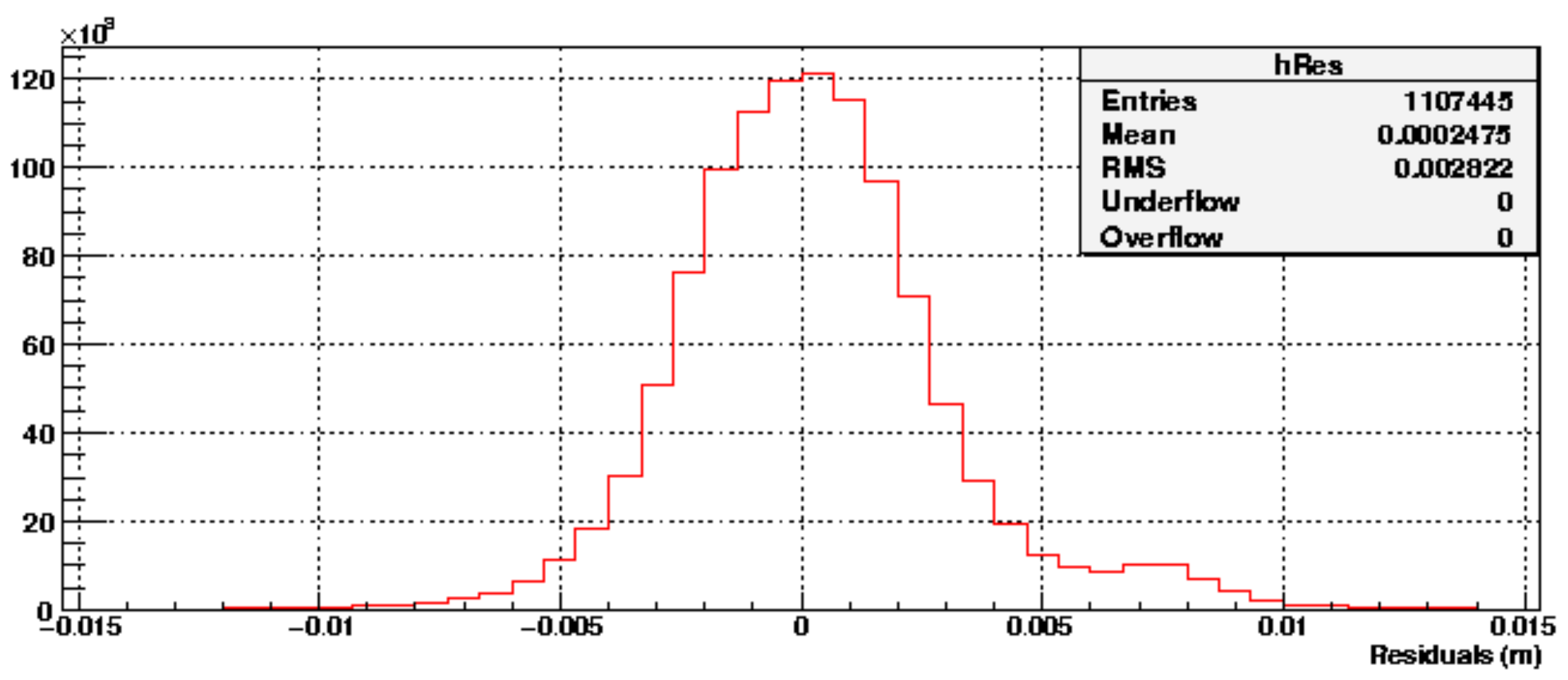}
  \caption{The difference between the distance calculated from the acoustic cycles and the distance re-calculated after triangulation, for Line 12, storey 1.}
  \label{apscompare}
 \end{figure}
 
The strength of the APS can be seen in Figure \ref{acou_tri}, which illustrates the horizontal movements of the line with respect to the BSS position, over a 6 month period. As illustrated in Figure \ref{acou_tri}, the top storeys of the detector line experience the largest amount of displacement due to the water current. Furthermore, this movement is mostly confined to an East-West heading, due to the dominant Ligurian current which flows at the ANTARES site. 
 
 \begin{figure}[!t]
  \centering
  \includegraphics[width=3.0in]{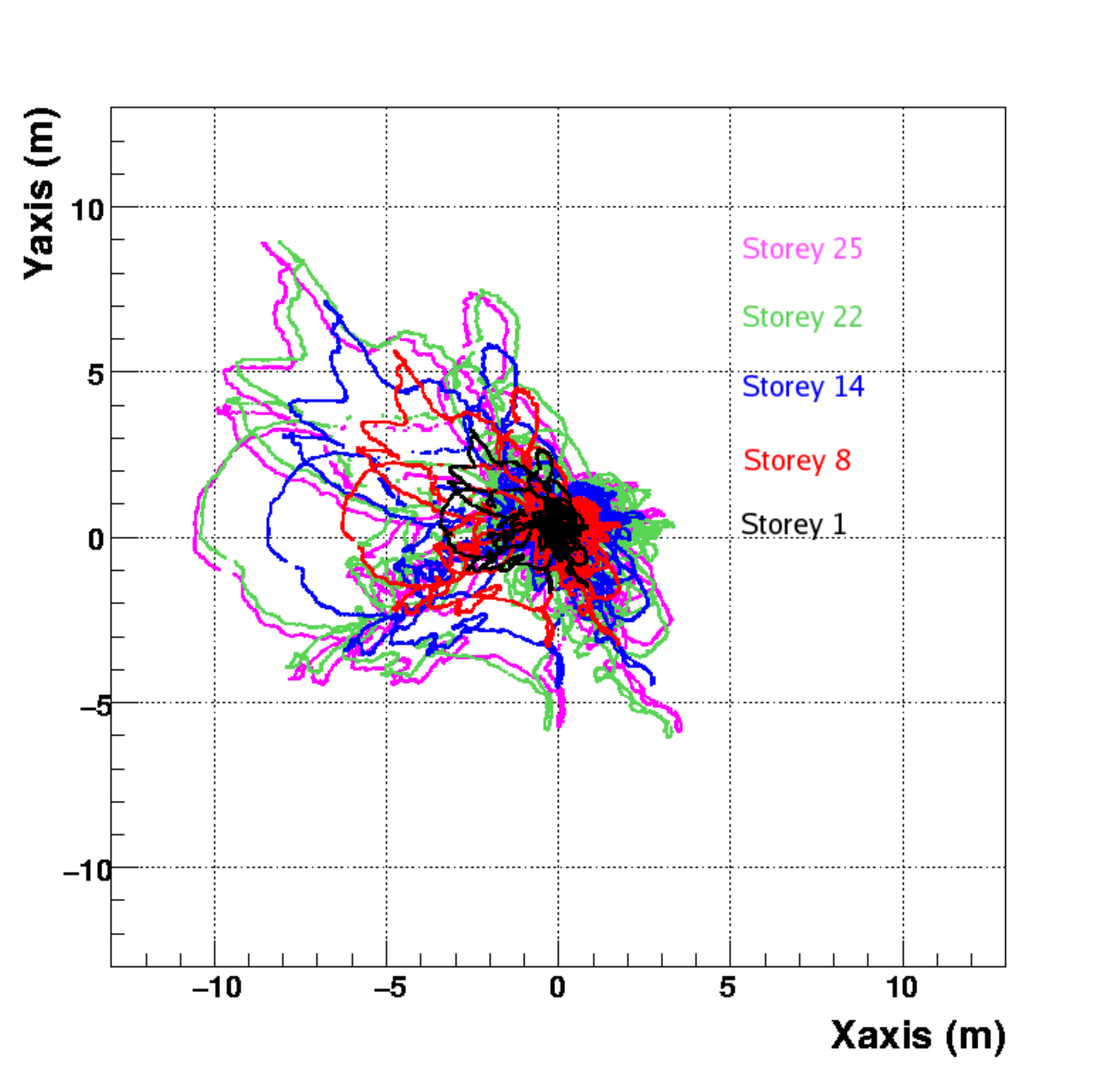}
  \caption{The horizontal movements from BSS, of all hydrophones on Line 11 for a 6 month period. The dominant East-West heading of the line movements is due to the dominant Ligurian current which flows at the ANTARES site.}
  \label{acou_tri}
 \end{figure}
 
To evaluate the compatibility of the data from the APS and TCS, in the overall line shape reconstruction, the line shape reconstructions, as based on the measurements of the APC and TCS alone, are compared. An illustration of such a comparison is shown in Figure \ref{compare}. As with Figure \ref{apscompare}, the spread in the distribution of this comparison is an indication of the compatibility of the two data sets, with a large spread implying a incompatibility of the data from the APS and TCS with regards to the overall line shape fit. As can be seen in Figure \ref{compare} the mean difference between the X position, as calculated using the two different data sets, is less than 1 cm. It should be noted that similar distributions are observed for the differences in the Y and Z positions of the hydrophones. 

 \begin{figure}[!t]
  \centering
  \includegraphics[width=3.0in]{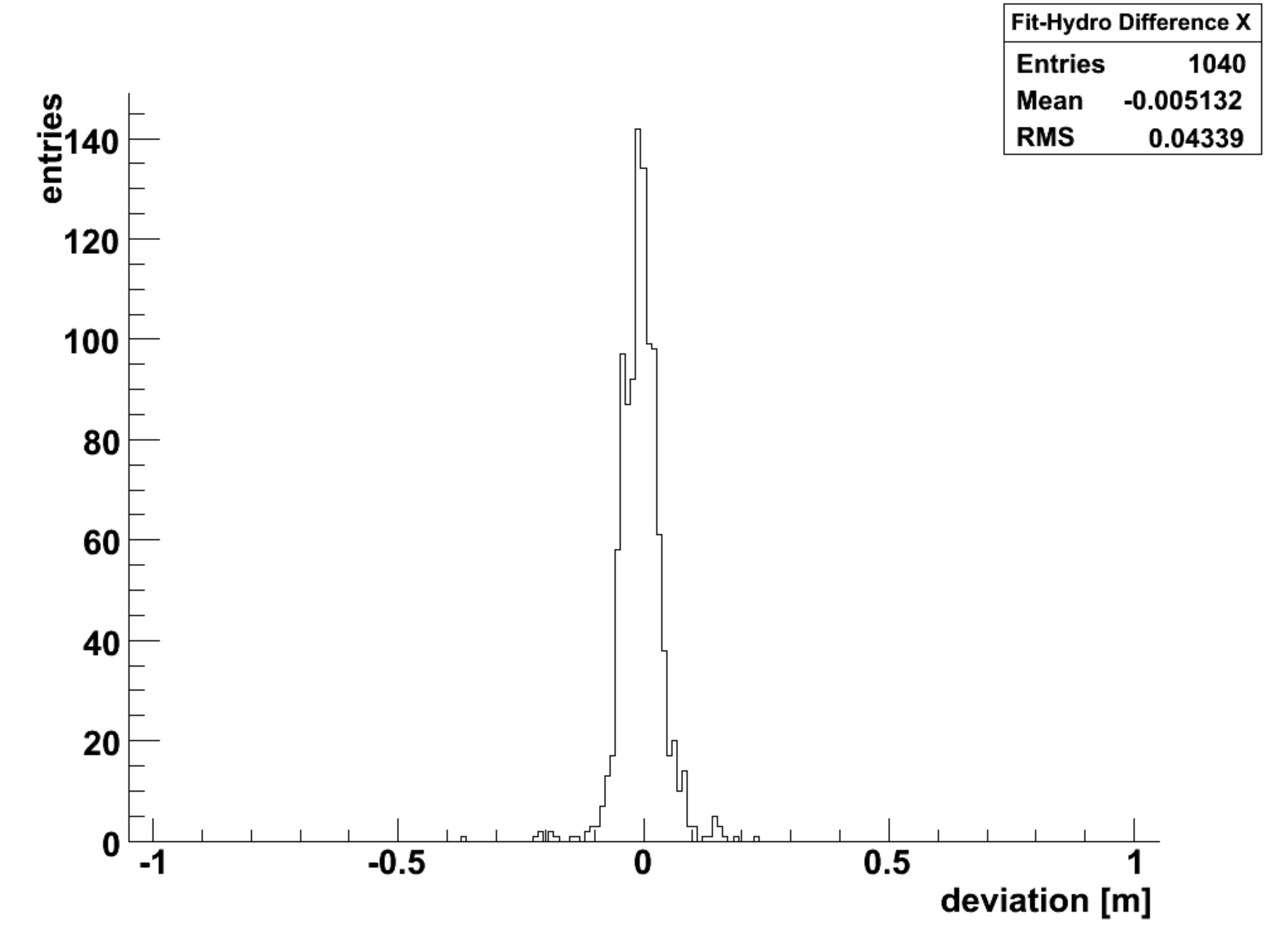}
  \caption{The difference between the X-position of storey 14 on Line 1 calculated by the hydrophone data and the X-position obtained by the line fit.}
  \label{compare}
 \end{figure}

\section{Conclusions}

ANTARES positioning system is stable and accurate in its operation for the complete 12 line configuration. Using the two independent systems of acoustic positioning and tiltmeters-compasses, we are able to reconstruct the 3 dimensional position of the OMs to an accuracy of 10 cm or less. Completing a 3 dimensional positioning of ANTARES every 2 minutes, the quasi-instantaneous knowledge of the line's position allows us to minimise the uncertainty in ANTARES angular resolution due to the fluid nature of our detecting medium. Furthermore, with its stability over a large detector volume, the current ANTARES positioning system is a sound starting point for the design of the positioning system for the future sea-based cubic kilometer sized neutrino telescope of KM3NET \cite{km3v2}.

 {\bf
 Acknowledgments}

Anthony Brown acknowledges the financial support of both the French funding agency `Centre National de la Recherche Scientifique' (CNRS) and the European KM3NET design study grant.

\label{icrc0178:end}

\setcounter{figure}{0}
\setcounter{table}{0}
\setcounter{footnote}{0}
\setcounter{section}{0}
\newpage





\hyphenation{abcdef-ghijklmnoprstuwxyz IEEEtran}

\title{Timing Calibration of the ANTARES Neutrino Telescope}

\author{\IEEEauthorblockN{Juan Pablo G\'omez-Gonz\'alez\IEEEauthorrefmark{1}\\
On behalf of the ANTARES Collaboration} \\
                            
\IEEEauthorblockA{\IEEEauthorrefmark{1} IFIC- Instituto de F\'isica Corpuscular, Edificios de Investigaci\'on de Paterna,\\ 
CSIC - Universitat de Val\`encia, Apdo. de Correos 22085, 46071 Valencia, Spain.}}

\shorttitle{Juan Pablo G\'omez-Gonz\'alez. Timing Calibration in ANTARES.}
\label{icrc0239:begin}
\maketitle

\begin{abstract}
On May 2008 the ANTARES collaboration completed the installation of a neutrino telescope in the Mediterranean Sea. This detector consists of a tridimensional array of almost 900 photomultipliers (PMTs) distributed in 12 lines. These PMTs can collect the Cherenkov light emitted by the muons produced in the interaction of high energy cosmic neutrinos with the matter surrounding the detector. A good timing resolution is crucial in order to infer the neutrino track direction and to make astronomy. 
In this presentation I describe the time calibration systems of the ANTARES detector including some measurements 
(made both at the laboratory and in-situ) which validate the expected performance.  
\end{abstract}

\begin{IEEEkeywords}
Neutrino Telescope, Timing Calibration.
\end{IEEEkeywords}
 
\section{Introduction}
The ANTARES Collaboration has completed the construction of the largest underwater neutrino telescope in the Northern hemisphere [1].  The final detector, located 40 km off the Toulon coast (France) at 2475 m depth, consist of an array of 884 photomultipliers (PMT) distributed along 12 lines separated by $\sim$74 m. Each line has 25 floors (or storeys) holding triplets of 10$"$ PMTs housed in pressure resistant glass spheres called Optical Modules (OM). The clock signal, slow control commands, HV supply, and the readout, arrive at the OMs via the electronic boards housed in the Local Control Module (LCM) container made of titanium. The Bottom String Socket (BBS) anchors each line to the seabed while a buoy at the top gives them vertical support. The whole detector is operated from a control room (shore station) via the electro-optical cable, linked to the Junction Box (JB) which splits the connection to the BBS of the 12 lines (see Fig. 1).
 
The aim of the experiment is to detect high energy cosmic neutrinos and to identify the source of emission by reconstructing the muon tracks. Since the track reconstruction algorithms are based on the arrival time of the Cherenkov photons in the PMTs and the position of the PMTs, both  timing and positioning in-situ  calibration are needed. In ANTARES we have developed several timing  and positioning calibration systems to ensure the expected performance, i.e., an angular resolution  better than 0.3 deg at energies larger than 10 TeV. This paper describes the timing calibration system of  ANTARES while details on the positioning system can be found in [2].\\

 \begin{figure}[!t]
  \centering
  \includegraphics[width=3.2in]{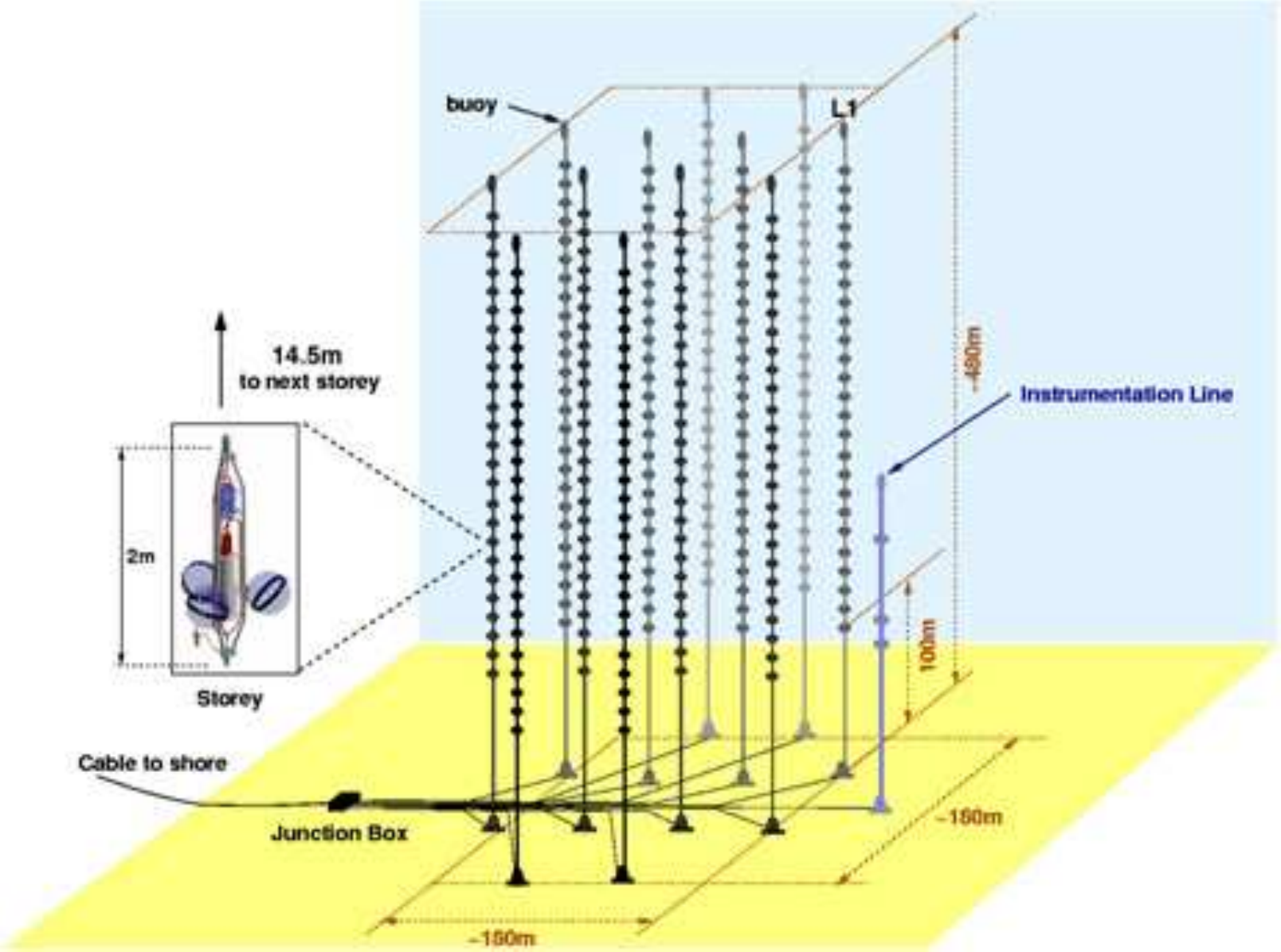}
  \caption{Schematic view of the ANTARES detector layout. The active part are the photomultiplier tubes grouped in triplets in each storey.}
  \label{simp_fig1}
 \end{figure}
 
\section{Timing Calibration}
Concerning to the time calibration we can distinguish between absolute and relative time resolution. The requirement on absolute temporal resolution has the main purpose of correlate detected signals with astrophysical transient events (GRBs, AGN flares, SGR bursts).
In this case the precision needed to measure the time of each event with respect to the universal time (to establish correlations) is of the order of milliseconds. The absolute timestamping  is  made by interfacing the shore station master clock and a card receiving the GPS time ($\sim$100 ns accuracy with respect to UTC). The main uncertainty comes from the electronic path between the electro-optical cable that links the junction box and the shore station.

The relative time resolution refers to the individual time offsets in the photon detection due to differences on the transit times and the front-end electronics among PMTs. The main uncertainties come from the transit time spread (TTS) of the signal in the PMTs ($\sim$1.3 ns) and the optical properties of the sea water (light scattering and chromatic dispersion), which means $\sim$1.5 ns considering 40 m distance.  Therefore, all electronics and calibration system are required to contribute less than 0.5 ns to the relative time resolution in order to guarantee the expected angular resolution. 

\section{Calibration systems}
Several techniques have been developed in ANTARES in order to perform the time calibration [3] of the detector: (A)
The on-shore calibration is performed at laboratory to check the individual time offsets and calibrate the electronics before the deployment in the sea. (B) The clock system enables the measurement of the signal time delay from the clock board located on each OM to the shore station. (C) The internal LED monitors the PMT transit time. (D) The Optical Beacons system, which consist of LED and laser sources of pulsed light with a well know emission time, are used for in-situ timing calibration. (E) The K40 calibration allows (together with the Beacons) the determination of the time differences among PMTs in the same floor, which is used as a check of the time offsets previously determined.
The main features of these method are summarized in this section.

\subsection{On-shore calibration systems}
The on-shore calibration is necessary in order to check that all detector components work properly before the deployment in the sea site. Moreover, they allow us to obtain the first calibration parameters wich will be confirmed and sometimes corrected by the in-situ systems. To this end each integration site of ANTARES has a setup consisting in a laser ($\lambda$ = 532 nm) sending light through an optical fiber to the PMTs and to the OBs placed in a dedicated dark room. Knowing the difference between the time emission of the laser light with respect to the time when the signal is recorded by the PMT we can compute the individual PMT offsets after correcting by the time delay of the fiber light transport. Taking one PMT as reference the relative time offsets can be corrected for the whole detector.

\subsection{Clock system}
The clock system consists of a 20 MHz clock generator on shore synchronized with the GPS, plus a clock distribution system and a signal transceiver on each LCM. The aim is to provide a common signal to all the PMTs. It works essentially by sending optical signals  from shore to each LCM, where they are sent back as soon as they arrive. The corresponding round trip time is twice the time delay due to the cables length for each individual LCM. The uncertainties when computing the time difference between the anchor of a line and one particular floor are of the order of 15 ps, good enough for our purposes in relative time calibration. Considering the time difference between the shore station and the JB we found that the fluctuations in the whole trip delay are around 200 ps, which fulfills the requirements for the absolute time calibration.

\subsection{Internal LED}
An internal blue ($\lambda$ = 472 nm) LED is glued to the back of every PMT in order to illuminate the photocathode from inside. The aim of this device is to monitor the transit time (TT) of the PMT measuring the difference between the time of the LED flash and the time when the flash light is recorded by the PMT. The results obtained with this system indicate that the average TT of the PMTs is stable within 0.5 ns.\\

	\begin{figure}[!t]
   \centering
   \includegraphics[width=3.0in]{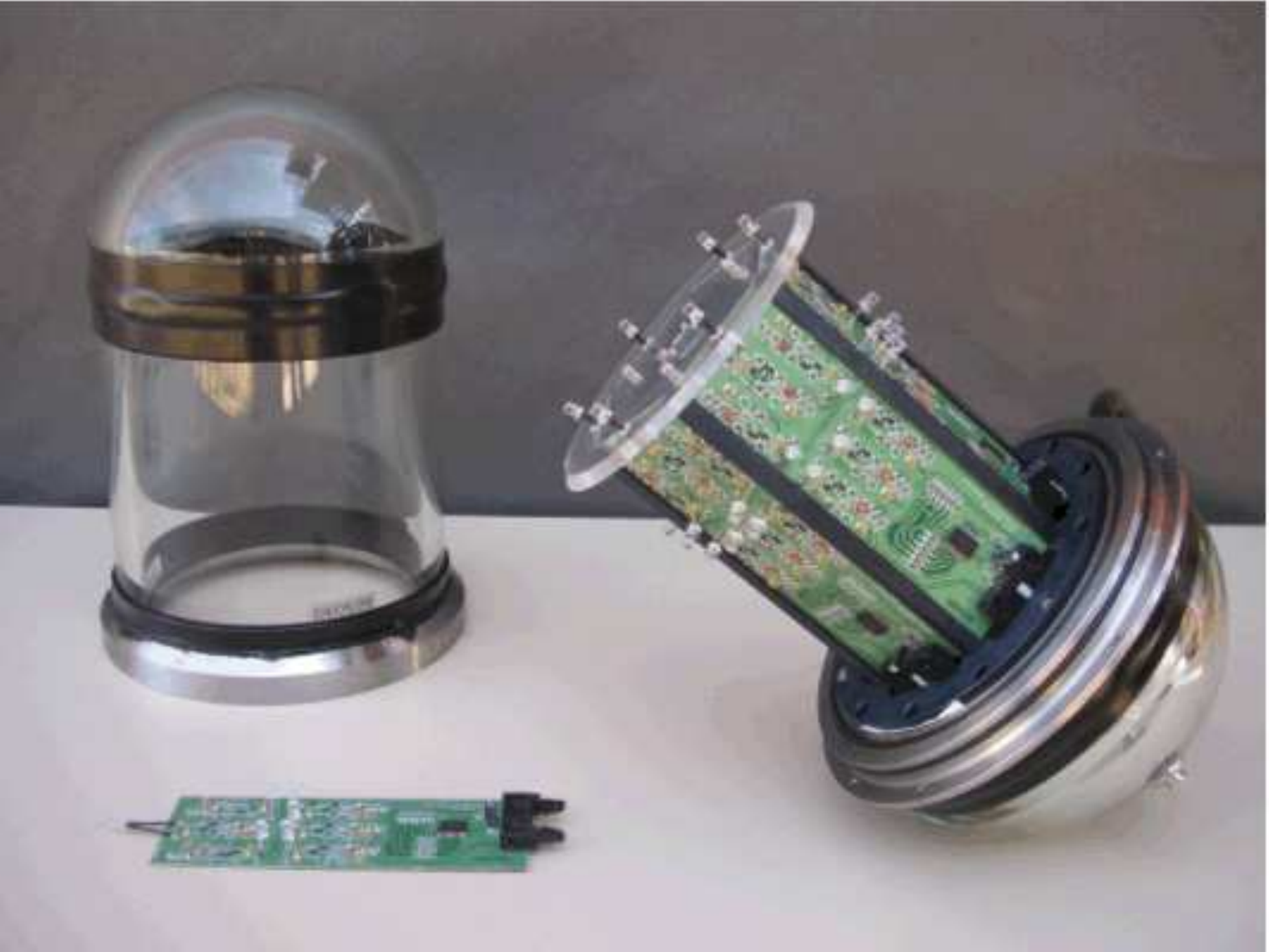}
   \caption{The LED Optical Beacon without the upper cap of it borosilicate container. The light emission time is measured with an  internal PMT. One of the board circuits is also shown separately. }
   \label{simp_fig2}
  \end{figure}

\subsection{Optical Beacons}
The in-situ calibration of the time offsets is performed with a system of external light emitters called Optical Beacons (OBs) [4]. This system comprises two kinds of devices: (1) LED Optical Beacons, and (2) Laser Beacon.  LED Beacons are conceived to calibrate the relative time offsets among OMs within each line. Moreover they can also be used with other purposes such as measuring and monitoring optical water properties, or studying possible PMT efficiency loss. Laser Beacon is being used for interline calibration and cross-checking of the time offsets of the lower floors.\\

\begin{figure}[!t]
  \centering
  \includegraphics[width=2.5in]{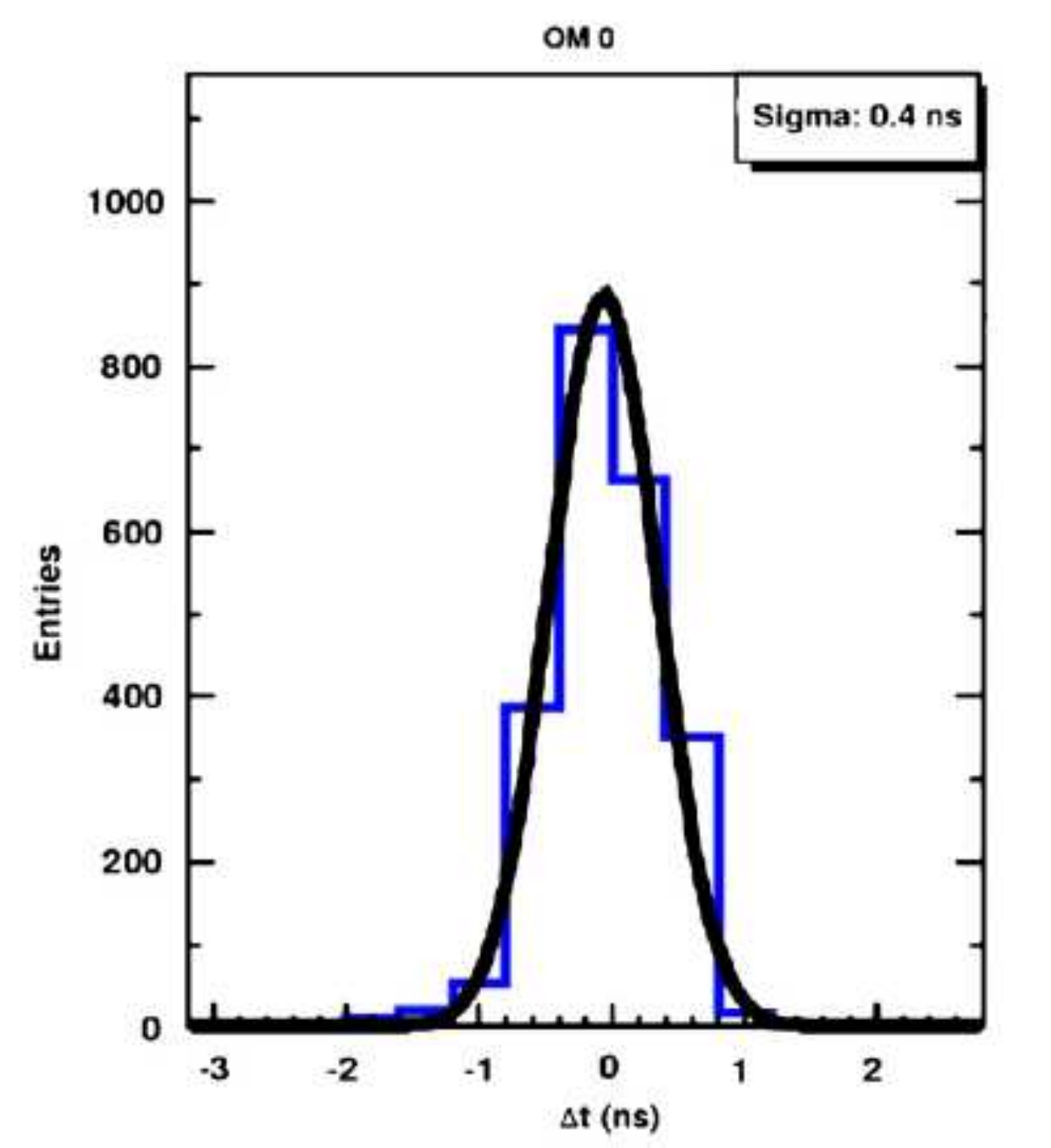}
  \caption{Time difference distribution between a PMT and a LED OB placed 14.5 m away. The standard deviation of 0.4 ns can be understood as an estimation of the ANTARES time resolution from the electronics. }
  \label{simp_fig3}
 \end{figure} 

\begin{enumerate}
	\item The LED Optical Beacon (LOB) is a device made up by 36 individual blue LEDs ($\lambda$ = 472 nm) arranged in groups of six on six vertical boards which are placed side by side forming an hexagonal cylinder (see Fig. 2).  The 36 LEDs in the Beacon can be flashed independently or in combination, and at different intensities (maximum intensity produces 160 pJ per pulse). In ANTARES each standard line contains four LOBs placed every 2, 9, 15 and 21 floors in order to illuminate the OMs located above in the same line.
The time offset computation is based on the time residuals defined as the difference between the time emission of the LED light with respect to the time when the flash is recorded by the OM. In order to know the time emission of the LED light, a small photocathode is placed inside the frame of the LOB. To check the validity of a set of time offset we have calculated the time difference by pairs of OMs in the same storey. The results obtained from the in-situ computation have shown that the changes of these offsets with respect to the measured values in the integration sites are not very large in general, and that only 15\% of the PMTs need a correction greater than 1 ns. The LOBs have also been used to check the ANTARES time resolution. One way to do that is by flashing nearby PMTs. In this case, the contribution of the TTS, which is inversely proportional to the square root of the number photo-electrons, is negligible due to the great amount of light collected by the PMT. The contribution of the photon dispersion is small, since the arrival time distribution is dominated by the first photons.  Finally, the contribution from the small internal OB PMT is also negligible. Therefore, one can safely say that the only important contribution comes from the electronics. We have found that this term takes values of around 0.4 ns (see Fig. 3), which is lower than the 0.5 ns required for the relative time calibration in order to reach the expected angular resolution.\\

 \begin{figure}[!t]
   \centering
   \includegraphics[width=3.0in]{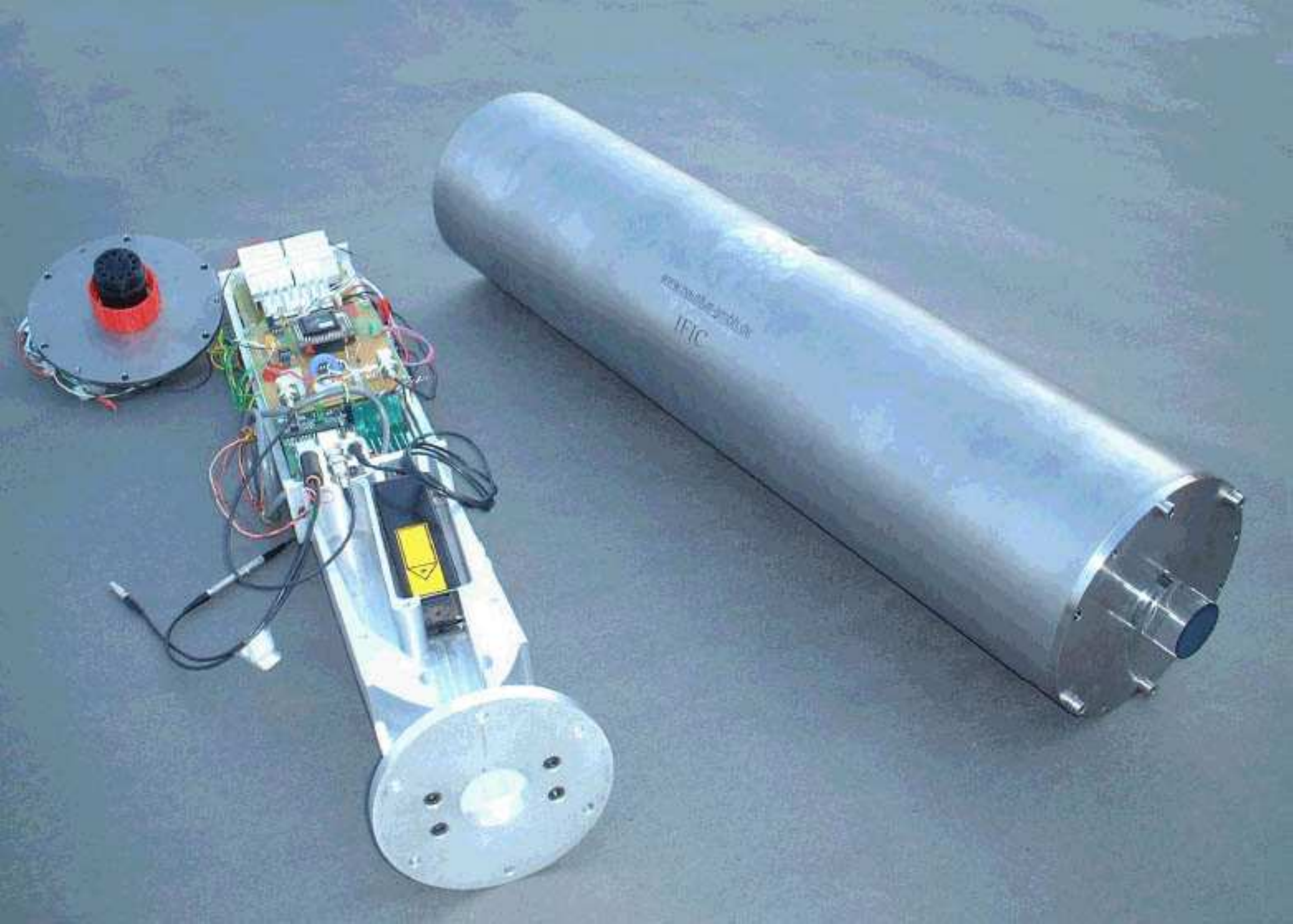}
   \caption{The LED Optical Beacon with it titanium container. The inner mechanics holding the laser head and its associated electronics are visible.}
   \label{simp_fig4}
 \end{figure}

	\begin{figure}[!t]
   \centering
   \includegraphics[width=2.5in]{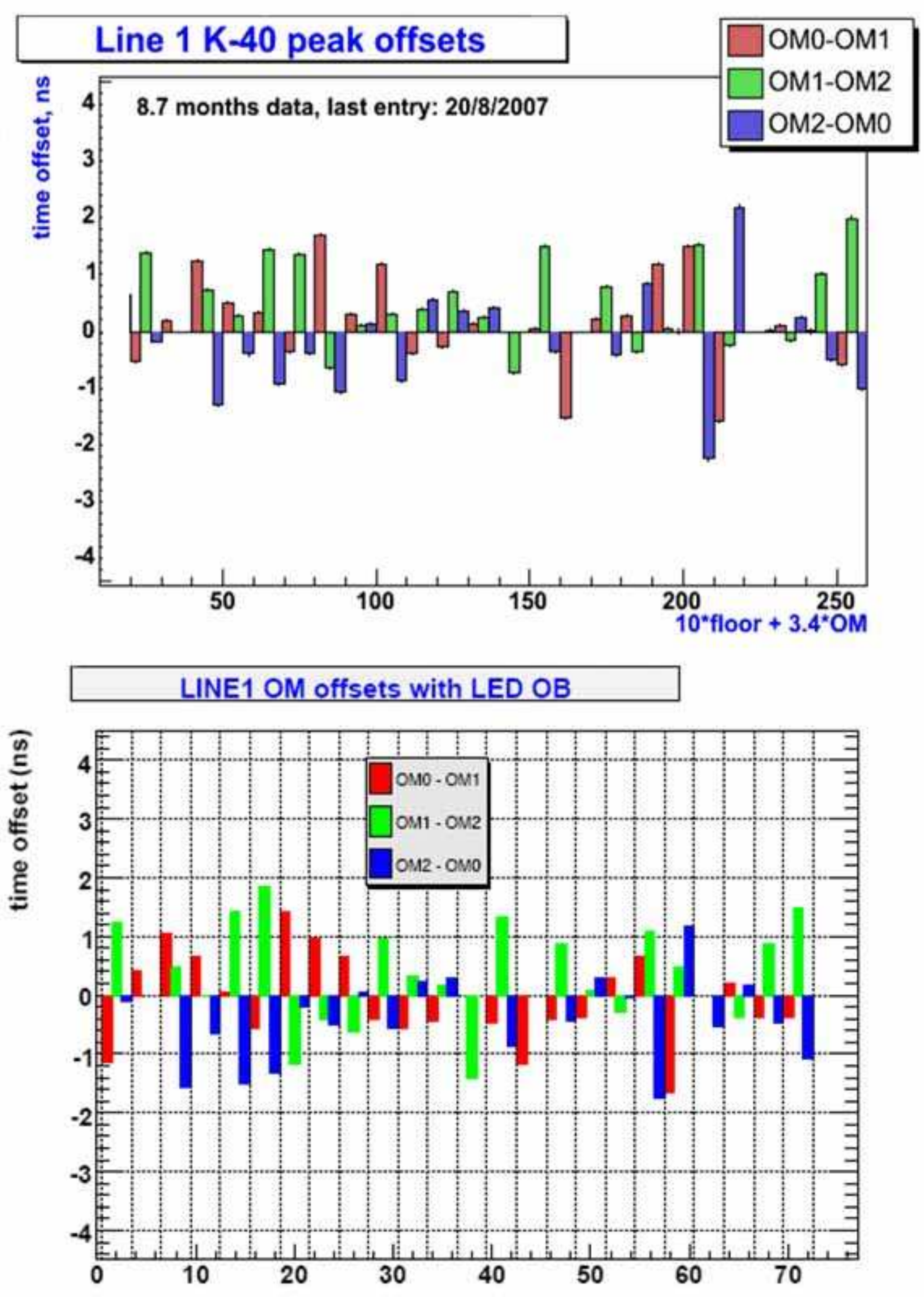}
   \caption{Time difference between three PMTs located in the same floor in line 1 computed with the K40 (up) and with the Optical Beacons (down), where X axis represents number of OM. In the ideal case (perfect calibration) the bars should be essentially zero.}
   \label{simp_fig5}
 	\end{figure}

\item The Laser Beacon system is a device capable of emitting intense ($\sim$ 1 $\mu$J) and short ($<$1 ns) light pulses (see Fig. 4). In ANTARES there are two LOB placed at the bottom of two central lines from where are able to illuminate most of the PMTs of the detector. This system can be used for interline time calibration of the OMs in the lower floors of the detector. In addition it can be used to perform cross-check with positioning systems of the detector by computing the time difference, for a given time period, between the LOB emission and the recording in the OMs of some particular floors on a line. Proceeding in this way we could see that the distribution of the values improves significantly when the real shape of the lines provided by the positioning system is considered (RMS $\sim$0.6 ns), rather than thinking in rigid straight lines (RMS $\sim$2.3 ns).

\end{enumerate} 	

\subsection{Potassium-40}
The potassium-40 is a $\beta$-radioactive isotope naturally present in the sea water. The decay of this substance produces an 1.3 MeV energy electron that will exceeds the Cherenkov threshold inducing a light cone capable of illuminate two OMs in coincidences if the emission occurs in the vicinity of a detector floor.  This event will result on a visible bump over the distribution of the relative time delays between hits in two PMTs of the same floor. Ideally, this coincidence peak must be centered at zero position.  Therefore, the K40 offers a completely independent method to compute the time offsets [5]. Experimental measurements show a small spread of the offsets in good agreement with those values obtained with the Optical Beacons systems (see Fig. 5), and confirm the  accuracy of the timing calibration in ANTARES.
\section{Conclusions} 
In this paper I have reviewed the different methods used in the ANTARES Neutrino Telescope to perform the timing calibration. The results from laboratory an in-situ measurements confirm their validity and the high \newpage level of accuracy reached. With regard to the absolute time resolution the demanded resolution of $\sim$1 ms could be reached by the precision of the clock system. The time resolution was found to be $<$0.5 ns by the Optical Beacons systems. This result was validated by the independently measuments from the K40 coincidences analysis.\\

\label{icrc0239:end}

\setcounter{figure}{0}
\setcounter{table}{0}
\setcounter{footnote}{0}
\setcounter{section}{0}
\newpage




\hyphenation{abcdef-ghijklmnoprstuwxyz IEEEtran}

\title{Measurement of the atmospheric muon flux \\ with the ANTARES detector}

\author{\IEEEauthorblockN{Marco Bazzotti \IEEEauthorrefmark{1} on the behalf of the ANTARES coll.
			                     \\
\IEEEauthorblockA{\IEEEauthorrefmark{1} University of Bologna and INFN Sezione di Bologna.}}}

\shorttitle{M. Bazzotti - Measurement of the atmospheric muon flux with the ANTARES detector}
\label{icrc_bazzotti:begin}
\maketitle

\begin{abstract}
ANTARES is a submarine neutrino telescope deployed in the Mediterranean Sea, at a depth of about 2500 m. It consists of a three-dimensional array of photomultiplier tubes that can detect the Cherenkov light induced by charged particles produced in the interactions of neutrinos with the surrounding medium. Down-going muons produced in atmospheric showers are a physical background to the neutrino detection, and are being studied.
In this paper the measurement of the Depth Intensity Relation (DIR) of atmospheric muon flux is presented. The data collected in June and July 2007,  when the ANTARES detector was in its 5-line configuration, are used in the analysis. The corresponding livetime is $724\,h$. A deconvolution method based on a Bayesian approach was developed, which takes into account detector and reconstruction inefficiencies. Comparison with other experimental results and Monte Carlo expectations are presented and discussed.
\end{abstract}

\begin{IEEEkeywords}
Cherenkov neutrino telescope, underwater muon flux.  
\end{IEEEkeywords}
 
\section{Introduction}
The largest event source in neutrino telescopes is \textit{atmospheric muons}, particles
created mainly by the decay of $\pi$ and $K$ mesons originating in the interaction of cosmic rays with atmospheric nuclei. 
Although ANTARES \cite{antares_web, antares_daq, antares_OM, antares_OM1, pcoyle} "looks downwards" in order to be less sensitive to signals due to downward going atmospheric muons, these represent the most abundant signal due to their high flux.
They can be a background source because they can be occasionally wrongly reconstructed as upward going particles mimicking muons from neutrino interactions.
On the other hand they can be used to understand the detector response and possible systematic effects. In this scenario the knowledge of the underwater $\mu$ intensity is very important for any Cherenkov neutrino telescope and the future projects \cite{icrc_bazzotti:km3net, nemo}. Moreover, it would also provide information on the primary cosmic ray flux and on the interaction models.  \\
 
\section{Data and simulation samples}\label{samples}
The considered data sample is a selection of June and July 2007 data: only runs with good background conditions\footnote{Good run: averaged baseline rate below $120\,kHz$, burst fraction (due to biological activity) below $20\%$, muon trigger rate more than $0.01\,Hz$ and less than $10 \,Hz$.} are considered in the analysis. The livetime of the real data sample corresponds to $724\,h$.

Atmospheric muons were simulated for the 5-line ANTARES detector. The equivalent livetime corresponds to $687.5\, h$. 
The Monte Carlo programs used in the analysis are the following: 
\begin{itemize}
\item Physics generator: MUPAGE program \cite{mupage0,mupage1}. It generates the muon kinematics on the surface of an hypothetical cylinder (\textit{can}) surrounding the detector instrumented volume (see Tab. \ref{tab-mupage}). 
\item Tracking and Cherenkov light generation: KM3 program \cite{km3-prog}. 
\end{itemize}
A dedicated program inserts the background in the simulation taking it from a real run. The Monte Carlo data are then processed by the trigger software, which requires the same trigger conditions as in the real data. 

Physical information is inferred from the triggered events (both Monte Carlo and real data) by a chi-square based track reconstruction program \cite{BBfit}. Each event is reconstructed as a single muon, even if it is a muon bundle.

\begin{table}[!t]
\begin{center} 
\begin{tabular}{|c|c|c|}
\hline
\multicolumn{3}{|c|}
{\textbf{MUPAGE generation parameters}} \\
\hline 
 & Min & Max \\
\hline 
Shower Multiplicity & 1 & 100 \\
\hline 
Shower Energy (TeV) & 0.02 & 500 \\
\hline 
Zenith angle (degrees) & 95 & 180\\
\hline
\end{tabular}
\caption{Generation parameters set in the MUPAGE simulation. \label{tab-mupage}}
\end{center}
\end{table}

\section{Cut selections based on the track reconstrution algorithm}\label{section_cut} 
A chi-square reconstruction strategy \cite{BBfit} is used in the analysis.
Different fits, based on a chi-square minimization approach, are applied by the tracking algorithm:\newline
- a linear rough fit whose extracted parameters are used as starting point for the next refined fits; \newline
- a track fit which looks for a muon track; \newline
- a bright point fit which looks for a point light source as for example electromagnetic showers originated by muon interactions with matter.\newline
Particular interest in the analysis is given to the following quality parameters:
\begin{itemize}
\item \textit{nline}: number of lines containing hits used in the track fit algorithm;
\item \textit{nhit}: number of hits (single or merged) used in the track fit algorithm;
\item $\chi^2_t$: normalized chisquare of the track fit. The smaller is its value the larger is the probability that the reconstructed track belongs to a muon event;
\item $\chi^2_b$: normalized chisquare of the bright point fit. The larger is its value the larger is the probability that the reconstructed track belongs to an electromagnetic shower and not to a muon. 
\end{itemize}
\noindent

Some cuts, based on these quality parameters, are necessary to improve the purity of the data sample.

The hits used in the track reconstruction may belong only to one line (Single Line-SL event) or to more than one line (Multiple Line-ML event). 
The events detected with a single line usually have a well reconstructed zenith angle but undefined azimuth angle (if the line is perfectly vertical, the hit informations are independent from the azimuth angle of the track).
The measurement of the Depth Intensity Relation is not strictly related with the azimuth angle and for this reason single line events are also considered here.

The cuts are performed in sequence on the quality parameters of the reconstruction program. The Efficiency\footnote{The Efficiency is defined as the 
fraction of events surviving the cuts over all the reconstructed events.} and the Purity\footnote{The Purity is defined as the fraction of events with 
a zenith reconstruction error $\Delta\theta \equiv |\theta_t-\theta_m|<5^o$ over the selected events. Applicable only to MC.} of the selected data set after each cut are presented in Tab. \ref{TOT}. 
The first cut is needed in order to remove the events for which the reconstruction algorithm does not converge toward a definite value of the fitting parameters.

\begin{table}[!t]
\begin{center}
\begin{tabular}{|c|c|c|c|}
\hline
            & Efficiency(\%) & Efficiency(\%)  & Purity(\%) \\
            &  Real data     & MC data &   MC data  \\
\hline
No cut      & 100               & 100                     & 62 \\
\hline
Reconstructed track         & 99                & 99                      & 63\\
\hline 
 $nhit>5$*       & 89                & 94                      & 64 \\
\hline
$\chi^2_t<3$    & 51                & 54                      & 77 \\
\hline
$\chi^2_b>2$ & 50                & 53                      & 78 \\
\hline 
\end{tabular}
\caption{Efficiencies and Purities. The cuts are performed in sequence. 
\newline *$nhit>5$ applied only on SL events.}\label{TOT}
\end{center}
\end{table}



\section{Depth Intensity Relation}\label{DIRsec} 
In the present section the quantities are given as functions of the zenith angle $\vartheta$ obtained from the unfolded real events.

One method to derive the DIR is to compute the muon flux $I_{h_{0}}(\cos\vartheta)$ as a function of the zenith angle $\vartheta$ at a fixed vertical depth $h_0$ in the sea. Once this distribution is known, it can be transformed into the DIR using the relation \cite{lipa,gaisserbook}:
\begin{eqnarray}\label{vertical}
I_V(h) & = & I_{h_{0}}(\cos\vartheta)\cdot |\cos\vartheta| \cdot \kappa(\cos\vartheta) \\
&&[s^{-1}\cdot cm^{-2}\cdot sr^{-1}]\nonumber
\end{eqnarray}
where the subscript "V" stands for "\textit{Vertical events}" (i.e. $\cos\vartheta=-1$) and $h=h_0/\cos\vartheta$ represents the \textit{slant depth} (i.e. the distance covered in the sea water by muons, to reach the vertical depth $h_0$ with zenith angle direction $\vartheta$). In the following $h_0=2000\,m$ corresponds to the sea depth of the highest ANTARES Photomultiplier Tubes (PMTs).  
The equation \ref{vertical} is referred to as \textit{"flux verticalization"}: it transforms the muon flux $I_{h_{0}}(\cos\vartheta)$ as a function of the zenith angle $\vartheta$ at the fixed sea depth $h_0$, into the DIR $I_V(h)$, i.e. the flux  of the vertical muons as a function of the sea depth $h$.
The $|\cos\vartheta|$ and the $\kappa(\cos\vartheta)$ factors are needed in order to take into account the zenith angle dependence of the atmospheric muon flux at sea level \footnote{The sea level flux has a zenith angle dependence $\propto 1/(\cos\vartheta \cdot \kappa(\cos\vartheta))$  where the corrective factor is needed to take into consideration the Earth curvature. $\kappa(\cos\vartheta)$ can be considered equal to 1 for $\cos\vartheta < -0.5$.} \cite{lipa,gaisserbook}. 


The measured zenith distribution $N^R_m(\cos\theta_{m})$ is obtained from the track reconstruction of the selected real events.
The deconvolution procedure is a method to derive a true distribution from a measured one. In this work the goal is to transform the measured real data distribution $N^{R}_m(\cos\theta_{m})$ into its parent angular distribution $N^{R}(\cos\vartheta)$, which represents the real events crossing the can surface during the considered experimental time:   
\begin{equation}\label{deco}
N^{R}_m(\cos\theta_m)\longrightarrow (deconvolution) \longrightarrow N^{R}(\cos\vartheta)
\end{equation}
This is possible using the Monte Carlo simulations of the detector response.

Several methods to unfold data exist. The approach that has been chosen consists in an iterative method based on Bayes' theorem proposed in \cite{DAGO}.

\begin{figure}[t!]
\begin{center}
\includegraphics[width=8.4 cm] {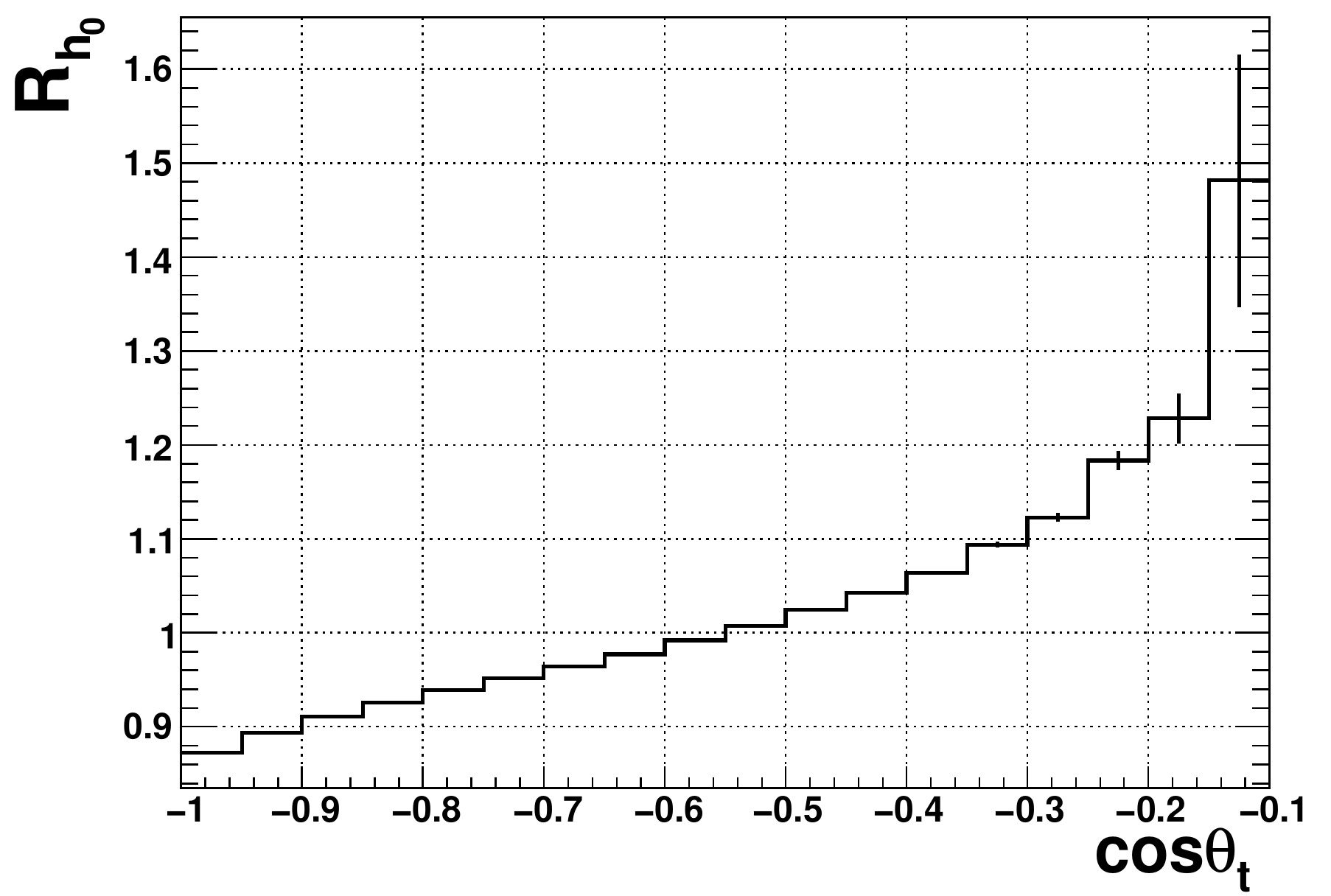}
\caption{\label{densR} $R_{h_{0}}(\cos\theta_t)$. \textbf{From Monte Carlo}. Only statistical errors are shown.}
\end{center}
\end{figure}

\begin{figure}[t!]
\begin{center}
\includegraphics[width=8.4 cm] {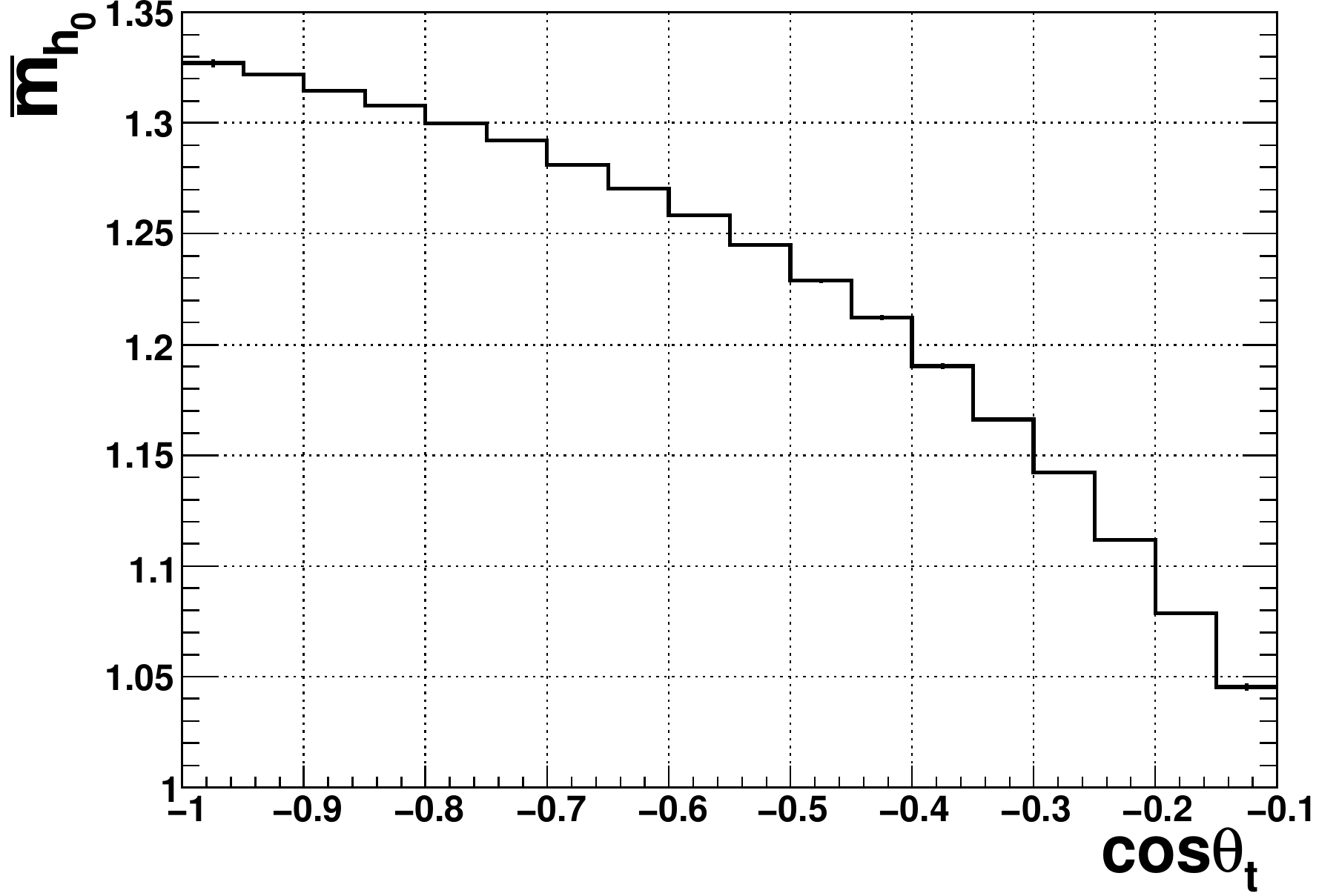}
\caption{\label{molt} Average muon event multiplicity $\overline{m}_{h_{0}}(\cos\theta_t)$ at the fixed sea depth $h_0 = 2000\,m$ for $E_\mu>20\,GeV$. \textbf{From Monte Carlo}. Only statistical errors are shown.}
\end{center}
\end{figure}

Once the distribution $N^{R}(\cos\vartheta)$ is known, it is possible to derive the atmospheric muon flux $I_{h_{0}}(\cos\vartheta)$ at the fixed depth $h_0$.
From relation \ref{vertical} the DIR can be finally written as in the following equation:
\begin{eqnarray}\label{DIRfinale}
 I_V(h) & = & \frac{N^R(\cos\vartheta)\cdot \overline{m}_{h_{0}}(\cos\vartheta)\cdot R_{h_{0}}(\cos\vartheta)}{\Delta T\cdot\Delta\Omega\cdot A_c(\cos\vartheta) } \cdot \\
 & & \cdot |\cos\vartheta|\cdot \kappa(\cos\vartheta) \;\;\;\;\; [s^{-1}\cdot cm^{-2}\cdot sr^{-1}]\nonumber
\end{eqnarray}
where the quantities in the equation are the followings:\newline
- $\Delta T=2.61\cdot10^6\,s$ is the livetime of the considered real data sample. \newline 
- $\Delta\Omega=2\pi\cdot0.05 \, sr$ is the solid angle subtended by two adjacent zenith angle bins as considered in the analysis. \newline
- $A_c(\cos\vartheta)$ is the generation can area as seen under the zenith angle $\vartheta$ (projection of a cylinder):
\begin{equation}\label{a_perp}
A_c(\cos\vartheta) = \pi R_c^2\cdot |\cos\vartheta| + 2R_c\cdot H_c\cdot |\sin\vartheta|
\end{equation}
$R_c=511\,m$ and $H_c=585\,m$ are the radius and the height of the generation can.\newline
- $N^R(\cos\vartheta)$ represents the number of muon events reaching the generation can surface during the considered experimental time $\Delta T$. \newline
- $R_{h_{0}}(\cos\vartheta)$ is a correction factor needed to get the event flux at the sea depth $h_0$ from the event flux averaged on the whole can area. $R_{h_{0}}(\cos\vartheta)$, computed from Monte Carlo ($\vartheta = \theta_t$, where $\theta_t$ is the -"true"- generated zenith angle of the Monte Carlo muon event), is shown in Figure \ref{densR}.\newline
- $m_{h_{0}}(\cos\vartheta)$ is the average muon bundle multiplicity at the fixed sea depth $h_0=2000\,m$. This quantity, computed from Monte Carlo ($\vartheta = \theta_t$), is shown in Figure \ref{molt}. \newline
- $\kappa(\cos\vartheta)\cdot \cos(\vartheta)$ are the correction factors \cite{lipa,gaisserbook} introduced in eq. \ref{vertical}.

\section{Estimation of systematic uncertainties}\label{syst}
During MC simulation several input parameters are required to define the environmental and geometrical characteristics of the detector. Some of
them are considered as sources of systematic uncertainties. 
In \cite{annarita} the effect of the variation of the following quantities on the muon reconstructed track rate is considered.
\begin{itemize}
\item Modifying by $\pm10\%$ the reference values of the sea water absorption length, an almost negligible effect on the shape of
the zenith distributions was noticed, while the absolute flux changed by $+18\%/-20\%$.
\item Decreasing and increasing the effective area of the ANTARES optical module (OM) by $10\%$ with respect to the values used in the analysis, a change of about $\pm20\%$ was observed in the muon flux.
\item The effect of the maximum angle between the OM axis and the Cherenkov photon direction allowing light collection was considered.
Moving the cut-off of this OM angular acceptance 
the rate of reconstructed tracks change of about $+35\%/-30\%$.
\end{itemize}
Summing in quadrature the different contributions, a global systematic effect of about $+45\%/-40\%$ can be
considered as an estimate of the errors produced by uncertainties on environmental and geometrical parameters.

\begin{figure}[!t]
\begin{center}
\includegraphics[width=8.4 cm] {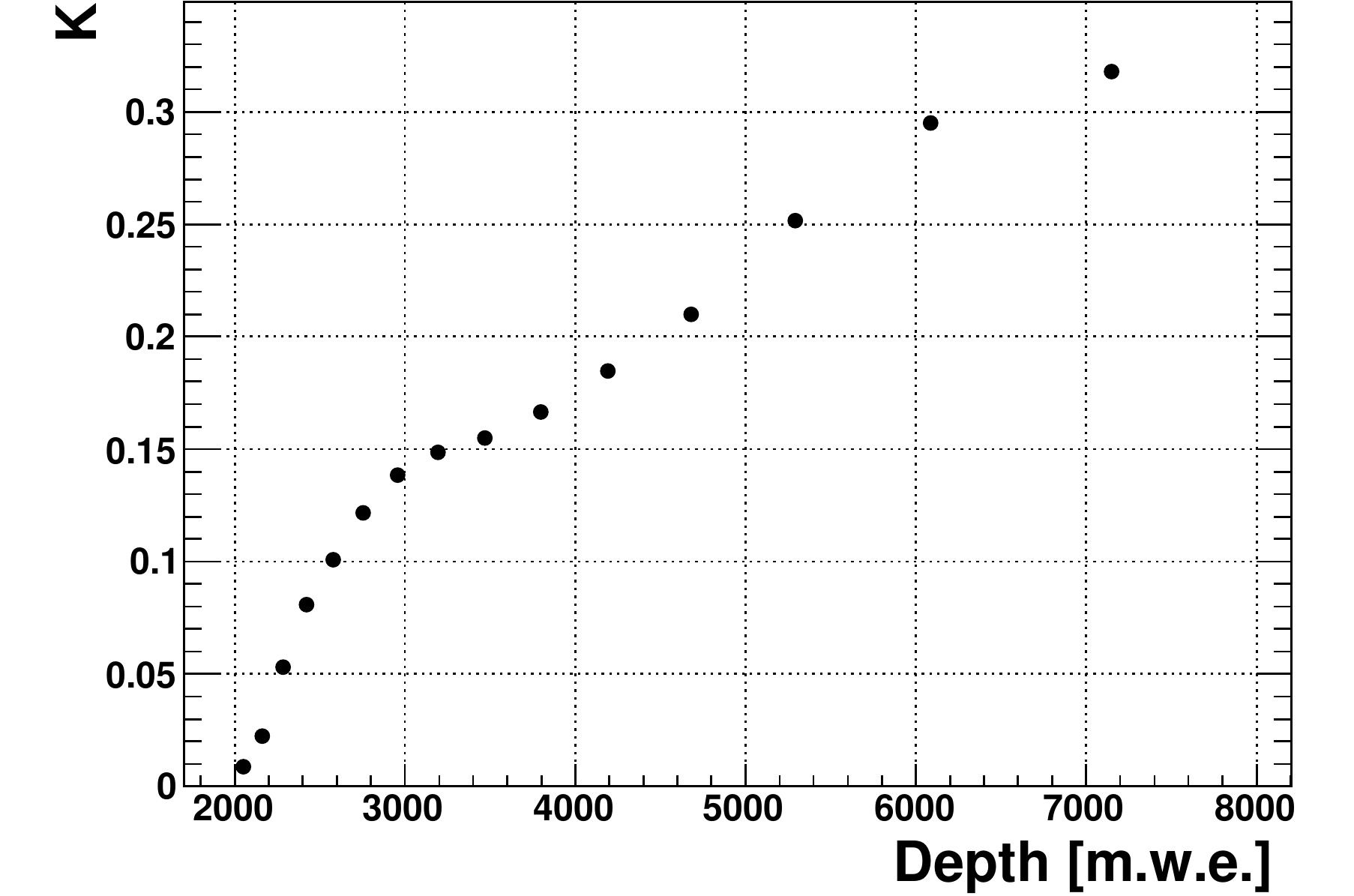}
\caption{\label{diff} Relative difference between the DIRs obtained with the defined quality cuts and without quality cuts (see eq. \ref{kfactor}).}
\end{center}
\end{figure}

The obtained results also depend on the quality cuts performed on the data set. The unfolding algorithm depends on the relative ratio of Monte Carlo and real reconstructed events used in the analysis. As seen in Tab. \ref{TOT} the selections applied to the events have slightly different effects on the two data sets.
In order to take into account this effect the unfolded DIR $I^*_V(h)$ 
has been obtained without considering any cut but the first one which eliminates not reconstructed tracks. The relative difference $K(h)$ between the two final DIRs is defined in the following equation:
\begin{equation}\label{kfactor}
K(h)=\frac{I^*_V(h) - I_V(h)}{I_V(h)}
\end{equation}
The quantity, considered as a systematic uncertainty, depends on the slant depth and is shown in Figure \ref{diff}.
This uncertainty is summed in quadrature with the systematic error estimated above. 

In Figure \ref{flussoNEWerr} the muon flux $I_{h_{0}}(\cos\vartheta)$ ($E_\mu>20\,GeV$) at $2000$ m depth is shown. The Monte Carlo simulation from MUPAGE is also displayed. The result, as the Monte Carlo simulation, takes into account only muons with energy higher than $20\,GeV$ because muons with lower energy are not able to trigger the detector.

In Figure \ref{DIRallNEW} the DIR $I_V(h)$ is shown together with other experimental results.
The Sinegovskaya parameterization ($E_\mu>20\,GeV$) \cite{sine} and the Monte Carlo simulation from MUPAGE are also shown. 
The ANTARES results are in reasonable agreement with the ones of the deep telescope prototypes DUMAND \cite{dumand} and NESTOR \cite{nestor} and with the data of the Baikal \cite{baikal1, baikal2} and AMANDA \cite{icrc_bazzotti:amanda} collaborations.

\begin{figure}[!t]	
\begin{center}
\includegraphics[width=8.4 cm] {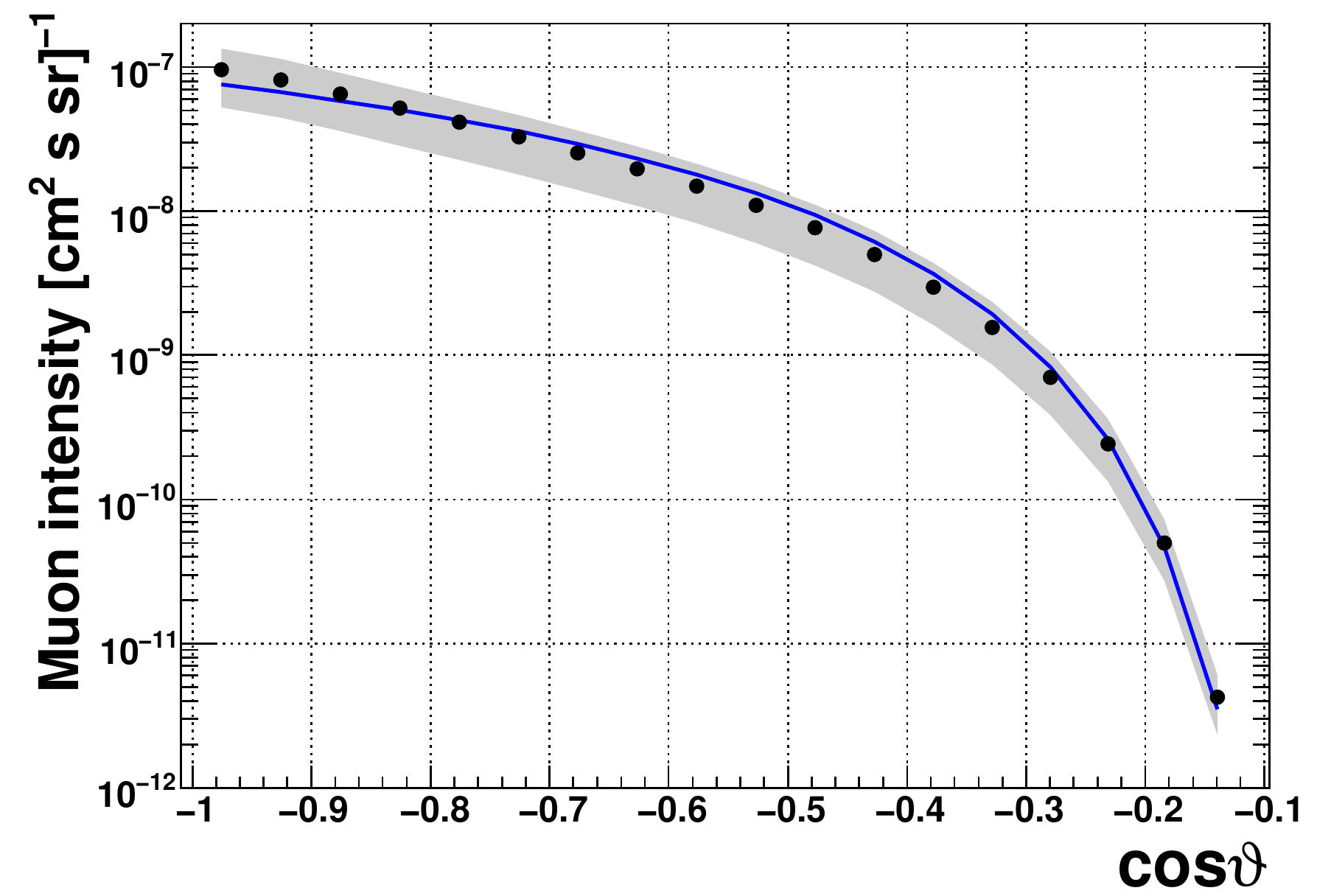}
\caption{\label{flussoNEWerr} \textbf{PRELIMINARY}. Flux of atmospheric muons for $E_\mu>20\,GeV$ at $2000\, m$ of sea depth ($I_{h_{0}}(\cos\vartheta)$) with systematic uncertainties (the statistical uncertainties are negligible). The MUPAGE simulation is superimposed.}
\end{center}
\end{figure}

\section{Conclusions\markboth{Conclusions}{}}
The aim of the presented analysis is the measurement of the muon flux at the depth of ANTARES and the derivation of the vertical component of the atmospheric muon flux as a function of the sea depth. The goal is also to assess the performance of ANTARES in detecting muons. 
The analysis has been performed on a selection of the experimental data of June and July 2007 when the ANTARES detector was in its 5-line configuration. 

Several quality cuts have been applied on the reconstructed events in order to improve their purity, in particular concerning the zenith angle reconstruction.

An unfolding algorithm, based on an iterative method, has been applied on the selected experimental data in order to retrieve back the flux of atmospheric muons with $E_\mu>20\,GeV$ at the fixed sea depth $h_0=2000\,m$. 
The experimental DIR was finally obtained. 

The results are in good agreement, within the uncertainties, with the experimental fluxes obtained by other Cherenkov telescopes.


\begin{figure}[!t]
\begin{center}
\includegraphics[width=8. cm] {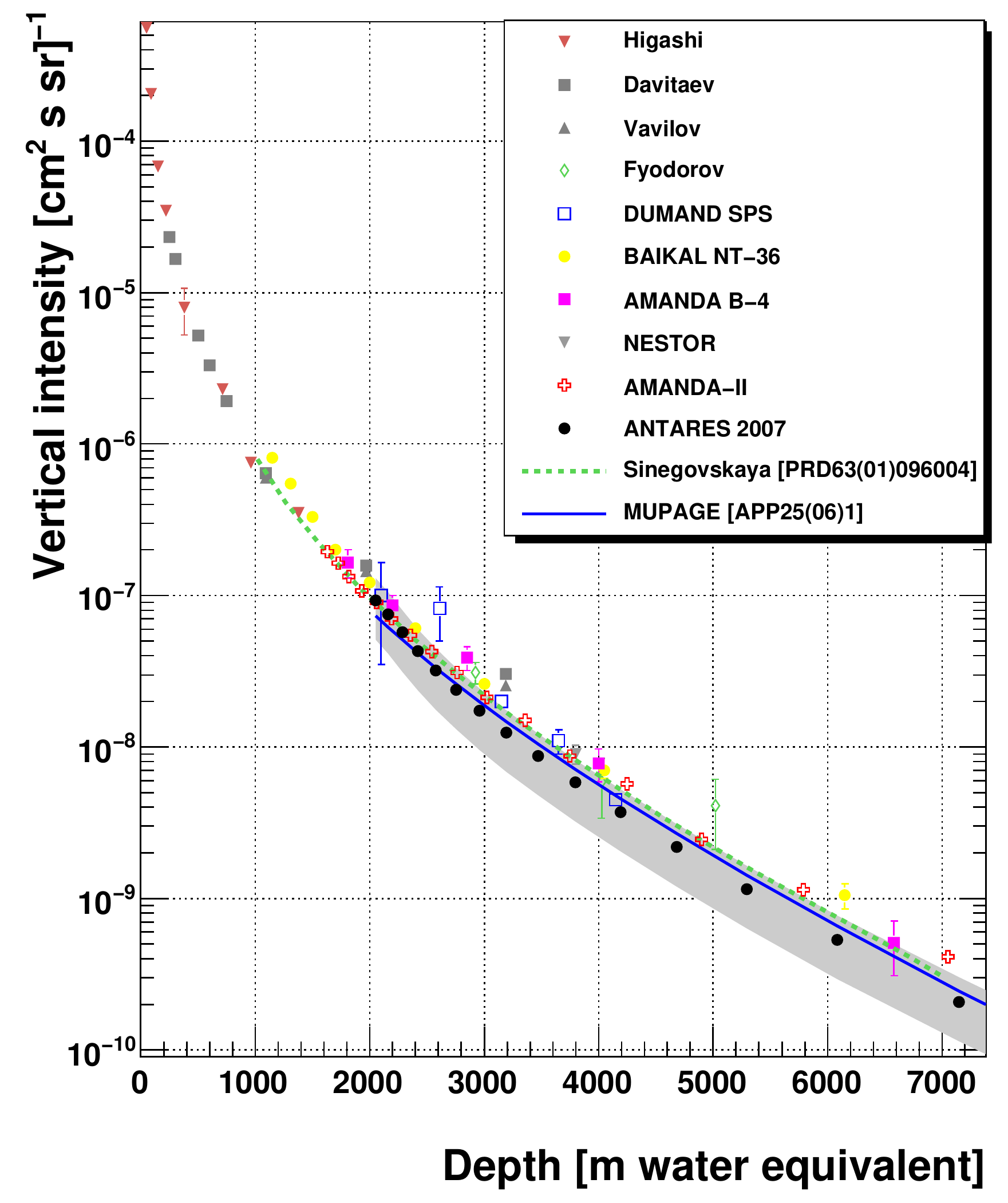}
\caption{\label{DIRallNEW} \textbf{PRELIMINARY}. Depth Intensity Relation of atmospheric muons for $E_\mu>20\,GeV$, with systematic uncertainties (the statistical uncertainties are negligible). The DIR obtained from other underwater measurements are also shown: Higashi \cite{d-higa}, Davitaev \cite{d-davitaev}, Vavilov \cite{d-vavilov}, Fyodorov \cite{d-fyodorov}, DUMAND-SPS \cite{dumand}, BAIKAL NT-36 \cite{baikal2}, NESTOR \cite{d-nestor}, AMANDA B-4 \cite{d-amandab4}, AMANDA-II \cite{d-amandaII}. The Sinegovskaya parameterization ($E_\mu>20\,GeV$) \cite{sine} and the MUPAGE simulation are superimposed.}
\end{center}
\end{figure}

\label{icrc_bazzotti:end}

\setcounter{figure}{0}
\setcounter{table}{0}
 \setcounter{footnote}{0}
\setcounter{section}{0}
\newpage





\hyphenation{abcdef-ghijklmnoprstuwxyz IEEEtran}

\title{Reconstruction of Atmospheric Neutrinos in Antares}

\author{\IEEEauthorblockN{Aart Heijboer\IEEEauthorrefmark{1}, for the Antares Collaboration}
\\
\IEEEauthorblockA{\IEEEauthorrefmark{1}Nikhef, Amsterdam.}}

\shorttitle{Aart Heijboer, Neutrino Reconstruction in Antares}
\maketitle
\label{icrc1045:begin}

\begin{abstract}

In May 2008, the Antares neutrino telescope was
completed at 2.5 km depth in the Mediterranean Sea;
data taking has been going on since. A prerequisite
for neutrino astronomy is an accurate reconstruction
of the neutrino events, as well as a detailed
understanding of the atmospheric muon and neutrino
backgrounds. Several methods have been developed to
confront the challenges of muon reconstruction in
the sea water environment, which are posed by e.g.
backgrounds due to radioactivity and bioluminescence.
I will discuss the techniques that allowed Antares
to confidently identify its first neutrino events, as
well as recent results on the measurement of atmospheric
neutrinos. 
\end{abstract}

\begin{IEEEkeywords}
 neutrino astronomy reconstruction
\end{IEEEkeywords}

\section{Introduction}

 The Antares collaboration is currently operating
 a 12-line neutrino telescope in the 
 Mediterranean Sea, at a depth of about 2500 m, 40 km from 
 the shore of southern France (see \cite{pas} for 
 a full status report). The full detector
 comprises 12 lines, each fitting 75 Optical
 Modules (OMs), arranged in 25 triplets, also called
 `floors', which are placed 14.5 metres apart along 
 the line. The OMs house a 10 inch photomultiplier
 tube, which is oriented downward at a 45 degree 
 angle to optimise the detection efficiency for 
 neutrino-induced, upgoing muons. The positions
 of the OMs are measured using the systems described
 in \cite{anthony}.

 The deployment of the first detector line took place
 at the beginning of 2006. This line was used to measure 
 the flux of atmospheric muons using a specialised
 reconstruction algorithm \cite{line1paper}. 
 By January 2007, five such detector lines were 
 operational, allowing the application of the methods 
 developed for the 3d reconstruction of muon 
 trajectories. This led to the identification of the 
 first neutrino events; see Fig. \ref{fig_evt}.
 The 12th line was deployed in May 2008, which 
 completed the construction of the detector. 
 
\section{Muon Reconstruction}

 The challenge of measuring muon neutrinos consists 
 of fitting the trajectory of the muon to the arrival
 times, and -to a lesser extent- to the amplitudes
 the Cherenkov light detected by the OMs.
 
 For a given muon position (at an arbitrarily chosen
 time $t^0$) and direction, the 
 expected arrival time $t^{\rm exp}$ of 
 the Cherenkov photons follows from the geometric
 orientation of the OM with respect to the muon path:
\begin{equation}
t^{\rm exp} = t^0 + {1 \over c } \bigl( l - {k \over {\tan \theta_C}} \bigr) +
                    {1 \over v } \bigl(     {k \over {\sin \theta_C}} \bigr),
\end{equation}
 where the distances $l$ and $k$ are defined in Fig. \ref{fig_schema},
 $\theta_C$ is the Cherenkov angle ($\sim 42^o$ in water) and $v$ is
 the group velocity of light in the water.
 The difference between
 $t^{\rm exp}$ and the measured arrival time of 
 the photon (i.e. the 'hit time) defines the time 
 residual: $r \equiv t^{\rm measured} - t^{\rm exp}$.

 Photons that scatter in the water and photons
 emitted by secondary particles 
 (e.g. electromagnetic showers created 
 along the muon trajectory) will arrive at the OM
 later than $t^{\rm exp}$, leading to positive 
 residuals. The residual distribution obtained 
 from in data is shown in Fig. \ref{fig_res}. 
 The tail on right due to late photons is 
 clearly visible. 

 The reconstruction algorithms attempt to find muon
 track parameters (i.e. three numbers for the position and two 
 for the direction) for which the residuals are small. This
 can be done by minimising a quantity like $\chi^2 = \sum_{i=1}^{N_{\rm hits}} r_i^2$
 or by maximising the likelihood function $\log L = \sum_{i=1}^{N_{\rm hits}} \log P(r_i)$. 
 Here, $P(r)$ is the probability density function (PDF) for the residuals.
 While relatively simple, the equation to compute the residuals
 is non-linear in the track parameters. As a consequence, iterative 
 methods are required for minimising a $\chi^2$-like 
 variable or maximising the likelihood.

 A complicating factor in the reconstruction process 
 is the presence of
 background hits, caused by decaying $K^{40}$ in the sea 
 water and by aquatic life (bioluminescence). If not
 accounted for in the muon reconstruction, the background
 light degrades both the angular resolution and the
 reconstruction efficiency. Antares has developed several
 strategies to deal with this problem. Two of them
 will be discussed in the following sections. We refer
 to these two as the 'full likelihood' and 'online' algorithms.
 They are currently widely used for data analysis. 

\begin{figure}[!ht]
  \centering
  \includegraphics[width=3.0 in]{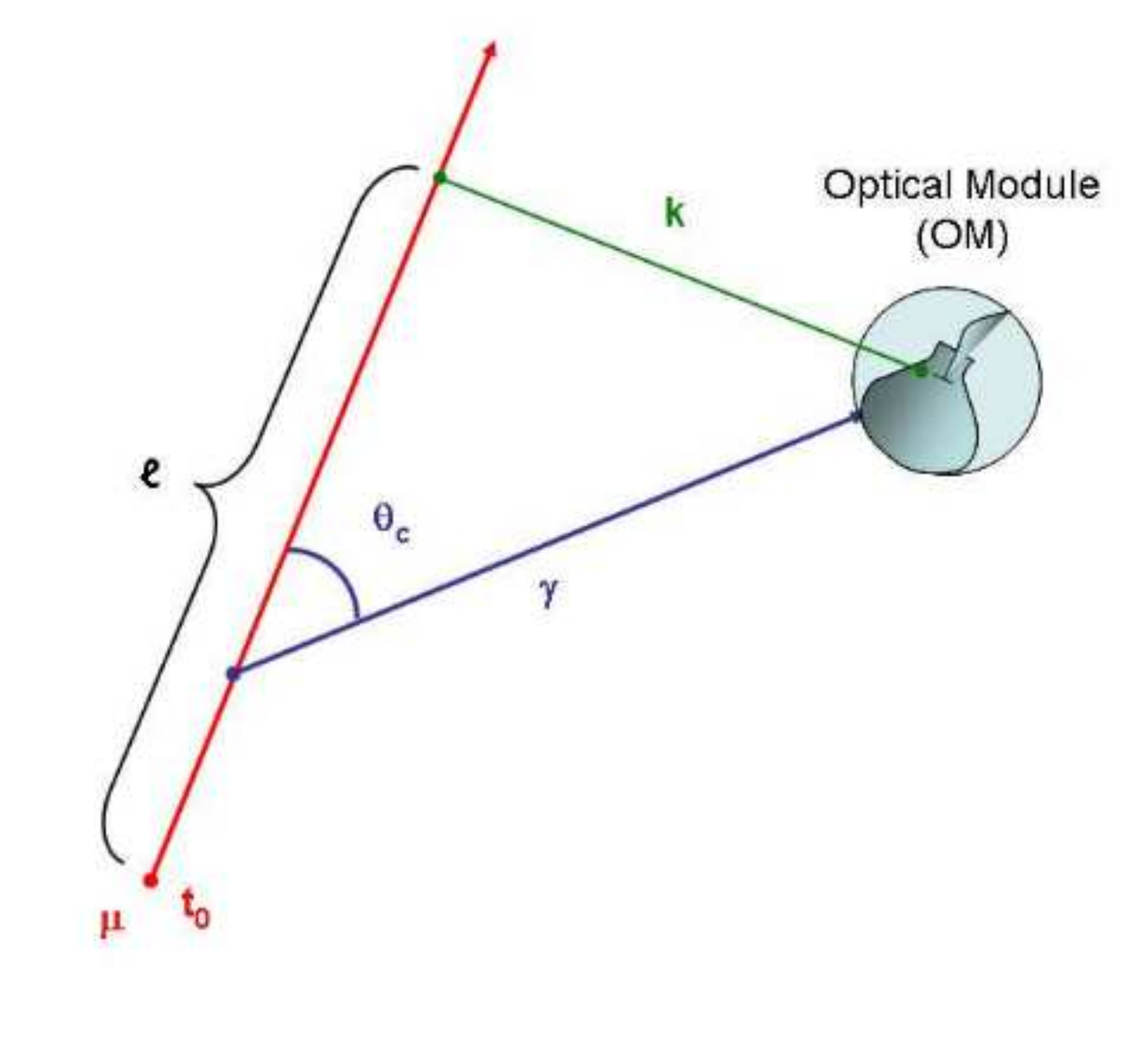}
  \caption{Schematic representation of the relation between the
           muon trajectory and the OM. The line labelled $\gamma$
           indicates the path travelled by a Cherenkov photon from 
           the muon
           to the OM.}
  \label{fig_schema}
\end{figure}

\begin{figure}[!t]
  \centering
  \includegraphics[width=3.5in]{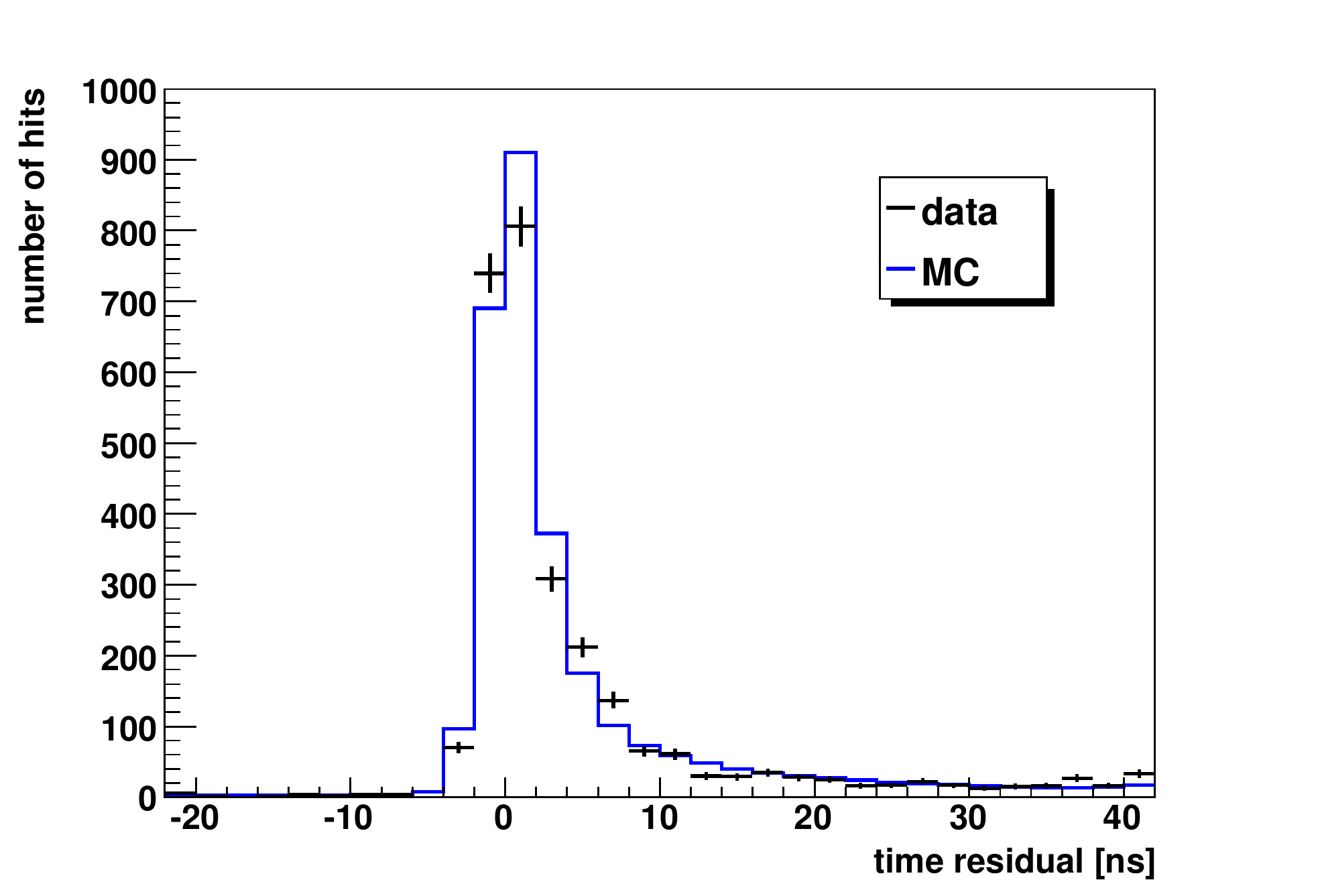}
  \caption{Time residuals of the hits with respect to the result
           of the full likelihood fit for selected, upgoing events
           (neutrino candidates). The data were taken in 2007 
            with 5 detector lines. The peak shows the (order 1 ns) intrinsic
            timing resolution of the OMs. The tail is due to light
            from secondary particles and to scattered photons.}
  \label{fig_res}
\end{figure}

\section{Full Likelihood Fit}

 The first algorithm was developed 
 several years before deployment of the detector and
 is described in detail in \cite{aartstrat}.
 This method is based on a likelihood fit that uses a
 detailed parametrisation, derived from simulation,
 for the PDF of the arrival time of the hits $P(r)$, which takes
 into account the probability of hits arriving
 late due to Cherenkov emission by secondary particles
 or light scattering. Moreover, the probability of a hit
 being due to background is accounted for as a function 
 of the hit amplitude and the orientation of the OM with 
 respect to the muon track. 
 It was found that the likelihood function has many local 
 maxima and that the likelihood fit is only successful if the
 maximisation procedure is started with 
 track parameters that are already a good approximation
 to the optimal solution. To obtain this approximate
 solution, the full likelihood fit is preceded by 
 a series of `prefit' algorithms of increasing sophistication.
 An important ingredient in the prefit stage is the use 
 of a so-called `M-estimator', which is a variant of a
 $\chi^2$-fit in which hits with large residuals are
 given less importance compared to a regular $\chi^2$. 
 This is crucial, as it allows the fit to converge
 to a solution relatively close (typically a few degrees) to 
 the true muon parameters, while being robust against the 
 presence of background hits at large residuals. 
 The M-estimate is followed by
 two different versions of the likelihood fit, the last
 of which fully accounts for the presence of background
 hits. The procedure is started at nine different starting 
 points to increase the probability of finding the global 
 minimum. To mitigate the associated loss in speed, 
 analytical expressions for the gradient of the 
 likelihood function are used in the min/maximisation 
 processes.

 The value of the final log-likelihood per degree
 of freedom that is obtained from the final fit
 is used as a measure of the goodness of fit. 
 This is combined with information on the number
 of times the repeated procedure converged to the
 same result, $N_{\rm comp}$ to provide a value 
 $\Lambda=\log(L)/N_{\rm dof} - 0.1(N_{\rm comp} -1 )$.
 The variable $\Lambda$ can be used to reject badly 
 reconstructed events; in particular atmospheric 
 muons that are reconstructed as upward-going. An
 example of the use of this algorithm for reconstructing
 and selecting neutrinos for a point source search is
 given in \cite{point}.

\subsection{Results}

 The full likelihood fit is optimised for the high energy
 neutrinos that are expected from astrophysical sources 
 ($E_{\nu}^{-2}$ spectrum, yielding muons in the 
 multi-TeV range). Simulations indicate that, 
 with this algorithm, Antares reaches an angular 
 resolution (defined as the median angle between the 
 true and reconstructed muon) smaller than 0.3 
 degrees for neutrino energies above 10 TeV. Below 
 this energy the scattering angle of the neutrino 
 interaction dominates.

 As the hit residuals are the main ingredient driving the
 angular resolution, the agreement observed between data
 and simulation in the residual distribution (see Fig. 2) 
 is good evidence that the algorithm is performing as 
 expected.

\begin{figure}[!ht]
  \centering
  \includegraphics[width=3.0in]{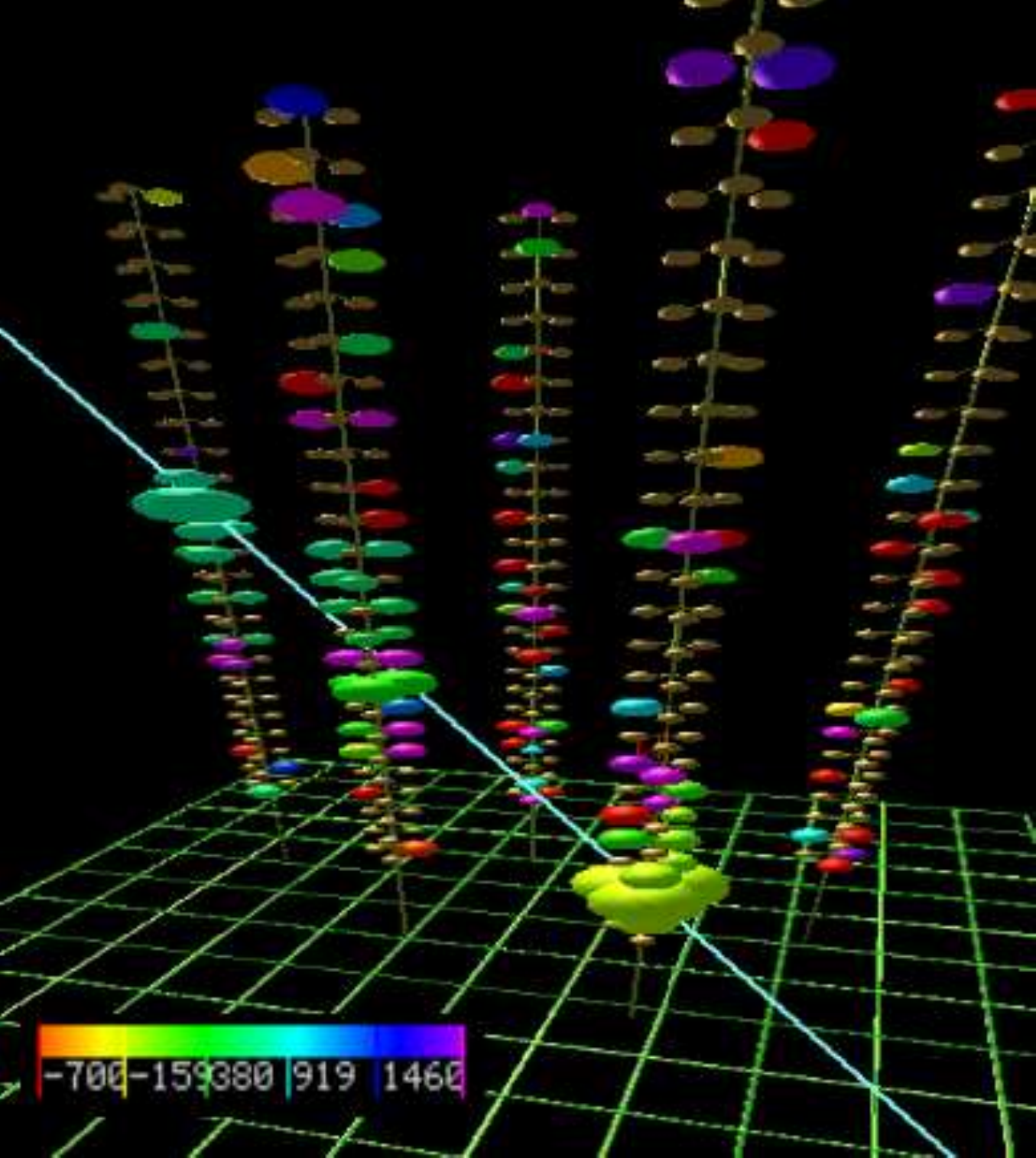}
  \caption{Event display of one of the first neutrino 
           candidates detected with Antares. This event was
           found and reconstructed using the full 
           likelihood fit. The colour of the hits indicates
           the time (according to the legend in the lower left
           corner), while the size indicates the collected charge.}
  \label{fig_evt}
\end{figure}

\section{Online Algorithm}

 The second approach to reconstructing tracks in the
 presence of background hits was developed 
 during the commissioning of the detector for the
 online event display that is used for monitoring the detector. We
 therefore refer to this fit as the `online algorithm', although
 it is now also frequently used for offline data analysis.

 Whereas the full likelihood fit described above attempts to incorporate
 all the background hits in the fit, the philosophy of the online
 algorithm is to select a very high purity sample
 of signal hits. This is followed by a fit of the muon 
 trajectory using a model that can be relatively simple. In this
 case a $\chi^2$-like fit is performed.

 The algorithm merges hits on the three OMs in a floor and 
 uses the centre of the triplets in the fit. While degrading
 the timing precision (and therefore, in theory the angular
 resolution) somewhat, this does make the algorithm independent
 of measurements of the azimuthal orientation of the triplets, which
 vary due to sea currents and which normally 
 need to be measured using compasses located on the floors.

\subsection{Hit Selection}

 The selection of hits starts by identifying floors
 that collected an amplitude (i.e. charge collected on the PMTs) 
 corresponding  to more than  2.5 photo-electrons (pe), 
 or 1.5 pe in case multiple hits 
 were detected within a time window 
 of 20 ns. Such configurations are rarely produced by optical
 backgrounds, but occur in most of the signal events.
 Doublets of such floors are identified, allowing for no
 more than one empty floor in between and requiring that
 the time difference of the hits is smaller 
 than $80 {\rm ns}$ per floor of separation. Only the 
 detector lines which at least one such doublet are 
 used in the fit. 
 Clusters of hits are then formed by iteratively 
 complementing the 
 doublets with adjacent hits that
 are close in time (within $80 {\rm ns}$ per floor) and 
 distance (no more than one empty floor) to 
 the already identified cluster.

\subsection{Fit}

 The fit is performed using a score function that is derived
 from the expression for the $\chi^2$ of the hit residuals, with
 an added term that promotes solutions where hits with large
 amplitudes pass the OM at close range. The minimised function is:
\begin{equation}
Q = \sum_i \bigr[ {1 \over {\sigma^2}} r_i^2 + \alpha q_i d_i \bigl],
\end{equation}
 where the sum is over all selected hits and where 
 $r_i$ is the residual of hit $i$, $q_i$ is the amplitude associated
 with that hit, and $d_i$ is
 the distance from the hypothesised track to the OM. The constants 
 $\sigma$ and $\alpha$ were set to $10 \rm ~ns$ and $50 ~\rm m^{-1} pe^{-1}$.

 For events with multiple selected lines, the position of
 the track is determined using the $Q$ fit to 
 minimise the hit residuals. An additional fit is performed
 using a `bright point' hypothesis corresponding to a single,
 localised flash of light. A comparison of the
 quality of the two fits is used to reject 
 events in which downgoing muons create a bright 
 electromagnetic shower that may mimic an upgoing track.

 Events with only one selected detector line do not carry information
 on the azimuth direction of the muon. Hence, for these events,
 a four-parameter fit is performed, yielding the zenith angle
 of the muon track. Also in this case, a bright point fit is
 done for comparison.


 As in the full likelihood fit, the value of $Q$ found by the fit is used to reject badly 
 reconstructed events;
 in particular atmospheric muons that are reconstructed 
 as upward-going. The track fit quality is required to be
 better (i.e. smaller) than 1.35 (1.5) for reconstruction
 with two (more than two) lines; Events with a bright-point
 fit quality better than 1.8 are vetoed.

 The fact that only a single minimisation is performed 
 makes the online algorithm about an order of magnitude
 faster than the algorithm used for the full fit, which
 typically performs 20 full minimisations.

 With the current trigger setup, both algorithms are 
 fast enough to run on all triggered events in real 
 time on a single CPU.

\subsection{Results}

 Simulations show that, at high energies, the online algorithm 
 achieves a typical angular resolution on the muon direction 
 of 2 degrees (1 degree for more stringent cuts), independently 
 of the energy. At high energies, 
 this is a factor 6 to 3 significantly worse than the resolution o
 f the full likelihood fit, which is not unexpected as the assumption of 
 Gaussianity of the time residuals that underlies the $\chi^2$-fit
 is known to be an approximation. The online algorithm is therefore
 not well suited for those neutrino astronomy studies that require 
 optimal angular resolution. On the other hand, the simplicity 
 of the algorithm and the (expected) robustness against inaccuracies in the 
 detector description, have made it a good alternative for the 
 initial studies of the atmospheric muon (see \cite{atm_mu}) and
 neutrino fluxes.  

 Figure \ref{fig_bbm} shows the elevation above
 the horizon for the data taken in 2008 for the 
 multi-line events in comparison to simulation. 
 The data were taken using a 9,10 and 12 line detector 
 and represent
 a total live time of 173 days.
 The atmospheric muons were simulated using Corsika
 with Horandel fluxes and the QGSJET hadronic model. 
 A combined theoretical and systematic uncertainty
 of 30 (50)\% on the expected number of 
 neutrinos (muons) accounts for uncertainties in the 
 (primary) flux and interaction model, and in detector 
 acceptance. 
 
 A total of 582 upward going multi-line events were reconstructed, whereas
 the simulation predicts 494(13) events due to atmospheric 
 neutrinos (muons). The difference, a factor of 0.87, is
 within the systematic  uncertainty on the simulation. 

\begin{figure}[!t]
  \centering
  \includegraphics[width=3.3in]{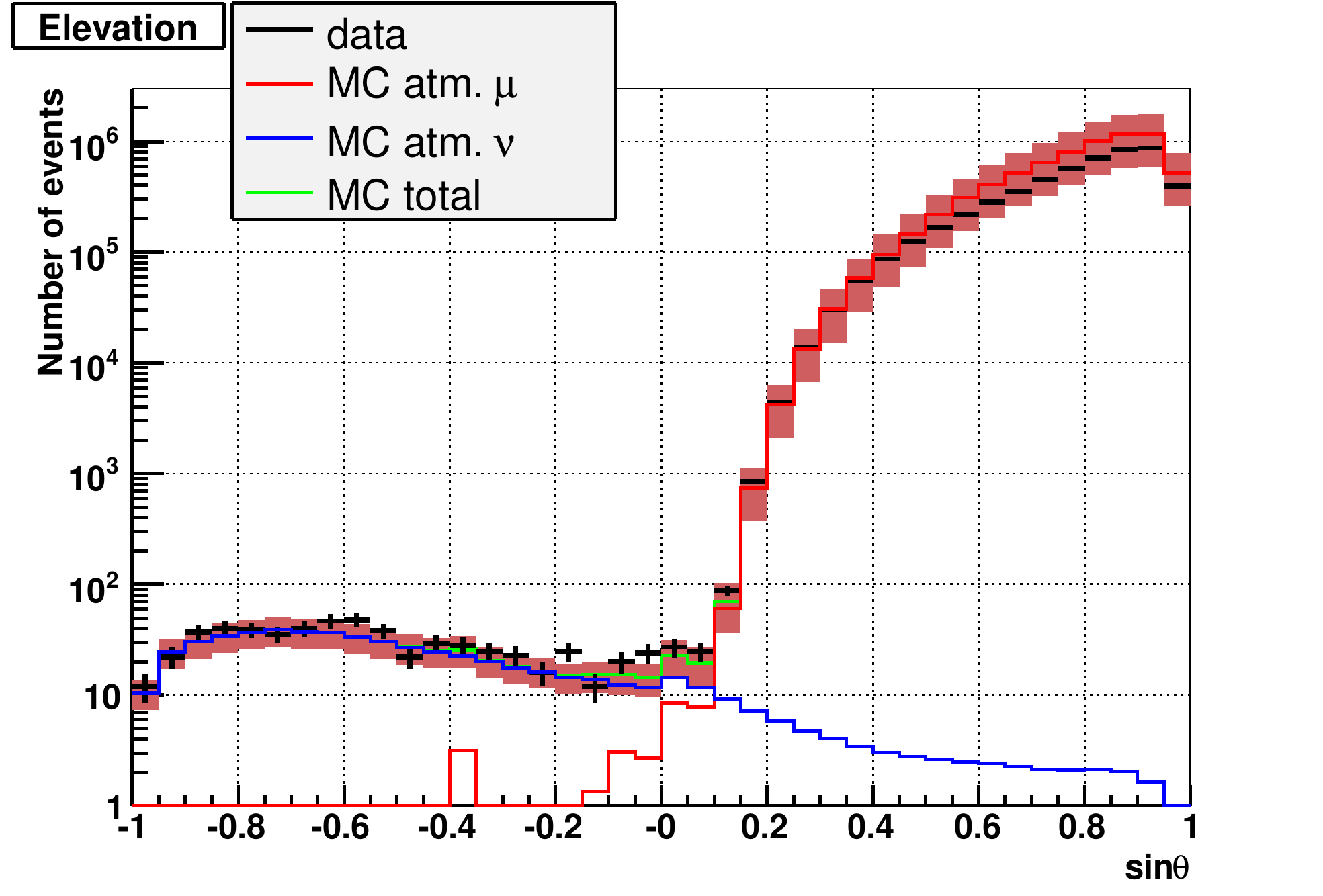}
  \caption{Distribution of the sine of the elevation angle for muons obtained             
           from the multi-line online reconstruction algorithm 
           for the 2008 data, i.e. 9-12 lines.
           }
  \label{fig_bbm}
 \end{figure}

\begin{figure}[!t]
  \centering
  \includegraphics[width=3.3in]{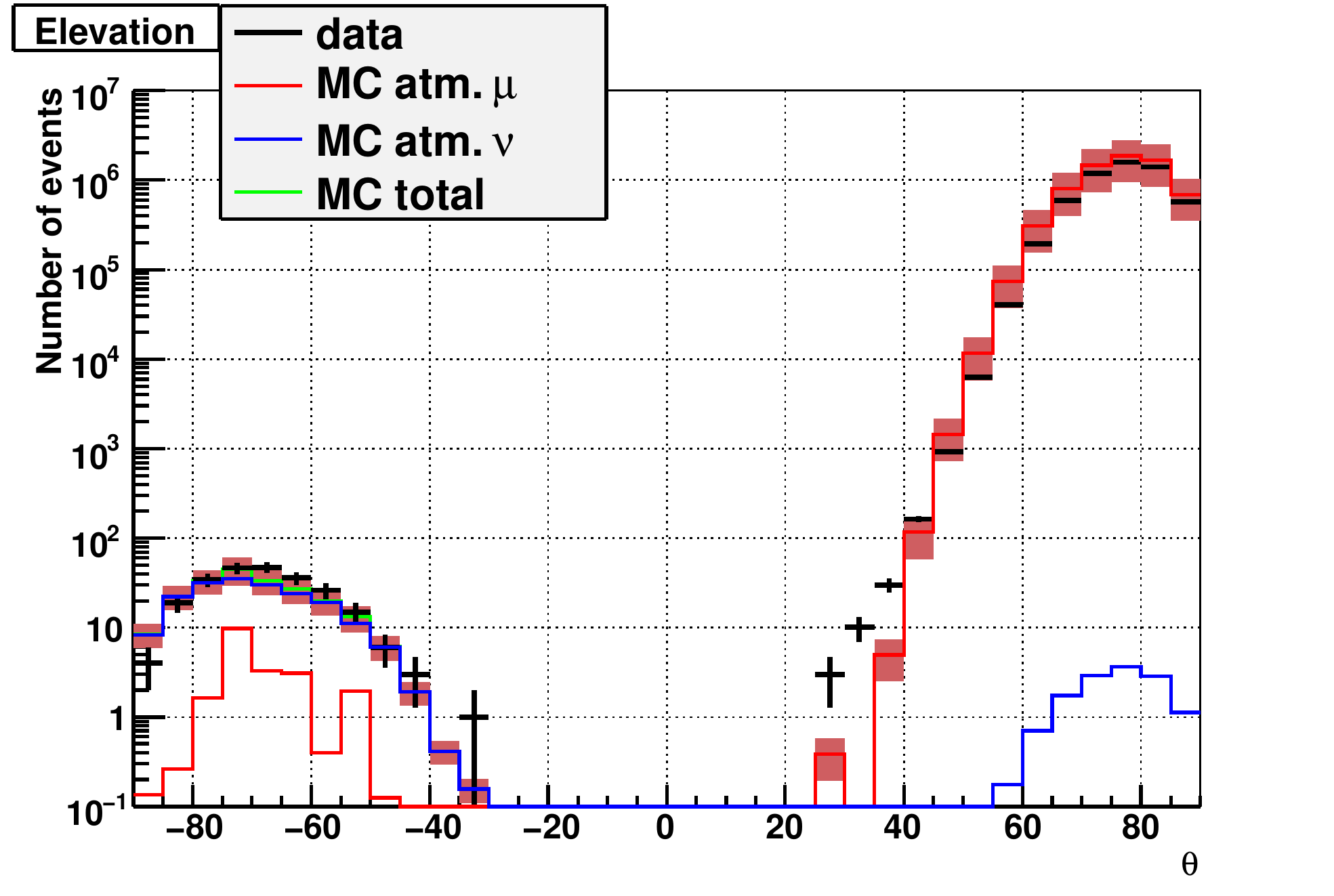}
  \caption{Distribution of the elevation angle f muons reconstructed
           on a single line using the online reconstruction algorithm for 2008 
           data, i.e. 9-12 lines.
           }
  \label{fig_bb1}
 \end{figure}

\vspace*{10cm}
\newpage

 Figure \ref{fig_bb1} shows the elevation angle for events
 fit on a single line using the online algorithm. The efficiency
 for such events is greatly enhanced for vertical (either upward, 
 or downward-going) tracks. The data agree well with the simulation,
 which is dominated by the atmospheric neutrino
 contribution for upward-going tracks. A total of 237 upward going 
 single-line events are found, while simulation predicts 190 (21) due to
 atmospheric neutrino (muon) events. The MC/data ratio is 0.89, which 
 is similar to the ratio reported above for multi-line events.

\section{Conclusion} 

 Since the deployment of the first five lines of the detector, 
 the Antares collaboration has been routinely detecting muons
 and atmospheric neutrinos. About five high-quality upgoing
 neutrino candidates are detected per day. The number of detected 
 neutrinos and their zenith and azimuth angle distributions
 agree well with simulations. 
 The observed time residuals of the hits with respect to reconstructed
 neutrino candidate tracks is in good agreement to the corresponding
 simulation. This further strengthens
 the confidence that Antares is performing as expected
 and that it is achieving sub-degree angular resolution.

\label{icrc1045:end}

\setcounter{figure}{0}
\setcounter{table}{0}
\setcounter{footnote}{0}
\setcounter{section}{0}





\hyphenation{abcdef-ghijklmnoprstuwxyz IEEEtran}

\title{Search for gamma-ray bursts with the \antares\ neutrino telescope}

\author{\IEEEauthorblockN{Mieke Bouwhuis\IEEEauthorrefmark{1} 
			on behalf of the \antares\ collaboration}
\IEEEauthorblockA{\IEEEauthorrefmark{1}National Institute for
Subatomic Physics (Nikhef), Amsterdam, The Netherlands}}

\shorttitle{Bouwhuis \etal GRB search \antares}
\maketitle
\label{icrc_grb:begin}

\begin{abstract}
Satellites that are capable of detecting gamma-ray bursts can trigger
the \antares\ neutrino telescope via the real-time gamma-ray bursts 
coordinates network.
Thanks to the ``all-data-to-shore" concept that is implemented
in the data acquisition system of \antares, the sensitivity to
neutrinos from gamma-ray bursts is significantly increased when
a gamma-ray burst is detected by these satellites.
The performance of the satellite-triggered data taking is shown,
as well as the resulting gain in detection efficiency.
Different search methods can be applied to the data taken in
coincidence with gamma-ray bursts.
For gamma-ray bursts above the \antares\ horizon, for which a neutrino 
signal is more difficult to find, an analysis method is
applied to detect muons induced by the high-energy gamma rays
from the source. 
  \end{abstract}

\begin{IEEEkeywords}
gamma-ray bursts; neutrino telescope
\end{IEEEkeywords}

\section{Introduction}

Several models predict the production of high-energy neutrinos by
gamma-ray bursts (GRBs)~\cite{grbs}.
The detection of these neutrinos would provide evidence for hadron
acceleration by GRBs. 
Such an observation would lead to a better understanding of the extreme
processes associated with these astrophysical phenomena.
In particular, it would give insight into the creation and
composition of relativistic jets.

One of the goals of the \antares\ neutrino telescope is to detect
high-energy neutrinos from GRBs.
The \antares\ telescope is situated in the Mediterranean Sea
at a depth of about 2500~m.
Neutrinos are detected through the detection of Cherenkov light
induced by the charged lepton that emerges from a neutrino
interaction in the vicinity of the detector.
Measurements are focused mainly on muon-neutrinos, since the muon 
resulting from a neutrino interaction can travel a distance of up to 
several kilometres.
Due to the transparency of the sea water (the absorption length is about
50~m), the faint Cherenkov light can be detected at relatively large
distances from the muon track.
A large volume of sea water is turned into a neutrino
detector by deploying a 3-dimensional array of light sensors in the
water.
The instrumented volume of sea water in the \antares\ telescope 
amounts to about 50 million cubic metres.
The track of the muon can be reconstructed from the measured arrival
times of the Cherenkov photons at the photo-multiplier tubes.
Since the muon and neutrino paths are approximately co-linear at high energies, 
the direction of the neutrino, and thus its origin, can be determined.

The GRB Coordinates Network (GCN)~\cite{gcn} announces the occurrence
of a GRB by distributing real-time alerts.
The data acquisition system of the \antares\ detector is designed such
that it can trigger in real time on these alerts.
This increases the detection efficiency for neutrinos from GRBs
significantly.
The \antares\ detector is currently the only neutrino telescope that
can trigger in real time on GRB alerts.

\section{Data taking with the \antares\ detector}
\label{s:data_taking}
The \antares\ telescope is operated day and night.
During operation, all signals from the photo-multiplier tubes are
digitised, and the raw data (containing the charge and time information of detected
Cherenkov photons) are sent to shore in a continuous data stream.
This is known as the all-data-to-shore concept~\cite{grb:daq}.
Although daylight does not penetrate to the depth of the \antares\
site, a ubiquitous background luminosity is present in the deep-sea
due to the decay of radioactive isotopes (mainly $^{40}$K) and
to bioluminescence.
This background luminosity produces a relatively high count rate of
random signals in the detector (60--150~kHz per photo-multiplier tube).
The total data rate is primarily determined by this background
luminosity, and amounts to about 1 GB/s.
On shore, the continuous data stream is divided over a farm of PCs.
Each of these PCs has a sophisticated filter program running, which
processes the data it receives in real time.
This filter scans the full sky, and finds the correlated photons that
are caused by a muon traversing the detector.
It triggers at a threshold of 10 such photons, which translates to a high
detection efficiency for muons, while preserving a high muon purity
(better than 90\%).
At the average background rate, the total trigger rate is 5--10~Hz
(depending on the trigger conditions).
The data are effectively reduced by a factor of 10$^4$.

\section{Satellite triggered data taking}

The data acquisition system of the \antares\ detector is linked with a socket 
connection to the GCN. 
The GCN network includes the Swift and Fermi satellites, both capable
of detecting GRBs.
When a GRB alert is received from the GCN, the standard data
processing continues (described in section~\ref{s:data_taking}), and in parallel 
to that the satellite triggered data taking is applied: all raw data 
covering a preset period (presently 2~minutes) are saved to disk for each GRB alert. 
There are about 1--2 GRB alerts per day, and half of them correspond to
a real GRB.

The buffering of the data in the data filter processors is used to store
the data up to about one minute before the actual alert.
The amount of data that can be kept in memory depends on the 
background rate in the sea water, the number of data processing PCs,
and the size of the RAM.
These data not only cover the delay between the detection of the GRB
by the satellite and the arrival time of the alert at the
\antares\ site, but also include data collected by the \antares\
detector before the GRB occurred.
These data therefore include a possible early neutrino signal that is
observable before the gamma rays. 

For each GRB that is detected by a satellite, and announced by
the GCN, all raw data collected by the \antares\ detector in
coincidence with the GRB are available on disk, as shown schematically
in Fig.~\ref{fig:fig01}.
\begin{figure}[t!]
\centering
\includegraphics[width=3.0in]{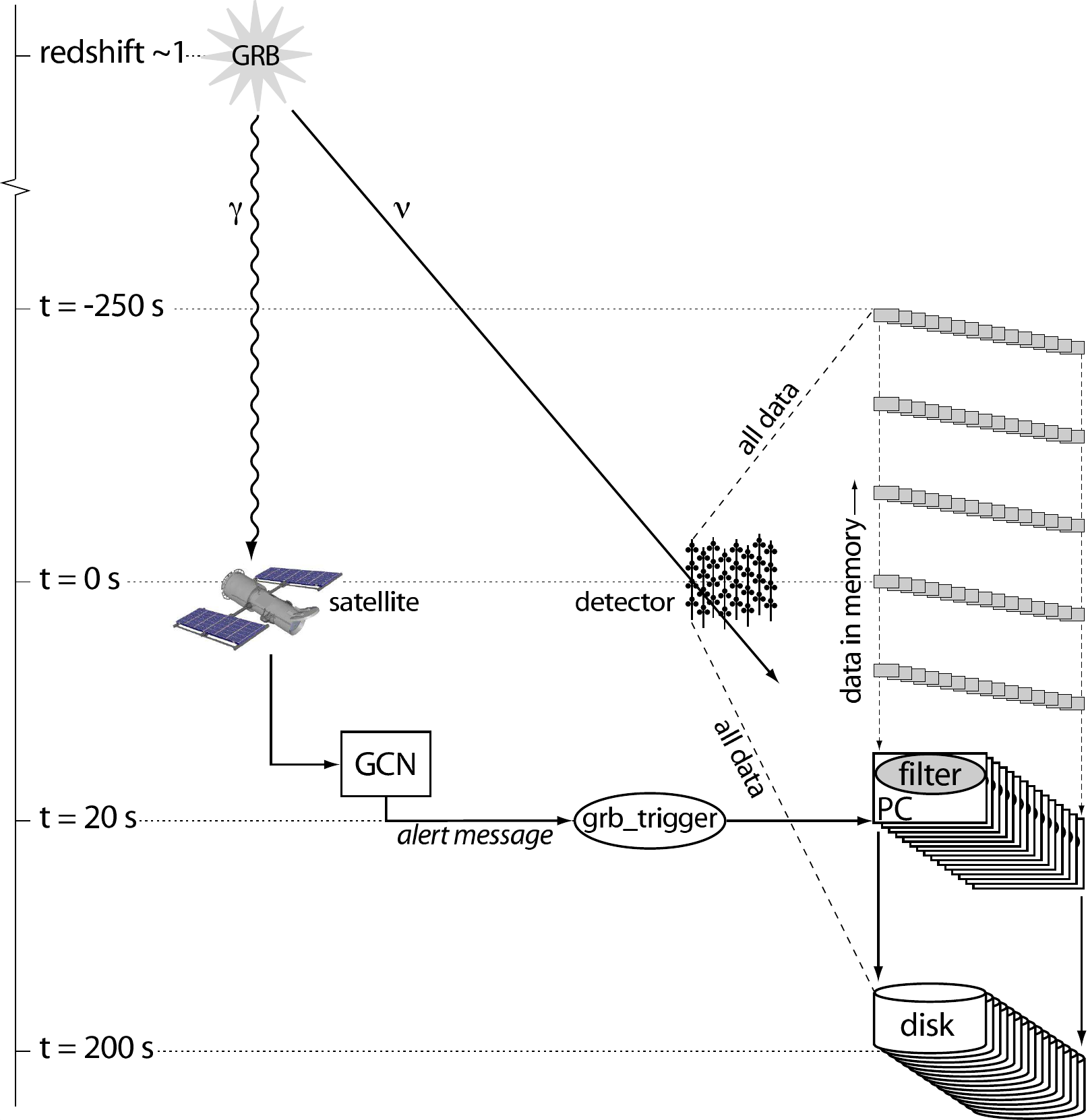}
\caption{Time line of the different events during a GRB. When an alert 
from the GCN is received by \antares\ (grb\_trigger program),
all raw data covering a few minutes are saved to disk, including all
data in memory. Any neutrino signal from the GRB (before, during, and
after the photon detection by the satellite) is stored on disk.}
\label{fig:fig01}
\end{figure}
The satellite triggered data taking period per GRB
 corresponds to a few times the typical duration of a GRB.
As a result, any time-correlated neutrino signal from the GRB
---before, during, and after the photon detection by the satellite---
is stored on disk.
Saving all raw data is only possible for transient sources like GRBs.
It can not be done for continuous sources because of the high data 
output rate of the detector.

\section{Satellite triggered data taking performance}

The satellite triggered data taking for all GRB alerts
distributed by the GCN is shown in Fig.~\ref{fig:fig02}.
The satellite triggered data taking system became operational in 
autumn 2006.  
The dashed line shows the number of GRB alerts from the GCN per
month as a function of time, and the solid line shows the
number of satellite triggered data taking sessions that were realised.
Although data taking with \antares\ is in principle continuous, 
inefficiencies can occur, for example, when a GRB alert is 
distributed during a calibration run, or due to power loss.  
As can be seen from Fig.~\ref{fig:fig02}, the typical efficiency of
the satellite triggered data taking is about 90\%. 
\begin{figure}[t!]
\setlength{\unitlength}{1cm}
\begin{center}
\begin{picture}(8,8.5)
\put(0,0){\scalebox{0.45}{\includegraphics{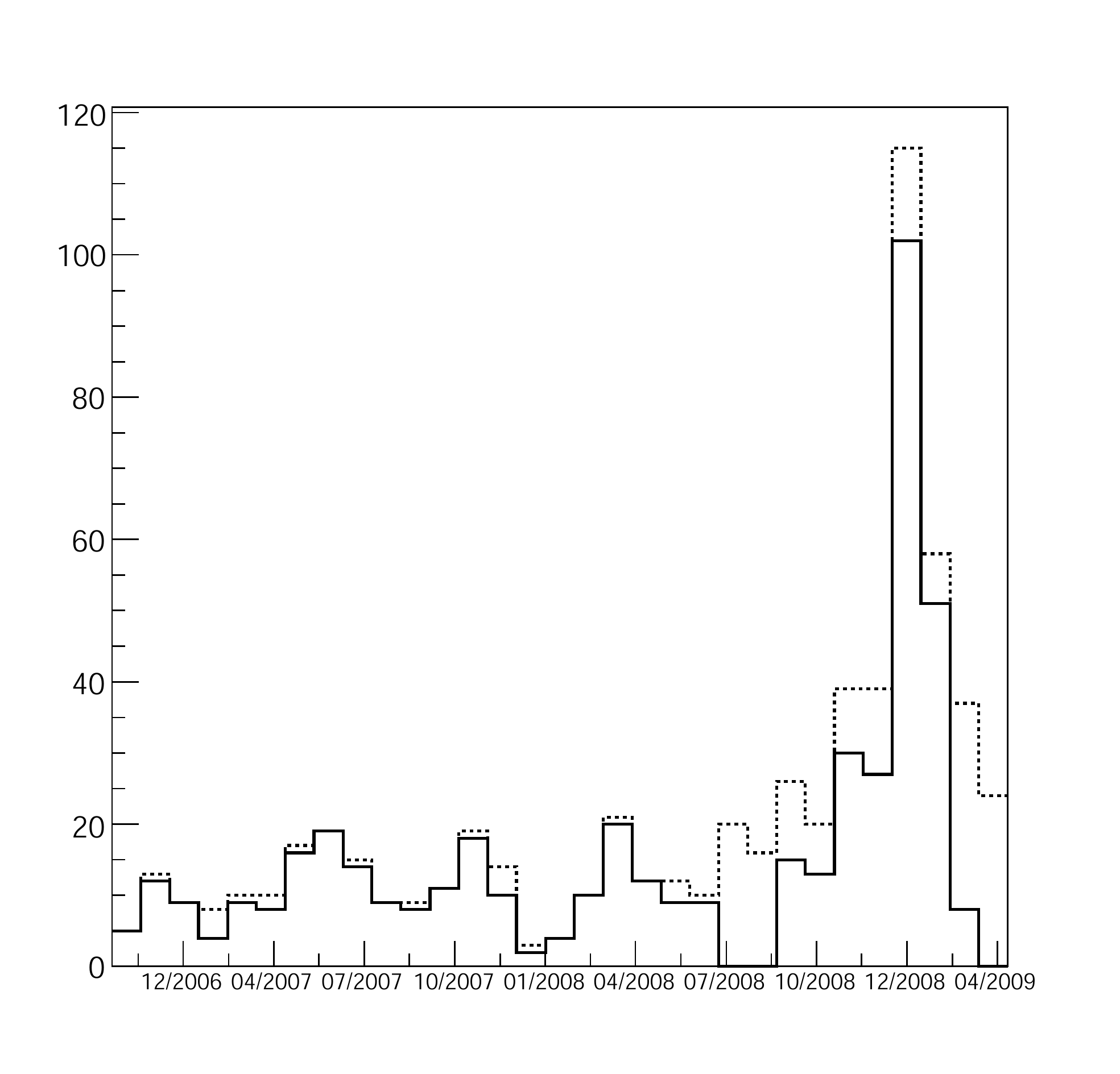}}}%
\put(4.5,0.1){\makebox(0,0)[b]{date (mm/yyyy)}}%
\put(0,4.5){\makebox(0,0)[l]{\rotatebox{90}{number of GRB alerts}}}%
\end{picture}
\end{center}
\caption{The satellite triggered data taking of \antares\
since the implementation of the satellite triggered data taking system.
The dashed line indicates the number of GRB alerts distributed
by the GCN per month. 
The solid line indicates the number of GRB alerts for which
the satellite triggered data taking was applied.
In November~2008, the GCN started the distribution of alerts from the 
Fermi satellite.
The relatively large amount of alerts received in January and 
February~2009 is due to the flaring activity of SGR~1550-5418.} 
\label{fig:fig02}
 \end{figure}

The response time of the \antares\ satellite triggered data taking to
the detection of the GRB by the satellite is shown in
Fig.~\ref{fig:fig03}.
\begin{figure}[t!]
\setlength{\unitlength}{1cm}
\begin{center}
\begin{picture}(8,5.5)
\put(0.5,0){\scalebox{0.45}{\includegraphics{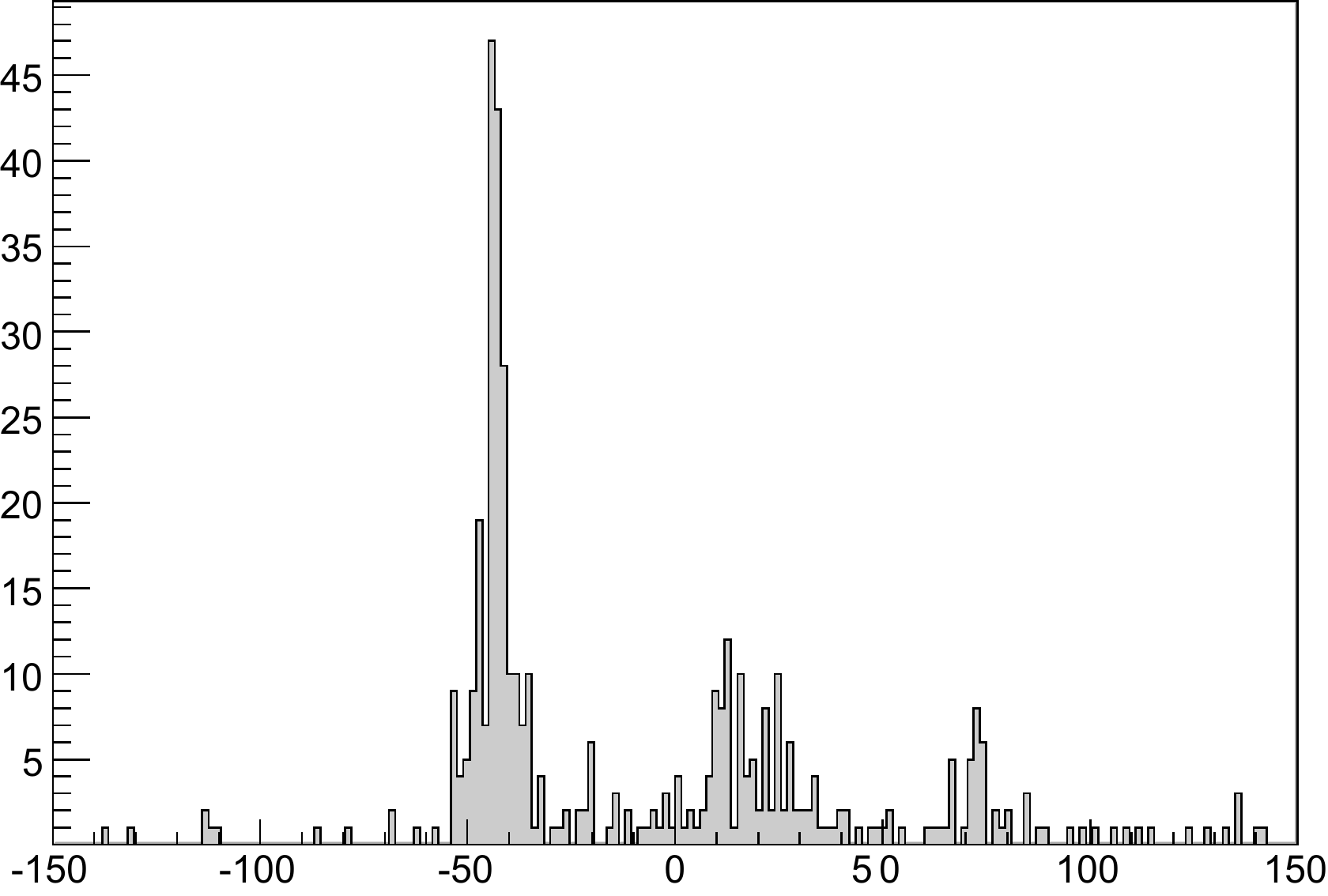}}}%
\put(4.5,-0.4){\makebox(0,0)[b]{response time (s)}}%
\put(0,3){\makebox(0,0)[l]{\rotatebox{90}{number of GRB alerts}}}%
\end{picture}
\end{center}
\caption{Earliest datum in the unfiltered GRB data set relative to the
detection time of the GRB by the satellite. At negative response
times, the unfiltered data set includes data prior to the detection of
the GRB by the satellite. Positive response times indicate a
substantial delay between the detection of the GRB by the satellite,
and the arrival time of the alert at the site.}
\label{fig:fig03}
\end{figure}
The response time is defined as the time difference between the GRB
time, as given in the GCN alert message, and the earliest datum in the
unfiltered data set available on disk. 
This indicates the amount of overlap of the unfiltered data set
with the observation period of the GRB by the satellite (the satellite
triggered data taking lasts for a fixed period of time).
At a response time of 0~seconds, the data in the unfiltered data set
completely cover the period during which the GRB was detected by the
satellite.
For positive response times, the delay between the detection of the GRB
by the satellite and the arrival time of the alert at the
\antares\ site plays a role.
As a result, the unfiltered data saved on disk do not fully cover the
period during which the GRB was detected by the satellite.
For negative response times, the buffering of the data filtering PCs
becomes apparent: the unfiltered data set on disk includes data
that were recorded before the GRB was detected by the satellite,
and could include an early neutrino signal.

\section{GRB data analyses} 

The GRB data analysis can be done in two alternative ways.
The standard way is based on real-time filtered data and
reconstruction of the muon trajectory. 
The reconstruction is based on a five parameter fit, including the two
direction angles~\cite{aart}.
 
The alternative method is based on the unfiltered data saved on disk
after a GRB alert. 
This analysis takes into account the position of the GRB on the sky, 
which is also provided by the GCN. 
Since these data do not need to be processed in real time, a much
lower detection threshold can be applied than is done for the standard data filtering. 
In the GRB analysis with unfiltered data, at least 6 time-position
correlated photons are required, compatible with a muon travelling in
the same direction as a neutrino from the GRB. 
In this way, the analysis method is only sensitive to a physics signal
from a specific GRB. 
The position of the GRB on the sky is also used to constrain the direction
angles in the fit. 
The same fit is repeated using many alternative directions, covering 
the opposite hemisphere and the downward hemisphere.
A cut on the likelihood ratio between 
the result of the first fit and 
the best result of all fits using the alternative directions  
is applied in order to select neutrinos coming from the GRB.
The whole analysis is thus reduced to a simple counting experiment.
In addition, the remaining background is low due to the short
duration of the GRB.  
The gain in detection efficiency over the standard method
is shown in Fig.~\ref{fig:fig04}.
\begin{figure}[t!]
\centering
\includegraphics[width=3.0in]{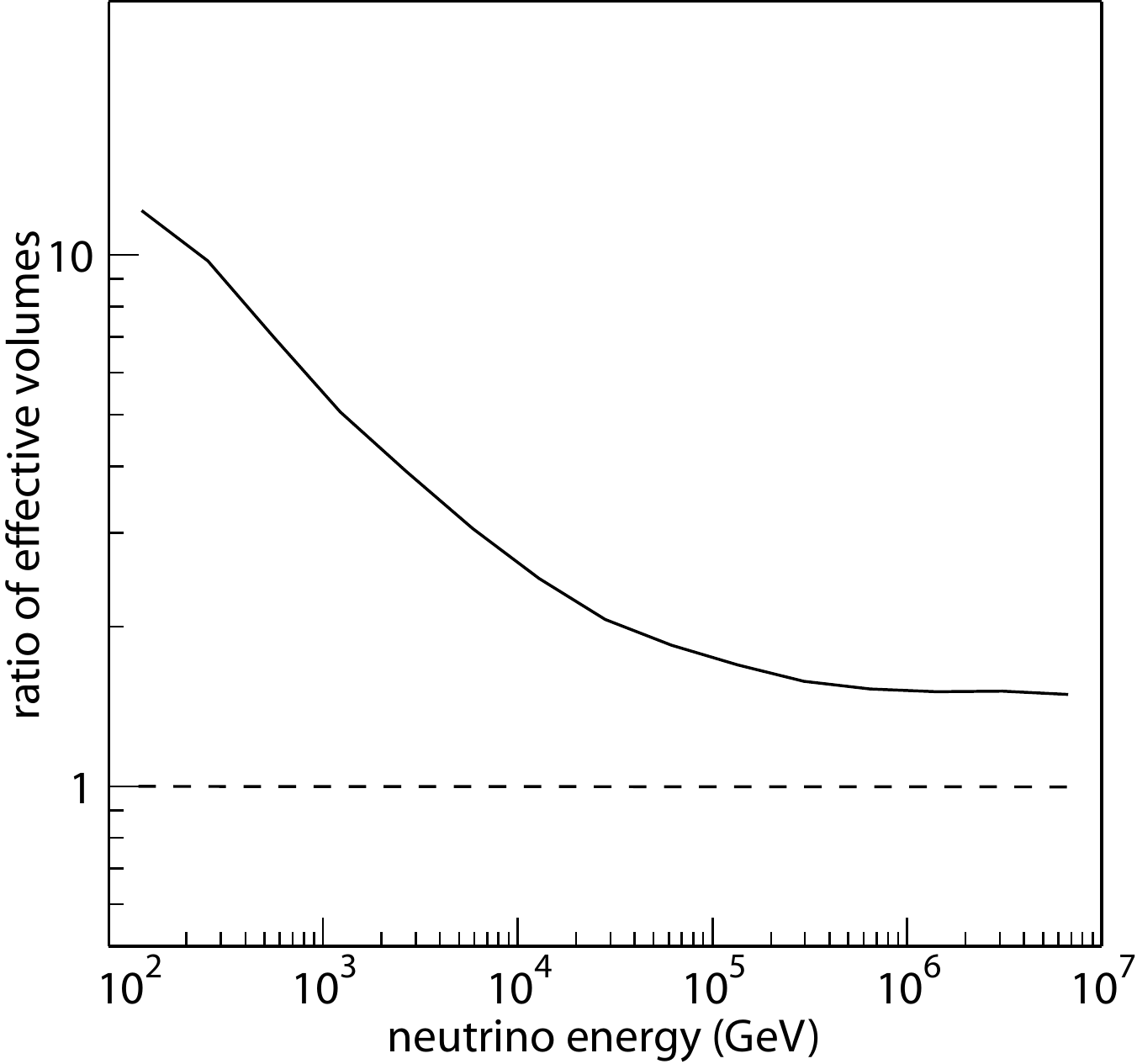}
\caption{The gain in detection efficiency for GRBs as a function of
neutrino energy (simulation). The solid line shows the result obtained
with the satellite triggered data taking, in combination with the
analysis method that uses the position of the GRB, relative to
the result obtained with the standard data taking and standard
analysis method (dashed line).}
\label{fig:fig04}
\end{figure}
This result is obtained using a simulation of the detector response to
muons, originating from neutrinos from a specific GRB.
The gain is expressed as the ratio of effective volumes 
(the volume in which a neutrino interaction produces a detectable
muon), and is shown relative to the result obtained with
the standard data taking and standard analysis method, using the same
simulated data.
The increased detection efficiency that is obtained with the unfiltered GRB data
is due to the lower detection threshold.
A higher threshold is required in the standard data taking method in order to
process the data in real time, which leads to an unavoidable detection inefficiency. 

For GRBs above the \antares\ horizon, for which a neutrino signal is
more difficult to find, an analysis method is applied to detect muons
induced by high-energy gamma rays from the source.  
The detection principle is presented in reference~\cite{goulven}.
A similar gain in detection efficiency can be expected when applying 
a specialised analysis to the unfiltered data saved on disk after a GCN
alert.

\section{Conclusions}  

The unique features of the \antares\ data acquisition system, 
in combination with the real-time distribution of GRB alerts
by the GCN, make it possible to trigger in real time on GRB alerts.
The \antares\ detector also has the possibility to buffer a large amount of data, 
resulting in very fast, and even negative, response times to GRB alerts. 
It is foreseen that the future, much larger, neutrino telescope KM3NeT~\cite{km3net}
will be designed such that it can trigger on GRB alerts in the same way.
These satellite triggered data lead to a significant increase in the sensitivity to neutrinos
from GRBs. 
Therefore the availability of networks like the GCN are very important
for neutrino telescopes. 
It is, however, imperative that the GRB alerts are distributed
within a few tens of seconds in order to maximise the discovery potential
for the detection of neutrinos from GRBs.

\label{icrc_grb:end}


\setcounter{figure}{0}
\setcounter{table}{0}
\setcounter{footnote}{0}
\setcounter{section}{0}






\hyphenation{abcdef-ghijklmnoprstuwxyz IEEEtran}

\title{Search for neutrinos from transient sources with the ANTARES telescope and optical follow-up observations}

\author{\IEEEauthorblockN{Damien Dornic\IEEEauthorrefmark{1},
			  St\'ephane Basa\IEEEauthorrefmark{2},
			  Jurgen Brunner\IEEEauthorrefmark{1},
			  Imen Al Samarai\IEEEauthorrefmark{1},
			  Jos\'e Busto\IEEEauthorrefmark{1},\\ 			  
			  Alain Klotz\IEEEauthorrefmark{3}\IEEEauthorrefmark{4},
			  St\'ephanie Escoffier\IEEEauthorrefmark{1},
			  Vincent Bertin\IEEEauthorrefmark{1},
			  Bertrand Vallage\IEEEauthorrefmark{5},		  
			  Bruce Gendre\IEEEauthorrefmark{2},\\
			  Alain Mazure\IEEEauthorrefmark{2}and
                          Michel Boer\IEEEauthorrefmark{4} \\ 
			  on behalf the ANTARES and TAROT Collaboration}
                            \\
\IEEEauthorblockA{\IEEEauthorrefmark{1}CPPM, CNRS/IN2P3 - Universit\'e de M\'editerran\'ee, 163 avenue de Luminy, 13288 Marseille Cedex 09, France}
\IEEEauthorblockA{\IEEEauthorrefmark{2}LAM, BP8, Traverse du siphon, 13376 Marseille Cedex 12, France}
\IEEEauthorblockA{\IEEEauthorrefmark{3}OHP, 04870 Saint Michel de l'Observatoire, France}
\IEEEauthorblockA{\IEEEauthorrefmark{4}CESR, Observatiore Midi-Pyr\'en\'ees, CNRS Universit\'e de Toulouse, BP4346, 31028 Toulouse Cedex04, France}
\IEEEauthorblockA{\IEEEauthorrefmark{5}CEA-IRFU, centre de Saclay, 91191 Gif-sur-Yvette, France}
}

\shorttitle{Damien Dornic \etal the TAToO project}
\maketitle
\label{icrc0055:begin}

\begin{abstract}
The ANTARES telescope has the opportunity to detect transient neutrino sources, such as gamma-ray bursts, core-collapse 
supernovae, flares of active nuclei... To enhance the sensitivity to these sources, we have developed a new detection method
based on the optical follow-up of "golden" neutrino events such as neutrino doublets coincident in time and space or single
neutrinos of very high energy.
\\
The ANTARES Collaboration has therefore implemented a very fast on-line reconstruction with a good angular resolution. These
characteristics allow to trigger an optical telescope network; since February 2009. ANTARES is sending alert trigger one or
two times per month to the two 25 cm robotic telescope of TAROT. This follow-up of such special events would not only give
access to the nature of the sources but also improves the sensitivity for transient neutrino sources.
\end{abstract}

\begin{IEEEkeywords}
neutrino, GRB, optical follow-up.
\end{IEEEkeywords}
 
\section{Introduction}
The ANTARES neutrino telescope \cite{BAntares} is located 40 km off shore Toulon, in the South French coast, at about 2500\,m below sea level. 
The complete detector comprises 12 detection lines, each equipped with up to 75 photomultipliers distributed in 25 storeys, which are the sensitive elements. 
Data taking started in 2006 with the operation of the first line of the detector. The construction of the 12 line detector was 
completed in May 2008. The main goal of the experiment is to detect high energy muon induced by neutrino interaction in the 
vicinity of the detector. The detection of these neutrinos would be the only direct proof of hadronic accelerations and so, the discovery of the ultra high 
energy cosmic ray sources without ambiguity.

Among all the possible astrophysical sources, transients offer one of the most promising perspectives for the detection of cosmic 
neutrinos thank to the almost background free search. The fireball model, which is the most commonly assumed, tells us how the GRBs operate but there are still remaining 
important questions such as which processes generate the energetic ultra-relativistic flows or how is the shock acceleration realized. The observation 
of neutrinos in coincidence in time and position with a GRB alert could help to constrain the models.

In this paper, we discuss the different strategies implemented in ANTARES for the transient sources detection. To detect transient 
sources, two different methods can be used \cite{BBasa}.

\section{Transient source detection strategies}
The first one is based on the search for neutrino candidates in conjunction with an 
accurate timing and positional information provided by an external source: the triggered search method. The second one is based 
on the search for high energy or multiplet of neutrino events coming from the same position within a given time window: 
the rolling search method. 

\subsection{The Triggered search}
Classically, GRBs or flare of AGNs are detected by gamma-ray satellites which deliver in real time an alert to the Gamma-ray bursts 
Coordinates Network (GCN \cite{BGCN}). The characteristics (mainly the direction and the time of the detection) of this alert are then distributed to 
the other observatories. The small difference in arrival time and position expected between photons and neutrinos allows a very 
efficient detection by reducing the associated background. This method has been implemented in ANTARES mainly for the GRB 
detection since the end of 2006. Today, the alerts are primarily provided by the Swift \cite{BSwift} and the Fermi \cite{BFermi} satellites. 
\\
Data triggered by more than 500 alerts (including the fake one) have been stored up to now. The "all data to shore" concept used in ANTARES allows to store 
all the data unfiltered during short periods. Based on the time of the external alert, in complement to the standard acquisition strategy, an on-line 
running program stores the data coming from the whole detector during 2 minutes without any filtering. This allows to lower the energy threshold of the event selection 
during the off-line analysis with respect to the standard filtered data. Due to a continuous buffering of data (covering 60s) and thanks to the 
very fast response time of the GCN network, ANTARES is able to recorded data before the detection of the GRB by the satellite \cite{BBouwhuis}. The 
analysis of the data relying on those external alerts is on-going in the ANTARES Collaboration.

Due to the very low background rate, even the detection of a small number of neutrinos correlated with GRBs could set a discovery. 
But, due to the relatively small field of view of the gamma-ray satellites (for example, Swift has a 1.4\,sr field of view), only 
a small fraction of the existing bursts are triggered. Moreover, the choked GRBs without photons counterpart can not be detected 
by this method.

\subsection{The Rolling search}
This second method, originally proposed by Kowalski and Mohr \cite{BKowalski}, consists on the detection of a burst of neutrinos in temporal and 
directional coincidence. Applied to ANTARES \cite{BDornic2}, the detection of a doublet of neutrinos is almost statistically significant. Indeed, 
the number of doublets due to atmospheric neutrino background events is of the order of 0.01 per year when a temporal window of 900\,s and a 
directional one of 3\,$^\circ$ x 3\,$^\circ$ are defined. It is also possible to search for single cosmic neutrino events by 
requiring that the reconstructed muon energy is higher than a given energy threshold (typically above a few tens of TeV). This 
high threshold reduces significantly the atmospheric neutrino background \cite{BDornic}. 
 
In contrary to the current gamma-ray observatories, a neutrino telescope covers instantaneously at least an hemisphere if only up-going events 
are analyzed and even $4\pi$\,sr if down-going events are considered. When the neutrino telescope is running, this method is almost 
100\% efficient. Moreover, this method applies whenever the neutrinos are emitted with respect to the gamma flash. More importantly no assumption is made on the nature 
of the source and the mechanisms occurring inside. The main drawback of the rolling search is that a detection is not automatically 
associated to an astronomical source. To overcome this problem, it is fundamental to organize a complementary follow-up program. The observation of any transient sources will require a quasi real-time analysis and an angular precision lower than a degree.

\section{ANTARES neutrino alerts}
Since the beginning of 2008, ANTARES has implemented an on-line event reconstruction. This analysis strategy contains a very efficient trigger based on local 
clusters of photomultiplier hits and a simple event reconstruction. The two main advantages are a very fast analysis (between 5 and 
10\,ms per event) and an acceptable angular resolution. The minimal condition for an event to be reconstructed is to contain a minimum of 
six storeys triggered on at least two lines. To select a high purity sample of up-going neutrino candidates, one quality cut is 
applied to the result of the $\chi^2$ minimisation of the muon track reconstruction based on the measured time and amplitude of 
the hits. In order to obtain a fast answer, the on-line reconstruction does not use the dynamic reconstructed geometry of the 
detector lines. This has the consequence that the angular resolution is degraded with respect to the one obtained with the standard 
off-line ANTARES reconstruction (of about 0.2 - 0.3\,$^\circ$) which includes the detector positioning.

 \begin{figure}[!t]
  \centering
  \includegraphics[width=2.8in]{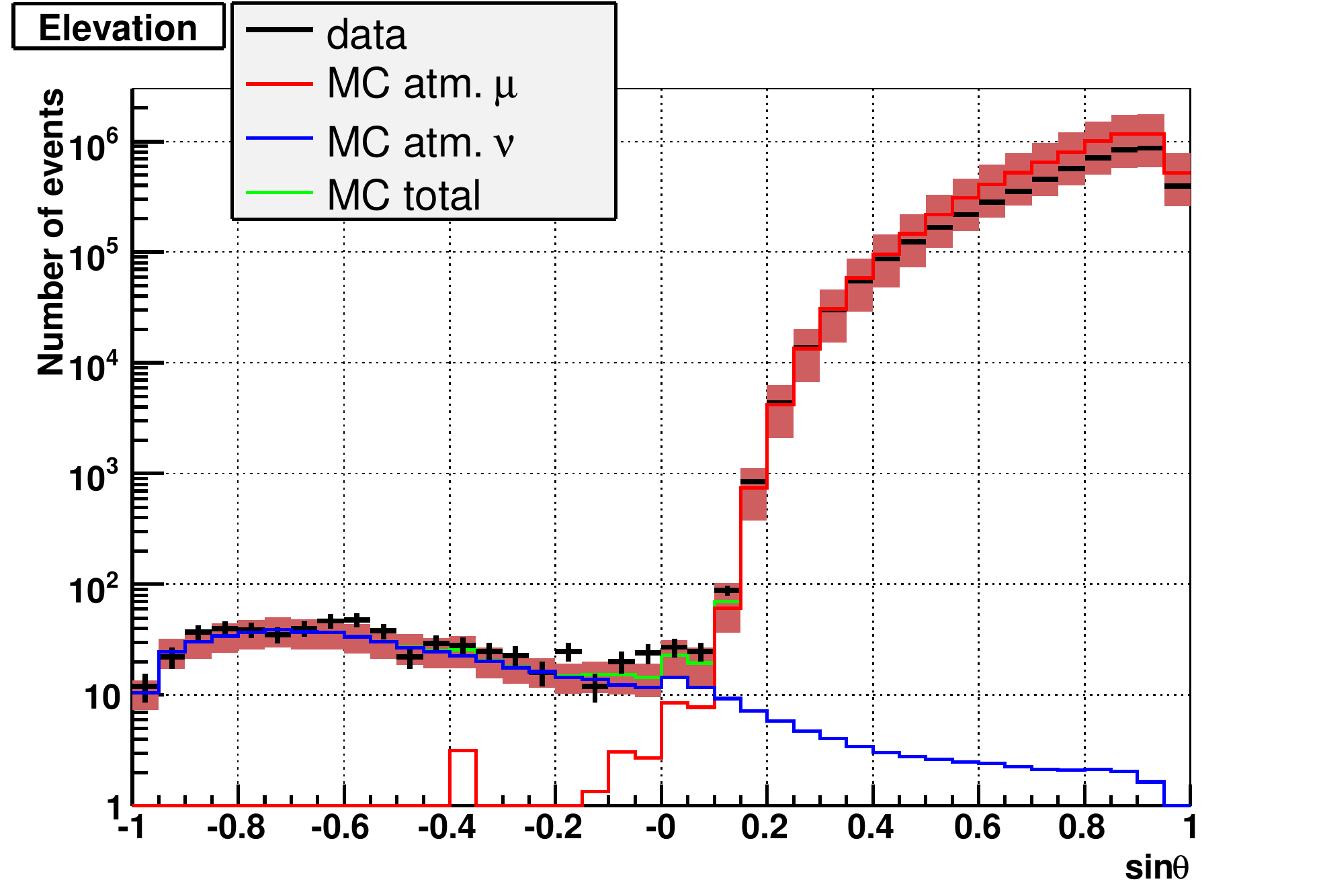}
  \caption{Elevation distribution of the well reconstructed muon tracks (black dots with error bar) recorded in 2008. The blue and red 
    lines represent the Monte-Carlo distribution for atmospheric neutrinos and atmospheric muons (the shaded band contains the systematic error).}
  \label{fig:elev}
 \end{figure}
 
 \begin{figure}[!t]
  \centering
  \includegraphics[width=2.8in]{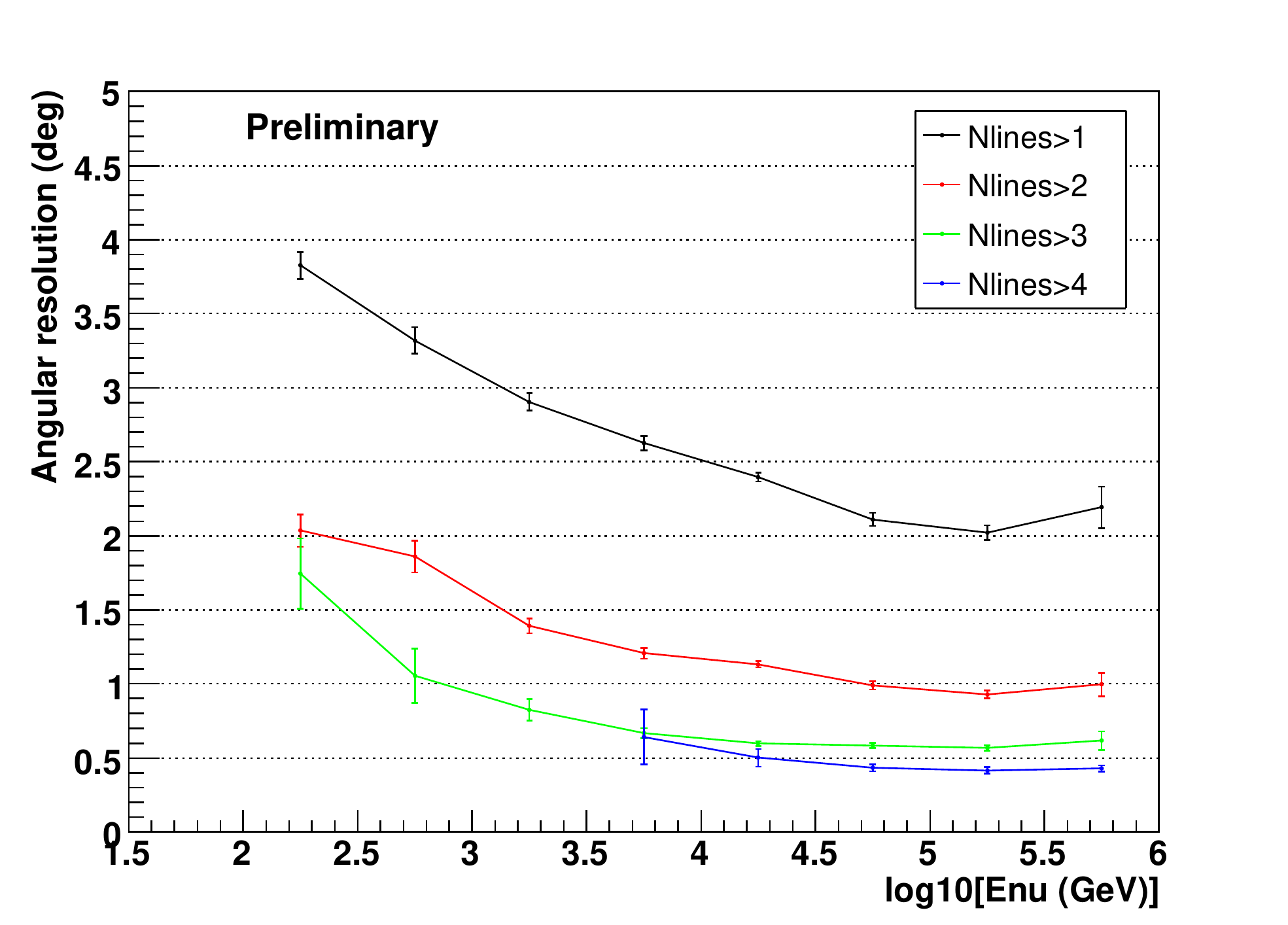}
  \caption{Angular resolution evolution with energy for the event reconstructed on-line with at least 2, 3, 4 and 5 lines.}
  \label{fig:angres}
 \end{figure}
 
In order to set the cuts used for our "golden" neutrino event selection, we have analysed the data taken in 2008 corresponding to 173 active 
days. During this period, around 582 up-going neutrino candidates were recorded. The figure~\ref{fig:elev} shows the elevation distribution of 
the well reconstructed muon events. This plot shows as the same time the distribution of the down-going atmospheric muons and the up-going 
neutrino candidates compare to the distribution predicted by the Monte-Carlo simulations. In order to obtain an angular resolution lower 
than the field of view of the telescope used for the follow-up (around 1\,$^\circ$ in radius), we select reconstructed events which trigger several hits 
on at least 3 lines. The dependence of this resolution with the number of lines used in the fit is shown in the figure~\ref{fig:angres}. For the 
highest energy events, this resolution can be as good as 0.5 degree. An estimation of the energy in the on-line reconstruction is indirectly 
determined by using the number of hits of the event and the total amplitude of these hits. In order to select events with an energy above around 5 TeV, 
a minimum of about 20 storeys and about 180 photoelectrons per track are required. These two different trigger logics applied on the 
2008 data period select around ten events. With a larger delay (few hours after the time of the burst), we are able to run the standard reconstruction tool 
which provides an angular resolution better than around 0.4\,$^\circ$ (still without the dynamic reconstructed geometry of the detector lines).

\subsection{Optical follow-up}
ANTARES is organizing a follow-up program in collaboration with TAROT (\textit{T\'elescope \`a Action Rapide pour les Objets Transitoires}, Rapid 
Action Telescope for Transient Objects, \cite{BTAROT}). This network is composed of two 25\,cm optical robotic telescopes located at Calern (South of France) and La 
Silla (Chile). The main advantages of the TAROT instruments are the large field of view of 1.86\,$^\circ$ x 1.86\,$^\circ$ and their very fast 
positioning time (less than 10\,s). These telescopes are perfectly tailored for such a program. Since 2004, they observe automatically the 
alerts provided by different GRB satellites \cite{BTAROTgrb}. 
\\
As it was said before, the rolling search method is sensitive to all transient sources producing high energy neutrinos. For example, a 
GRB afterglow requires a very fast observation strategy in contrary to a core collapse supernovae for which the optical signal will appear 
several days after the neutrino signal. To be sensitive to all these astrophysical sources, the observational strategy is composed of a real 
time observation followed by few observations during the following month. For each observation, six images integrated over a period of 3 minutes are taken by the telescope. 
We are adapting an image substraction program coming from Supernovae search (\cite{BImage}) in order to look for transient objects in the large field of view. Such a program 
does not require a large observation time. Depending on the neutrino trigger settings, an alert sent to TAROT by this rolling search program would be issued at a 
rate of about one or two times per month.

\subsection{Example of alerts}
In the beginning of 2009, with the final run control, 3 alerts were recorded by ANTARES during a test period. Unfortunetly, none of them has been followed by TAROT. 
The figure~\ref{fig:event} shows as example the event display of the first alert recorded in 2009 with ANTARES during this 
\newpage
test period. An \textit{a posteriori} search in astronomical
catalogues has shown not interesting objet in a one degree search window correlated with the position of this first alert.

 \begin{figure}[!t]
  \centering
  \includegraphics[width=2.8in]{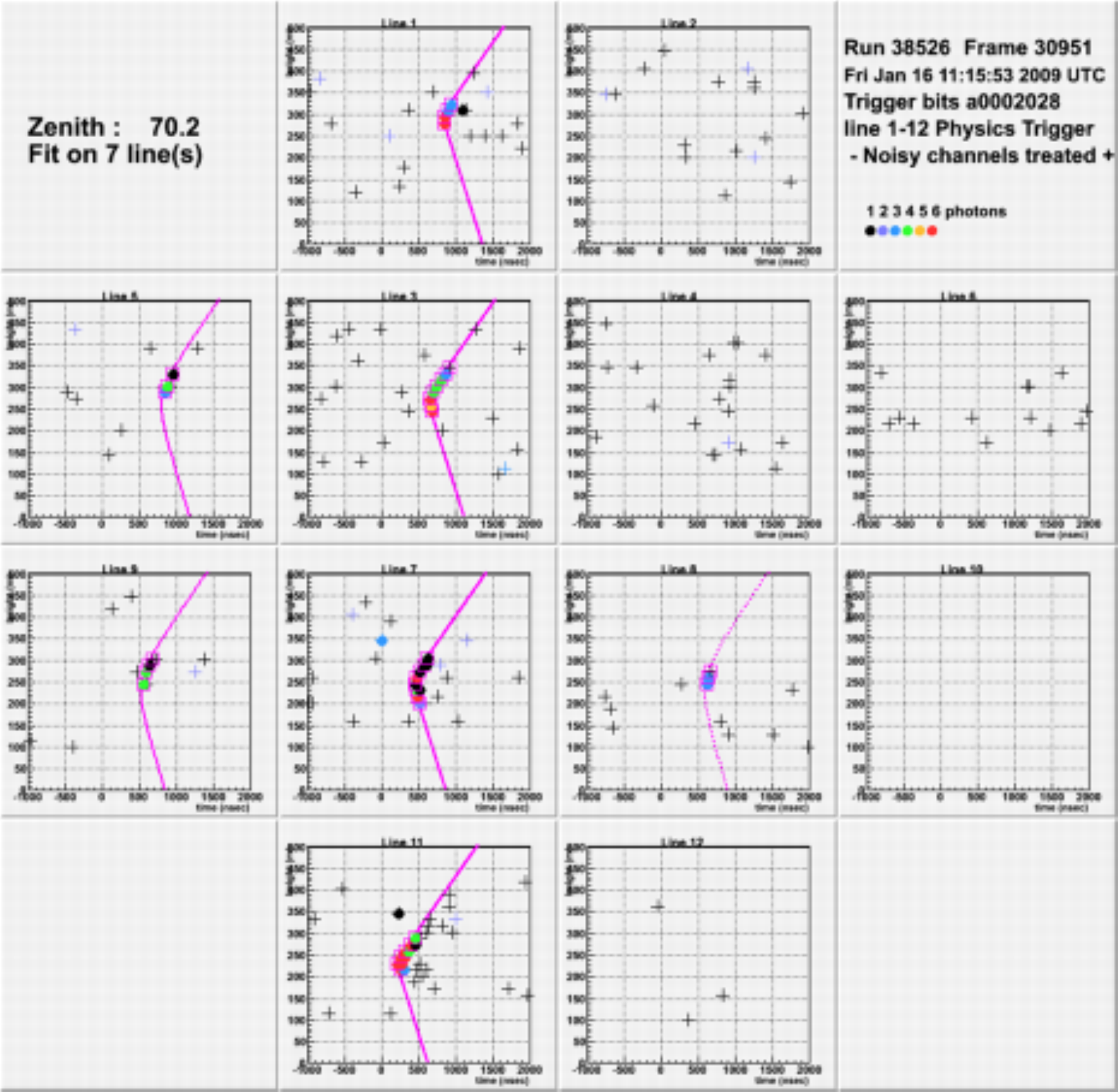}
  \caption{Two dimensional event display of the first neutrino alert reconstructed with 7 lines (not sent to TAROT). Each box represents for each line the height of all the
  hits acquired during a 2 $\mu$s time window versus their times of the detection.}
  \label{fig:event}
 \end{figure}
 
\section{Summary}
The follow-up of golden events would improve significantly the perspective for neutrino detection from transient sources. The most 
important point of the rolling search method is that it is sensitive to any transient source. A confirmation by an optical telescope 
of a neutrino alert will not only give the nature of the source but also allow to increase the precision of the source direction 
determination in order to trigger other observatories (for example very large telescopes for the redshift measurement). The program for 
the follow-up of ANTARES golden neutrino events is operational with the TAROT telescopes since February 2009. It would be also 
interesting to extend this technique to other wavelength observation such as X-ray or radio.

\label{icrc0055:end}


\setcounter{figure}{0}
\setcounter{table}{0}
\setcounter{footnote}{0}
\setcounter{section}{0}





\hyphenation{abcdef-ghijklmnoprstuwxyz IEEEtran}

\title{Gamma ray astronomy with \textsc{Antares}}

\author{\IEEEauthorblockN{Goulven Guillard\IEEEauthorrefmark{1}, for the \textsc{Antares} Collaboration}\\
\IEEEauthorblockA{\IEEEauthorrefmark{1}Institut Pluridisciplinaire Hubert Curien (CNRS/IN2P3)\\\& Universit\'e Louis Pasteur, Strasbourg, France}}

\shorttitle{Goulven Guillard - Gamma ray astronomy with \textsc{Antares}}
\maketitle
\label{icrc0503:begin}

\begin{abstract}
It has been suggested that underwater neutrino telescopes could detect muons from gamma ray showers. \textsc{Antares}' ability to detect high energy muons produced by TeV photons is discussed in the light of a full \textsl{Monte Carlo} study. It is shown that currently known sources would be hardly detectable.
\end{abstract}

\begin{IEEEkeywords}
\textsc{Antares} TeV gamma rays
\end{IEEEkeywords}
 
\section{Introduction}
The last decade has been fruitful in terms of high energy astronomy. More than 80 gamma rays sources have been found to emit in the TeV range~\cite{catalog}, thanks to Imaging Atmospheric \v{C}erenkov Telescopes (IACTs) such as \textsc{Cangaroo}, \textsc{HEGRA}, H.E.S.S, \textsc{Magic} or \textsc{Veritas}/Whipple~\cite{IACTs}.

There have also been many discussions about the possibility to detect muons produced by high energy gamma rays in underground, underice or underwater neutrino telescopes~\cite{gammarays}\cite{gammaraysok}\cite{Kudryavtsev}. In contrast to upward-going muons from neutrinos, which are the primary purpose of such a telescope, downward-going muons from gamma rays suffer from a high atmospheric muon background. Therefore the sensitivity of a neutrino telescope to gamma ray induced muons is quite lower than IACTs'. However, it has the advantage of monitoring continuously all directions. In addition to their physics potential, muons from gamma rays may also offer calibration benefits in terms of pointing accuracy and angular resolution.

Gamma ray showers are believed to be muon poor, but there are at least three processes by which a photon can produce muons\,: photoproduction, muon pair production and charm decay. The first process involves the (semi)leptonic decay of a pion or a kaon produced by the interaction of the photon with an atmospheric nucleus. Such muons are said to be conventional. The second process is self-explanatory, and its final particles are referred to as direct muons. The final case corresponds to the (semi)leptonic decay of a photoproduced charm meson, and secondary muons are called prompt muons. The prompt muon production was not taken into account in this work since it was not implemented in the software used for the \textsl{Monte Carlo} production. The charm production involves QCD processes that are not fully understood, but measurements at HERA have shown that at photon energies of several TeV charm production is significant~\cite{charm}.

Some calculations have estimated that the muon flux from gamma ray sources could be sufficient for neutrino telescopes to detect them. However, most attempts to estimate this muon flux rely on one-dimensional analytic models, and do not take into account the muon propagation from sea level to the detector and the detector sensitivity. A first attempt to estimate the underwater flux using a \textsc{Monte Carlo} simulation, without considering detection efficiency, has found gamma ray sources to be hardly detectable by a neutrino telescope~\cite{Kudryavtsev}.

In this paper, a full \textsl{Monte Carlo} simulation, including \v{C}erenkov light detection in realistic background conditions and track reconstruction, is presented, within the \textsc{Antares} framework. The expected number of events from the main sources of interest are presented.

\section{The \textsc{Antares} detector}

The Mediterranean sea currently houses the first operational undersea neutrino telescope, and also the largest neutrino telescope in the Northern hemisphere, namely \textsc{Antares} (Astronomy with a Neutrino Telescope and Abyss environmental RESearch)~\cite{ICRC0503:ANTARES}. Its full configuration has been completed in May 2008, though data has been taken with partial detector configurations since the first line was in water, in March 2006.

\textsc{Antares}' main focus is to detect astrophysical neutrinos, thanks to the \v{C}erenkov light produced in water by muons resulting from the interactions of neutrinos with the Earth. Because of the atmospheric muon background, \textsc{Antares} field of view is the Southern hemisphere, and in particular the Galactic center.

Installed at 40\,km off Toulon, in France (40$^{\circ}$50$^{\prime}$\,N, 6$^{\circ}$10$^{\prime}$\,E), \textsc{Antares} comprises 12 vertical detection lines positioned on the Mediterranean sea bed, at about 2500\,m depth. Each line hosts up to 25 floors of three 10$^{\prime\prime}$ photomultiplier tubes (PMTs) regularly distributed over 350\,m, the lowest floor being 100\,m above the sea bed. On a given floor, the three PMTs are placed symmetrically around the vertical string and look downwards at 45$^{\circ}$ in order to optimize the collection of \v{C}erenkov photons from upgoing muons rather than from downgoing muons~\cite{OMs}.

The lines are separated from each other by approximately 70\,m, and set as a regular octagon surrounding a square. An instrumented line intended to monitor the environmental conditions completes the apparatus.

The sea current induced displacements of the lines with regard to their vertical axis do not exceed a few meters, and are monitored in real time using compasses, tiltmeters and hydrophones hosted on each line. A position accuracy of about 10\,cm for each PMT is obtained.

An electro-optical cable transfers the electronic readout of the whole detector to shore, where digitized informations are processed in a computer farm.

\section{Monte Carlo simulation}

Extensive Air Shower have been simulated using Corsika v6.720~\cite{ICRC0503:CORSIKA}. High energy hadronic interactions are modeled through QGSJET01~\cite{hadronic}, while electromagnetic interactions are processed through EGS4~\cite{EGS4}. QGSJET01 is found to be the most conservative model in comparison to SIBYLL, VENUS and QGSJET-II~\cite{hadronic}, regarding the number of photons creating high energy muons (using VENUS leads to a 7\% rise, assuming a E$_\gamma^{-1}$ flux, in the [1;100]\,TeV range). However, the effect of this increase at the depth of \textsc{Antares} still has to be investigated, the muon range being energy dependent. The energy range considered in the present work goes from 1\,TeV to 100\,TeV.

MUSIC has been used for the propagation of muons in water~\cite{MUSIC}. The \textsc{Antares} \textsl{Monte Carlo} simulation chain then allows for simulation of \v{C}erenkov light in the detector, taking into account, in particular, the water properties and the PMTs angular acceptance~\cite{MC}. It also allows for the addition of realistic bioluminescence background using real data streams.

In this work the data used for the bioluminescence background corresponds to golden running conditions\,: runs are selected where the baseline of raw counting rates and the fraction of bursts\footnote{The burst fraction is defined as the ratio between the time when the counting rate is higher than 250\,kHz and the overall time.} are low (about 60\,kHz and less than 20\%, respectively).

Finally, the events are reconstructed using \textsc{Antares} standard reconstruction strategy. It has to be noticed that this strategy is optimized for upgoing events. The results presented here might thus be slightly enhanced using a dedicated strategy. On the other hand, the cut made on the reconstruction quality is very loose, so the effect of hardening the quality cut may compensate the effect of improving the reconstruction strategy.

\section{Sources of interest}

In order to have a reasonable probability to reach the depth of the \textsc{Antares} detector, a downgoing vertical muon must be more energetic than 1\,TeV at sea level\,: the muon probability to survive to a 2200\,m depth in water is less than 70\% for a 1\,TeV muon (13\% at 700\,GeV). Hence only TeV gamma ray sources may be seen by \textsc{Antares}. More than 80 gamma ray sources have been detected in the TeV range by IACTs~\cite{catalog}. However, not all of them are good candidates for \textsc{Antares}\,: most of them are located in the galactic plane, which is not in \textsc{Antares} field of view\footnote{The gamma rays field of view being the Northern hemisphere, as opposed to the neutrinos field of view.}. Moreover, weak and/or soft fluxes are not likely to produce enough muons.

Fortunately, several of the most powerful sources are visible by \textsc{Antares}, including the so-called ``standard candle'', the Crab pulsar. Characteristics of the most interesting candidates in terms of fluxes and visibility are summarized in table~\ref{ICRC0503:tab:sources}. In addition to the Crab, three extragalactic sources have been selected.

Though most of these sources are variable or flaring sources, they are known to have long periods of high activity, which make them more promising over a long period than most steady sources~\cite{1ES}\cite{Mkn421}\cite{Mkn501}.

\begin{table}[!h]
\centering
\begin{tabular}{lccc}\hline
source		& visibility	& mean zenith	& type	\\\hline\hline
Crab		&	62\%	&	51.7	& PWN	\\\hline
1ES 1959+650	&	100\%	&	49.7	& HBL	\\\hline
Mkn 501		&	78\%	&	49.4	& HBL	\\\hline
Mkn 421		&	76\%	&	49.2	& HBL	\\\hline
\end{tabular}
\caption{\textsl{Characteristics of \textsc{Antares} best gamma ray sources. HBL stands for High frequency peaked BL Lac object, and PWN for Pulsar Wind Nebula.}}
\label{ICRC0503:tab:sources}
\end{table}

\section{Simulation results}

\begin{table*}[!th]
\centering
\begin{tabular}{lcccccc}\hline
source		& $\mathrm{F_{\gamma}^{atm}}$ 	& $\mathrm{F_{\gamma}^{sea}(\times10^{-3})}$	& $\mathrm{N^{det}_{5{\scriptscriptstyle +}}}$& $\mathrm{N_{10{\scriptscriptstyle +}}^{det}}$&  $\mathrm{N^{reco}}$\\[0.1cm]\hline\hline
Crab		&	5-8	&	0.2-0.8		&	30-70		&	20-45		&	1-4	\\
		&		&	\emph{0.4-1}	&	\emph{20-50}	&	\emph{15-35}	&	\emph{1-3}	\\\hline
1ES 1959+650	&	0.8-30	&	0.1-2		&	3-150		&	2-100		&	0.2-8	\\
		&		&	\emph{0.1-2}	&	\emph{2-100}	&	\emph{1-70}	&	\emph{0.1-5}	\\\hline
Mkn 421		&	1.5-45	&	0.1-4		&	5-330		&	3-230		&	0.2-20	\\
		&		&	\emph{0.1-5}	&	\emph{2-260}	&	\emph{2-190}	&	\emph{0.1-15}	\\\hline
Mkn 501		&	1.5-40	&	0.1-15		&	6-1350		&	4-950		&	0.3-90	\\
		&		&	\emph{0.1-15}	&	\emph{3-1200}	&	\emph{2-880}	&	\emph{0.1-80}	\\\hline
\end{tabular}
\caption{\textsl{Number of photons/muons produced by several gamma ray sources at different levels, during one year, assuming a 100\% visibility, for primaries in the [1;100]\,TeV energy range. When relevant, straight font corresponds to a 10 degrees zenith angle, while italic stands for a 40 degrees zenith angle. Fluxes of photons at the top of the atmosphere ($\mathrm{F_{\gamma}^{atm}}$) and at sea level ($\mathrm{F_{\gamma}^{sea}}$) are expressed in m$^{-2}$. $\mathrm{N^{det}_{X{\scriptscriptstyle +}}}$ is the number of photons which produce more than $X$ hits on the detector PMTs, and $\mathrm{N^{reco}}$ corresponds to the number of reconstructed events in realistic bioluminescence conditions.}}
\label{tab:results}
\end{table*}

The number of detected and reconstructed events depends on several parameters, such as the precise level of background, the trigger strategy, the reconstruction strategy and the source flux parametrization. Therefore only range estimates are given. They are reported in table~\ref{tab:results}, assuming a 100\% visibility over one year. Most excentric parametrizations have been omitted.

It is found that only a few photons can be expected to be seen by \textsc{Antares}\,: less than ten events per year are reconstructed for the Crab in realistic conditions, though a few tens produce hits on the detector.

In comparison, a rough estimate on data with similar bioluminescence conditions gives $1.1\times10^5$ (resp. $4.1\times10^4$) reconstructed background events (atmospheric muons) within a one degree cone of the 10 degrees zenith angle direction (resp. 40 degrees), and $2.1\times10^6$ (resp. $7.3\times10^5$) background events within a 5 degrees cone.

It seems thus not reasonable to expect \textsc{Antares} to extract any gamma signal from the background under these conditions for any known flux. Though Markarian 501 may seem promising, the upper limit actually corresponds to high state fluxes parametrizations on dayscale variations~\cite{Mkn501}. A more precise selection of the fluxes and a study of the significance of the expected number of muons are still to be done.

However, these estimates are not so bad as they seem. First, the simulation is conservative in terms of photoproduction cross-section~\cite{sigmaph} and does not take muon production from charm decay into account. In addition, background discrimination has not yet been investigated. In particular, the muon poorness of gamma ray showers may help to reduce the atmospheric background\,: by rejecting multimuon events, one can improve both the signal to noise ratio and the angular resolution. If achievable, the muon pair tagging may also improve the background rejection. Moreover, a dedicated reconstruction strategy could increase the number of detected photons. Finally, galactic sources such as the Crab are not subject to the universe opacity above 100\,TeV, and the extension of their spectra to higher energy could lead to reasonable numbers of detectable photons. Short and powerful bursts are not to be excluded either, the associated background being in such cases almost negligible.

\section{Conclusion}

A complete \textsl{Monte Carlo} study has been processed in order to estimate \textsc{Antares}' ability to detect downgoing muons from gamma ray sources.

It has been found that \textsc{Antares} is not likely to detect any of the currently known sources, unless they show some unexpected behaviour. However the conservative estimates computed in this work show that the gamma ray astronomy field is not completely out of reach of underwater neutrino telescopes, at least for the next generation of detectors.

This study will be refined and extended to the km$^3$-scale successor of \textsc{Antares}, namely KM3NeT, which is currently being designed~\cite{KM3NET}.

\label{icrc0503:end}


\setcounter{figure}{0}
\setcounter{table}{0}
\setcounter{footnote}{0}
\setcounter{section}{0}





\hyphenation{abcdef-ghijklmnoprstuwxyz IEEEtran}

\title{First results on the search for dark matter in the Sun\\ with the ANTARES neutrino telescope}

\author{
        \IEEEauthorblockN{G.M.A. Lim\IEEEauthorrefmark{2} on behalf of the ANTARES collaboration}\\
        \IEEEauthorblockA{\IEEEauthorrefmark{2}NIKHEF, Science Park 105, P.O. Box 41882, 1009 DB Amsterdam, The Netherlands}
       }

\shorttitle{G.M.A. Lim - ANTARES dark matter results}
\maketitle
\label{icrc0031:begin}

\begin{abstract}

The ANTARES collaboration is currently operating the largest neutrino detector in the Northern Hemisphere. One of the goals of ANTARES is the search for dark matter in the universe. In this paper, the first results on the search for dark matter in the Sun with ANTARES in its 5 line configuration, as well as sensitivity studies on the dark matter search with the full ANTARES detector and the future cubic-kilometer neutrino telescope studied by the KM3NeT consortium are presented.

\end{abstract}

\begin{IEEEkeywords}

Dark matter, neutrino telescopes, supersymmetry

\end{IEEEkeywords}
 
\section{Introduction}
\vspace*{1mm}

Observational evidence shows that the majority of the matter content of the universe is of non-baryonic nature, befittingly called dark matter. Various extensions of the Standard Model of particle physics provide a well-motivated explanation for the constituents of dark matter: the existence of hitherto unobserved massive weakly interacting particles. A favorite amongst the new particle candidates is the neutralino, the lightest superpartner predicted by supersymmetry, itself a well-motivated extension of the Standard Model. 

In the supersymmetric dark matter scenario, neutralinos have been copiously produced in the beginning of the universe. The expansion of the Universe caused a relic neutralino density in the universe today analogous to the cosmic microwave background. These relic neutralinos could accumulate in massive celestial bodies in the Universe like the Sun, thereby increasing the local neutralino annihilation probability. In the annihilation process new particles would be created, amongst which neutrinos. This neutrino flux could be detectable as a localised emission with earth-based neutrino telescopes like ANTARES.\\[-3mm]

\section{The ANTARES neutrino telescope}
\vspace*{1mm}
 
\begin{figure}[!t]
  \centering
  \includegraphics[width=0.45\textwidth]{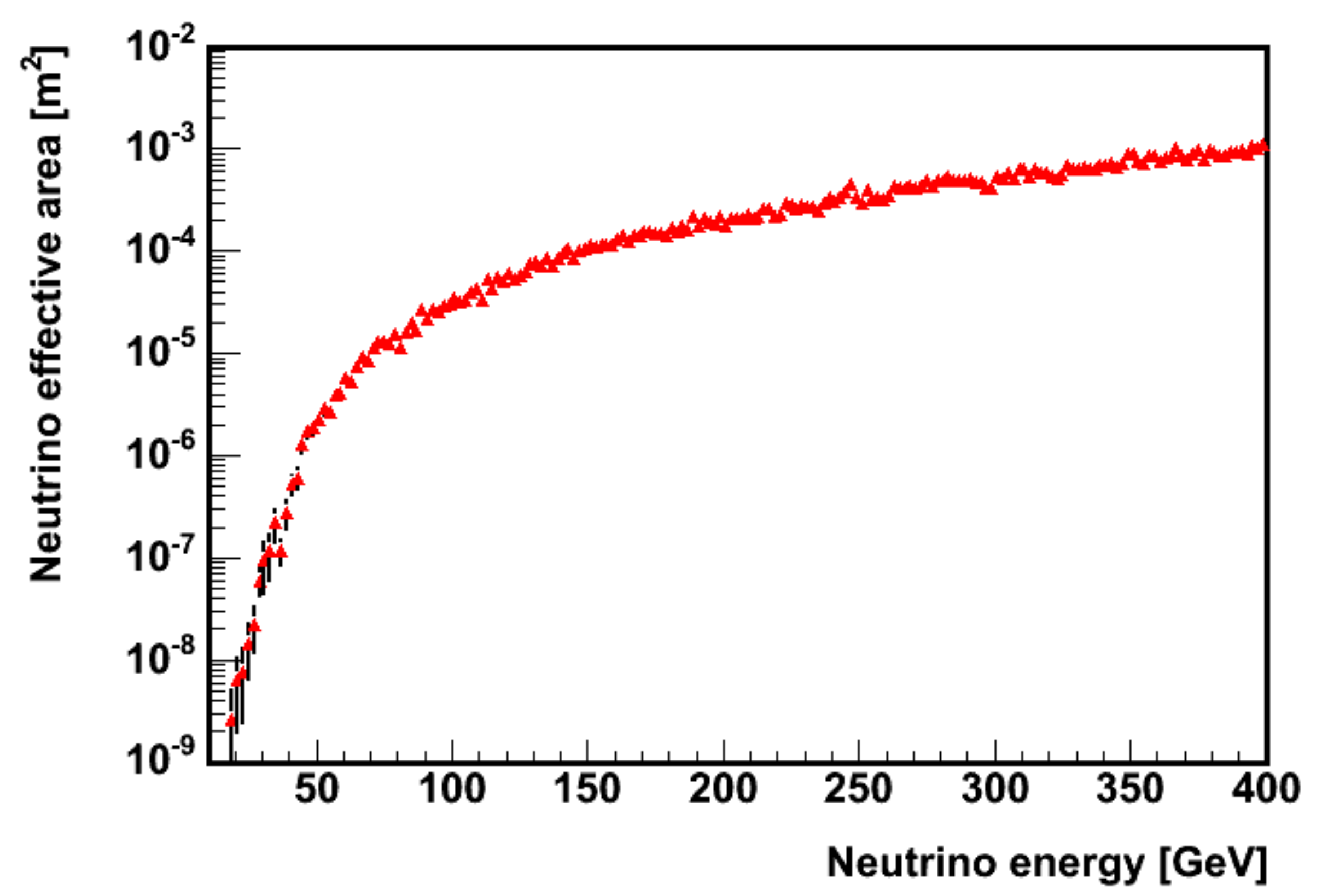}
  \caption{The 12~line ANTARES effective area.}
  \label{NEAnew_colour}
\end{figure} 

\begin{figure*}[!t]
  \centering
  \includegraphics[width=0.8\textwidth]{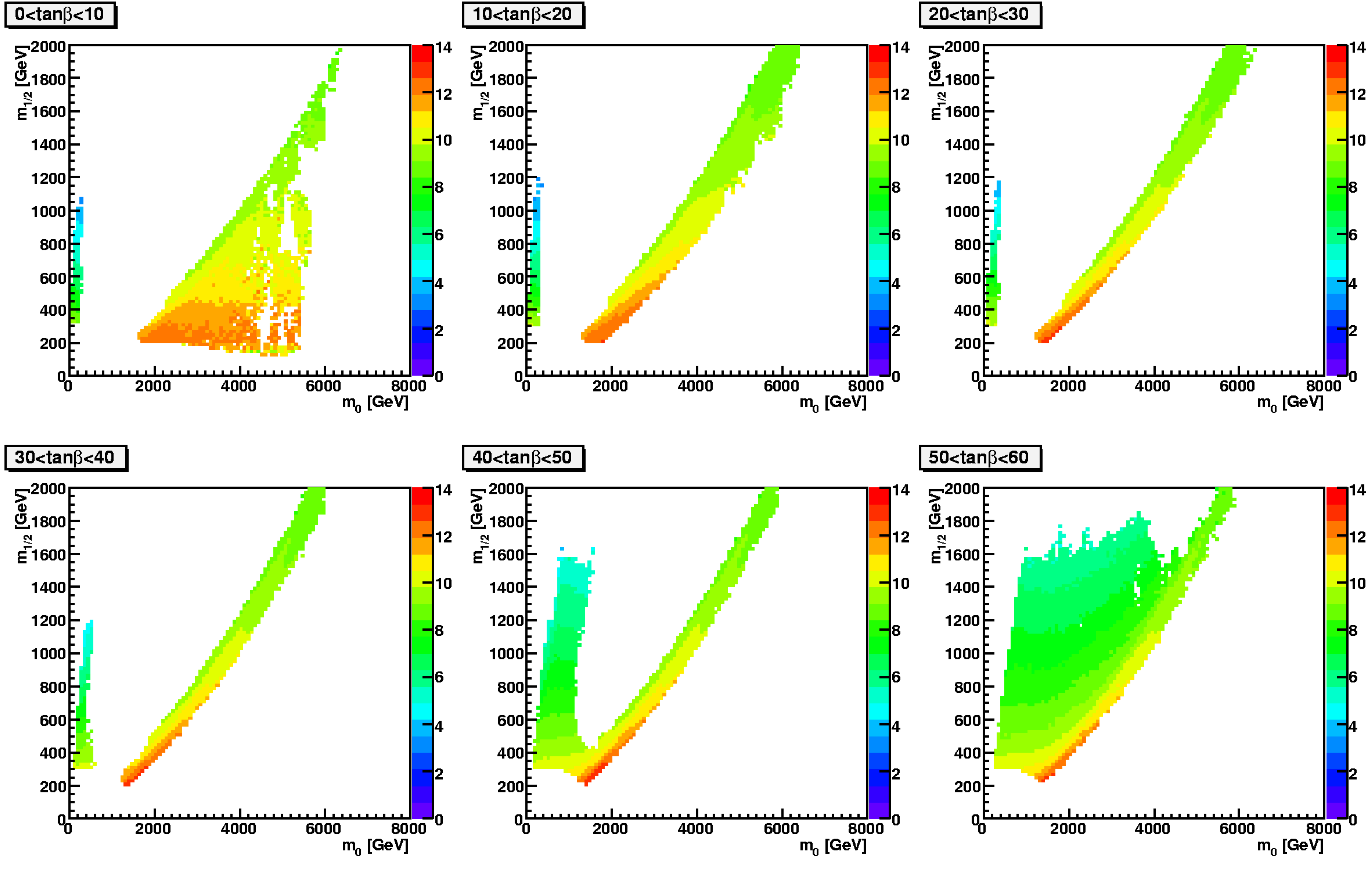}
  \caption{The $\log_{10}(\nu_\mu+\bar{\nu}_\mu)$ flux from neutralino annihilation in the Sun $[\textrm{km}^{-2}\textrm{year}^{-1}]$ in $m_0$-$m_{1/2}$ mSUGRA parameter space, for six different $\tan(\beta)$ intervals. $A_0$ varies between $-3m_0$ and $3m_0$. The flux was integrated above \mbox{$E_\nu=10$~GeV}.}    
  \label{psflux}
\end{figure*}
 
\begin{figure*}[!t]
  \centering
  \includegraphics[width=0.8\textwidth]{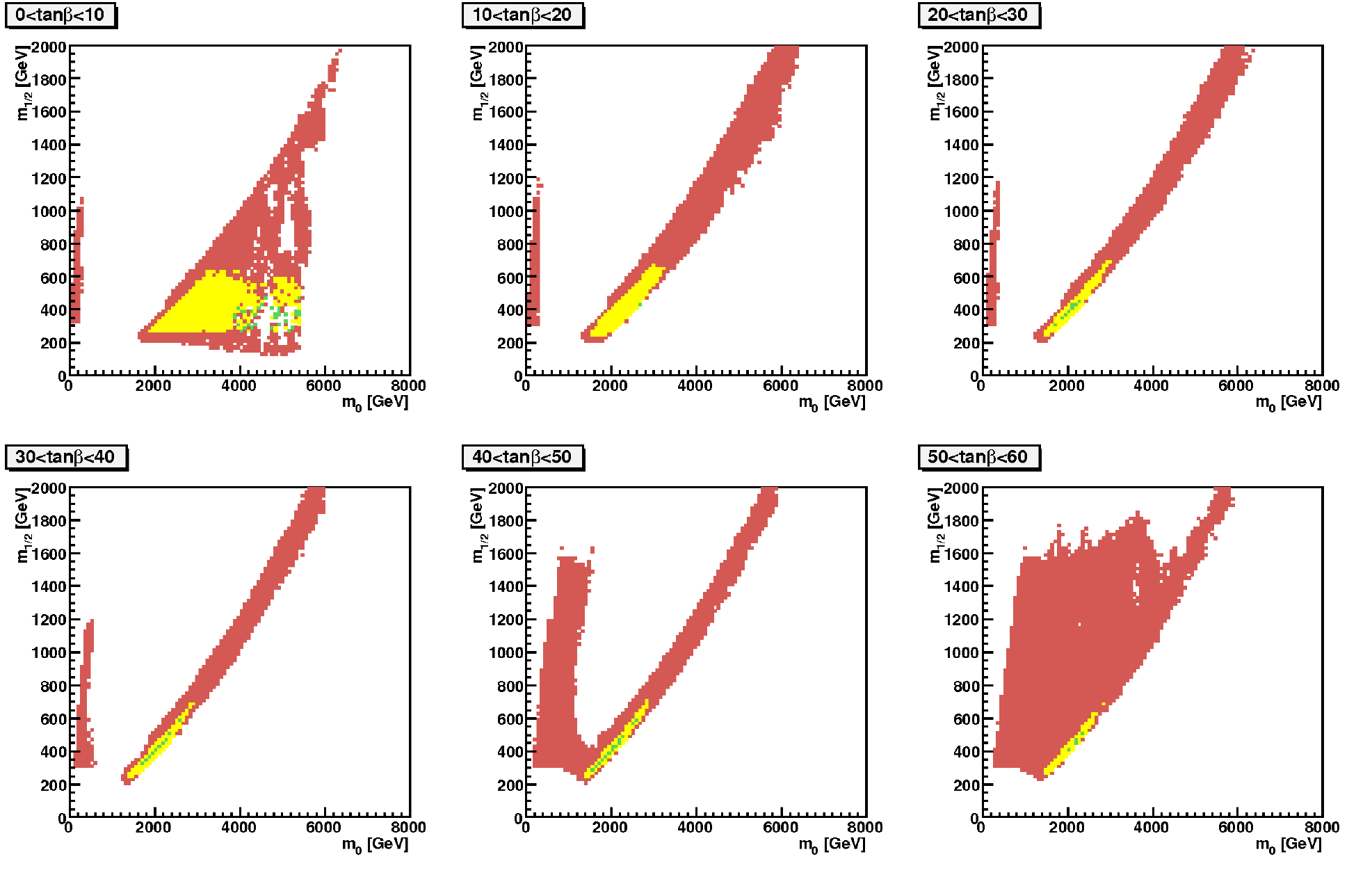}
  \vspace*{-1mm}
  \centerline{\footnotesize (a) ANTARES (12~lines) }\\[4mm]
  \includegraphics[width=0.8\textwidth]{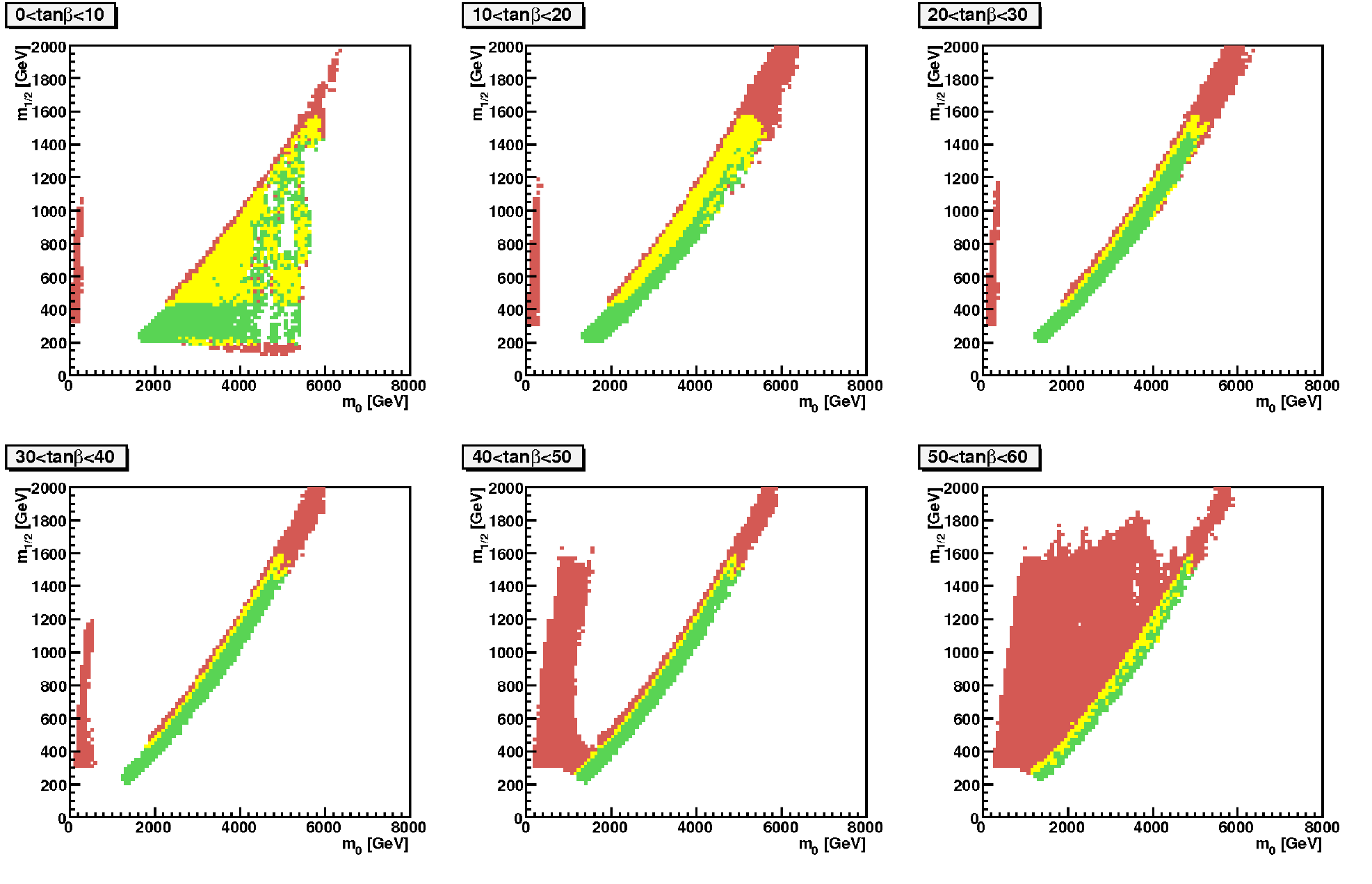}
  \centerline{\footnotesize (b) KM3NeT}\\[2mm]
  \caption{Sensitivity of ANTARES (Fig.~\ref{exclfig}a) and KM3NeT (Fig.~\ref{exclfig}b) in $m_0$-$m_{1/2}$ mSUGRA parameter space, for six different $\tan(\beta)$ intervals. $A_0$ varies between $-3m_0$ and $3m_0$. Green/yellow/red indicates that all/some/no mSUGRA models (depending on~$A_0$ and~$\tan(\beta)$) can be excluded at 90\%~CL after 3~years.}
  \label{exclfig}
\end{figure*}

\addtocounter{figure}{1}

\begin{figure*}[!t]
  \centerline{
    \subfloat[Upper limit on the neutrino flux from the Sun]{\includegraphics[width=0.45\textwidth]{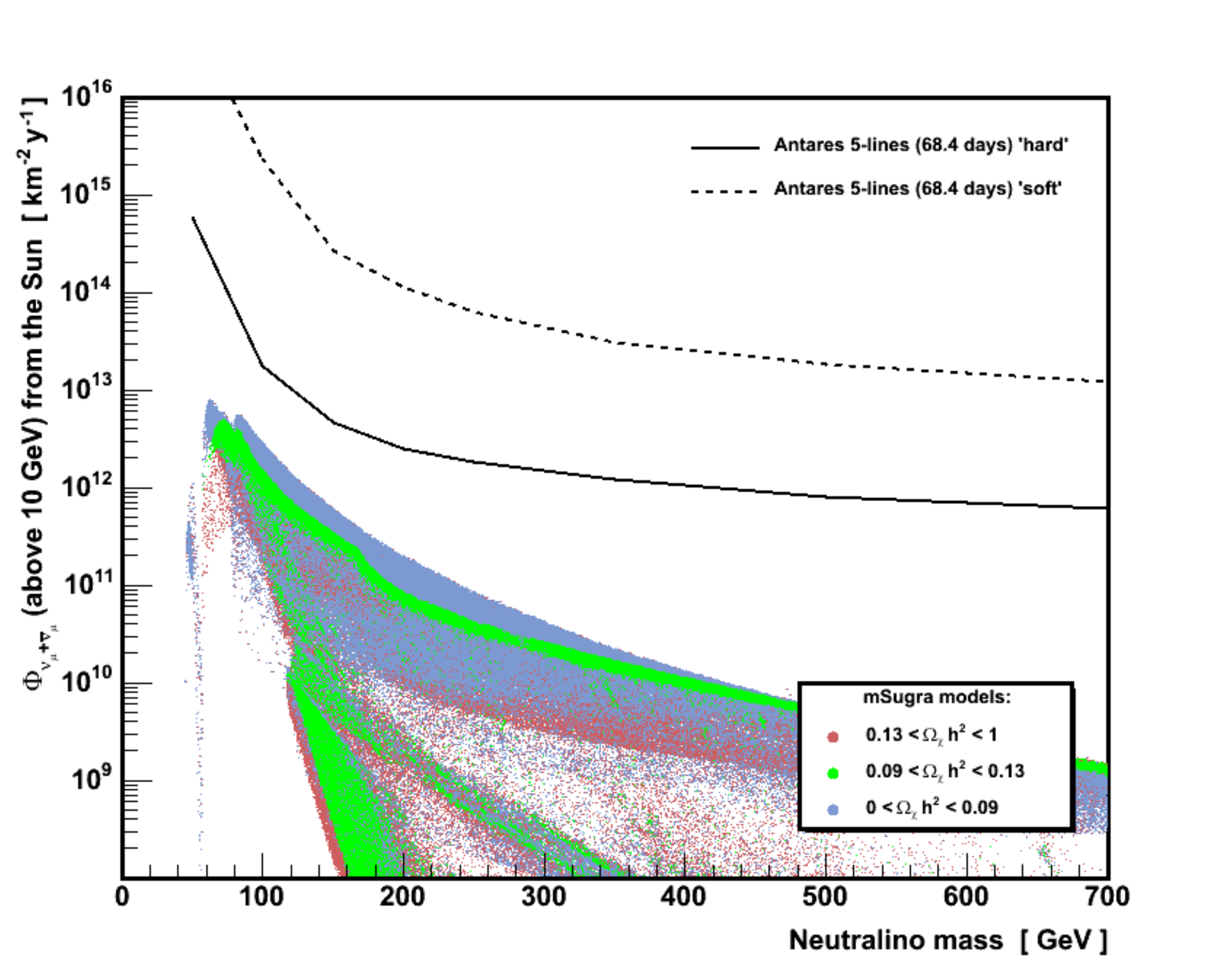} \label{sub_fig1}}
    \hfil
    \subfloat[Upper limit on the muon flux from the Sun]{\includegraphics[width=0.45\textwidth]{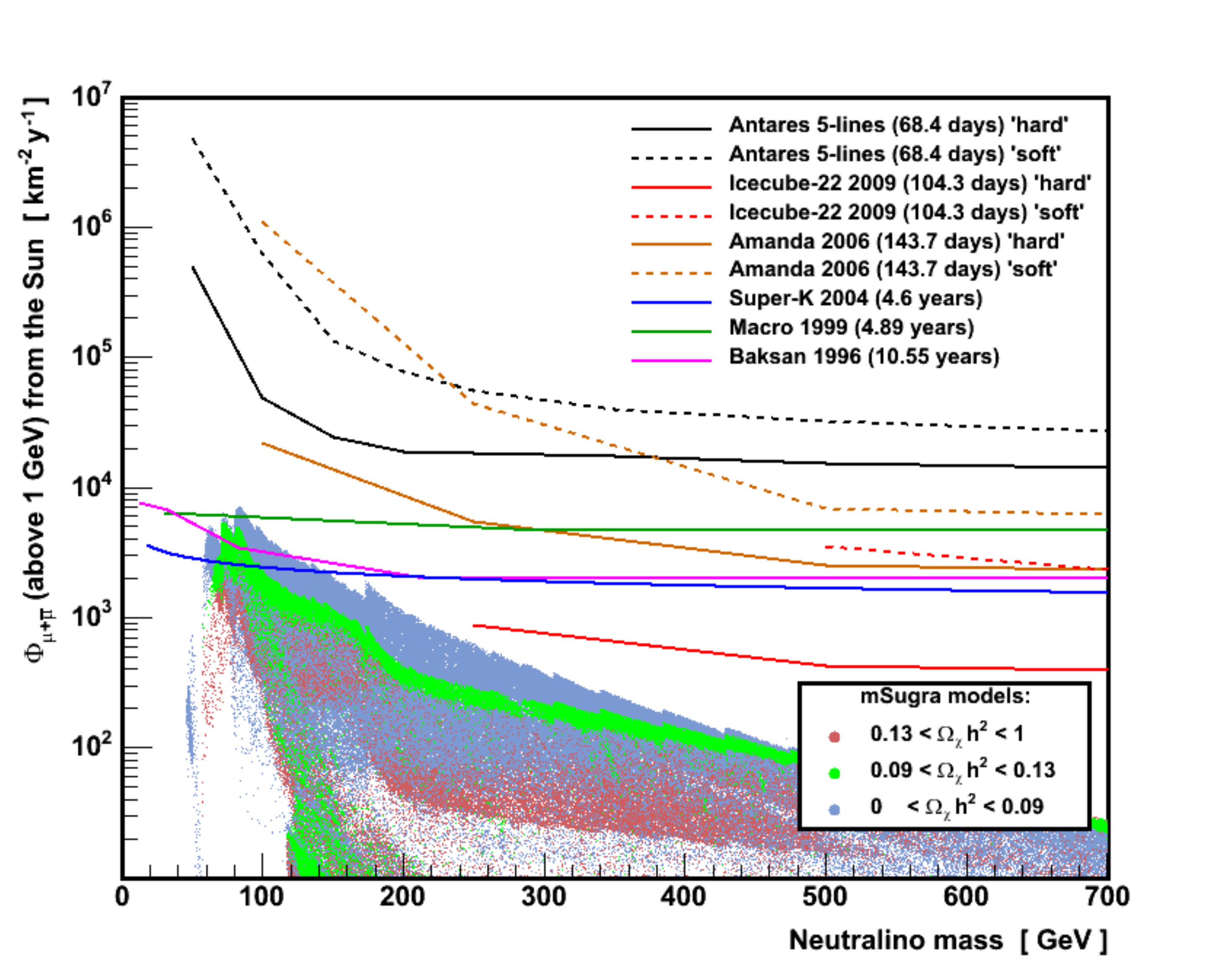} \label{sub_fig2}}
  }
  \vspace*{3mm}
  \caption{Upper limit on the neutrino flux from the Sun above $E_\nu=10$~GeV (Fig.~\ref{sub_fig1}) and the corresponding muon flux above $E_\mu=1$~GeV (Fig.~\ref{sub_fig2}) for the 5-line ANTARES period as a function of the neutralino mass, in comparison to the expected flux from the mSUGRA models considered in Sect. \ref{theory} and other experiments.}
  \label{nflmfl}
\end{figure*}

\addtocounter{figure}{-2}
 
\begin{figure}[!b]
  \centering
  \includegraphics[width=0.45\textwidth]{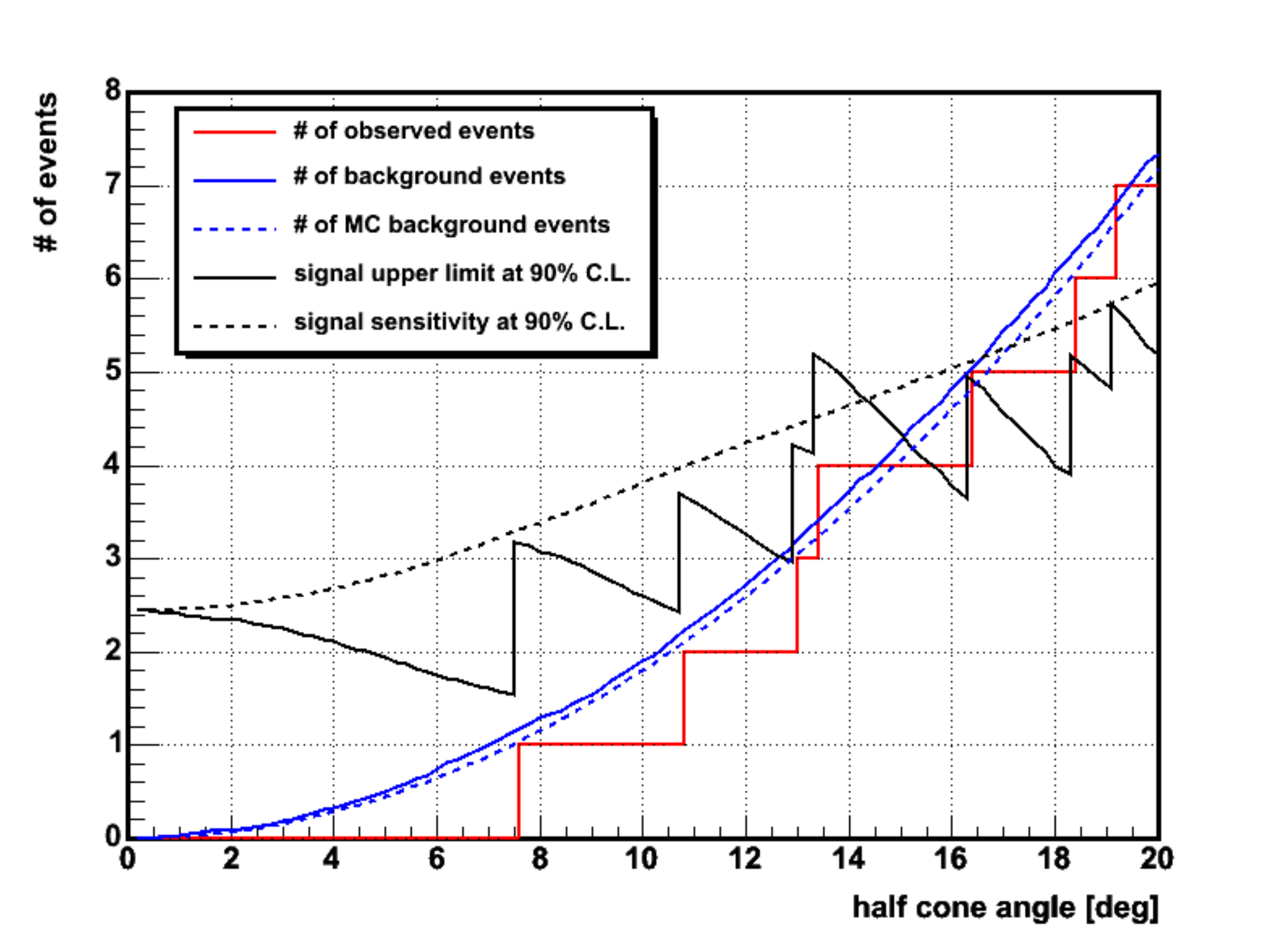}
  \caption{Number of observed neutrinos and expected background events as a function of the search cone radius around the Sun.}
  \label{N_vs_cone_official}
\end{figure}

ANTARES is currently the largest neutrino detector in the Northern Hemisphere~[\ref{antareslabel},\ref{antareslabel2},\ref{antareslabel3}]. Located at a depth of about 2.5~km in the Mediterranean Sea offshore from France, ANTARES is a Cherenkov neutrino telescope comprising 12~detection lines in an approximately cylindrical layout. Each line is comprised of up to 25~storeys, each storey contains \mbox{3~ten-inch~photomultiplier} tubes. The distance between detection lines is 70\,m, and vertically between adjacent storeys 14.5\,m, resulting in an instrumented detector volume of about 0.02~km$^3$. The detector was completed in May~2008. Prior to its completion, ANTARES has been taking data in intermediate configurations of 5~and 10~detector lines for more than one year. The angular resolution of the telescope depends on the neutrino energy~$E_\nu$ and is of the order of one degree at low energy ($E_\nu<$~1~TeV), relevant to dark matter searches.

The sensitivity of a neutrino telescope is conventionally expressed by the neutrino effective area $A_{\textrm{\footnotesize eff}}^{\nu}$. It is defined as

\vspace*{-3mm}
\begin{equation}
  R(t) \;=\; \Phi(E_\nu,t)\;A_{\textrm{\footnotesize eff}}^{\nu}(E_\nu)
  \label{nea}
\end{equation}
\vspace*{-3mm}

\noindent where $R(t)$ is the detection rate and  $\Phi(E_\nu,t)$ is the incoming neutrino flux. The ANTARES effective area (12~lines) for upgoing \mbox{$\nu_\mu+\bar{\nu}_\mu$'s} is shown as a function of the neutrino energy in the low energy regime in Fig.~\ref{NEAnew_colour}. The increase of the effective area with neutrino energy is mainly due to the fact that the neutrino-nucleon cross section as well as the muon range in water are both proportional to the neutrino energy.\\[-3mm]

\section{KM3NeT}
\label{km3net}
\vspace*{1mm}

The KM3NeT consortium aims to build a cubic-kilometer scale neutrino telescope in the Mediterranean Sea~[\ref{KM3NeTlabel},\ref{KM3NeT2label}]. During the Design Study phase of the project, several detector designs are explored. In this study (see Sect.~\ref{sensitivitysect}) we assume one of the possible detector configurations, the so-called ``reference detector''. This homogeneous cubic configuration consists of 225~(15$\times$15)~detection lines, each line carrying 37~optical modules. Each optical module contains \mbox{21~three-inch~photomultiplier} tubes. The distance between detection lines is 95\,m, and vertically between adjacent optical modules 15.5\,m, resulting in an instrumented detector volume of 1~km$^3$.\\[-3mm]

\section{Neutralino annihilation in the Sun} 
\label{theory}
\vspace*{1mm}

We calculated the $\nu_\mu+\bar{\nu}_\mu$ flux resulting from neutralino annihilation in the centre of the Sun. Instead of the general supersymmetry scenario, we used the more constrained approach of minimal supergravity (mSUGRA) in which models are characterized by four parameters and a sign:~$m_{1/2}$, $m_0$, $A_0$, $\tan(\beta)$ and sgn$(\mu)$. The calculation was done using the DarkSUSY simulation package~\cite{DarkSUSY} in combination with the renormalisation group evolution package ISASUGRA~\cite{Isasugra}. We assumed a local neutralino halo density of~0.3~GeV/cm$^3$. To investigate specifically those mSUGRA models that possess a relic neutralino density~$\Omega_\chi$ that is compatible with the cold dark matter density~$\Omega_{\chi,\textrm{\footnotesize WMAP}}$ as measured by WMAP~\cite{WMAP}, the neutrino flux was calculated for approximately four million mSUGRA models using a random walk method in mSUGRA parameter space based on the Metropolis algorithm where~$\Omega_\chi$ acted as a guidance parameter~\cite{MarkovChain}. We considered only \mbox{sgn$(\mu)=+1$} models within the following parameter ranges: \mbox{$0<m_0<8000$~GeV,} \mbox{$0<m_{1/2}<2000$~GeV,} \mbox{$-3m_0<A_0<3m_0$} and \mbox{$0<\tan(\beta)<60$.}

The resulting $\nu_\mu+\bar{\nu}_\mu$ flux from the Sun, integrated above a threshold energy of \mbox{$E_\nu=10$~GeV}, can be seen in the \mbox{$m_0$-$m_{1/2}$~plane} for six different $\tan(\beta)$ ranges in Fig.~\ref{psflux}. Models in the so-called focus point region produce the highest solar neutrino flux. In this region of mSUGRA parameter space the neutralino has a relatively large higgsino component and therefore a large neutralino vector-boson coupling. This enhances the neutralino capture rate in the Sun as well as the neutralino annihilation to vector-bosons, resulting in a relatively high neutrino flux with a relatively hard energy spectrum \cite{Nerzi}.\\[-3mm]

\section{Expected detection sensitivity} 
\label{sensitivitysect}
\vspace*{1mm}

The ANTARES detection sensitivity for neutralino annihilation in the Sun was determined by considering the irreducible background from atmospheric neutrinos and an additional 10\% of that flux due to misreconstructed atmospheric muons in a search cone of 3~degree radius around the Sun. Based on the ANTARES effective area in Fig.~\ref{NEAnew_colour}, the average background prediction after 3~years of effective data taking is \mbox{$\sim\!\!$~7} neutrinos. Assuming that only the average background rate will be measured, a 90\%~CL upper limit on the neutrino flux from the Sun for 3~years of effective data taking was derived, according to~\cite{FeldmanCousins}. This can be compared to the ANTARES detection rate from the expected neutrino flux from neutralino annihilation in the Sun. 

The resulting ANTARES detection sensitivity in the $m_0$-$m_{1/2}$ mSUGRA parameter space for six different $\tan(\beta)$ intervals is shown in Fig.~\ref{exclfig}a. The results of the analogous calculations for the KM3NeT detector outlined in Sect.~\ref{km3net} are shown in Fig.~\ref{exclfig}b. Green/yellow/red colours indicate respectively that all/some/no mSUGRA models (depending on~$A_0$ and~$\tan(\beta)$) can be excluded at 90\%~CL after 3~years of effective data taking by the considered experiment. The sensitivity of ANTARES is sufficient to put constraints on parts of the focus point region of mSUGRA parameter space, while KM3NeT would be sensitive to most of this region.\\[-3mm]

\section{Data analysis} 
\vspace*{1mm}

The data taken during the operation of the first 5~lines (Jan-Dec '07) were used to search for a possible excess in the neutrino flux from the Sun. The effective livetime of this period corresponds to 68.4~days, reduced from the full 164~days due to detector dead-time and the condition that the Sun has to be below the horizon. The number of observed neutrinos in a search cone around the Sun is shown in Fig.~\ref{N_vs_cone_official} as a function of the search cone radius. The expected number of background events from Monte Carlo simulation is in good agreement with the estimation obtained by randomising the direction and arrival time of the observed events. The upper limit and sensitivity on the number of signal events according to~\cite{FeldmanCousins} is also shown.

The corresponding upper limit on the neutrino flux from the Sun was calculated as a function of the neutralino mass assuming two extreme annihilation cases: Pure annihilation into vector-bosons and into $b\bar{b}$~quarks only, referred to as ``hard'' and ``soft'' annihilation respectively. The neutrino energy spectra at Earth for both cases were calculated as a function of the neutralino mass using the simulation package WimpSim~\cite{wimpsim}, assuming the standard oscillation scenario. The search cone used in both cases was optimized as a function of the neutralino mass by Monte Carlo simulation before data analysis. The resulting upper limit at 90\%~CL on the neutrino flux from the Sun, integrated above a threshold energy of \mbox{$E_\nu=10$~GeV}, can be seen for both cases in Fig.~\ref{sub_fig1}. Also shown is the expected neutrino flux from the mSUGRA models considered in Sect.~\ref{theory}, divided into three categories according to the compatibility of their $\Omega_{\chi}$ to $\Omega_{\chi,\textrm{\footnotesize WMAP}}$. Models indicated in green lie within $2\sigma$ of the preferred WMAP value, while models in red/blue have a higher/lower relic density.

The corresponding upper limits on the muon flux above a muon energy threshold of \mbox{$E_\mu=1$~GeV} for both annihilation channels is shown in Fig.~\ref{sub_fig2}, as well as the expected muon flux from the mSUGRA models considered in Sect.~\ref{theory} and upper limits determined by other indirect detection experiments~[\ref{icecubelabel},\ref{amandalabel},\ref{superklabel},\ref{macrolabel},\ref{baksanlabel}].\\[-3mm]

\section{Conclusion} 
\vspace*{1mm}

The first upper limits of the ANTARES detector in its intermediate 5~line configuration on the neutrino and muon flux from neutralino annihilation in the Sun have been obtained. Monte Carlo simulations based on the full ANTARES detector and a possible configuration of the future KM3NeT detector show that these experiments are sensitive to the focus point region of the mSUGRA parameter space.\\[-3mm]

\label{icrc0031:end}

\setcounter{figure}{0}
\setcounter{table}{0}
\setcounter{footnote}{0}
\setcounter{section}{0}





\hyphenation{abcdef-ghijklmnoprstuwxyz IEEEtran}

\title{Skymap for atmospheric muons at TeV energies measured in deep-sea neutrino telescope ANTARES}

\author{\IEEEauthorblockN{Salvatore Mangano\IEEEauthorrefmark{1}, for the ANTARES collaboration}
                            \\
\IEEEauthorblockA{\IEEEauthorrefmark{1}IFIC - Instituto de F\'isica Corpuscular, Edificio Institutos de Investigati\'on, \\
                                 Apartado de Correos 22085, 46071 Valencia, Spain}}

\shorttitle{S. Mangano \etal Skymap for atmospheric muons in ANTARES}
\maketitle
\label{icrc1131:begin}

\begin{abstract}
Recently different experiments mention to have observed a large
scale cosmic-ray anisotropy at TeV energies, e.g. Milagro, Tibet and Super-Kamiokande.
For these energies the cosmic-rays are expected to be nearly isotropic. Any
measurements of cosmic-rays anisotropy could bring some information about
propagation and origin of cosmic-rays.

Though the primary aim of the ANTARES neutrino telescope is the
detection of high energy cosmic neutrinos, the detector measures
mainly down-doing muons, which are decay products of cosmic-rays
collisions in the Earth's atmosphere. This proceeding describes
an anlaysis method for the first year measurement of down-going atmospheric muons at
TeV energies in the ANTARES experiment, when five out of the final
number of twelve lines were taking data.
\end{abstract}

\begin{IEEEkeywords}
 Underwater neutrino telescope, Skymap, Atmospheric muons, Cosmic rays
\end{IEEEkeywords}
 
\section{Introduction}

Spatial distributions of the muon flux have been measured by the Milagro
observatory \cite{Milagro}, Super-Kamiokande \cite{SuperK1, SuperK2}
as well as by the Tibet Air Shower Array \cite{Tibet}.  
These experiments report large-scale
anisotropy at primary cosmic-ray energies in the TeV range. 
The measured deviation from an isotropic distribution is of the order of $0.1\%$
and the excess region and deficit region have the size of several tens degrees. 

The ANTARES neutrino telescope \cite{Antares}, located in the Northern hemisphere, can detect
down-going muons from the North as the
above mentioned detectors. 
IceCube has also mentioned that an
investigation on a high statistic down-going muons data is ongoing.

ANTARES is located on the bottom of the
Mediterranean Sea, \mbox{40 km} off the French coast at $42^o 50'$ N,
$6^o 10'$ E.  The main objective is to detect high energy neutrinos
from galactic or extragalactic sources.  Neutrinos are detected
indirectly through the detection of Cherenkov light produced by
relativistic muons emerging from charged-current muon neutrino
interactions in the surroundings.

The detector has been successfully deployed between March 2006 and May
2008. In its full configuration the 
detector comprises 12 vertical detection lines, each of about 450 m height, 
installed at a depth of about \mbox{2500 m}. The lines
are set from each other at a distance of 60 m to 70 m. They are
anchored at the sea floor and held taut by buoys. The instrumented
part of the line starts at \mbox{100 m} above the sea floor.
Photomultipliers are grouped in triplets (up to 25 floors on each line) for a
better rejection of the optical background. They are oriented with
their axis pointing downward at an angle of $45^{\circ}$ with respect
to the vertical in order to maximize the effective area for upward-going
tracks.  ANTARES is operated in the so called all-data-to-shore mode,
which means that all photomultiplier digitized information is sent to
shore and treated in a computer farm at the shore station.  These data
are mainly due to background light caused by bioluminesence and
$^{40}K$ decay. The background light varies in time and can cause
counting rates of the photomultiplier tubes that varies between \mbox{50 kHz} 
and 500 kHz. The data flow rate at the shore station are reduced
by fast processing of the events looking for interesting physics,
which is a challenge because of the high background rates. The main
idea of the on-shore handling of these data is to take into account the
causal connection between photomultiplier tubes signals which are
compatible with the light produced by a relativistic muon.  The
reconstruction of muon tracks is based on the measurements of the
arrival times of Cherenkov photons at photomultipliers and their
positions.

Although ANTARES is optimized for upward-going particle
detection, the most abundant signal is due to the atmospheric
down-going muons. They are produced in air showers induced by
interactions of primary cosmic-rays in the Earth's atmosphere. The
muons are the most penetrating particles in such air showers. Muons
with energies above around 500 GeV can reach the detector, producing
enough amount of Cherenkov light to reconstruct the direction of the
muon.  At larger zenith angle the minimum muon energy increases. 
The muons represent a high statistic data
set that can be used for calibration purposes as well as to check the
simulation of the detector response.
The atmospheric muons will also provide information about primary cosmic rays
at energies above few TeV. For these energies the cosmic-rays
are mostly of Galactic origin and are expected to be nearly isotropic due to interactions 
with the Galactic magnetic field.   

The cosmic-ray muon spatial distribution at these energies may be not isotropic for different reasons, like the
instabilities related to temperature and pressure variations in the
Earth's atmosphere. The temperature in the upper atmosphere is related
to the density of the atmosphere and thus to the interaction of the
particles in the air showers which will affect the muon flux.
This time variations of the cosmic-ray muon flux could mimic a flux  anisotropy.
This effect is canceled out when the time scale of data taking is larger
than the time of thermodynamic changes in the Earth's atmosphere.  At
cosmic-ray energies lower of around a TeV the movements of solar
plasma through the heliosphere may change the Earth's magnetic field,
causing a modulation of the cosmic-ray anisotropy. Furthermore the
Compton-Getting effect predicts a dipole effect due to the moving of
the Earth with respect to an isotropic cosmic-ray rest system. If the
Earth is moving in the rest system, the cosmic ray flux from the
forward direction becomes larger.

\section{Skymap for down going muons}
The data used in this analysis cover the period from 
February 2007 until the middle of December 2007, during which only five
out of twelve detector lines were installed.  More than $10^7$ events
were collected, most of which are atmospheric muons. The tracks are
reconstructed with a linear fit algorithm which uses hits selected
using the time information. The algorithm used for this analysis has
the advantage to be fast, robust and has a high purity for atmospheric
muons.  The disadvantage is that this algorithms has not the ultimate
angular resolution which can be achieved by ANTARES with
a more sophisticated reconstruction algorithm.
 
Only down-going tracks detected by at least two lines and six floors
are considered for the analysis.  The angular resolution for this
selection cuts is around seven degrees.

 \begin{figure}[!t]
  \centering
  \includegraphics[width=2.5in]{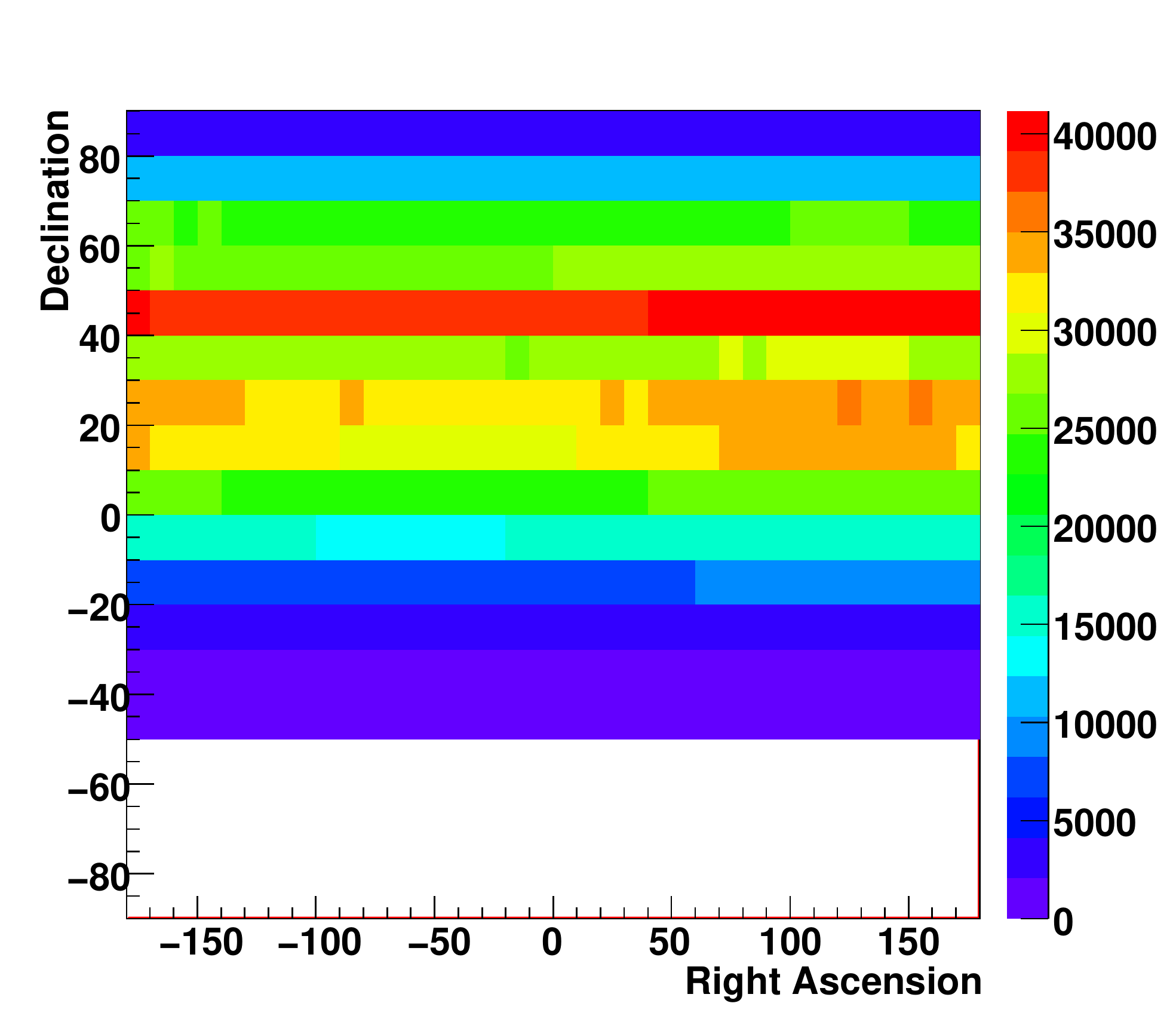}
  \caption{Muon events in equatorial coordinates for the period with
  five operating detection lines. The sky is divided into $10^{\circ}
  \times 10^{\circ}$ cells. Declinations below $-47^{\circ}$ are
  always below the horizon and are thus invisible to the detector.}
  \label{simp1_fig}
 \end{figure}

 \begin{figure}[!t]
  \centering
 \includegraphics[width=2.5in]{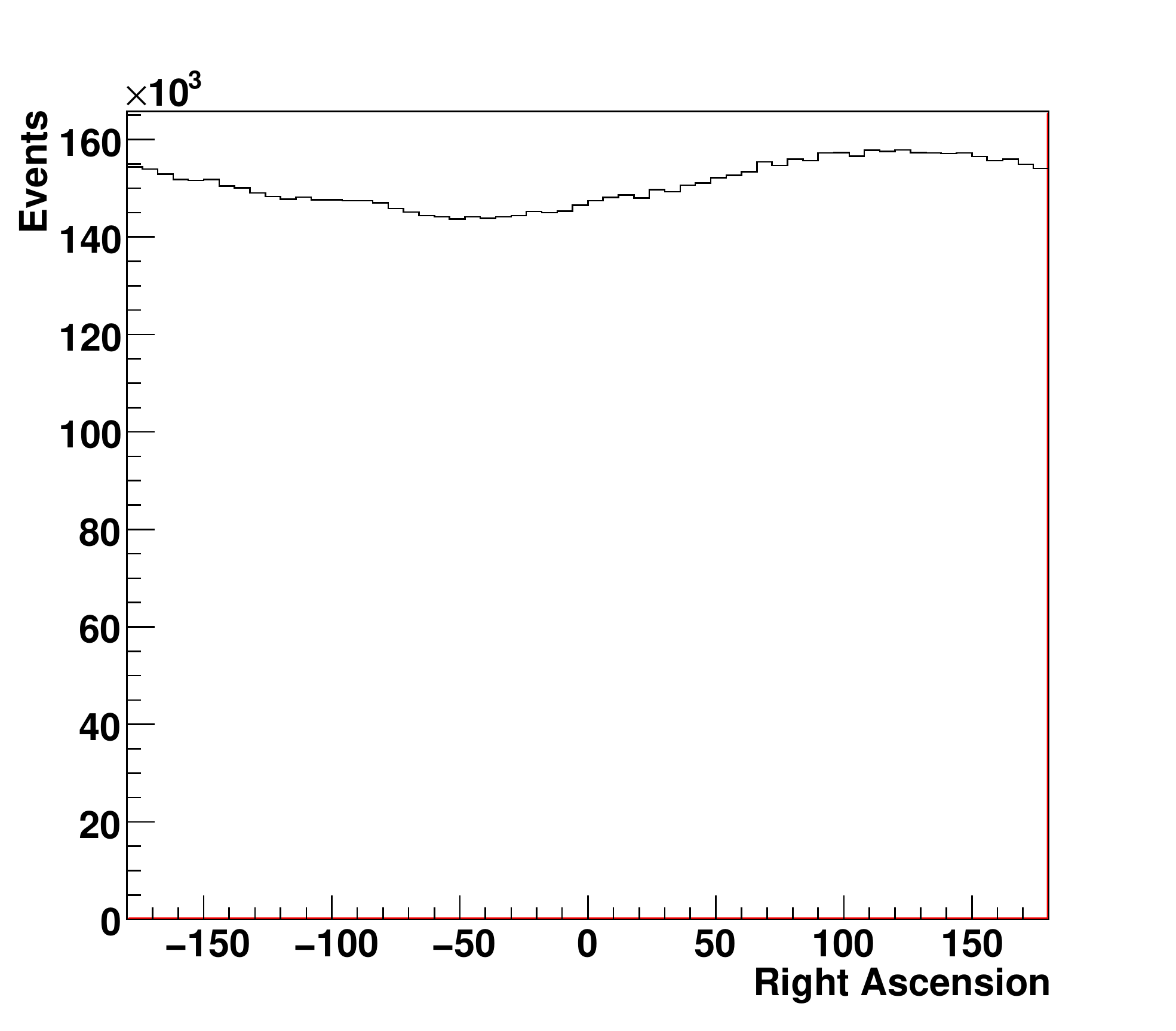}  
 \caption{Muon events as function of the right ascension without any corrections.}
  \label{simp2_fig}
 \end{figure}

The track directions are given through the zenith and azimuth angle of
the ANTARES local coordinates. Considering the time of the
reconstructed tracks, the local coordinates are translated into
equatorial coordinates: right ascension and declination. Figure
\ref{simp1_fig} shows the results of the corresponding data sample in
equatorial coordinates. The variation in declination is affected by
the visibility and by the propagation length of the muons in the
atmosphere.  The number of events as a function of the
right ascension should be uniform, because the
distribution of cosmic rays with a given zenith direction in the ANTARES local
coordinates travel a line in the declination coordinate.  A given
direction in the ANTARES local coordinates 
returns to the same right ascension after one sidereal day
because of the Earth's rotation.  The right ascension distribution is
shown in figure \ref{simp2_fig}, which is not uniform.  

The exposure for different directions is not
uniform because the detector was not constantly taking data. Several
days with high background rates have been removed for the
analysis while during normal operation the detector is   
halted regularly for calibration purposes. 
This makes the data-taking not completely uniform along the day. The
right ascension coordinates can be corrected for the introduced
fluctuations in the exposure time.  Taking the exposure time for the
analyzed runs as well as the measured direction of the track in local
coordinates, the modulation in right ascension can be calculated. The
calculated amplitude of this modulation matches well the data shown in
figure \ref{simp2_fig}. Taking these values the data right ascension
can be normalized by exposure. 

After this correction, fluctuations,
above or below the statistical expectation can be calculated. For a
given declination band with $m$ bins in a right ascension, the
probability to observe the number of events in each particular bin is
calculated.  The probability for the average expectation to fluctuate
to the observed number of events or more in this bin is calculated
with the equation
$$P(n \geq n_{obs} | \nu_b)=
1-\sum_{n=0}^{n_{obs}-1}\frac{\nu_b^n}{n!} e^{-\nu_b},$$ where $\nu_b$
is the number of expected events which is estimated from the average
background in the declination band containing that bin and $n_{obs}$ is
the observed number of events in that bin. Finally, the significance
of the cosmic ray signal of one bin is defined as
$$S=-log_{10}(P).$$

The statistics will be enlarged 
by using data of the following years. 
Assuming that the detector effects are under control, 
then the expected sensitivity to measure cosmic-ray anisotropy is 
only constrained by the number of detected muons.

\section{Conclusion}
Large scale cosmic ray anisotropies at TeV energies have been observed
by the Milagro observatory, the Tibet Air Shower Array and
Super-Kamiokande. ANTARES has a high
statistics of TeV down-going muons available, 
\newpage
which allows to search
for possible anisotropies in the primary flux. A first attempt to
reproduce a TeV energy muon skymap is ongoing, with the one year
ANTARES data, where only five lines were deployed.

\label{icrc1131:end}

\setcounter{figure}{0}
\setcounter{table}{0}
\setcounter{footnote}{0}
\setcounter{section}{0}




\hyphenation{abcdef-ghijklmnoprstuwxyz IEEEtran phe-no-me-na o-pe-ra-ted pro-per-ly tra-ver-sing}

\title{Search for Exotic Physics with the ANTARES Detector}

\author{\IEEEauthorblockN{
			  Gabriela Pavalas\IEEEauthorrefmark{1} and
                          Nicolas Picot Clemente\IEEEauthorrefmark{2},
                          \\
			  on behalf of the ANTARES Collaboration}
                            \\
\IEEEauthorblockA{\IEEEauthorrefmark{1}Institute for Space Sciences, Bucharest-Magurele, Romania}
\IEEEauthorblockA{\IEEEauthorrefmark{2}CNRS / Centre de Physique des Particules de Marseille, Marseille, France}}

\shorttitle{G. Pavalas \etal Exotic Physics with ANTARES}
\maketitle
\label{icrc0695:begin}

\begin{abstract}
Besides the detection of high energy neutrinos, the ANTARES telescope offers an opportunity
to improve sensitivity to exotic cosmological relics.  
In this article we discuss the sensitivity of the ANTARES detector to relativistic monopoles and slow nuclearites. 
Dedicated trigger algorithms and search strategies are being developed to search for them. The data filtering, background rejection selection criteria are described, as well as the expected sensitivity of ANTARES
to exotic physics.

\end{abstract}
\begin{IEEEkeywords}
ANTARES, magnetic monopoles, nuclearites
\end{IEEEkeywords}
\section{Introduction}
The ANTARES neutrino telescope is aimed to observe high energy cosmic neutrinos through
the detection of the Cherenkov light produced by up-going induced muons. However, the
ANTARES detector is also sensitive to a variety of exotic particles, and can provide an
unique facility for the search of magnetic monopoles and nuclearites.

\section{The ANTARES detector}

The ANTARES detector has reached its nominal size in May 2008. The 884 Optical Modules (OM) are deployed on
twelve vertical lines in the Western Mediterranean, at depths between 2050 and 2400 meters.
The OMs, consisting of a glass sphere housing a 10'' Hamamatsu photomultiplier (PMT)~\cite{Ant_PMT}, are arranged by triplet per storey. Each detector line, made of 25 storeys, is connected via interlinks to a Junction Box, itself connected to the 
shore station at La Seyne-sur-Mer through a 40 km long electro-optical cable. The strategy 
of the ANTARES data acquisition is based on the ``all-data-to-shore'' concept~\cite{Ant_Acq}.
This implementation leads to the transmission of all raw data above a given threshold to shore, where different
triggers are applied to the data for their filtering before storage. 
\\
For the analysis presented here, only the two general trigger logics operated up to now have been considered. 
Both are based on local coincidences. A local coincidence (L1 hit) is defined either 
as a combination of two hits on two OMs of the same storey within 20 ns, or as a single hit 
with a large amplitude, typically 3 pe. The first trigger, a so-called directional trigger, 
requires five local coincidences anywhere in the detector but causally connected, within a 
time window of 2.2 $\mu$s. The second trigger, a so-called cluster trigger, requires two
T3-clusters within 2.2 $\mu$s, where a T3-cluster is a combination of two L1 hits in adjacent 
or next-to-adjacent storeys. When an event is triggered, all PMT pulses are recorded over
2.2 $\mu$s.
\\
The ANTARES observatory was build gradually, 
giving rise to various detector layouts used for physics analysis. The 5-line, 10-line and 
12-line detector configurations match with data taken from January 2007, from December 2007 
and from May 2008, respectively. 

\section{Magnetic monopoles}
\subsection{Introduction}

Most of the Grand Unified Theories (GUTs) predict the creation of magnetic monopoles in the early Universe. Indeed, in 1974, 't Hooft \cite{thooft} and Polyakov \cite{polyakov} showed independently that each time a compact and connected gauge group is broken into a connected subgroup, elements caracterising well a magnetic charge, as introduced by Dirac in 1931 \cite{Dirac}, appear.\\
These particles are topologically stable and carry a magnetic charge defined as a multiple integer of the Dirac charge $g_D=\frac{\hbar c}{2e}$, where e is the elementary electric charge, c the speed of light in vacuum and $\hbar$ the Planck constant. Depending on the group, the masses inferred for magnetic monopoles can take range over many orders of magnitude, from $10^8$ to $10^{17}$ GeV.

As magnetic monopoles are stable, and so would survive until now, they should have been very diluted in the Universe, as predicted by numerous theoretical studies which set stringent limit on their fluxes, like the Parker flux limit \cite{Parker}. More stringent limits were set recently by different experiments (MACRO \cite{MACRO}, AMANDA \cite{AMANDA}, ...).

The development of neutrino astronomy in the last decade led to the construction of huge detectors, which allow new hopes in the search for magnetic monopoles.
Actually, the ANTARES detector seems to be well designed to detect magnetic monopoles, or at least to improve limits on their fluxes, as described below.
\subsection{Signal and background simulations}

Since fast monopoles have a large interaction with matter, they can lose large amounts of energy in the terrestrial environment. The total energy loss of a relativistic monopole with one Dirac charge is of the order of $10^{11}$ GeV \cite{EnergyLoss} after having crossed the full diameter of the Earth. Because magnetic monopoles are expected to be accelerated in the galactic coherent magnetic field domain to energies of about $10^{15}$ GeV \cite{CosmicFlux}, some could be able to cross the Earth and reach the ANTARES detector as upgoing signals.

The monopole's magnetic charge $g=n g_D$ can be expressed as an equivalent electric charge $ g = 68.5 n e $, where n is an integer. Thus relativistic monopoles with $\beta\geq0.74$ carrying one Dirac charge will emit a large amount of direct Cherenkov light when traveling through the ANTARES detector, giving rise to $\sim 8500$ more intense light than a muon. The number of photons per unit length ($cm^{-1}$) emitted on the path of a monopole is shown in figure \ref{nberofphotons}, as a function of the velocity of the monopole up to $\gamma = 10 $ ($\beta=0.995$).
\begin{figure}[!h]
  \centering
  \includegraphics[width=2.5in]{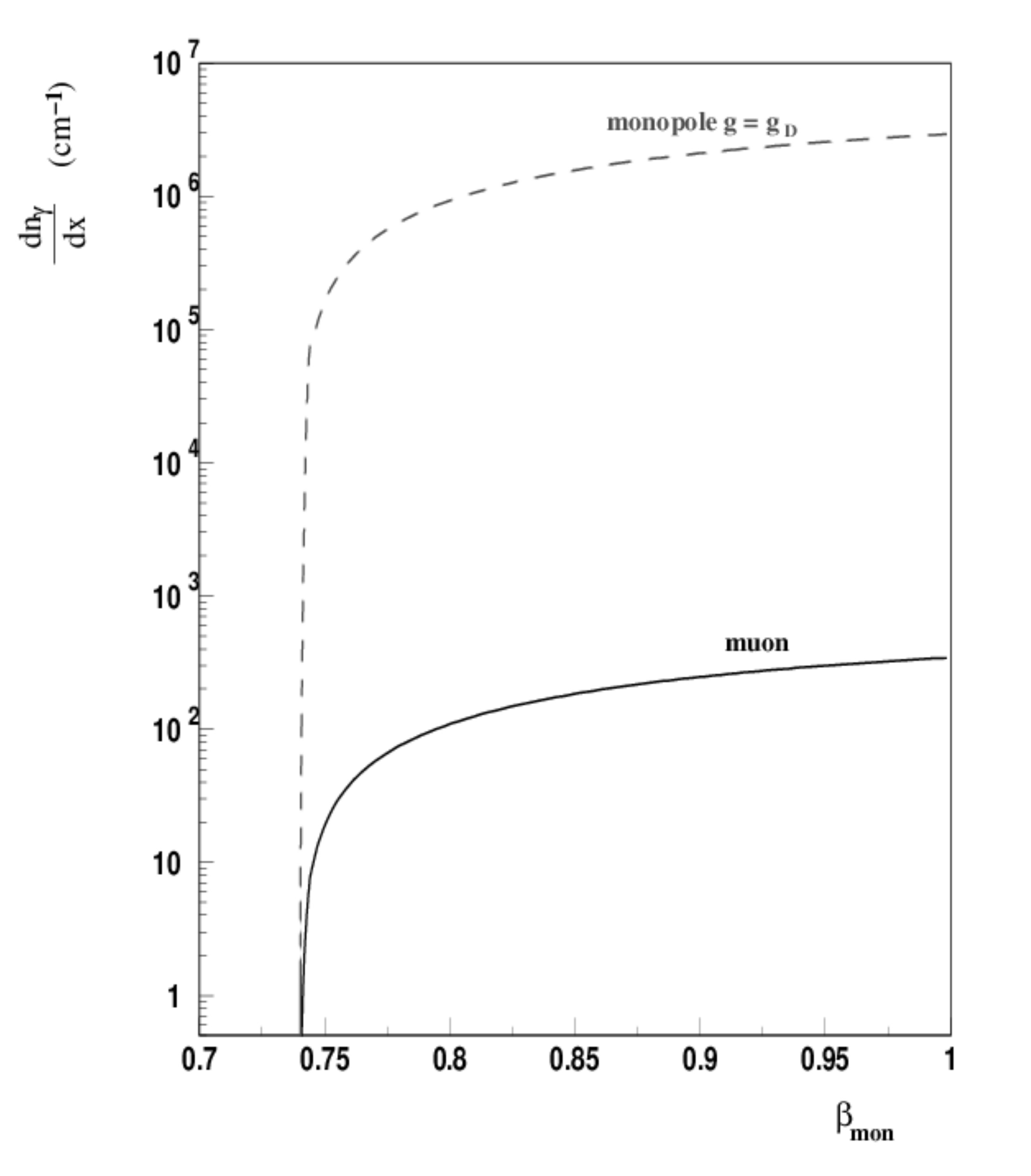}
  \caption{Number of emitted photons per unit length ($cm^{-1}$) by a magnetic monopole with a charge $g=g_D$ through direct Cherenkov emission (dashed line) compared to the number of photons emitted by a muon (black line), as a function of their velocities.}
  \label{nberofphotons}
 \end{figure}
 
For the analysis, monopoles have been simulated inside an optimized volume containing the 12-line detector, for six ranges of velocities between $\beta=0.74$ and $\beta=0.995$.
In addition, downgoing atmospheric muons have been simulated using the CORSIKA package \cite{Corsika}, as well as upgoing and downgoing atmospheric neutrinos according to the Bartol flux \cite{Bartol1,Bartol2}. Optical background from $^{40}K$ decay has been added to both magnetic monopole signal and atmospheric muon and neutrino background events.
\subsection{Search strategy}

The 12-line detector data are triggered by both trigger logics, the directional and the cluster triggers (see section II). A comparison of efficiency was therefore performed on magnetic monopoles and restricted only to upward-going events. 
As the efficiency of the directional trigger was found to be lower than for the cluster trigger, it was decided to perform searches for upgoing magnetic monopoles with the cluster trigger only.

The standard reconstruction algorithm, developed in ANTARES for upward-going neutrino selection, and mainly based on a likelihood maximization, was applied. In order to select upgoing particles, only reconstructed events with a zenith angle lower than 90$^{\circ}$ were selected.
However, muon bundles are difficult to reconstruct properly and some of them can be reconstructed as upward-going events.\\
As it is shown in figure \ref{DistribT3Norm}, a magnetic monopole traversing the detector will emit an impressive quantity of light, compared to atmospheric muons or muons induced by atmospheric neutrinos.
\begin{figure}[!h]
  \centering
  \includegraphics[width=2.5in]{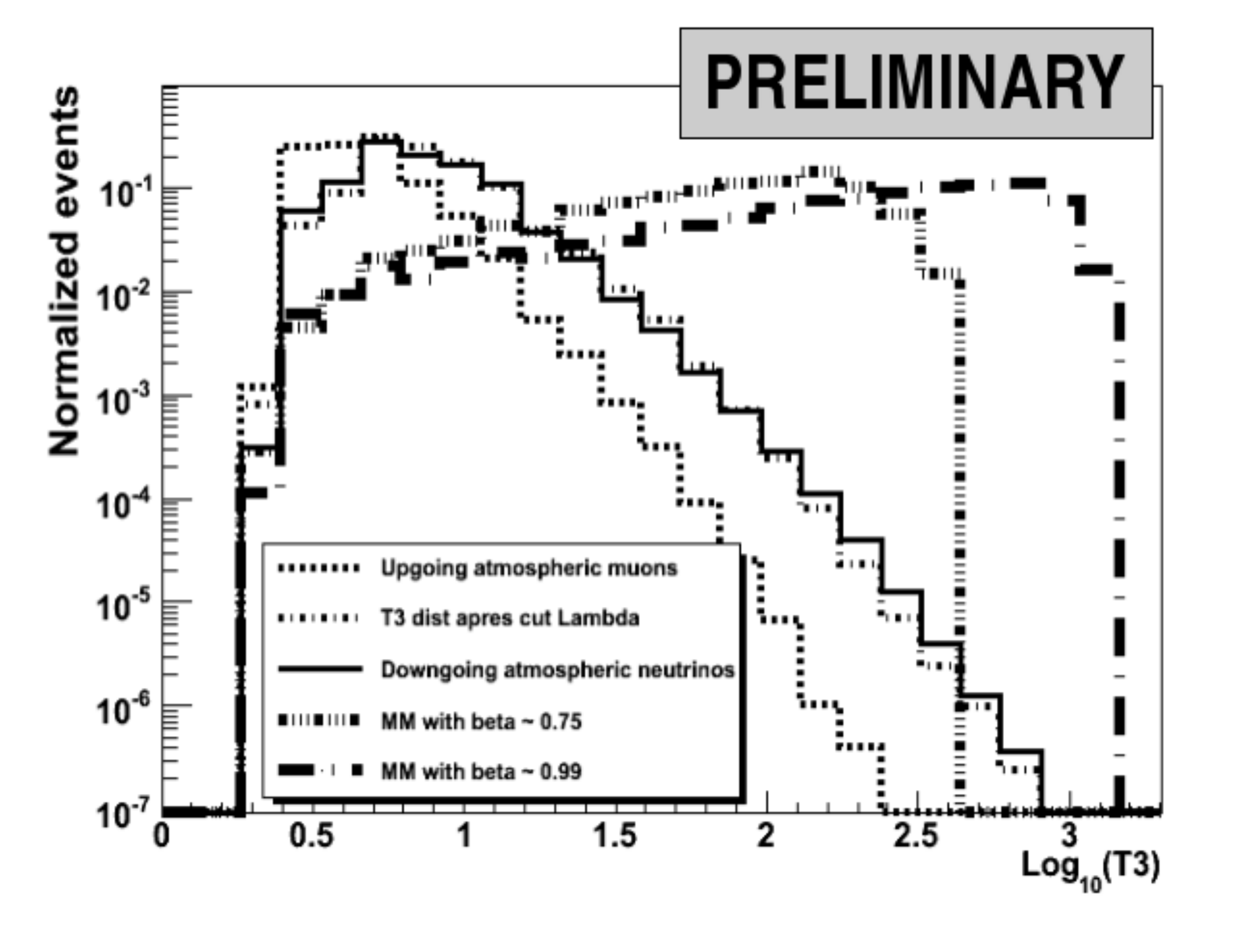}
  \caption{Normalized events as a function of the number of T3 clusters for downgoing atmospheric muons, upgoing and downgoing atmospheric neutrinos, and upgoing magnetic monopoles with $\beta \sim 0.75$ and $\beta \sim 0.99$.}
  \label{DistribT3Norm}
 \end{figure}
The large amount of induced hits in the detector, more precisely the number of T3 clusters, is therefore used as a criteria to remove part of the atmospheric background.

Before applying a supplementary cut to reduce the remaining background, 10 active days of golden\footnote{Golden data assumes experimental data complying with certain selection criteria like low baserate and burstfraction.} data were taken as reference to check the data Monte Carlo agreement. The comparison of T3 distributions between the data and the background simulation for 10 days is shown in figure \ref{DistribT3_10days}.
\begin{figure}[!h]
  \centering
  \includegraphics[width=2.5in]{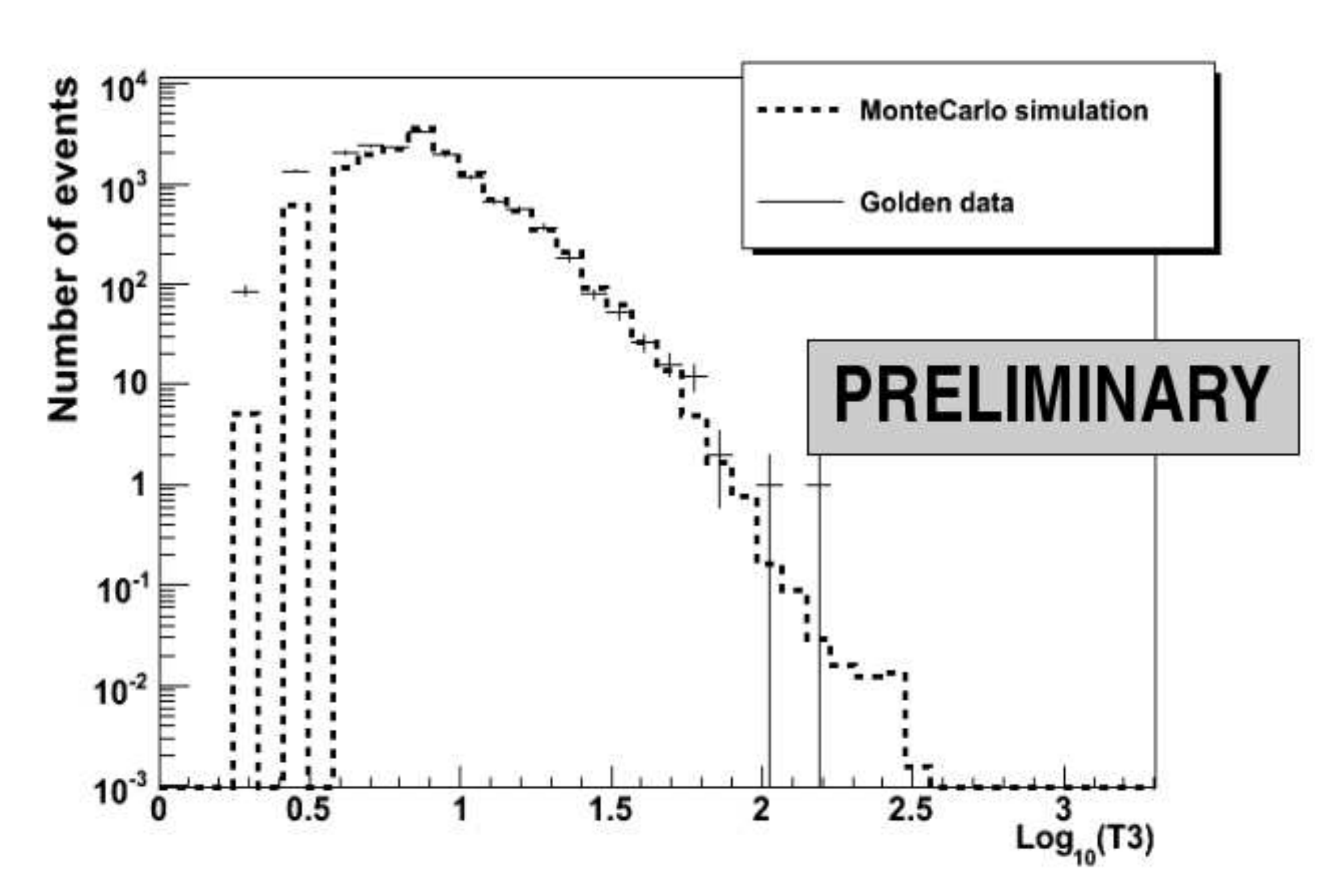}
  \caption{Comparison of T3 distributions between data and Monte Carlo simulations for 10 days of data taking.}
  \label{DistribT3_10days}
 \end{figure}

We optimized the cuts on the number of T3 clusters to maximize the 90$\%$ C.L. sensitivity, calculated with the usual Feldman-Cousins formula\cite{Feld}, for magnetic monopoles after one year of data taking.
In the optimization process the same selection criteria have been applied to calculate the sensitivity to magnetic monopole events over the whole velocity range $0.74 \leq \beta \leq 0.995$.

Finally the 90$\%$ C.L. sensitivity for this range was found, for a cut of at least 140 T3 clusters, for which about 1.7 background events are expected.
The 90$\%$ C.L. sensitivity for ANTARES after one year of data taking is of the order of $\sim 1\cdot10^{-17}cm^{-2}s^{-1}sr^{-1}$.

\section{Nuclearites}
\subsection{Introduction}
    Nuclearites are hypothetical nuggets of strange quark matter that could be present in
cosmic radiation. Their origin is related to energetic astrophysical phenomena
(supernovae, collapsing binary strange stars, etc.). Down-going nuclearites could reach
the ANTARES depth with velocities $\sim$ 300 km/s, emitting blackbody radiation at visible
wavelengths while traversing sea water.

    Heavy nuggets of strange quark matter ($M\geq10^{10}$GeV), known as nuclearites, would be electrically neutral; the small positive electric charge of the quark core would be neutralized by electrons forming an electronic cloud or found in week equilibrium inside the core. The relevant energy loss mechanism is represented by the elastic collisions with the atoms of the traversed media, as shown in ref. \cite{ruj}:
  \begin{equation}
    \frac{dE}{dx} = -\sigma\rho{v^2},
    \label{en_loss}
   \end{equation}  
where $\rho$ is the density of the medium, $v $ is the nuclearite velocity and $\sigma$ its geometrical cross section:
 
   \begin{displaymath}
        \sigma = \left\{ \begin{array}{ll}
                        \pi(3M/4\pi\rho_N)^{2/3} & \mbox{for $M\geq8.4*10^{14}$ GeV};\\
                       \pi\times10^{-16} \mbox{cm}^2 &  \mbox{for lower masses}.
                    
	             \end{array}
	             \right. 
	\label{sigma_cross}
     \end{displaymath}
with $\rho_N=3.6\times10^{14}$ g cm$^{-3}$. The mass limit in the above equation corresponds to a radius of the strange quark matter of 1\AA. Assuming a nuclearite of mass $M$ enters the atmosphere with an initial (non-relativistic) velocity $v_0$, after crossing a depth L it will be slowed down to

 \begin{equation}
    v(L) = v_0e^{-\frac{\sigma}{M}\int_0^L\rho dx}   
    \label{velocity}
   \end{equation} 
where $\rho$ is the air density at different depths. Considering the parametrization of the standard atmosphere from \cite{shi}:

 \begin{equation}
    \rho(h) = ae^{-\frac{h}{b}}=ae^{-\frac{H-l}{b}}, 
    \label{density}
   \end{equation} 
where a=$1.2\times10^{-3}$ g cm$^{-3}$ and b$\simeq8.57\times10^5$ cm, H$\simeq$50 km is the total height of the atmosphere, the integral in Eq.\ref{velocity} may be solved analytically.

\begin{figure}
\centering
\includegraphics[width=2.5in]{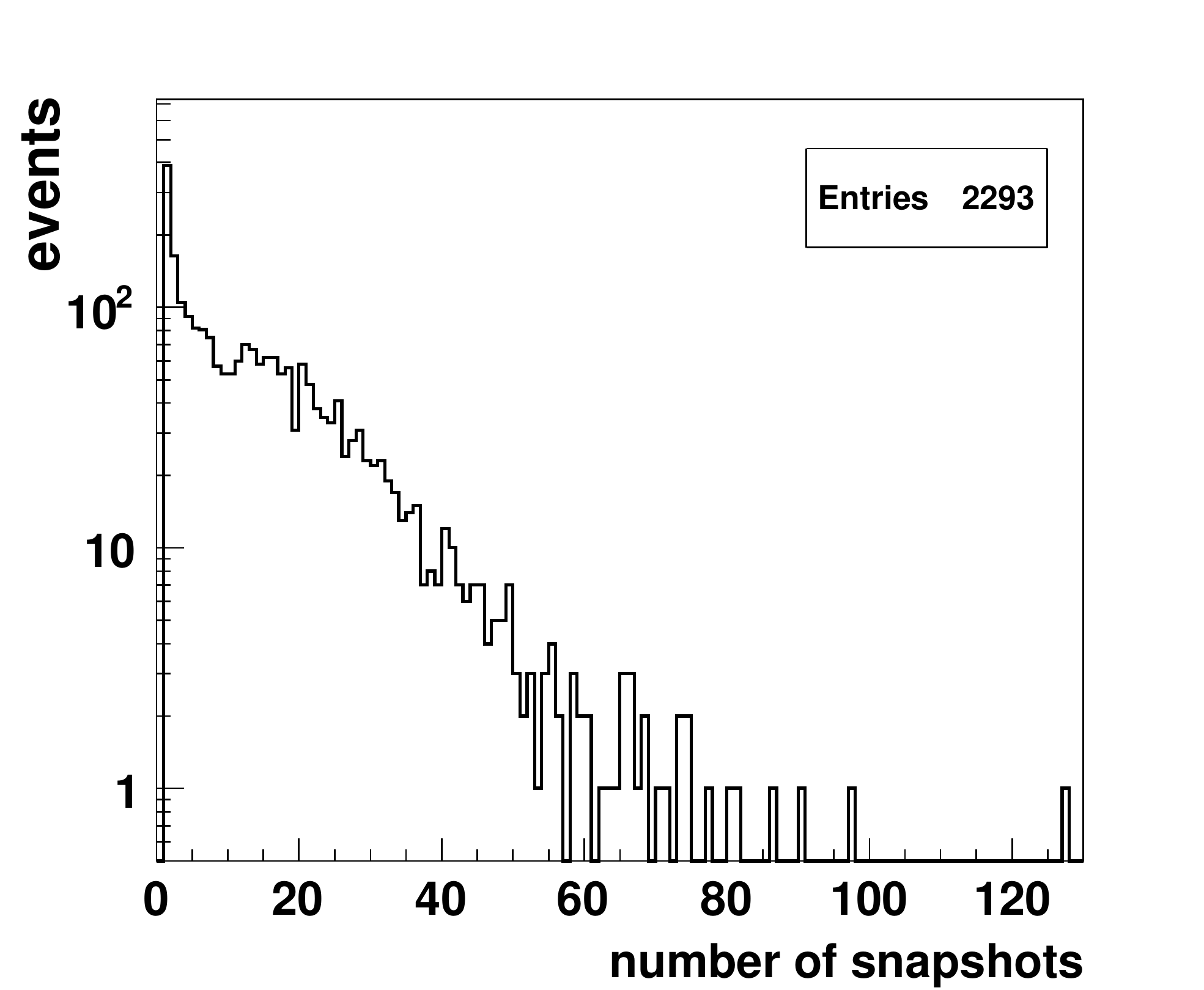}
\caption{Snapshot distribution obtained for simulated nuclearite events with masses 3$\times10^{16}$, 10$^{17}$ and 10$^{18}$ GeV.}
\label{mult}
\end{figure}

The propagation of nuclearites in sea water is described also by Eq. \ref{velocity}, assuming $\rho=1$ g cm$^{-3}$ and substituting $v_0$ with the speed value at the Earth surface. Nuclearites moving into the water could be detected because of the black-body radiation emitted by the expanding cylindrical thermal shock wave \cite{ruj}. The luminous efficiency (defined as the fraction of dissipated energy appearing as light) was estimated, in the case of water, to be $\eta\simeq3\times10^{-5}$ \cite{ruj}. The number of visible photons emitted per unit path length can be computed as follows:
 \begin{equation}
    \frac{dN_{\gamma}}{dx}=\eta\frac{dE/dx}{\pi(eV)},   
    \label{en_eta}
   \end{equation} 
assuming the average energy of visible photons $\pi$ eV.

\subsection{Search strategy and results}

 The Monte Carlo simulation of nuclearite detection in ANTARES assumes only the down-going part of an isotropic flux of nuclearites and an initial velocity (before entering the atmosphere) of $\beta=10^{-3}$. A typical nuclearite event would cross the Antares detector in a characteristic time of $\sim$ 1 ms, producing a luminosity that would exceed that of muons by several orders of magnitude. Down-going atmospheric muons represent the main background for nuclearite events.  
This analysis refers to simulated nuclearite and muon events using the 5-line detector configuration and 84 days of data taken from June to November 2007.   
\\
	Simulated nuclearite and muon events have been processed with the directional trigger, that operated during the 5-line data acquisition. Background was added from a run taken in July 2007, at a baserate of 63.5 kHz. We found for nuclearites a lower mass limit detectable with the directional trigger of $3\times10^{16}$ GeV. 
Nuclearite events were simulated for masses of $3\times10^{16}$ GeV, $10^{17}$ GeV and $10^{18}$ GeV.
The atmospheric muons were generated with the MUPAGE code \cite{MUP}. The parameters used in our analysis comprised the number of L1 triggered hits (see Section II), the number of single hits (L0 hits, defined as hits with a threshold greater than 0.3 photoelectrons), the duration of the snapshot (defined as the time difference between the last and the first L1 triggered hits of the event) and total amplitude of hits in the event. Data were reprocesed with the directional trigger.
We obtained a good agreement between simulated muon events and data.

The algorithm of the directional trigger selects from all the hits produced by a nuclearite only those that comply to the signal of a relativistic muon. These hits can be contained in a single snapshot or multiple snaphots for a single event. 

Because nuclearites are slowly moving particles, multiple snapshots belonging to a single nuclearite event may span into intervals from tens of $\mu$s up to $\sim$1ms. The multiplicity of snapshots in the simulated nuclearite events is presented in Fig. \ref{mult}.     
The majority of the snapshots produced by nuclearites from the studied sample are of short duration ($<$500 ns), see Fig. \ref{dt}.  
Long duration snapshots (up to tens of $\mu$s) are also presented, due to the large light output of events with masses $\geq 10^{17}$ Gev.

Snapshots of nuclearites passing through the detector would also be characterised by a larger number of single hits than for atmospheric muons.  
In the following, the selection criteria for nuclearite signal were obtained considering only the data sample as background.
The distributions for data and simulated nuclearite events in the 2-dimensional plot L1 triggered hits vs L0 single hits show a clear separation, that was optimized using the linear cut presented in Fig. \ref{L10_hits}. This cut reduces the data by 99.998$\%$.  

\begin{figure}
\centering
\includegraphics[width=2.5in]{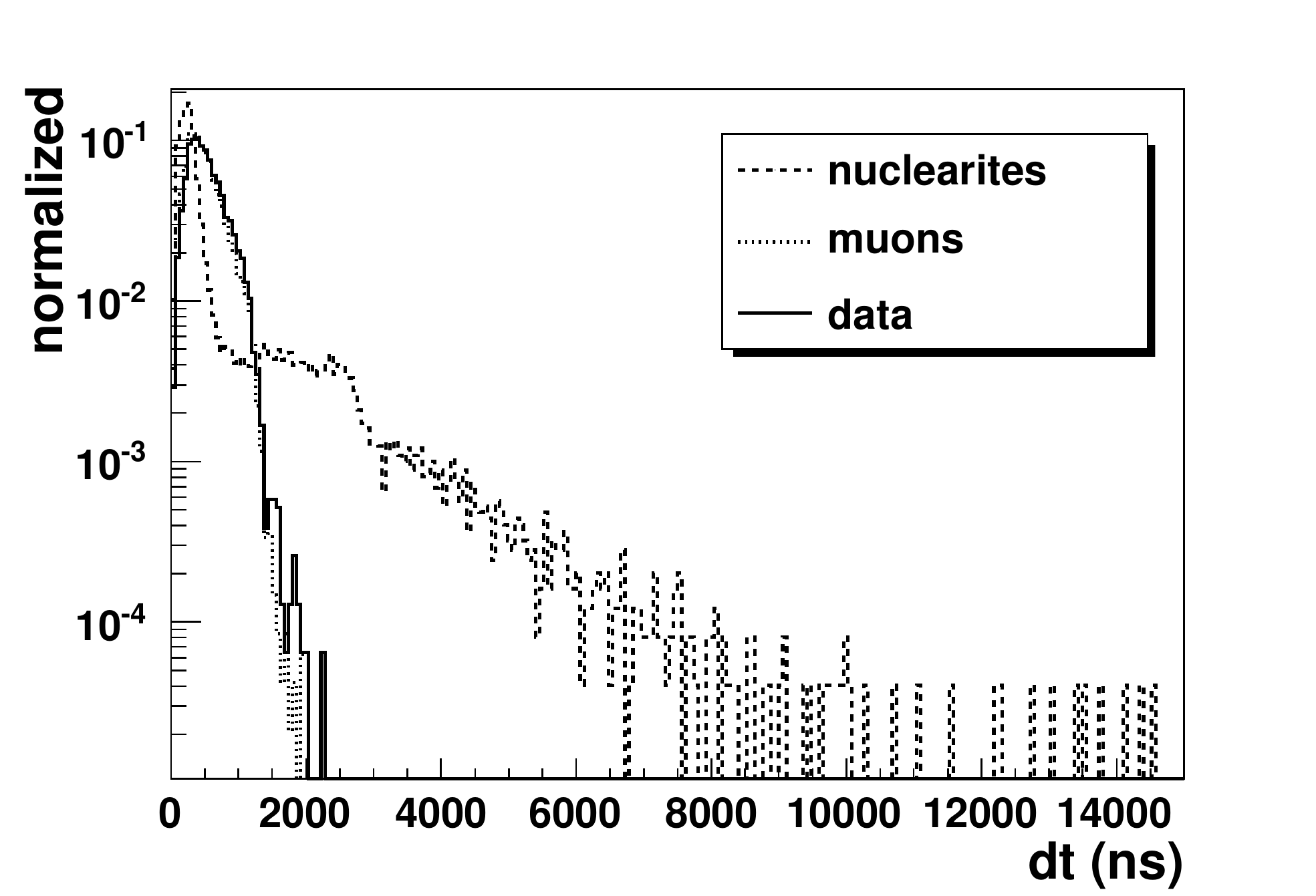}
\caption{Normalized distributions as a function of the duration of snapshot. Comparison between data (continuous line), simulated muon (dotted line) and nuclearite (dashed line) events is shown.}
\label{dt}
\end{figure}

A second cut has been applied to the remaining events, by requiring the multiple snapshot signature within a time interval of 1 ms, characteristic to the crossing time of the detector by a nuclearite event. Three "events" with a double snapshot have passed the second cut. 

The percentage of simulated nuclearite events remained after applying the cuts is given in Table \ref{table_nucl}.
The events below the linear cut are characterized by snapshots with a low number of L1 hits and a large number of single hits.
\begin{figure}
\centering
\includegraphics[width=2.5in]{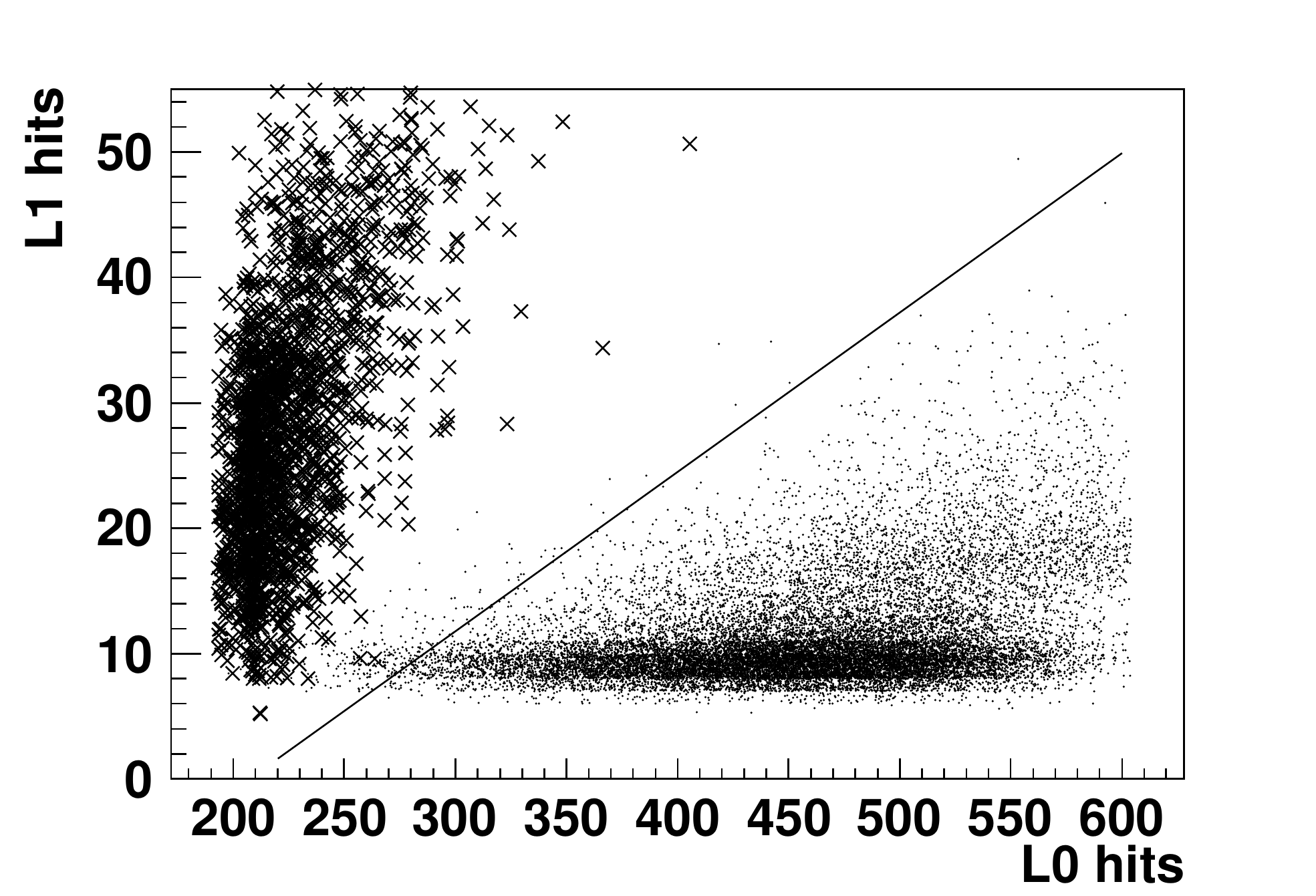}
\caption{Data (crosses) and simulated nuclearite events (points) in a L1 triggered hits vs. L0 single hits plot. The points represent the corresponding values per snapshot. The linear cut was optimized for signal selection.}
\label{L10_hits}
\end{figure} 

\begin{table}[!h]
  \caption{Percentage of nuclearite events after cuts}
  \label{table_nucl}
  \centering
  \begin{tabular}{|c|c|c|}
  \hline
   Nuclearite mass (GeV)  &  linear cut & multiple snapshot cut \\
   \hline 
    3$\times10^{16}$ & 72.5\% & 16.3\% \\
    1$\times10^{17}$ & 96.8\% & 89.1\% \\
    1$\times10^{18}$ & 98.7\% & 96.2\% \\
  \hline
  \end{tabular}
  \end{table} 
 
The sensitivity of the ANTARES detector in 5-line configuration after 84 days of data taking to nuclearite events with masses $\geq 10^{17}$ GeV is of the order of $\sim 10^{-16}cm^{-2}s^{-1}sr^{-1}$. 
We are currently investigating the possibility of implementation in the data acquisition program of a trigger that uses the characteristics of nuclearite events in order to keep all data sent to shore for a time interval of about 20 ms. Also, test runs on simulated nuclearite data using the cluster trigger show a better efficiency and a lower detectable mass limit than for directional trigger.  

\section{Conclusions}
We presented search strategies and expected sensitivities for monopoles and non-relativistic nuclearites with the ANTARES detector in 12-line and 5-line configurations. The sensitivities obtained for both upward magnetic monopoles and nuclearites are preliminary. The sensitivity for nuclearites can be improved by using data taken in nominal configuration and the cluster trigger.

\label{icrc0695:end}

\setcounter{figure}{0}
\setcounter{table}{0}
\setcounter{footnote}{0}
\setcounter{section}{0}





\hyphenation{abcdef-ghijklmnoprstuwxyz IEEEtran}

\title{Underwater acoustic detection of UHE neutrinos with the ANTARES experiment}

\author{\IEEEauthorblockN{Francesco Simeone, on behalf of the ANTARES Collaboration.\IEEEauthorrefmark{1}}
                            \\
\IEEEauthorblockA{\IEEEauthorrefmark{1}University "La Sapienza" and INFN Sez. Roma.}}

\shorttitle{Simeone \etal Acoustic neutrino detection}
\maketitle
\label{icrc0471:begin}

\begin{abstract}
 The ANTARES Neutrino Telescope is a water Cherenkov detector composed of an array of approximately 900 
 photomultiplier tubes in 12 vertical strings, spread over an area of about 0.1 km$^{2}$ with an 
 instrumented height of about 350 metres. ANTARES, built in the Mediterranean Sea, is the biggest 
 neutrino telescope operating in the northern hemisphere. Acoustic sensors (AMADEUS project) have been 
 integrated into the infrastructure of ANTARES, grouped in small arrays, to evaluate the feasibility 
 of a future acoustic neutrino telescope in the deep sea operating in the ultra-high energy regime.
 In this contribution, the basic principles of acoustic neutrino detection will be presented. 
 The AMADEUS array of acoustic sensors will be described and the latest results of the project 
 summarized.\\
\end{abstract}

\begin{IEEEkeywords}
 Acoustic neutrino detection, Beam-Forming, AMADEUS
\end{IEEEkeywords}
 
\section{Introduction}
 Almost everything we know about the Universe came from its observation by means of electromagnetic radiation. 
 Using photons as observation probe it has been possible to discover very energetic sources. 
 However, photons are highly absorbed by matter and so their observation only allows us to directly 
 obtain information of the surface process at the source. Moreover energetic photons interact with 
 the infrared photon background and are attenuated during their travel from the source toward us.\\
 Observation of the proton component of cosmic rays can give information about the sources but, 
 since they are charged, low energy protons are deflected by the magnetic galactic fields and loose 
 the directional information that would allow us to point back to their source. Ultra high energy protons 
 are slightly deflected by magnetic fields and in principle could be a good probe for the high energy 
 Universe. Unfortunately, as pointed out by Greisen-Zatsepin-Kuz'min \cite{GZK1} \cite{GZK2}, the 
 proton interactions with the CMBR(GZK effect) will reduce the proton energy and make them not useful as 
 astroparticle probes for the high energy Universe.\\
 In order to directly observe the physical mechanism of distant and energetic sources we need 
 to use a neutral, stable and weakly interacting messenger: the neutrino. The interest in studying 
 such high energy sources arises from the fact that much of the classical astronomy is related to 
 the study of the thermal radiation, emitted by stars or dust, while the non thermal energy density 
 in the Universe is roughly equal to the thermal one and it is assumed to play a relevant role in its 
 evolution.\\  
 The experimental techniques proposed to identify the cosmic neutrino signatures are mainly three: 
 the detection of Cherenkov light originating from charged leptons produced by neutrino interactions 
 in water or ice; the detection of acoustic waves produced by neutrino induced energy deposition in water, 
 ice or salt; the detection of radio pulses following a neutrino interaction in ice or salt.\\

 \begin{figure*}[ht]
  \centering
  \includegraphics[width=0.7\textwidth]{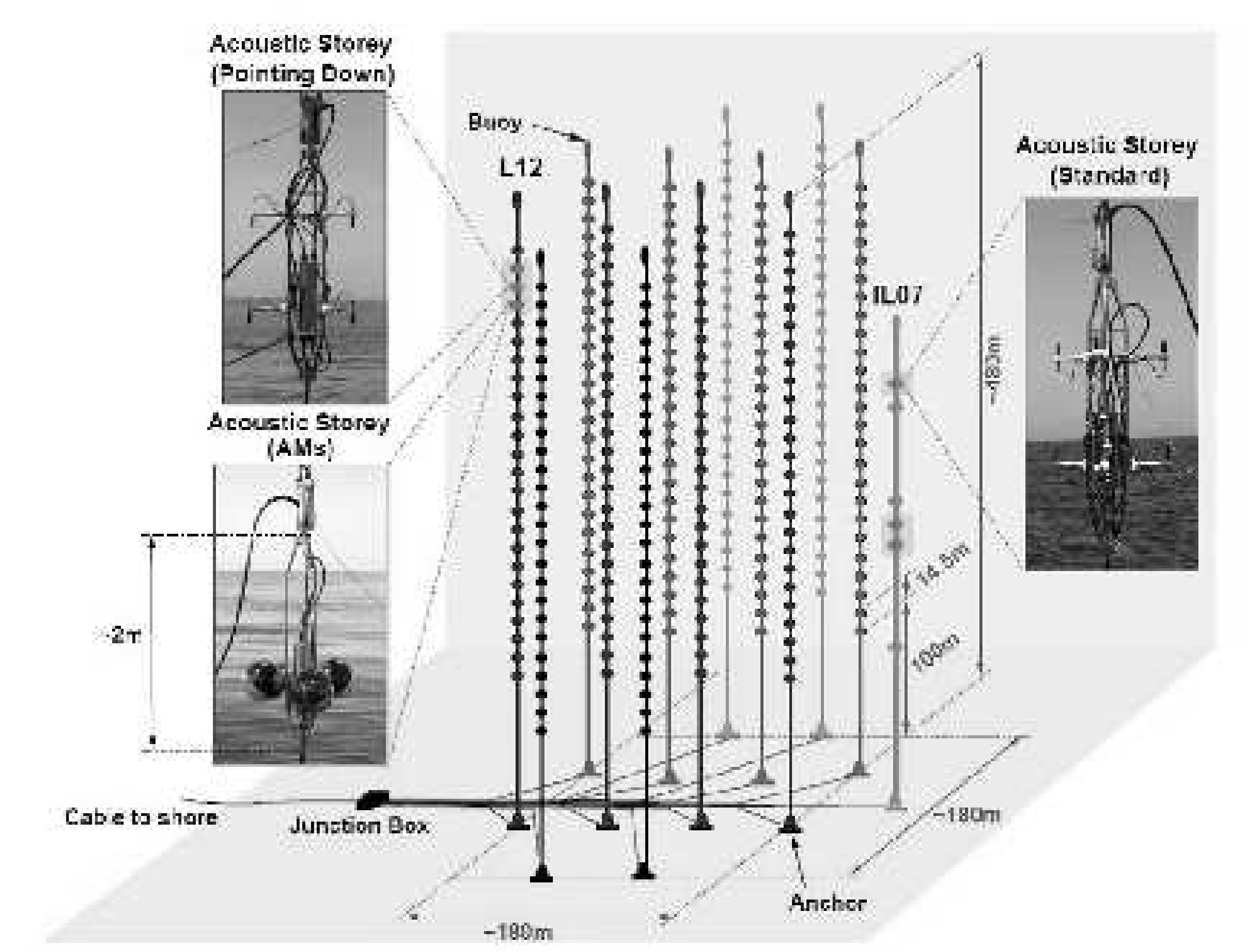}
  \caption{Acoustic storeys are present in four different configurations; two of
 them contain commercial hydrophones (one upward and one downward
 looking), one contains custom built hydrophones and the last is made of
 sensors inside three glass spheres. }
  \label{ANT_LAY}
 \end{figure*}
 
\section{Acoustic neutrino detection}
 The acoustic detection technique of neutrino-induced cascades, in water or ice, is based on the 
 thermo-acoustic effect \cite{TERMO1} \cite{TERMO2}. The cascade energy is deposited in a narrow region of the medium, 
 inducing a local heating and resulting in a rapid expansion of the water (or ice). The simultaneous 
 expansion of the medium along the shower leads to a coherent sound emission in the plane 
 perpendicular to the shower axis.\\
 Simulations performed by many authors \cite{SIMUL1} \cite{SIMUL2} show that the bipolar acoustic neutrino pulse is tens 
 of microseconds long and has a peak-to-peak amplitude of about 100~mPa at 1km distance if originated 
 by a shower of 10$^{20}$~eV. The signal is largely collimated in the plane perpendicular to the 
 shower axis and its amplitude decreases by almost two order of magnitude in few degrees.\\
 The acoustic detectors are composed of many acoustic sensors distributed in a wide instrumented 
 volume; by measuring the acoustic induced neutrino pulse with several sensors it will be possible 
 to infer the shower direction. The interest in this technique is related to the high attenuation 
 length ($\sim$km) of the sound in water (or ice). Consequently it is possible to instrument a large volume 
 using a relatively low number of sensors.\\

\section{AMADEUS project}
 The AMADEUS \cite{AMADEUS} (Antares Modules for Acoustic DEtection Under the Sea) project is fully integrated into 
 the ANTARES \cite{ANTARES} \cite{ANTARES1} \cite{ANTARES2} Cherenkov neutrino telescope (Fig.~\ref{ANT_LAY});
 Its main goal is to evaluate the feasibility 
 of a future acoustic neutrino telescope in the deep sea operating in the ultra-high energy regime.\\
 ANTARES is located in the Mediterranean Sea about 40~km south of Toulon (France) at a depth of about 
 2500m. It is composed of 12 vertical structures, called lines and labeled L1 - L12. An additional 
 line (IL07) is equipped with several instruments used for environmental monitoring. Each detection 
 line holds up to 25 storeys, each of them contains three photomultipliers (PMTs) and the electronics needed to 
 acquire the PMT signals and send them to shore. The storeys are vertically separated by 14.5~m 
 starting at a height of about 100m above sea floor. Each line is fixed to the sea floor by an 
 anchor and held vertically by a buoy.\\ 
 Three special storeys (Acoustic Storeys) are present on both L12 and IL7, each storey contains six 
 acoustic sensors and the electronics used to acquire, pre-process and send the samples to the onshore laboratory. The 
 hydrophones signals are amplified up to a sensitivity of 0.05V/Pa, filtered using a bandpass filter 
 from 1~kHz to 100~kHz and sampled at 250~ksps with a resolution of 16~bit. The acquisition system is 
 able to produce up to 1.5~TB per day; in order to select interesting signals and reduce the data 
 rate as well as the storage requirements, three online triggers are implemented at level of the 
 acoustic storey:
 \begin{itemize}
  \item Minimum bias filter, that stores 10~s of samples every hour. 
  \item Threshold filter, that store signals of amplitude greater than the selected threshold. 
  \item Matched filter, that store signals of cross correlation with the expected bipolar signal 
        greater than the selected threshold.
 \end{itemize}
 These filters reduce the data sent to the onshore laboratory by two orders of magnitude. The AMADEUS project is fully
 integrated into the ANTARES data acquisition system, in particular all the samples acquired by AMADEUS are tagged with
 an absolute time, common to the whole experiment, with a precision better than 1ns. This allows to correlate acoustic
 signals acquired in different parts of the apparatus and represents, at the moment, an unique underwater hydrophone
 array acquiring synchronously.\\
 \begin{figure*}[ht]
  \setcounter{figure}{2}
  \centering
  \includegraphics[width=0.75\textwidth]{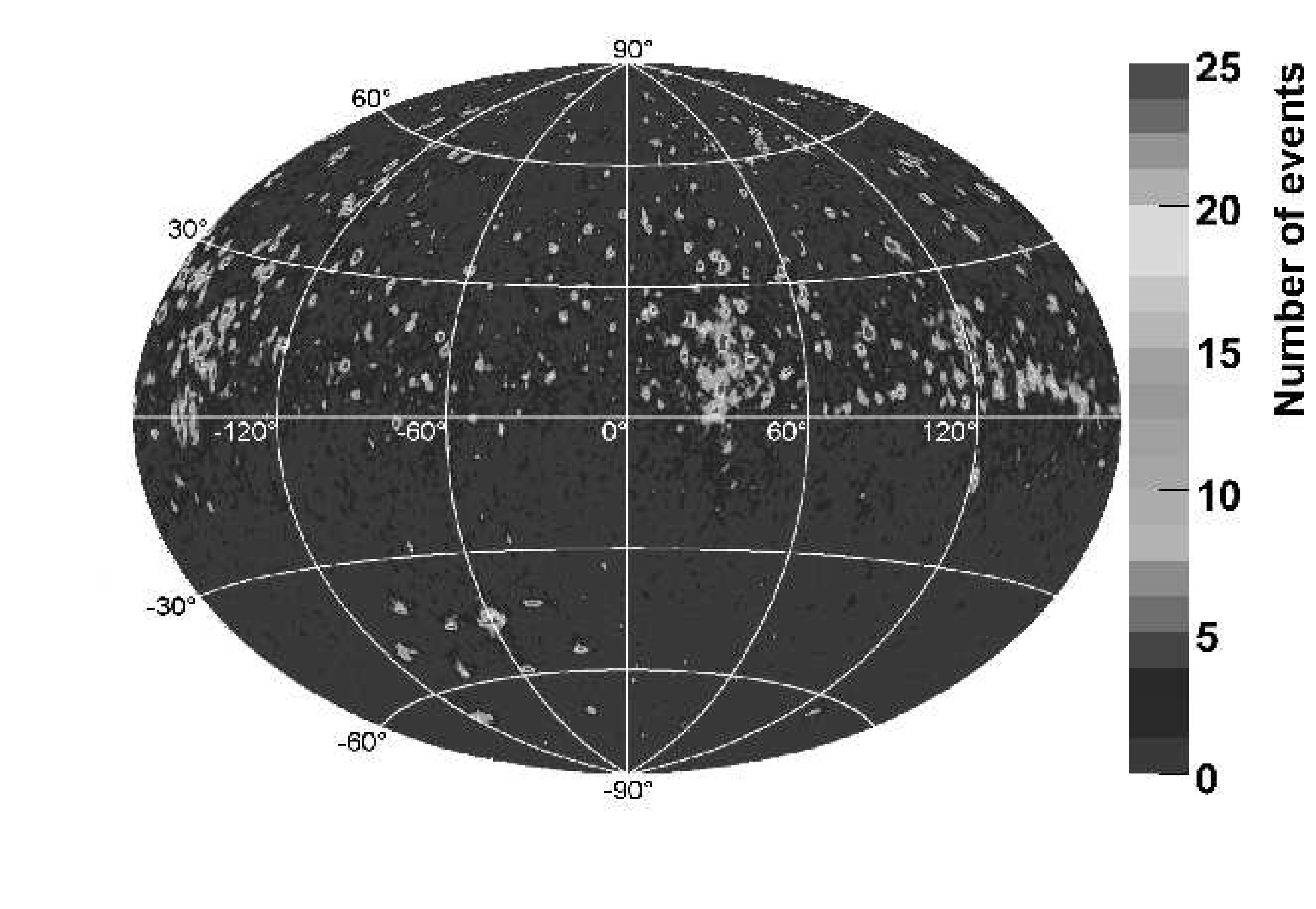}
  \caption{Mapping of the arrival directions of transient signals. The map is produced using the data acquired by the
 second storey on IL07 about 180m above the sea-bed. The origin is defined by the westward direction with respect to
 the horizontal of that storey.
 90$^{\circ}$ (-90$^{\circ}$) in longitude corresponds to north (south), 90$^{\circ}$ (-90$^{\circ}$) in
 latitude to vertically downward-going acoustic signals (upward-going). }
  \label{MAP_DIR}
 \end{figure*}

\section{First AMADEUS results}
 The underwater enviroment is an highly noisly ambient: thus the characterization of this background is of
 fundamental importance to develop detection algorithms able to identify neutrino induced acoustic signals.
 One of the main opportunity given by the AMADEUS system, is to study this background.\\
 The noise sources can be classified in two main group: 
\begin{itemize}
  \item Transient signals (like the ones expected by the UHE neutrinos). These signals contains a finite amount of energy and have a finite duration in time.
  \item Stationary random signals. These signals have statistical properties that are invariant with respect to a
 translation in time and are often characterized by their power spectral density (PSD).
 \end{itemize} 
 In the following subsections the contribution of these two kinds of signals will be discussed.

 \subsection{Transient Signals}
 Each storey of the AMADEUS project is an array of hydrophones; taking advantage of these geometry, 
 it is possible to reconstruct the arrival direction of the pressure wave on the storey using a 
 technique called beam-forming.\\
 The Beam-forming is a MISO (Multi Input Single Output) technique \cite{BEAM} originally developed to 
 passively detect submarines with an underwater array of hydrophones. This technique analyses the 
 data acquired by an array of sensors to compute the arrival direction of the incident waves. The 
 array of sensors samples the wave-front in space and time and the samples of different sensors are 
 combined, using the proper time delays, to sum up coherently all the waves arriving on the array 
 from a specific direction; this will lead to an increase of the signal to noise ratio (SNR) since the
 spatially white noise 
 will be averaged out. Since we don't know the arrival direction of the incident wave we need to 
 compute the time delays for many different arrival directions and produce an angular map of the 
 incident pressure wave as the one  reported in figure~\ref{MAP_ANG}. The aliases that are present in this map 
 are due to delays (directions) for which two or more of the six hydrophones sum coherently. The only 
 direction for which all the hydrophones sum up coherently is the true one.\\
\begin{figure}[htb]
  \setcounter{figure}{1}
  \centering
  \includegraphics[width=0.4\textwidth]{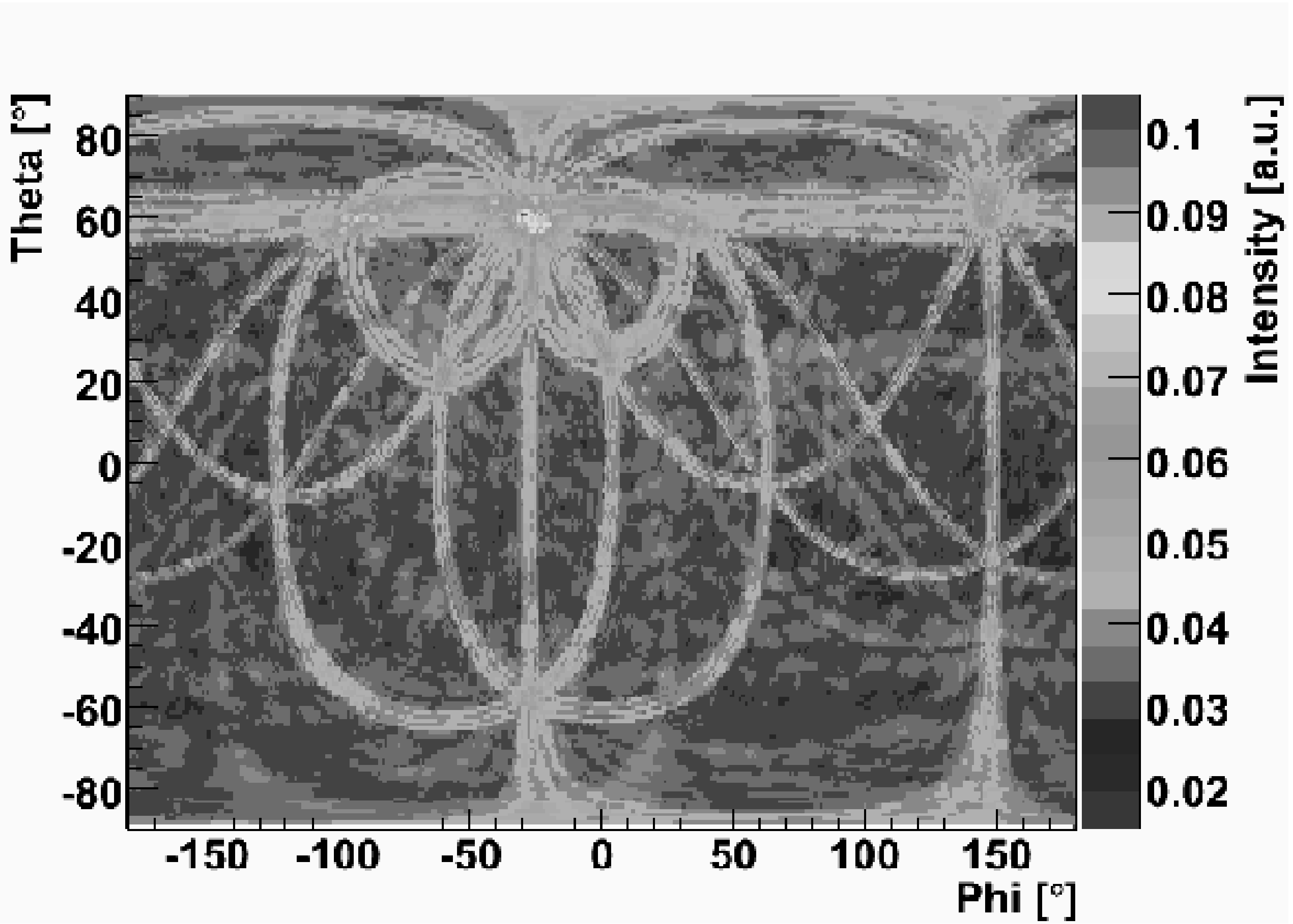}
  \caption{Example of beamforming output used to reconstruct a transient signal direction. The beamforming output
 is a function of the number of hydrophones whose signals sum up coherently, assuming that the transient signals
arrive from that specific direction.}
  \label{MAP_ANG}
 \end{figure}
 Using the directional information provided by three or more arrays it is possible to reconstruct the 
 acoustic source position. The directional information can be used in conjunction with ray-tracing 
 techniques that allows us to easily take into account the sound propagation in water.\\ 
 In figure~\ref{MAP_DIR} qualitative mapping of the arrival directions of transient acoustic signals are shown.
 The data sample has been collected with the minimum bias filter during a time period of about 6 
 months; the figure shows the directions of all reconstructed signals which have an amplitude greater 
 than eight times the standard deviation of the ambient noise. The majority of the reconstructed 
 acoustic signals are received from directions in the upper hemisphere; this is consistent with the expectations 
 since the major sources of transient noise are due to biological and anthropological activities. 
 The few souces visible in the lower left part of the sphere resemble the layout of the ANTARES strings.
 Those sources are the pingers of the ANTARES acoustic positioning system,
 emitting acoustic signals at the bottom of each line.\\
 \subsection{Correlation between underwater noise and weather condition.}
 An analysis correlating the ambient acoustic noise level, measured with AMADEUS and the surface weather data was
 performed for the whole year 2008. The data used in this analysis are the minimum bias ones and the weather condition
 were measured by 5 station around the ANTARES site:
 \begin{itemize}
  \item Station 1: Toulon.
  \item Station 2: Cap Cepet.
  \item Station 3: Hy$\grave{e}$res.
  \item Station 4: Porquerolles.
  \item Station 5: Toulon/Ile du Levant.
 \end{itemize}
 For each station a daily average of the wind speed and other weather observables are available; in this analysis only the
 mean wind speed was used. The noise level for a data sample acquired by AMADEUS is evaluated by integrating over the
 frequency range 1-50~kHz, the mean PSD is calculated using samples of 8.4 seconds duration.\\
\begin{figure}[htb]
  \setcounter{figure}{3}
  \centering
  \includegraphics[width=0.41\textwidth]{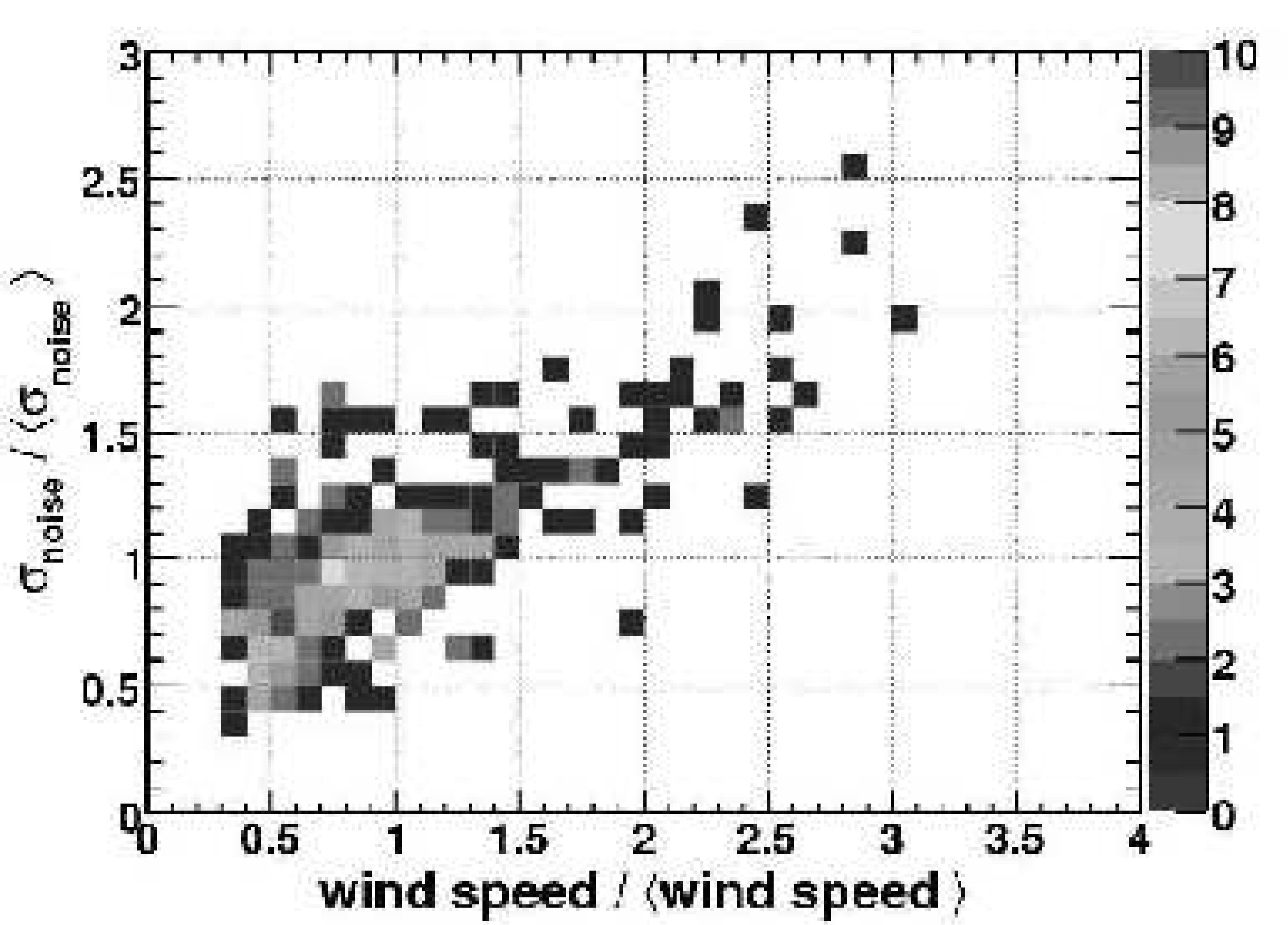}
  \caption{The relative variation of the wind speed, measured by the weather station 2, and the noise level measured by
 the sensor 17 (IL07). This weather station is located near the sea so its wind measurements are well suited for this
 kind of analysis.}
  \label{CORR_2}
 \end{figure}
 For better comparability between wind speed and noise level, the daily variation with respect to the annual mean is used.
 The correlation seems to became more evident for high wind speed (see figure~\ref{CORR_2}).\\
\begin{figure}[htb]
  \centering
  \includegraphics[width=0.4\textwidth]{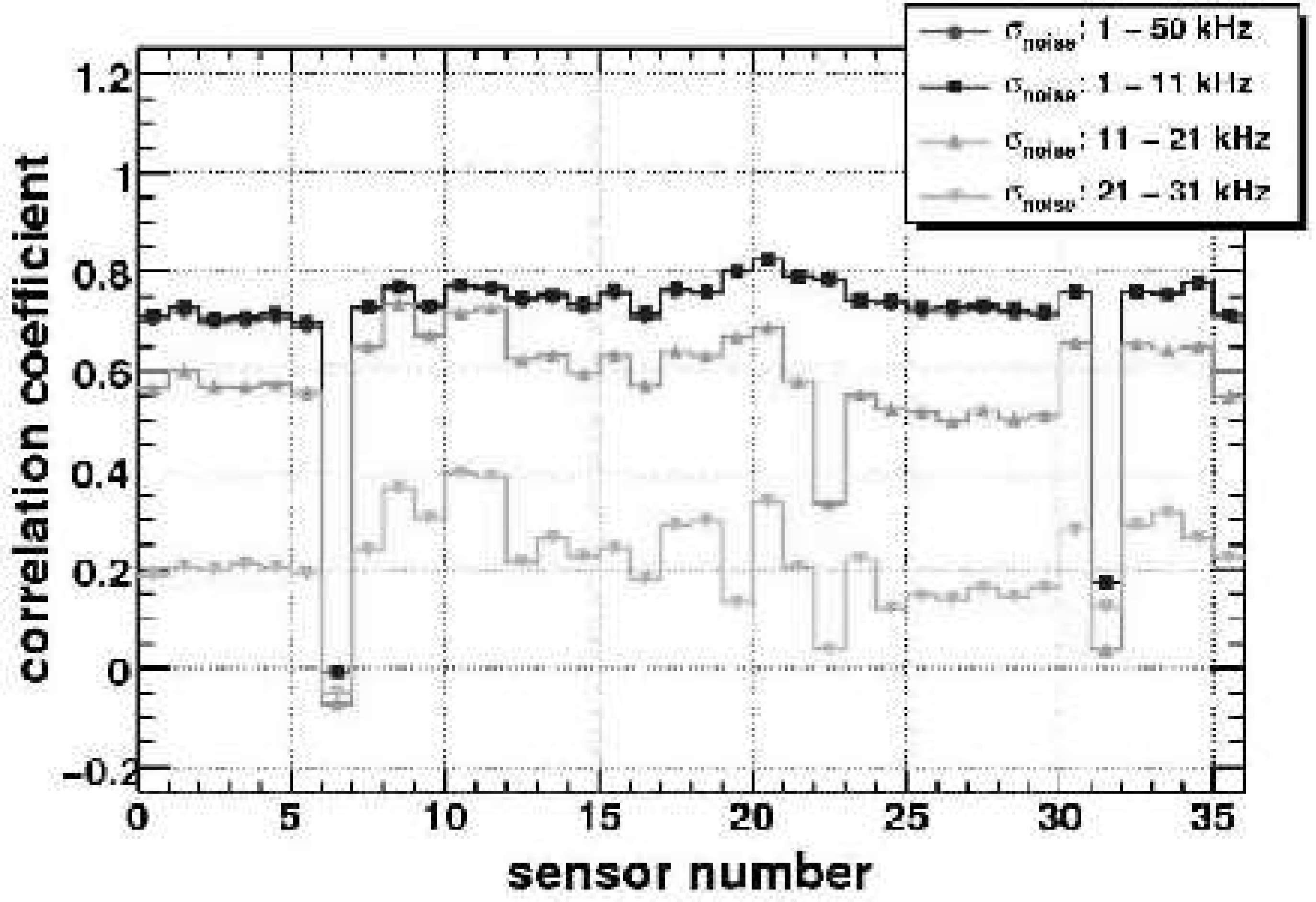}
  \caption{Correlation coefficient evaluated using different frequency bands. Sensors 6 and 31 have been defective from
 the beginning of the data taking and thus show no correlation with the weather conditions. Sensor 22 has a non-ideal
 coupling with the glass sphere and is not sensitive to the higher frequency range.}
  \label{CORR_3}
 \end{figure}
 For frequencies below 21~kHz a strong correlation ($\geq$50\%) is found; for higher frequency the ambient noise decreases
 and the system noise starts to becomes significant as a consequence the correlation with the weather condition
 decreases (see figure~\ref{CORR_3}).
 No systematic different observation is found using the other weather station data; the significance of the correlation
 is lower as the stations are situated in-land and typically measure lower wind speed.\\

\section{Conclusions} 
 The AMADEUS project is fully operating and the data analysis is going on; this system offers a unique opportunity
 to study the underwater noise in order to develop and test acoustic triggers, able to discriminate the neutrino-induced
 signals from underwater background.\\
 The matched filter is the optimum linear filter to discriminate a signal of known shape from a 
 stochastic background of known spectrum. The beam-forming technique allows us to further increase 
 the SNR by a factor that is equal to the number of hydrophones used in the phased array. Moreover 
 the beam-forming technique allows to reconstruct the arrival direction of the incident pressure wave. 
 The directional information, combined with a ray-tracing technique, permits an easy reconstruction 
 of the interaction point. The combination of these techniques could extend the energy range of 
 acoustic detectors and may increase the possibility to measure combined events in an hybrid 
 (optic-acoustic) detector.\\

  \label{icrc0471:end}


\setcounter{figure}{0}
\setcounter{table}{0}
\setcounter{footnote}{0}
\setcounter{section}{0}





\hyphenation{abcdef-ghijklmnoprstuwxyz IEEEtran}

\title{Point source searches with the ANTARES neutrino telescope}
\author{\IEEEauthorblockN{Simona Toscano\IEEEauthorrefmark{1}, for the ANTARES Collaboration}

                            \\
\IEEEauthorblockA{\IEEEauthorrefmark{1} IFIC- Instituto de F\'isica
                            Corpuscular, Edificios de Investigaci\'on de
                            Paterna, CSIC - Universitat de Val\`encia,\\
                            Apdo. de Correos 22085, 46071 Valencia, Spain.}}

\shorttitle{S.Toscano \etal Point source searches with the ANTARES.....}
\maketitle
\label{icrc0127:begin}

\begin{abstract}
With the installation of its last two lines in May 2008, 
ANTARES is currently the largest neutrino detector in the Northern Hemisphere. The detector comprises 12 detection lines, carrying 884 ten-inch photomultipliers, at a depth of about 2500 m in the Mediterranean Sea, about 40 km off shore Toulon in South France. 
Thanks to its exceptional angular resolution, 
better than 0.3$^\circ$ above 10 TeV, and its favorable location with the 
Galactic Center visible 63\% of time, ANTARES is specially suited for the search 
of astrophysical point sources. Since 2007 ANTARES has been taking data 
in smaller configurations with 5 and 10 lines. 
With only 5 lines it already has been possible to set the most restrictive 
upper limits in the Southern sky. In this contribution we present the search 
of point sources with the 5-line data sample.    
\end{abstract}

\begin{IEEEkeywords}
High energy neutrinos, Point source search, Neutrino telescope. 
\end{IEEEkeywords}
 
\section{Introduction}
The ANTARES (Astronomy with a Neutrino Telescope and Abyss environmental RESearch) detector~\cite{ANT09}~\cite{ANT08} has been completed in May 2008, and it is currently the major neutrino telescope in the Northern Hemisphere. The full detector consists on a 3$-$dimensional array of photomultipliers set out in 12 lines deployed in the Mediterranean Sea at a depth of about 2500~m and about 40~km from the south coast of France. The 10$^{\prime\prime}$ photomultipliers~\cite{AMR02} detect the Cherenkov light induced by the relativistic muons
generated in the high energy neutrino interactions with the surrounding material.\\
During the year 2007 ANTARES was operated in a smaller configuration of only 5 lines. Nevertheless, the good angular resolution of the 5 line configuration ($<$~0.5$^\circ$ at 10 TeV) made it possible to start with the physics analysis and perform a search for neutrino sources in the visible sky of ANTARES. 
In this contribution we present the results obtained by an unbinned analysis for the search of point sources with the 5-line data. \\
The description of the data is given in section~\ref{data} while the methodology is explained in section~\ref{method}.
The results are compatible with a background fluctuation and hence, the first upper limits on the cosmic neutrino flux set for ANTARES are given in section~\ref{results}.

\section{Data processing and detector performance}
\label{data}
Data used in the analysis correspond to a sample of data taken from February to December 2007\footnote{To exclude periods with high bio-activity only runs with a baseline~$<$~120 kHz and a burst fraction $<$~40\% have been used.}, in which ANTARES was operated in a smaller configuration of only 5 of its 12 lines. Due to several detector operations including the deployment and connexion of new lines, the actual live-time for the 5-line configuration is 140 days.

The reconstruction of the muon track is achieved by using the information given by the time and amplitude of the signal of the Cherenkov photons arriving to the photomultiplier. The reconstruction algorithm is based on a maximum likelihood method~ \cite{AART} where the quality parameter is defined by the maximum log-likelihood per degree of freedom plus a term that takes into account the number of compatible solutions, N$_{comp}$, found in the algorithm:
\begin{equation}
\Lambda = log(L)/N_{DOF} + 0.1(N_{comp}-1).
\label{eq:lambda}
\end{equation}
\noindent
Fig.\ref{fig:LAMBDA} shows the complementary cumulative distribution\footnote{It is the number of events above any given $\Lambda_{th}$.} of $\Lambda$ for up-going events with a cut in elevation $< -10^\circ$ in order to eliminate a possible atmospheric muon contamination near the horizon. 
	
	\begin{figure}[!h]
	  \centering
    \includegraphics[width=2.5in]{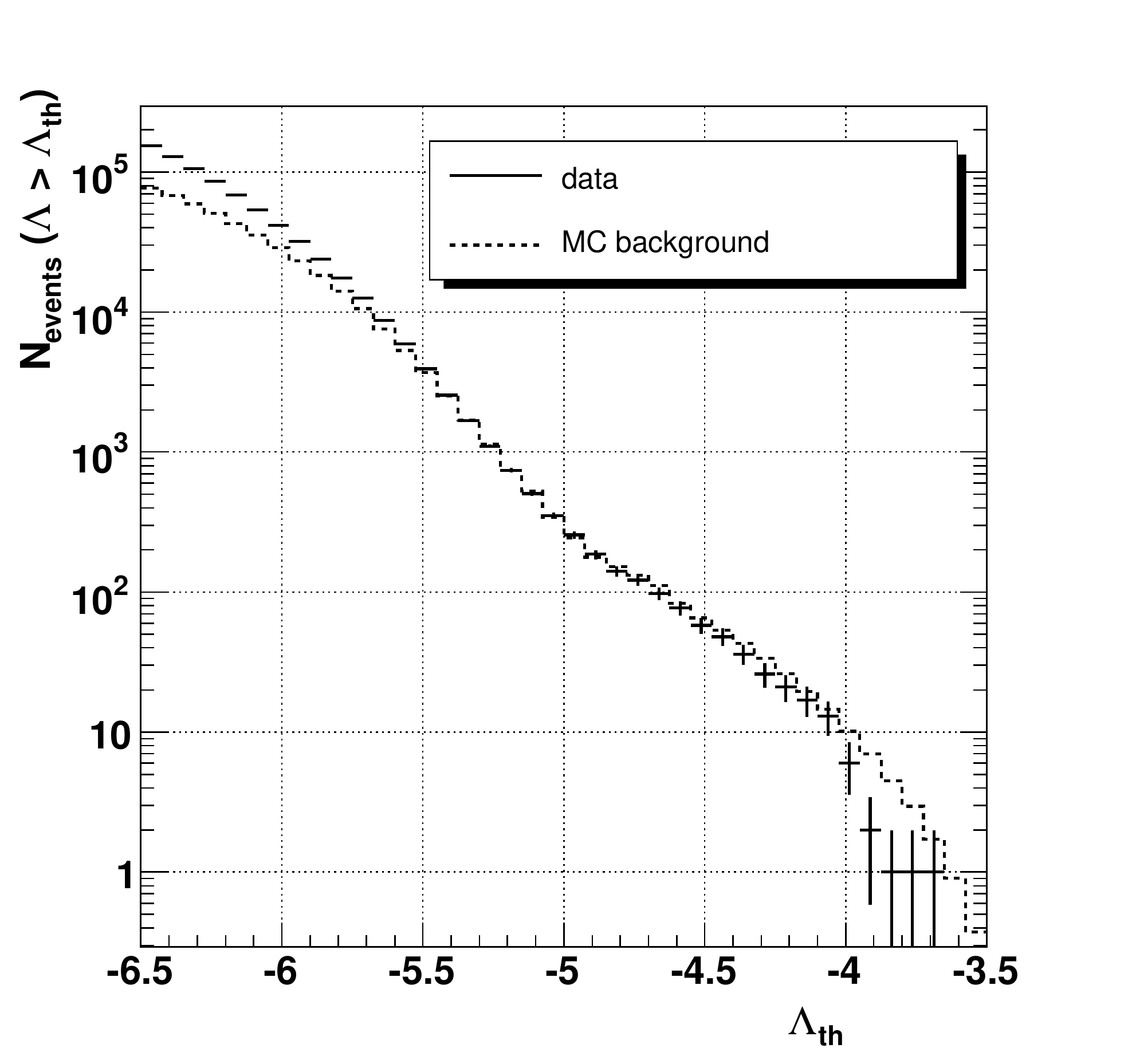}
    \caption{\small{Comparison between data (point) and MC (line) for the complementary cumulative distribution of $\Lambda$. Only up-going tracks with an elevation $< -10^\circ$ are shown.}}
    \label{fig:LAMBDA}
  \end{figure}
\noindent

In this analysis, the contributions from atmospheric muons and neutrinos coming from the cosmic ray interactions in the Earth$^\prime$s atmosphere are simulated using Monte Carlo techniques. Atmospheric muons are simulated with CORSIKA~\cite{CORSIKA}, with QGSJET~\cite{QGSJET} models for the hadronic interactions and Horandel~\cite{Horandel} for the cosmic ray composition; neutrinos are generated with the GENNEU~\cite{GENNEU} package and the Bartol~\cite{Bartol} model.\\
In fig.\ref{fig:DECL} the declination distribution for both Monte Carlo and real data is shown. 
After applying a cut in $\Lambda < -4.7$ and elevation$< -10^\circ$ a total number of 94 events are selected with a contamination\footnote{The ratio between the number of atmospheric muons and the total number of events expected from Monte Carlo simulations.} of 20\% due to misreconstructed atmospheric muons.  
This distribution is used in the analysis to estimate the background density function (see next section).

	\begin{figure}[!h]
	  \centering
    \includegraphics[width=2.5in]{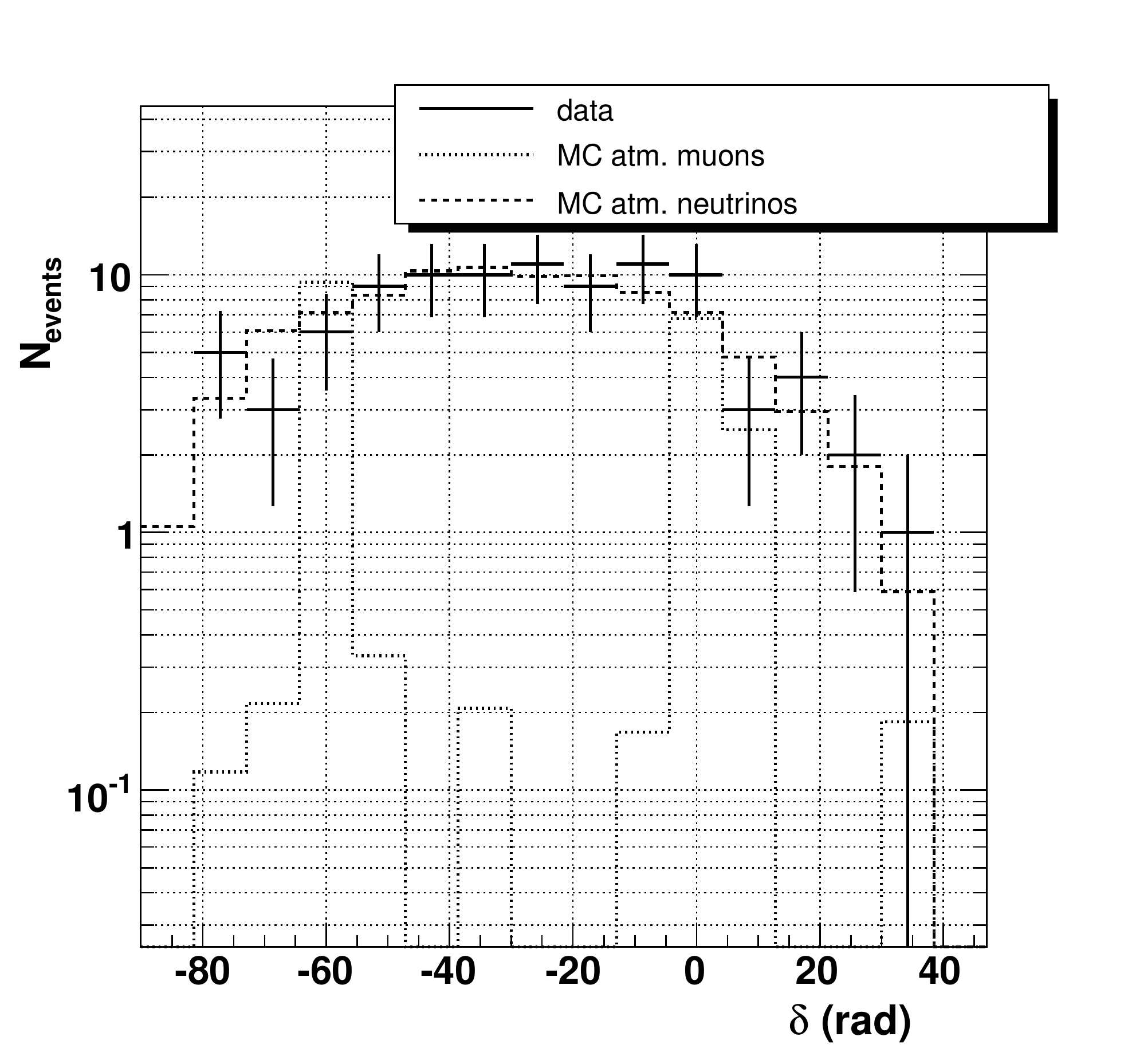}
    \caption{\small{Declination distribution for up-going events. Data (point) and MC (lines) are compared for the quality cuts used in the analysis ($\Lambda > -4.7$, elevation $< -10^\circ$).}}
    \label{fig:DECL}
  \end{figure}
  
\noindent 
The two plots show a good agreement between data and Monte Carlo.
    
The detector performance is usually done in terms of its effective area and angular resolution. In fig.\ref{fig:PERFORMANCE} the expected performance for the 5-line detector is shown as a function of the neutrino generated energy.
At 10 TeV the effective area (fig.\ref{fig:EffA}), averaged over the neutrino angle direction, has a value of about $4\times 10^{-2}~$m$^2$. The decrease of the effective area above $\sim$ 1 PeV is due to the opacity of the Earth for neutrinos at these energies. The angular resolution (fig.\ref{fig:AngRes}) for high energy neutrinos is limited by intrinsic detector and environmental characteristics (i.e. PMT transit time spread, dispersion and scattering of light) and it is better than $0.5^\circ$ for energies above 10~TeV. 

\begin{figure*}[!th]
   \centerline{\subfloat[Effective area]{\includegraphics[width=2.5in]{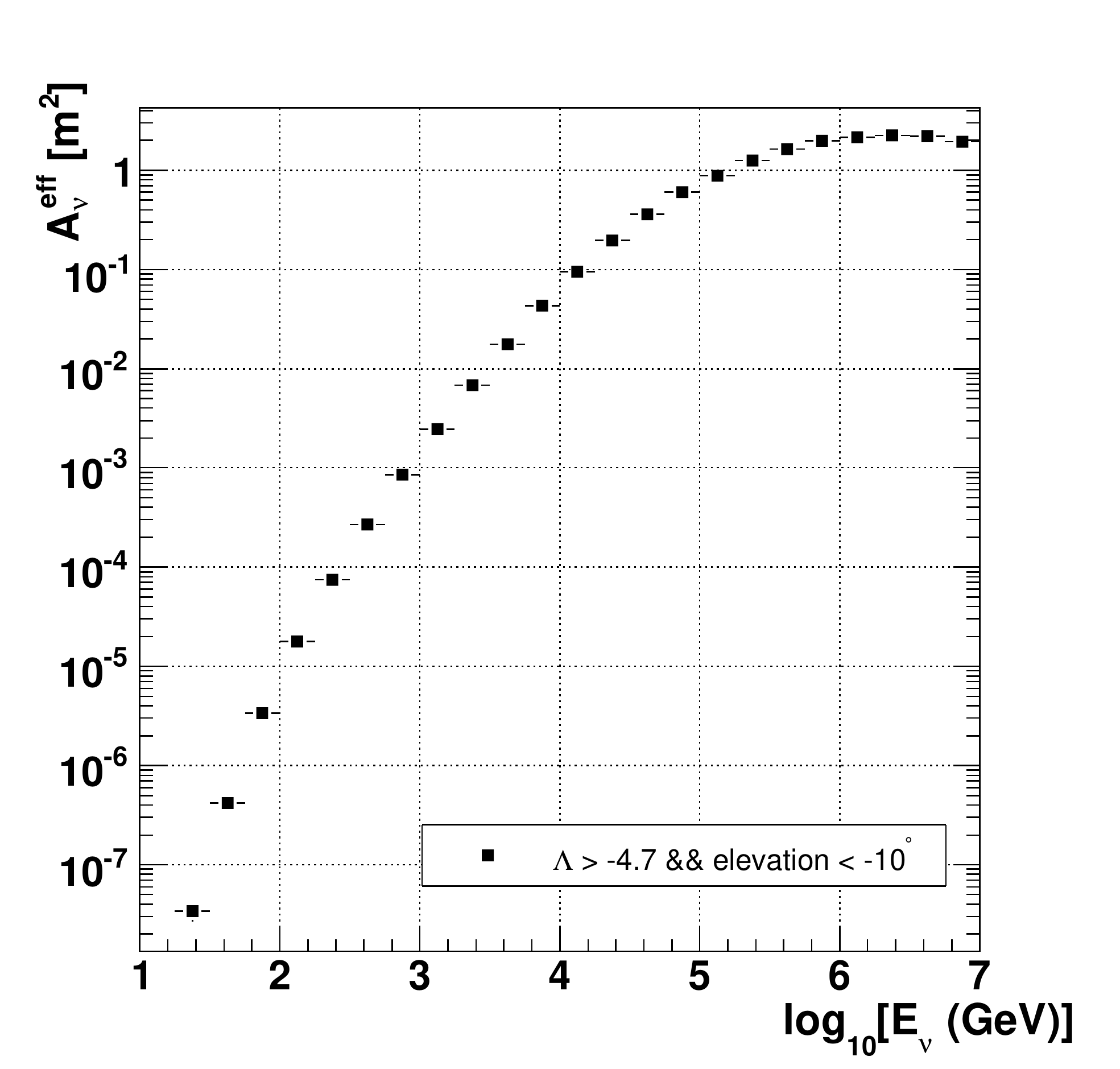} \label{fig:EffA}}
              \hfil
              \subfloat[Angular resolution]{\includegraphics[width=2.5in]{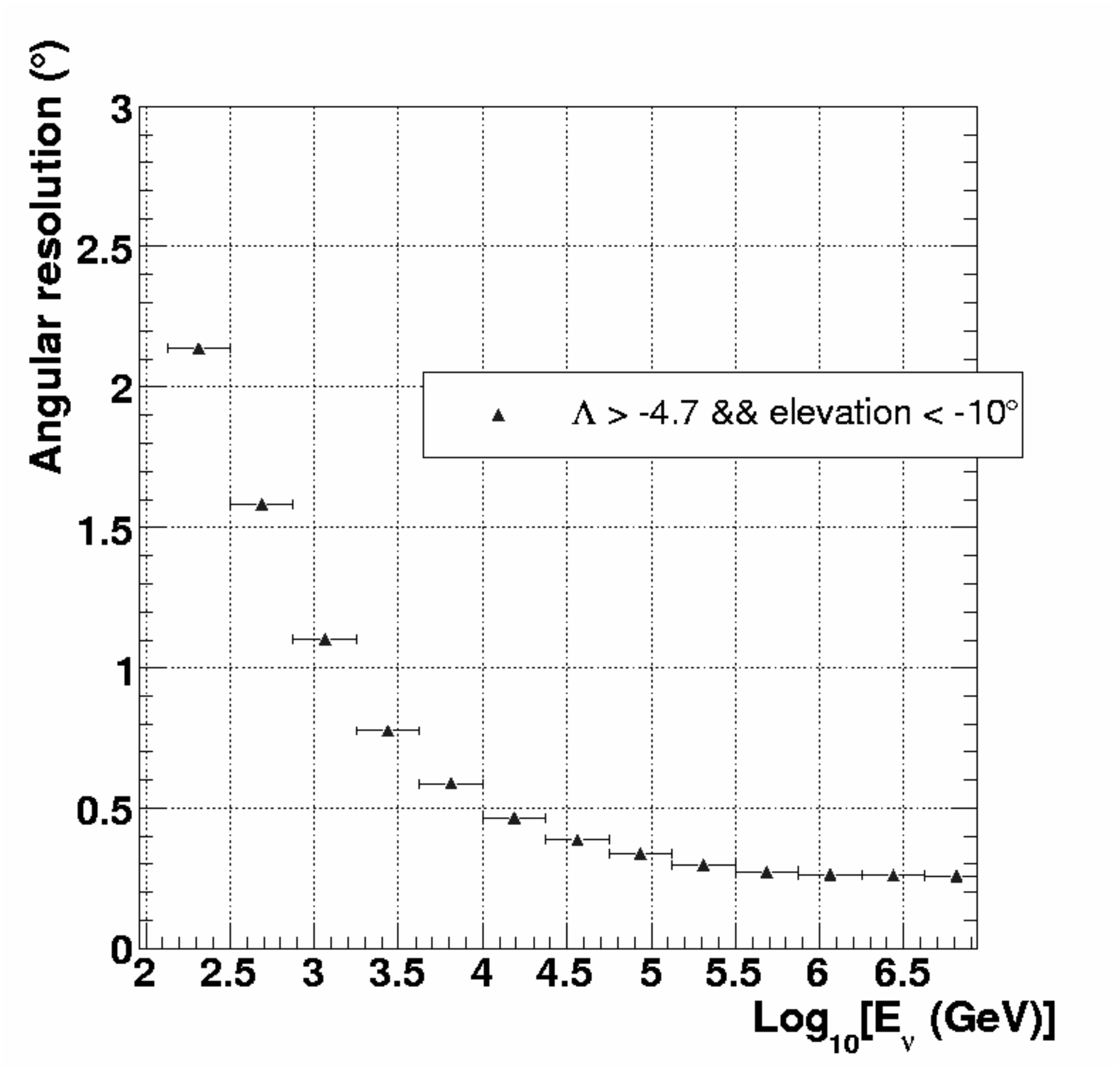} \label{fig:AngRes}}
             }
   \caption{\small{Performance of the ANTARES 5-line detector in terms of effective area (left) and angular resolution (right) as a function of the neutrino energy.}}
   \label{fig:PERFORMANCE}
 \end{figure*}

\section{Searching algorithms}
\label{method}
Two methods have been developed to perform the search for point sources. 
The first one is an unbinned method based on the expectation-maximization algorithm~\cite{EM} applied to the point source search in a neutrino telescope~\cite{JUANAN}. The second method is a standard binned method only used to cross-check the results obtained with the unbinned method.

\subsection{Unbinned method}

The expectation-maximization method is a clustering algorithm that analytically maximizes the likelihood in finite mixture problems which are described by different density components (pdf). In our case the mixture problem can be expressed as the sum of two density distributions: the background distribution of atmospheric neutrinos and the signal of cosmic neutrinos. 
Since the detector response is uniform in right ascension the background only depends on the declination. A fit of the declination distribution of real data (fig.\ref{fig:DECL}) gives the background density used in the algorithm. The signal pdf model is assumed to follow a two-dimensional Gaussian.
The EM algorithm assumes that the set of given observations (in our case the equatorial coordinates of the observed neutrino events) forms a set of incomplete data vectors. The unknown information is whether or not the observed event belongs to the background or the signal pdf. A class indicator vector or weight is added to each event taking the value 0 for background and 1 for signal. The observed events with this additional associated weight forms the complete data set with a new complete likelihood. The EM procedure consists of two main steps: in the expectation step (E-step) an estimate of the initial parameters ($\Psi^{(0)}$), including the value of the associated weights, is given and the expectation value of the complete log-likelihood is evaluated for this current set of parameters; in the Maximization step (M-step) a new set of parameters $\Psi = {\Psi^{(m)}}$ is found that maximize the complete data log-likelihood. Successive maximizations of the complete likelihood lead to the maximization of the likelihood of the incomplete data set.\\
After the likelihood maximization the model testing theory is used to confirm or reject the existence of a point source. 
In this analysis the Bayesian Information Criterion (BIC) is used as a test statistic:

\setlength{\arraycolsep}{0.0em}
  \begin{eqnarray}
 BIC=2\log p(\{x\}|\Psi_{1}^{ML},M_{1}) \nonumber\\
  - 2\log p(\{x\} | M_{0})- \nu_1 \log(n), 
\label{eq:BIC}
  \end{eqnarray}              %
  \setlength{\arraycolsep}{5pt}


\noindent
which is the maximum likelihood ratio of the two models we are testing ($M_1 =$ background + signal, over $M_0 =$ background) plus a factor that takes into account the degrees of freedom in our testing model ($\nu_1$) and the number of events $n$ in our sample; $\Psi_k^{ML}$ indicates the estimate obtained by the algorithm for the set of parameters that define each model. \\
Since there is no analytical probability distribution associated to our problem $10^4$ experiments have been simulated to infer the BIC distribution. The background is simulating by randomizing real events uniformly in right ascension. The signal simulation has been inferred from Monte Carlo using the corresponding angular error distributions for neutrino spectral index of 2.0. 

In case of an observation a $BIC_{obs}$ value is selected and the probability that this $BIC_{obs}$ is not due to a source is computed. When the probability reaches the 90\% it can be ensured that there is no source emitting a flux that would yield a BIC value higher than the one observed and an upper limit is set.  

\subsection{Binned method}

Since it is well known that binned methods are not as sensitive as unbinned analysis, in this work the binned method has been used only to cross-check the results obtained with the EM method.
The point source search analysis presented here consists in the optimization of the size of the search cone in order to maximize the probability of finding a cluster of events incompatible with background (at a 90\% C.L.). The approach followed in this analysis is the minimization of the Model Rejection Factor (MRF). The MRF is defined as the ratio between the average upper limit, which depends on the expected background inside the bin, and the signal contained in the searching bin. \\
The background is estimated from real data and the fraction of the contained signal is computed using the angular error distribution inferred from Monte Carlo simulations for a neutrino flux with a spectral index of 2.0.
The angular radius that minimizes the MRF is considered the optimum bin size for point source search.

\section{Results}
\label{results}
Two different implementations of a point source analysis have been applied. In the first one a search for a signal excess in pre-defined directions in the sky corresponding to the position of neutrino candidate sources is performed. The second is a full sky survey in which no assumptions about the source position is made.\\

The ANTARES collaboration follows a strict blinding policy. This means that the analyses have been optimized using real data scrambled in right ascension. Once the parameters have been fixed the results are obtained by looking at real data.

\subsection{Fixed source search}
The first analysis performed to search for a point-like signal consists in looking at a list of possible neutrino emitters.
Taking into account the ANTARES visibility (for up-going events) and the lack of statistics at declinations higher than $10^\circ$, 24 sources have been selected among the most promising galactic and extragalactic neutrino source candidates (supernova remnants, BL Lac objects, etc.). The hot spot reported by the IceCube collaboration in the analysis of the 22-line data \cite{IC22} has also been included in the ANTARES list. 
No significant excess has been found in the 5-line data sample. The results are shown in table~\ref{tab:sources}. The P-values, i.e. the probability of the background to produce the observed BIC value, has been reported for each candidate source in the list. The lowest P-value (a 2.8 $\sigma$ excess, pretrial) corresponds to the location ($\delta=$ -57.76$^\circ$, RA=155.8$^\circ$), which is expected in 10\% of the experiments when looking at 25 sources (post-trial probability).
The results are compatible with the binned method.

\begin{table}
 \begin{center}
{\footnotesize
\begin{tabular}{|l|c|c|c|c|} \hline 

Source name &    $\delta~(^\circ)$ &    RA ($^\circ$) &     P-value &    $\phi_{90}$ \\
            &                      &                  &             &                 \\ 
\hline
PSR B1259-63      &    -63.83 &    195.70  &     1  &    3.1  \\
RCW 86            &    -62.48 &    220.68  &     1  &    3.3   \\
ESO 139-G12       &    -59.94 &    264.41  &     1  &    3.4   \\
HESS J1023-575    &    -57.76 &    155.83  &  0.004 &    7.6   \\
CIR X-1           &    -57.17 &    230.17  &     1  &    3.3   \\
HESS J1614-518    &    -51.82 &    243.58  &  0.088 &    5.6   \\
PKS 2005-489      &    -48.82 &    302.37  &     1  &    3.7   \\
GX 339            &    -48.79 &    255.70  &     1  &    3.8   \\
RX J0852.0-4622   &    -46.37 &    133.00  &     1  &    4.0   \\
Centaurus A       &    -43.02 &    201.36  &     1  &    3.9   \\
RX J1713.7-3946   &    -39.75 &    258.25  &     1  &    4.3   \\
PKS 0548-322      &    -32.27 &    87.67   &     1  &    4.3   \\
H 2356-309        &    -30.63 &    359.78  &     1  &    4.2   \\
PKS 2155-304      &    -30.22 &    329.72  &     1  &    4.2   \\
Galactic Centre   &    -29.01 &    266.42  &  0.055 &    6.8   \\
1ES 1101-232      &    -23.49 &    165.91  &     1  &    4.6   \\
W28               &    -23.34 &    270.43  &     1  &    4.8   \\
LS 5039           &    -14.83 &    276.56  &     1  &    5.0   \\
1ES 0347-121      &    -11.99 &    57.35   &     1  &    5.0   \\
HESS J1837-069    &    -6.95  &    279.41  &     1  &    5.9   \\
3C 279            &    -5.79  &    194.05  &   0.030&    9.2   \\
RGB J0152+017     &     1.79  &    28.17   &     1  &    7.0   \\
SS 433            &     4.98  &    287.96  &     1  &    7.3   \\
HESS J0632+057    &     5.81  &    98.24   &     1  &    7.4   \\
IC22 hotspot      &    11.00  &    153.00  &     1  &    9.1   \\ \hline

\end{tabular}
}
 \caption{Flux upper limits for 25 neutrino source candidates. Together with the source name and
 location in equatorial coordinates, the P-value and the 90\% confidence level upper limit for $\nu_{\mu}$ flux with $E^{-2}$ spectrum (i.e. $E^2d\phi_{\nu_{\mu}}/dE~\le~\phi_{90}~\times~10^{-10}~TeV~cm^{-2}~s^{-1}$) over the energy range 10 GeV to 100 TeV are provided.}
 \label{tab:sources}
 \end{center}
\end{table}
\noindent
The corresponding flux limits have been also reported in the last column and shown in fig.\ref{fig:LIMIT}. As can be seen, the first limits from ANTARES are already competitive with the previous experiments looking at the Southern sky.

	\begin{figure}[!h]
	  \centering
    \includegraphics[width=3.0in]{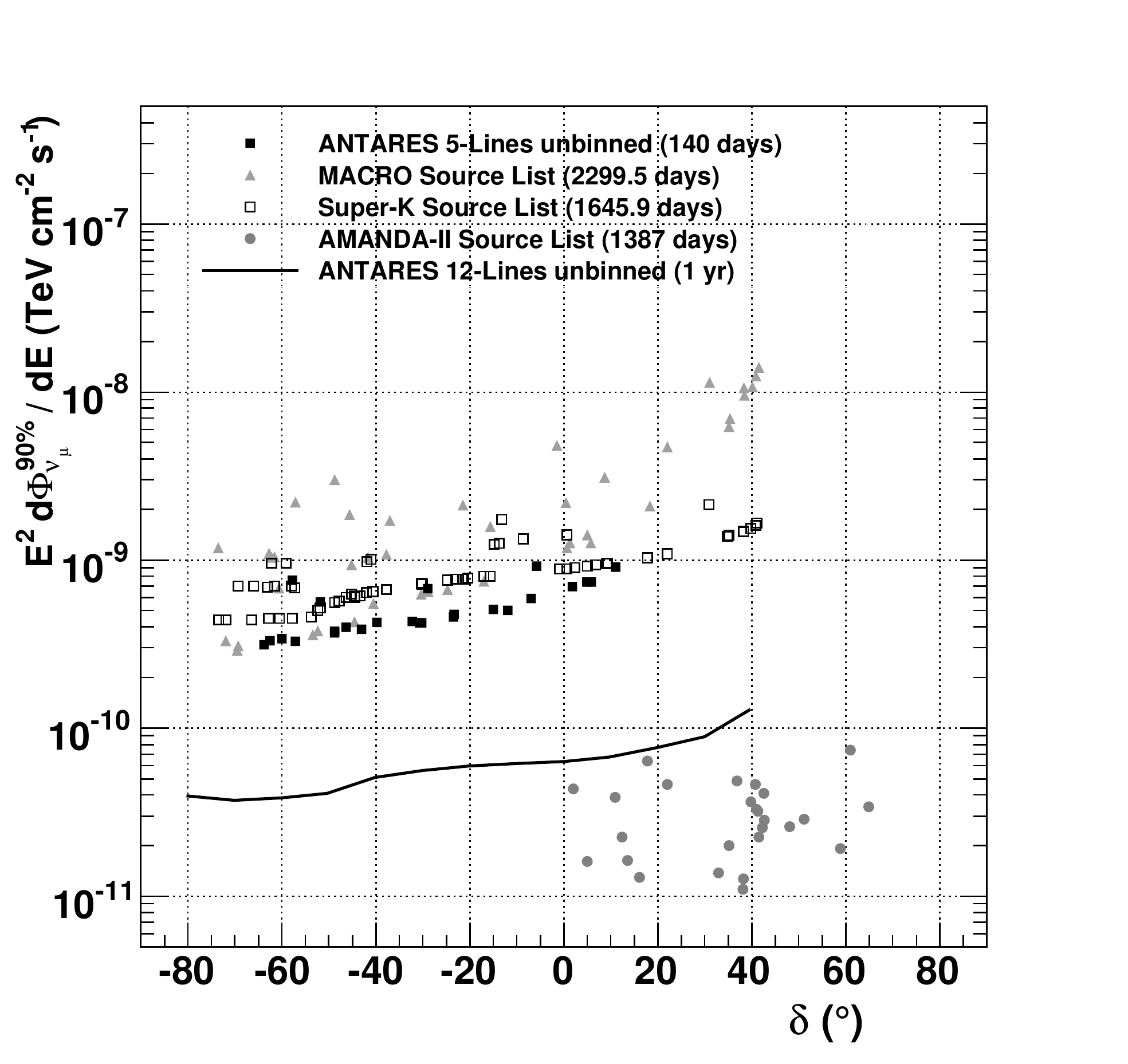}
    \caption{\small{Upper Limits for ANTARES 5-line data (black-filled squares) compared with the results published by other neutrino experiments (MACRO \cite{ICRC0127:MACRO}, AMANDA \cite{ICRC0127:AMANDA} and SuperKamiokande \cite{SuperK}). The predicted sensitivity of ANTARES for 365 days (line) is also shown.}}
    \label{fig:LIMIT}
  \end{figure}

\subsection{All sky search}
In the all sky survey no assumption about the source location is done. 
In the EM method a pre-clustering algorithm selects few cluster candidates and only the cluster with the highest significance is considered.  
The highest BIC value is found at a position ($\delta=$ -63.7$^\circ$, RA=243.9$^\circ$), corresponding to a P-value of 0.3 (1$\sigma$ excess). In this case no trial corrections has to be done because the method looks directly to the most significant cluster.  \\

\section{Conclusions}
ANTARES is the first undersea neutrino telescope and the largest apparatus in the Northern Hemisphere. It reached its full configuration in May 2008. Due to its exceptional angular resolution better than 0.3$^\circ$ at high energy it is especially suited for the search for neutrino point sources. In this contribution the analysis of the data taken with 5 lines during the year 2007 have been presented together with the first limits for a point source search. Two independent methods have been used: a more powerful unbinned method based on the Expectation-Maximization algorithm and a binned method to cross-check the results. 
The two methods have been applied to search for a signal excess at the locations of a list of candidate sources (fixed source search). No statistically significant excess have been found in the data. 
The lowest P-value is a 2.8 $\sigma$ excess (pre-trial) corresponding to the location ($\delta=$ -57.76$^\circ$, RA=155.8$^\circ$), expected in 10\% (post-trial) of the experiments when looking at 25 sources.\\
Moreover a \emph{blind} search over the full sky has been also performed, without making any assumption about the source position. Again, no excess has been found and the lowest P-value corresponds to $1\sigma$ excess at ($\delta=$ -63.7$^\circ$, RA=243.9$^\circ$).\\
Although no evidence of neutrino sources have been found in the data, with only 5 lines and 140 days of livetime, ANTARES has set the best upper limits for neutrino point sources in the Southern Hemisphere. \\
The analysis of new data is already underway. The new data sample corresponds to data taken during the year 2008, in which detector was taking data in different configurations (due to the deployment of new lines). The results will be presented soon.

\label{icrc0127:end}

\setcounter{figure}{0}
\setcounter{table}{0}
\setcounter{footnote}{0}
\setcounter{section}{0}




\hyphenation{abcdef-ghijklmnoprstuwxyz IEEEtran}

\title{Searching for high-energy neutrinos in coincidence with gravitational
waves with the ANTARES and VIRGO/LIGO detectors}

\author{\IEEEauthorblockN{V\'eronique Van Elewyck\IEEEauthorrefmark{1},
			  for the ANTARES Collaboration\IEEEauthorrefmark{2}}
                            \\
\IEEEauthorblockA{\IEEEauthorrefmark{1} AstroParticule et Cosmologie (UMR 7164) \& Universit\'e Paris 7 Denis Diderot, \\Case 7020, F-75205 Paris Cedex 13, France \\
email: elewyck@apc.univ-paris7.fr}
\IEEEauthorblockA{\IEEEauthorrefmark{2} http://antares.in2p3.fr}}

\shorttitle{V. Van Elewyck \etal Coincident searches ANTARES/VIRGO-LIGO}
\maketitle
\label{icrc_elewyck:begin}

\begin{abstract}
 Cataclysmic cosmic events can be plausible sources of both
gravitational waves (GW) and high-energy neutrinos (HEN). Both GW and HEN are
alternative cosmic messengers that may escape very dense media and
travel unaffected over cosmological distances, carrying information
from the innermost regions of the astrophysical engines. For the same
reasons, such messengers could also reveal new, hidden sources that
were not observed by conventional photon astronomy.

Requiring the consistency between GW and HEN detection channels shall
enable new searches as one has significant additional information
about the common source. A neutrino telescope such as ANTARES can
determine accurately the time and direction of high energy neutrino
events, while a network of gravitational wave detectors such as LIGO
and VIRGO can also provide timing/directional information for
gravitational wave bursts. By combining the  information from these
totally independent detectors, one can search for cosmic events that
may arrive from common astrophysical sources.

  \end{abstract}

\begin{IEEEkeywords}
 neutrinos; gravitational waves; multi-messenger astronomy.
\end{IEEEkeywords}

\section{Introduction}

Astroparticle physics has entered an exciting period with the recent development of experimental techniques that have opened new windows of observation of the cosmic radiation in all its components. In this context, it has been recognized that not only a multi-wavelength but also a multi-messenger approach was best suited for studying the high-energy astrophysical sources.

In this context, and despite their elusive nature, both high-energy neutrinos (HENs) and gravitational waves (GWs) are now  considered seriously as candidate cosmic messengers. Contrarily to high-energy photons (which are absorbed through interactions in the source and with the extragalactic background light) and cosmic rays (which are deflected by ambient magnetic fields, except at the highest energies), both HENs and GWs may indeed escape from dense astrophysical regions and travel over large distances without being absorbed, pointing back to their emitter.

It is expected that many astrophysical sources produce both GWs, originating from the cataclysmic event responsible for the onset of the source, and HENs, as a byproduct of the interactions of accelerated protons (and heavier nuclei) with ambient matter and radiation in the source. Moreover, some classes of astrophysical objects might be completely opaque to hadrons and photons, and observable only through their GW and/or HEN emissions. The detection of coincident signals in both these channels would then be a landmark event and sign the first observational evidence that GW and HEN originate from a common source. GW and HEN astronomies will then provide us with important information on the processes at work in the astrophysical accelerators. A more detailed discussion on the most plausible GW/HEN sources will be presented in Section \ref{sec:sources}, along with relevant references.

Furthermore, provided that the emission mechanisms are sufficiently well known, a precise measurement of the time delay between HEN and GW signals coming from a very distant source could also allow to probe quantum-gravity effects and possibly to constrain dark energy models\cite{QG}. Gamma-ray bursts (GRBs) appear as good candidates, {\it albeit} quite challenging\cite{QGlim}, for such time-of-flight studies.

Common HEN and GW astronomies are also motivated by the advent of a new generation of dedicated experiments. In this contribution, we describe in more detail the feasibility of joint GW+HEN searches between the ANTARES neutrino telescope\cite{icrc_elewyck:antares} (and its future, $km^3$-sized, successor KM3NeT\cite{icrc_elewyck:km3net}) and the GW detectors VIRGO\cite{virgo} and LIGO\cite{ligo} (which are now part of the same experimental collaboration). The detection principle and performances of these detectors are briefly described in Section \ref{sec:det}, along with their respective schedule of operation in the forthcoming years. Section \ref{sec:ana} presents an outlook on the analysis strategies that will be set up to optimize coincident GW-HEN detection among the three experiments.

It should be noted that similar studies involving the Ice Cube neutrino telescope\cite{ice3}, currently under deployment at the South Pole and looking for cosmic neutrinos from sources in the Northern Hemisphere, are under way\cite{aso}.

\begin{figure*}[!t]
   \centerline{{\includegraphics[width=5in]{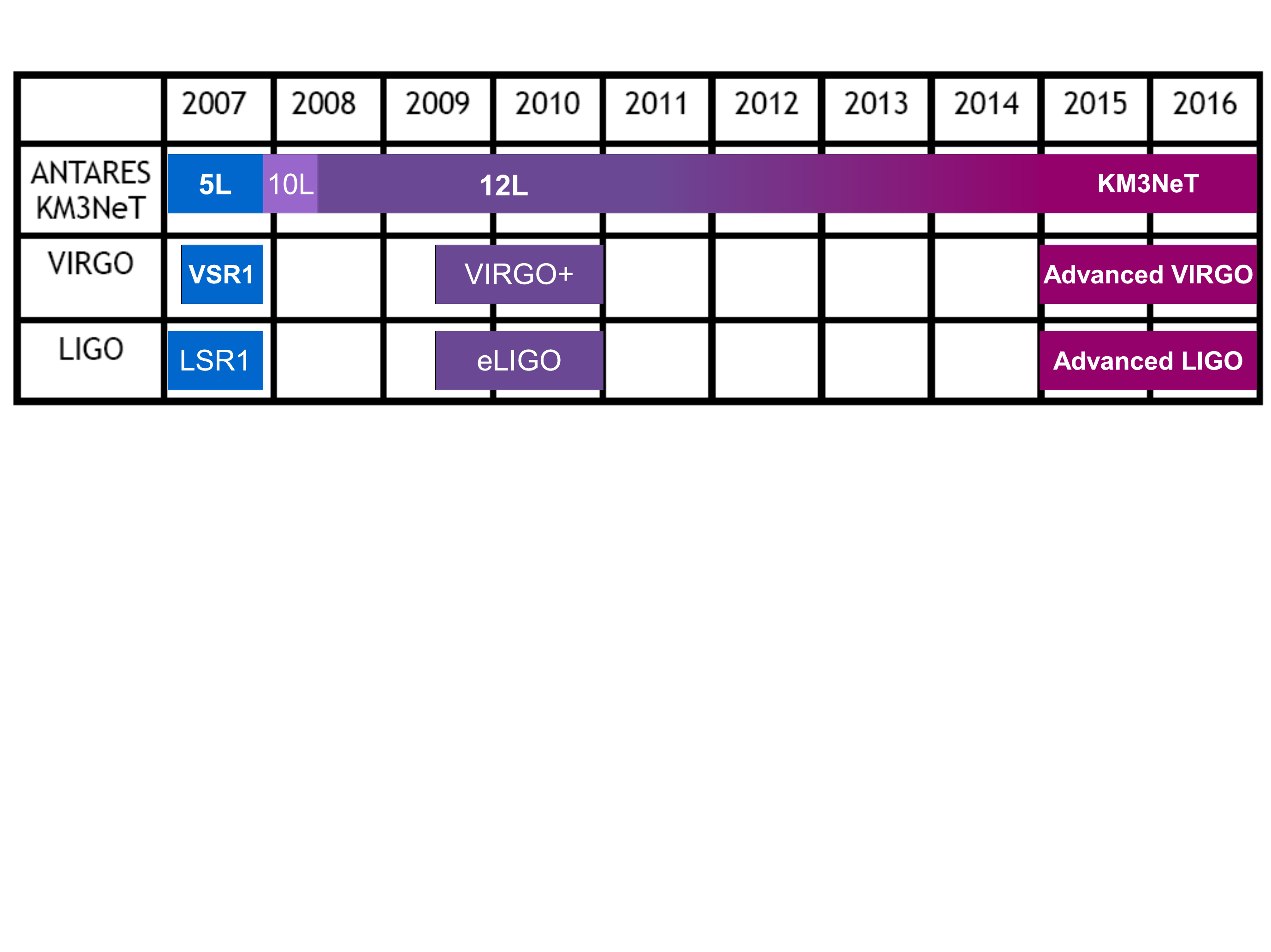} \label{icrc_elewyck:sub_fig1}}}
\vspace*{-5.5cm}
   \caption{\footnotesize Time chart of the data-taking periods for the ANTARES, VIRGO and LIGO experiments, indicating the respective upgrades of the detectors (as described in the text). The deployment of the KM3NeT neutrino telescope is expected to last three to four years, during which the detector will be taking data with an increasing number of PMTs before reaching its final configuration.\cite{icrc_elewyck:km3net}.}
   \label{fig:tab}
 \end{figure*}

\section{Potential common sources of Gravitational waves and high-energy neutrinos}
\label{sec:sources}
\subsection{Galactic sources}
 Several classes of galactic sources could feature both GW and HEN emission mechanisms potentially accessible to the present generation of GW interferometers and HEN telescopes, among which :
\begin{itemize}
 \item {\bf Microquasars} are associated with X-ray binaries involving a compact object (neutron star or black hole) that accretes matter from a companion star and re-emits it in relativistic jets associated with intense radio (and IR) flares. Such objects could emit GWs during both accretion and ejection phases\cite{pradierVLVNT}; and the latter phase could be correlated with a HEN signal if the jet has a hadronic component\cite{mq}.
\item {\bf Soft Gamma-ray Repeaters (SGRs)} are X-ray pulsars with a soft $\gamma$-ray bursting activity. The {\it magnetar} model described them as highly magnetized neutron stars whose outbursts are caused by global star-quakes associated with rearragements of the magnetic field\cite{magnetar}. The deformation of the star during the outburst could produce GWs, while HENs could emerge from hadron-loaded flares\cite{ioka}. The fact that both the AMANDA neutrino telescope and the LIGO interferometer have issued relevant limits on the respective GW and HEN emissions from the 2004 giant flare of SGR 1806-20\cite{sgrlim} indicates that a coincident GW+HEN search might further constrain the parameter space of emission models.
\end{itemize}

\subsection{Extragalactic sources}
Gamma-Ray Bursts (GRBs) are probably the most promising class of extragalactic sources. In the prompt and afterglow phases, HENs ($10^5 - 10^{10}$ GeV) are expected to be produced by accelerated protons in relativistic shocks and several models predict detectable fluxes in km$^3$-scale detectors\cite{GRBnus}.
\begin{itemize}
 \item {\bf Short-hard GRBs} are thought to originate from coalescing binaries involving  black holes and/or neutron stars; such mergers could emit GWs detectable from relatively large distances\cite{nakar}, with significant associated HEN fluxes.
\item {\bf Long-soft GRBs}, as described by the collapsar model, are compatible with the emission of a strong burst of GWs during the gravitational collapse of the (rapidly rotating) progenitor star and in the pre-GRB phase; however this population is distributed over cosmological distances so that the associated HEN signal is expected to be faint\cite{kotake}.
\item{\bf Low-luminosity GRBs}, with $\gamma$-ray luminosities a few orders of magnitude smaller, are believed to originate from a particularly energetic, possibly rapidly-rotating and jet-driven population of core-collapse supernovae. They could produce stronger GW signals together with significant high- and low-energy neutrino emission; moreover they are often discovered at shorter distances\cite{gupta}.
\item {\bf Failed GRBs} are thought to be associated with supernovae driven by mildly relativistic, baryon-rich and optically thick jets, so that no $\gamma$-rays escape. Such ``hidden sources'' could be among the most promising emitters of GWs and HENs, as current estimations predict a relatively high occurrence rate in the volume probed by current GW and HEN detectors\cite{ando}.
\end{itemize}

\section{The detectors}
 \label{sec:det}

\subsection{ANTARES}
  The ANTARES detector is the first undersea neutrino telescope; its deployment at a depth of 2475m in the Mediterranean Sea near Toulon was completed in May 2008. It consists in a three-dimensional array of 884 photomultiplier tubes (PMTs) distributed on 12 lines anchored to the sea bed and connected to the shore through an electro-optical cable. Before reaching this final (12L) setup, ANTARES has been operating in various configurations with increasing number of lines, from one to five (5L) and ten (10L); the respective periods are indicated on the time chart of Fig.~\ref{fig:tab}.
  
   ANTARES detects the Cherenkov radiation emitted by charged leptons (mainly muons, but also electrons and taus) induced by cosmic neutrino interactions with matter inside or near the instrumented volume. The knowledge of the timing and amplitude of the light pulses recorded by the PMTs allows to reconstruct the trajectory of the muon and to infer the arrival direction of the incident neutrino and to estimate its energy.
   
   Since the Earth acts as a shield against all particles but neutrinos, the design of a neutrino telescope is optimized for the detection of up-going muons produced by neutrinos which have traversed the Earth and interacted near the detector. The field of view of ANTARES is therefore $\sim\, 2 \pi\, \mathrm{sr}$ for neutrino energies $100\ \mathrm{GeV} \lesssim E_\nu \lesssim 100\  \mathrm{TeV}$. Above this energy, the sky coverage is reduced because of neutrino absorption in the Earth; but it can be partially recovered by looking for horizontal and downward-going neutrinos, which can be more easily identified as the background of atmospheric muons is much fainter at these energies. With an effective area $\sim 0,1\ \mathrm{km^2}$), ANTARES is expected to achieve an unprecedented angular resolution (about $0.3^\circ$ for neutrinos above 10 TeV) as a result of the good optical properties of sea water\cite{opt}.
  
  The data acquisition system of ANTARES is based on the "all-data-to-shore" concept, which allows to operate different physics triggers to the same data in parallel, each optimized for a specific (astro)physics signal\cite{miekedata}. In particular, satellites looking for GRBs can trigger the detector in real time via the GCN (Gamma-Ray Burst Coordinate Network) alert system. About 60s of buffered raw data (sometimes even including data recorded before the alert time) are then written on disk and kept for offline analysis, allowing for a significant gain in detection efficiency\cite{miekegrb}. 
  
  Another interesting feature recently implemented in ANTARES is the possibility to trigger an optical telescope network on the basis of "golden" neutrino events (such as neutrino doublets coincident in time and space or single neutrinos of very high energy) selected by a fast, online reconstruction procedure\cite{damien}.

   All these characteristics make the ANTARES detector especially suited for the search of astrophysical point sources\cite{simona} - and transients in particular.
  
\subsection{VIRGO and LIGO}

The GW detectors VIRGO\cite{virgo} (one site in Italy) and LIGO\cite{ligo} (two sites in the United States) are Michelson-type laser interferometers that consist of two light storage arms enclosed in vacuum tubes oriented at $90^\circ$ from each other. Suspended, highly reflective mirrors play the role of test masses. Current detectors are sensitive to relative displacements (hence GW amplitude) of the order of $10^{20}$ to $10^{22}$   Hz$^{-1/2}$. Their current detection horizon is about 15 Mpc for standard binary sources.

Their sensitivity is essentially limited at high frequencies by laser shot noise and at low frequencies by seismic noise ($<\ \sim O(50\ \mathrm{Hz})$) and by the thermal noise of the atoms in the mirrors (up to few 10 Hz). To reduce the sources of noise, GW interferometers rely on high-power, ultrastable lasers and on sophisticated techniques of position and alignment control and of stabilization of the mirrors.

Both detectors had a data-taking phase during 2007 (Virgo Science Run 1 and Ligo S5), which partially coincided with the ANTARES 5L configuration. They are currently upgrading to VIRGO+ and eLIGO, to improve their sensitivity by a factor of 2 - and hence the probed volume by a factor of 8. The collaborations have merged and are preparing for a common science run starting mid-2009 (Virgo Science Run 2 and LIGO S6), i.e. in coincidence with the operation of ANTARES 12L (see Fig.~\ref{fig:tab}). Another major upgrade for both classes of detectors is scheduled for the upcoming decade: the Advanced VIRGO/Advanced LIGO and KM3NeT projects should gain a factor of 10 in sensitivities respect to the presently operating instruments. 

The VIRGO/LIGO network monitors a good fraction of the sky in common with HEN telescopes: as can be seen from Figure~\ref{fig:map} the overlap of visibility maps with each telescope is about 4~sr ($\sim 30\%$ of the sky), and the same value holds for Ice Cube.

\begin{figure}[!t]
  \centering
  \includegraphics[width=3in]{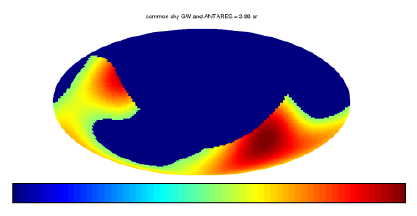}
  \caption{\footnotesize Instantaneous common sky coverage for VIRGO + LIGO + ANTARES in geocentric coordinates. This map shows the combined antenna pattern for the gravitational wave detector network (above half-maximum), with the simplifying assumption that ANTARES has 100\% visibility in its antipodal hemisphere and 0\% elsewhere. The colour scale is from $0\%$ (left, blue) to $100\%$ (right, red)\cite{ecm}.}
  \label{fig:map}
 \end{figure}

\section{Outlook on the analysis strategies}
\label{sec:ana}

GW interferometers and HEN telescopes share the challenge to look for faint and rare signals on top of abundant noise or background events. The GW+HEN search methodology involves the combination of independent GW/HEN candidate event lists, with the advantage of significantly lowering the rate of accidental coincidences.

The information required about any GW/HEN event consists of its timing, arrival direction and associated angular uncertainties (possibly under the form of a sigificance sky map). Each event list is obtained by the combination of  reconstruction algorithms specific to each experiment, and quality cuts used to optimize the signal-to-background ratio.

GW+HEN event pairs within a predefined, astrophysically motivated (and possibly source- or model-dependent), time interval can be selected as time-coincident events. Then, the spatial overlap between GW and HEN events is statistically evaluated, e. g. by an unbinned maximum likelihood method, and the significance of the coincident event is obtained by comparing to the distribution of accidental events obtained with Monte-Carlo simulations using data streams scrambled in time (or simulated background events).

Preliminary investigations of the feasibility of such searches have already been performed and indicate that, even if the constituent observatories provide several triggers a day, the false alarm rate for the combined detector network can be maintained at a very low level ($\sim (600\ \mathrm{yr})^{-1}$)\cite{aso,pradierVLVNT}.

\section{Conclusions and perspectives}
\label{sec:concl}

A joint GW+HEN analysis program could significantly expand the scientific reach of both GW interferometers and HEN telescopes. The robust background rejection arising from the combination of two totally independent sets of data results in an increased sensitivity and the possible recovery of cosmic signals. The observation of coincident triggers would provide strong evidence for the detection of a GW burst and a cosmic neutrino event, and for the existence of common sources. Beyond the benefit of a high-confidence discovery, coincident GW/HEN (non-)ob\-servation shall play a critical role in our understanding of the most energetic sources of cosmic radiation and in constraining existing models. They could also reveal new, ``hidden'' sources unobserved so far by conventional photon astronomy.
A new period of concurrent observations with upgraded experiments is expected to start mid-2009. Future schedules involving next-generation detectors with a significantly increased sensitivity (such as KM3NeT and the Advanced LIGO/Advanced VIRGO projects) are likely to coincide as well, opening the way towards an even more efficient GW+HEN astronomy.

\section{Acknowledgements}
 V. V. E.  thanks E. Chassande-Mottin for many fruitful discussions and for his help in preparing this manuscript. She acknowledges financial support from the European Community 7th Framework Program (Marie Curie Reintegration Grant) and from the French Agence Nationale de la Recherche (ANR-08-JCJC-0061-01).

\label{icrc_elewyck:end}


\end{twocolumn}
\end{document}